%% file: Thesis.tex
\RequirePackage[l2tabu, orthodox]{nag}

\documentclass[a4paper,12pt,twoside,openright,dalthesis]{report}

\makeatletter
\title{Adventures {\magi i}n Andromeda: The Interplay of Interstellar Dust {\magi a}nd Gas in our Big Neighbour} \let\Title\@title
\author{Gayathri Eknath} \let\Author\@author
\newcommand{\Shortauthor}{G. Eknath}
\date{26th September 2023} \let\Date\@date
\makeatother

\usepackage[T1]{fontenc}
\usepackage{fancyhdr}
\usepackage[document]{ragged2e}
\usepackage{Style/New_Cardiff_Thesis}
\usepackage{Style/deluxetable}
\usepackage{Style/aas_macros}
\usepackage[british]{babel}
\usepackage{aecompl}
\usepackage{graphicx} 
\usepackage{natbib}
\usepackage{url}
\usepackage{multirow}   
\usepackage{amssymb}	
\usepackage{amsmath}    
\usepackage{latexsym}
\usepackage{titlesec}
\usepackage{microtype}
\usepackage{enumitem}
\usepackage{textcomp}
\usepackage{longtable}
\usepackage{setspace}
\usepackage{float}
\usepackage{makecell}
\usepackage{bold-extra}
\usepackage{booktabs}   
\usepackage{relsize}
\usepackage[breaklinks]{hyperref}
\usepackage[labelsep=period,labelfont=bf,size=normalsize]{caption}
\usepackage{lscape}
\usepackage{breakurl}
\usepackage{mathtools}
\usepackage{bigdelim}
\usepackage{xcolor}

\usepackage{savesym}
\savesymbol{tablenum}
\usepackage{siunitx}
\restoresymbol{SI}{tablenum}

\usepackage[a4paper, inner=20mm, outer=20mm, top=20mm, bottom = 20mm, includehead, includefoot]{geometry}

\usepackage{pdflscape}

\usepackage{subcaption}  
\usepackage{todonotes}  

\newcommand*\magi{\color{black}}

\DeclareRobustCommand{\VAN}[3]{#2}
\let\VANthebibliography\thebibliography
\def\thebibliography{\DeclareRobustCommand{\VAN}[3]{##3}\VANthebibliography}

\newcommand{\chapquote}[3]{\begin{quotation} \textit{#1} \end{quotation} \begin{flushright}  #2 \textit{#3}\end{flushright} }

\graphicspath{Chapters/Plots_Tables/}

\usepackage{enumitem}
\setlist{listparindent=\parindent}

\begin{document}


\fontfamily{qtm}\selectfont   

\FlushLeft 
\input{journals}

\defcitealias{2018Coogan}{C18}
\phd
\title{\Title}
\submitdate{\Date}
\author{\Author}
\university{Cardiff University}
\twosupervisors
\supervisor{Stephen A. Eales}
\firstreader{Matthew W. L. Smith}

\noserif 
\noacknowledgementspage

\titleformat{\chapter}[display]{}{\chapter}{}{\huge \textbf \textsc}[\titlerule\vspace{2pt}\titlerule]

\dedicate{\vspace{6.35in}\textit{``Women don't need to find a voice - they have a voice. They need to feel empowered to use it and people need to be encouraged to listen.''}\\ - Meghan Markle}

\tablespagetrue
\figurespagetrue

\beforepreface
\prefacesection{Acknowledgements}
\input{Chapters/acknowledgements}

\prefacesection{Abstract}

\input{Chapters/abstract}

\prefacesection{Publications}

\input{Chapters/publications}

\afterpreface
 
\input{Chapters/dedication}

\onehalfspacing

\titleformat{\chapter}[display]{}{}{15pt}{{\textbf \huge \textsc{Chapter} \thechapter}\\ \huge \textbf \textsc}[\titlerule\vspace{2pt}\titlerule]
\addtolength{\parindent}{0.5in}
\addtolength{\parskip}{0.1in}
	
\chapter{Introduction}
\label{chapter:Introduction}
\input{Chapters/intro}

\chapter{Investigating variations in the dust emissivity index in the Andromeda galaxy}
\label{chapter:betavar}
\input{Chapters/chapter_betavar}

\chapter{Searching for excess dust in the Andromeda galaxy}
\label{chapter:sedfit}
\input{Chapters/chapter_sed}

\chapter{A dusty cloud catalogue - estimating gas using dust in the Andromeda galaxy}
\label{chapter:dustmass}
\input{Chapters/chapter_dustmass}

\chapter{The star formation efficiencies of clouds in the Andromeda galaxy}
\label{chapter:sfe}
\input{Chapters/chapter_starformation}

\chapter{A first look at testing PPMAP for extragalactic scales}
\label{chapter:ppmaptest}
\input{Chapters/chapter_ppmaptest}

\chapter{Conclusions and future work}
\label{chapter:Conclusion}
\input{Chapters/Conclusion}

\appendix

\titleformat{\chapter}[display]{}{}{0pt}{{\textbf \huge \textsc{Appendix} \thechapter}\\ \huge \textbf \textsc}[\titlerule\vspace{2pt}\titlerule]

\input{Chapters/An_appendix_betavar}

\label{ap:cloud_catscarmappmap}

\input{Chapters/An_appendix_ppmaptest}

\input{Chapters/An_appendix_dustmass}

\label{ap:hashtagcloudcat}

\titleformat{\chapter}[display]{}{}{0pt}{\huge \textbf \textsc}[\titlerule\vspace{2pt}\titlerule]

\bibliographystyle{mnras}
\bibliography{Thesis.bib}

\end{document}

%% file: journals.tex
\def\araa{{\em ARAA}}
\def\aj{{\em AJ}}
\def\apj{{\em ApJ}}
\def\apjl{{\em ApJ}}
\def\apjs{{\em ApJS}}
\def\aap{{\em A\&A}}
\def\apss{{\em Astrophys.\ Space Science}}
\def\baas{{\em Bull.\ Amer.\ Astron.\ Soc.}}
\def\bain{{\em Bull.\ Astron.\ Inst.\ Netherlands}}
\def\fcp{{\em Fund.\ Cosm.\ Phys.}}
\def\jcam{{\em J.\ Comput.\ Appl.\ Math.}}
\def\jcp{{\em J.\ Comput.\ Phys.}}
\def\jfm{{\em J.\ Fluid Mech.}}
\def\mnras{{\em MNRAS}}
\def\nat{{\em Nature}}
\def\pta{{\em Phil.\ Trans.\ A.}}
\def\ptp{{\em Prog.\ Theo.\ Phys.}}
\def\prd{{\em Phys.\ Rev.\ D}}
\def\pre{{\em Phys.\ Rev.\ E}}
\def\prl{{\em Phys.\ Rev.\ Lett.}}
\def\prsa{{\em Proc.\ R.\ Soc.\ London A}}
\def\pasj{{\em Pub.\ Astron.\ Soc.\ Japan}}
\def\pasp{{\em PASP}}
\def\pfl{{\em Phys.\ Fluids}}
\def\ppl{{\em Phys.\ Plasmas}}
\def\qjras{{\em Quarterly\ Journal\ of\ the\ Royal\ Astronomical\ Society}}
\def\rpp{{\em Rep.\ Prog.\ Phys.}}
\def\rmp{{\em Rev.\ Mod.\ Phys.}}
\def\zp{{\em Z.\ Phys.}}
\def\za{{\em Z.\ Astrophys.}}
\def\physrep{{\em Phys.\ Repts.}}

%% file: Chapters/acknowledgements.tex

\subsection*{Funding Bodies and Affiliations}
This work would not have been possible without the funding provided by the Bell Burnell Graduate Scholarship Fund,  the Science \& Technology Facilities Council and Cardiff University.  I use observations from the following telescopes and wholeheartedly thank all the people involved in collecting and providing the data used:
\begin{itemize}
\item Herschel Space Observatory
\item Combined Array for Research in Millimeter-wave Astronomy
\item Westerbork Synthesis Radio Telescope (WSRT)
\item James Clerk Maxwell Telescope
\item Spitzer Space Telescope
\item 30-m telescope at the Institut de Radioastronomie Millim\'etrique
\item Planck Space Telescope
\end{itemize}

\noindent I am grateful for the administrative support provided behind-the-scenes by the Physics \& Astronomy postgraduate office team,  for the IT engineering support provided by Richard Frewin,  and for the excellent computing training resources and supercomputing facilities provided by the Advanced Research Computing at Cardiff (ARCCA) team.

\subsection*{Software distribution}
I make use of the following software in this work:
\begin{itemize}
\item Python,  version 3.6+ (packages: \texttt{numpy} (\citealt{Harris2020}),  \texttt{astropy} (\citealt{astropy2013},  \citealt{astropy2018}, \citealt{astropy2022}),  \texttt{matplotlib} (\citealt{Hunter2007}),  \texttt{multicolorfits},  \texttt{lmfit} (\citealt{Newville2014}),  \texttt{colorcet} (\citealt{Bednar2022}),  \texttt{cmocean} (\citealt{Thyng2016}),  \texttt{reproject},  \texttt{scipy} (\citealt{SciPy1.0Contributors2020}),  \texttt{mpl\_toolkits},  \texttt{h5py} (\citealt{Collette2014}),  \texttt{astrodendro} (\citealt{Rosolowsky2008}),  \texttt{tqdm} (\citealt{DaCosta-Luis2023}),  \texttt{ipython} (\citealt{Perez2007}))
\item LaTeX
\item DS9 (\citealt{Joye2003})
\item CARTA (\citealt{Comrie2021})
\end{itemize}

\subsection*{Personal Thanks}
To say that this PhD has been `quite a journey' would be an understatement.  This PhD has brought out all of the calibre in my being,  a determination in me that I did not know I had,  and all of the monsters wriggling in the dark that I wanted to ignore.  While doing cutting-edge research defining new boundaries comes with its own level of pressure and isolation,  the COVID-19 pandemic added another layer of uncertainty that I had never faced before.  Despite this,  here I attempt to show my utmost gratitude to a support system which kept me going.

\hspace*{10mm} First of all,  \textit{ammumma},  thank you for the amazing way in which you hold yourself no matter what life throws at you and for believing in me whenever I was distressed by the PhD.  \textit{Amma},  thank you for knowing what I need before I do and for everything that you've done and overcome to get me to where I am today (and being right on time with the most delicious home-cooked food!) \textit{Achan},  thank you for the solid examples and advice that helped me solve problems independently in a world that seems constantly overwhelming and full of disconnect,  despite technology connecting us together now more than ever.  \textit{Aunty Doreen},  you've been a great friend - thank you for keeping me company during the pandemic and beyond.  \textit{Glory aunty \& family},  thank you for your unwavering support while watching me grow through the academic years.

\hspace*{10mm} \textit{Prof Paul Roche},  without you I probably would not have made it through the MSc let alone the PhD.  So thank you for all of the support,  adjustments and hours and hours spent `looking after' my MSc and research education.  \textit{Dr Richard Lewis},  thanks for always having the door open during my MSc days and being the incredibly impactful Director of Postgraduate Research that you were,  saving me from all sorts of problems.  \textit{Jess Phillips},  you were the first person I spoke to while I started my journey at Cardiff and to this day I think about what would have happened if you had not been the super efficient superhuman that you are to get my admissions sorted.  A massive thanks.

\hspace*{10mm} To my primary PhD supervisor,  \textit{Prof Stephen Eales},  firstly thanks for adopting me onto this PhD project in the early, challenging years.  Thanks also for picking up on all of the dodgy thought trails that I had and stopping me going down any rabbit holes just at the right time over the last four years.  I look forward to seeing what kind of science ideas you're living and breathing well into the distant future! My secondary supervisor,  \textit{Dr Matthew Smith},  thanks for all of the practical help,  support (and rebuttals to Steve!) and bringing the most sophisticated ideas into our discussions.  It's been great to be around your super sciency brain and to hear the stories of your academic journey over the years! To my mentor,  \textit{Dr Sam Ladak},  it's been great to have a dedicated space to chat about the PhD work outside of the supervision team and I thank you for your words of encouragement.  Thank you also to all of the academics who piped in during the ISM \& star formation meetings,  seminars,  larger group meetings,  journal club sessions and one-to-one meetings and played a big role in helping me grow as an astrophysicist over the last five years.

\hspace*{10mm} To my examiners,  \textit{Prof Stephen Serjeant \& Dr Ana Duarte Cabral},  thank you for taking the time to read through my thesis (or rather,  150+ page novel!) Steve,  you made the viva voce one of the best feedback-receiving experiences that I have had throughout my PhD and Ana,  the rigour that you brought to helping me present my best possible scientific work has not gone unnoticed.  I hope I did both of your examining experiences justice.

\hspace*{10mm} To \textit{Prof Dame Jocelyn Bell Burnell,  Prof Helen Gleeson,  the Bell Burnell fund panel and scholars},  thank you for your kindness and exemplar role modelling for people from minoritised backgrounds and creating a wonderful community for scientists to progress in when this is so often missing in the usual academic spaces.  \textit{Wendy Sadler},  \textit{Dr Ana Ros Camacho} and the \textit{Diversity in STEM network},  you've all been brilliant at creating a safe space for me to dive into the intricacies of systemic injustices alongside my PhD and without this,  I would not have made it very far. Thank you,  I am incredibly grateful.

\hspace*{10mm} To my long time best friend, \textit{Catherine Brayford},  you have been amazing.  Thank you for all of the late night calls,  life hacks and empathetic chats whenever I needed you and for challenging me to think from different perspectives even when I didn't want to.  \textit{South Wales Storm and Ultimate Frisbee friends},  it's been great (slowly!) getting back into Ultimate post-pandemic and thanks for giving me a space to run/throw/hammer out my problems,  away from the world of Astrophysics.   \textit{Dr Virginia D'Emilio},  thank you for your honesty and the critical,  lengthy discussions that we've had about all sorts of stimulating subjects.  Your friendship has been huge for me in this PhD journey.  \textit{Annie Mathew \& Reem Okasha},  you made my post-pandemic social life so much better - thank you for your continued support through all the ups and downs and for making sure that I celebrate my achievements.  \textit{Dr Ina Sander},  thank you for the long distance phone calls full of support,  keeping up our shared love of board games and teaching me a thing or two about data justice, despite a pandemic getting in the way! \textit{Dr Zoltan Sztranyovszky},  I still remember the days we went over those maths equations in the MSc - thanks a lot for the help and listening ear over the years.  \textit{Dr Jenifer Millard},  \textit{Dr Rosie Beeston},  \textit{Dr Elizabeth Watkins},  \textit{Dr Hannah Chawner},  \textit{Zoe Ballard} and \textit{Rhiannon Lunney},  you've kept me going through the tough times and I am ever so thankful to you for repeatedly and tirelessly answering my many questions.  \textit{Dr Nikki Zabel},  \textit{Dr James Dawson},  \textit{Dr Thomas Williams},  \textit{Dr Phil Cigan},  \textit{Dan Lewis},  \textit{Dr Vasileios Skliris}, \textit{Dr Rhys Green},  thank you for sharing your experiences in the early days and being great guides through various parts of my PhD.  

\hspace*{10mm} To the \textit{Planning Inspectorate team},  you've created a super supportive and inclusive environment while I juggled the last stages of my PhD and ventured into Operational Research.  I'm incredibly grateful to you for making me feel like I belong,  and very excited to see what we will do together.  \textit{Ian Palfreyman},  you have been so inspirational and such a good teacher since my A-Level days and I very much appreciate all of your social media cheerleading throughout this PhD.  To \textit{my therapist},  thank you for helping me to work things out during the toughest of times and to become more comfortable with my emotions.

\hspace*{10mm}  \textit{Lis Hughes Jones},  thank you for welcoming me into your house during the pandemic,  the numerous nature trips and weekend breaks spent catering for me and creating a space for me to look after my wellbeing during the PhD.  Finally,  last but not at all the least,  \textit{Dr Michael Anderson},  I am sure that these words cannot possibly describe the level of support that you've given me over this PhD and suffice to say,  it has been A LOT.  From all the research method ideas,  countless office hours figuring out why something isn't working and boosting my confidence in all sorts of difficult situations,  to almost always watching me bring my work home on every level.  Thank you for being the most patient,  tolerant and altruistic person/scientist/colleague that you are.  I think it's safe to say that we have grown on our PhD journeys both as individuals and together.  It has been amazing.  Let's never do it again.

\hspace*{10mm}  While my love for space threw me into this work,  down on Earth,  I have toured through the zen nature spots of Quy Nhon,  presented at the International Centre for Interdisciplinary Science and Education,  caught a tram or two in Gothenburg,  and experienced the winter grounds of Princeton \& New York - all of which has only been possible because of the opportunities awarded by this PhD.  That said,  I am mindful that this is very much an idyllic social-media-post version of a PhD and I leave you to read this thesis with the awareness that there is still a lot to be done in making the academic landscape more inclusive,  equal and safe.

%% file: Chapters/abstract.tex
Past studies of dust in the Andromeda galaxy (M31) have shown radial variations in the dust emissivity index ($\beta$).  Understanding the astrophysical reasons behind these radial variations may give clues about the chemical composition of dust grains, their physical structure, and the evolution of dust.  In Chapter \ref{chapter:betavar},  we use $^{12}$CO(J=1-0) observations and dust mass surface density measurements derived from \textit{Herschel} observations to produce two cloud catalogues.  We use these catalogues to investigate whether there is evidence that $\beta$ is different inside and outside molecular clouds.  Our results confirm the radial variations of $\beta$ seen in previous studies.  We find little difference between the average $\beta$ inside molecular clouds compared to outside molecular clouds, in disagreement with models which predict an increase of $\beta$ in dense environments.  We find some clouds traced by dust with very little CO which may be either clouds dominated by atomic gas or clouds of molecular gas that contain little CO. 

\hspace*{10mm} Finding excess emission at submillimetre wavelengths beyond 500 $\mu$m (coined "sub-mm excess") might imply that there is a lot of very cold dust in a galaxy or that the dust grains have unusual emission properties.  In Chapter \ref{chapter:sedfit},  we use new submillimetre observations of M31 at 450 and 850 $\mu$m to search for any excess emission from dust at these wavelengths.  We do not find strong evidence for the presence of a sub-mm excess.  We present the first results of the HASHTAG large programme showing the spatial distribution of dust temperature,  $\beta$ and dust mass surface density in M31. 

\hspace*{10mm} In Chapter \ref{chapter:dustmass},  we produce a dust-selected cloud catalogue using archival \textit{Herschel} observations and new JCMT observations of M31 from the HASHTAG large programme.  We then examine how much CO-traced molecular gas and atomic hydrogen is in our dust-selected clouds.  We show that dust is a good tracer of the ISM gas mass at the scale of individual molecular clouds but with an offset from the combined CO-traced + HI gas masses.  We propose that the offset could be due to variations in the ratio of gas to dust,  or due to missing CO-dark molecular gas which is not being traced. 

\hspace*{10mm} Star formation is an inefficient process.  What is driving this inefficiency is still a mystery.  In Chapter \ref{chapter:sfe},  we measure the star formation efficiency (SFE) of individual clouds for the clouds extracted in Chapter 4.  We investigate if the SFE varies in clouds across different positions in the galaxy.  We use FUV+24 $\mu$m emission to trace both the massive star formation and dust-obscured star formation.  We also study whether any observational dust properties influence the SFE of clouds.  We do not find any systematic trends of SFE with radius but do find strong correlations of the star formation rate with both atomic gas and CO-traced molecular gas.  We find that SFE is also correlated with dust temperature and $\beta$. 

\hspace*{10mm} In Chapter \ref{chapter:ppmaptest},  we use a simulated galaxy to test and optimise the Bayesian algorithm \textsc{PPMAP} for application on observations of external galaxies.  We find that there is an offset between the input simulation dust mass surface density values and the \textsc{PPMAP} output. 

%% file: Chapters/publications.tex
\subsection*{}
\noindent{\scshape \Large {Representative publications; published:}
\begin{itemize}
  \item[] \large {\textbf{Athikkat-Eknath G.}, Eales S. A., et al., 2022.  \textit{Investigating Variations in the Dust Emissivity Index in the Andromeda Galaxy}.  Monthly Notices of the Royal Astronomical Society, 511, 5287.}
\end{itemize}

\noindent{\scshape \Large {In prep:}
\begin{itemize}
  \item[] \large {\textbf{Athikkat-Eknath G.},  Eales S. A.,  et al.,  \textit{Searching for a sub-mm excess in Andromeda}.  In Prep.}  
  \item[] \large {\textbf{Athikkat-Eknath G.},  Eales S. A.,  et al.,  \textit{A dusty cloud catalogue - estimating gas using dust in Andromeda}.  In Prep.} 
   \item[] \large {\textbf{Athikkat-Eknath G. },  Eales S. A.,  et al.,  \textit{Star formation efficiencies of clouds in Andromeda at 68 pc spatial scales}.  In Prep.}  
      \item[] \large {Koch E.W.,  Armante M.,  \textbf{Athikkat-Eknath G.,} et al.,  \textit{ALMA Observations of GMCs in M33}.  In Prep.}
\end{itemize}  

\noindent{\scshape \Large {Other co-author publications:}}
\begin{itemize}
  \item[] \large {Smith M. W. L.,  Eales S. A et al.,  2021.  \textit{The HASHTAG Project: The First Submillimeter Images of the Andromeda Galaxy from the Ground}. The Astrophysical Journal Supplement Series,  257, 52}
\end{itemize}

%% file: Chapters/dedication.tex

%% file: Chapters/intro.tex
\chapquote{``Science progresses best when observations force us to alter our preconceptions.''}{Vera Rubin}{}

\noindent This chapter provides the necessary context of what we know about the dust and the interstellar medium of galaxies as relevant for the work in this thesis. 

\section{The interstellar medium}
The interstellar medium (ISM) is the space in between stars which frequently interacts with the star-forming process.  The ISM is made of various states of hydrogen gas (forming $\sim 70$\% of the ISM {\magi mass}),  helium gas (forming $\sim 28$\% of the ISM {\magi mass}) and a small blend of heavier elements termed `metals' (forming $\sim 2$\% of the ISM {\magi mass}),  a fraction of which can be locked up in interstellar dust grains (e.g.  \citealt{James2002},  \citealt{Roman-Duval2022}).  In our own galaxy,  there is evidence that about 50\% of metals in the ISM can be locked up in dust grains (\citealt{James2002},  \citealt{Draine2003},  \citealt{Jenkins2009}).  The gas in the interstellar medium can be segregated into many phases: molecular gas,  atomic gas and ionised gas.  While this gives a simplistic picture of what is predominantly in the ISM,  the ISM is a complex space with many different astronomical sources,  including supernova remnants,  reflection nebulae,  ejected stellar shells,  cloud-like and filamentary gaseous structures,  compact and diffuse ionised regions,  shocks from exploding massive stars and many different types of dust particles.  Table \ref{tab:ismprops} shows the different phases of the ISM and its properties.

\begin{table*}
\caption{Table showing different phases of the ISM and their properties.}
\label{tab:ismprops}      
\centering                                      
\begin{tabular}{c c c c}          
\hline       
ISM phase & Type of gas & \makecell{Number \\ density (cm$^{-3}$)} & Temperature (K)  \\
\hline
 \\
\makecell{Cold Molecular Medium \\ (CMM)} & \makecell{Gas is primarily molecular; \\ hydrogen exists as H$_2$} & $10^2 - 10^6$ & 10 - 30 \\
\hline
\makecell{Cold Neutral Medium \\ (CNM)} &  \makecell{Gas is primarily atomic; \\ hydrogen exists as H} & $1 - 100$ & 30 - 100 \\
\hline
\makecell{Warm Neutral Medium \\ (WNM)} &  \makecell{Gas is primarily atomic; \\ hydrogen exists as H} & < 1 & 6000 - 10,000 \\
\hline
HII regions &  \makecell{Molecular clouds ionised by \\ massive stars; hydrogen exists as H$^{+}$} & $10^2 - 10^6$ & 8000 - 10,000 \\
\hline
\makecell{Hot Ionised Medium \\ (HIM)} & \makecell{Diffuse gas surround galaxies \\ or in regions of recent supernova \\ explosions; hydrogen exists as H$^{+}$} & < $10^{-2}$ & $10^{6} - 10^{7}$ \\
\hline                                   
\end{tabular}
\end{table*}

Different phases of the ISM can be traced using atoms or molecules which are abundant in that particular medium or from direct emission from the gas,  using telescopes on the ground and in space.  Below we describe a selection of tracer molecules and direct emission which are used to observe the ISM and are relevant for this thesis.
\begin{itemize}
\item{\textbf{Tracing the cold molecular medium with CO emission}} - Molecules can emit line radiation by changing their rotational and vibrational states (ro-vibrational transitions).  Molecular hydrogen (H$_2$) is a symmetric molecule with no permanent electric dipole moment.  This means that it is very hard to detect this molecule in emission unless it is in regions with a high temperature ($\gtrsim 500$ K) where there is enough energy to change its rotational state and the molecule is excited.  For this reason,  it is near impossible to detect H$_2$ directly in the cold ISM where the temperatures are $\lesssim 30$ K.  As such,  molecular hydrogen gas is commonly traced using the carbon monoxide (CO) molecule which is the next most abundant molecule.  CO has a permanent dipole moment and is easily detectable at colder temperatures ($\sim 5$ K).  The $^{12}$CO(J=1-0) rotational transition line can be detected at a wavelength of 2.6 mm.  The line intensity of the radiation from the CO molecule can be converted into the mass surface density of molecular hydrogen using:
\begin{equation}
\label{eq:intro1}
\Sigma_\mathrm{H_2} =  X_\mathrm{CO} \times m(\mathrm{H_2}) \times I_\mathrm{CO}
\end{equation}
where $m(\mathrm{H_2})$ is the mass of a hydrogen molecule and $I_\mathrm{CO}$ is the intensity of emission in K km s$^{-1}$,  $X_{\mathrm{CO}}$ is a constant of proportionality {\magi given in units of cm$^{-2}$[K km s$^{-1}$]$^{-1}$.  {\magi In this thesis,  we use the terms `mass of molecular hydrogen in the ISM' and `molecular gas mass' interchangeably (we do not include helium and heavier elements in the definition of `molecular gas mass').} We discuss the uncertainties associated with assuming a constant $X_{\mathrm{CO}}$ factor in Section \ref{sec:xfactor}.}

\item{\textbf{Tracing the cold neutral medium and cold molecular medium with dust emission}} - Dust grains absorb ultraviolet light emitted by stars and re-emit the absorbed energy at infrared wavelengths.  In the optically thin limit,  dust continuum emission follows a modified blackbody function (see Section \ref{ssec:dustmbb} for full details).  Dust can be used as a tracer for the total ISM gas mass (e.g.  \citealt{Eales2012}).  The dust mass surface density ($\Sigma_{\mathrm{dust}}$) can be converted to ISM gas mass surface density ($\Sigma_{\mathrm{gas}}$) using:
\begin{equation}
\Sigma_{\mathrm{gas}} = \Sigma_{\mathrm{dust}} \times \mathrm{GDR}
\end{equation}
where GDR is a gas-to-dust ratio.  Since dust is thought to be well mixed with molecular gas in the ISM,  {\magi cold dust and} its infrared emission can also be used {\magi to probe regions of} molecular gas {\magi (e.g.  \citealt{Scoville2016},  \citealt{Liang2018}, \citealt{Privon2018})}. {\magi Typically,  in the Galaxy,  a constant GDR of $\sim$100 is often assumed (\citealt{Bohlin1978},  \citealt{Hildebrand1983}) but studies have shown variations in GDR with Galactic position (e.g. \citealt{Giannetti2017}).  For external galaxies with similar metallicities to the Milky Way,  we can assume a similar GDR.  But there too,  this ratio has been shown to vary radially (e.g. \citealt{Munoz-Mateos2009},  \citealt{Smith2012}) and with gas density (e.g. \citealt{Roman-Duval2017},  \citealt{Clark2023}).  The GDR also varies with metallicity (e.g.  \citealt{Remy-Ruyer2014},  \citealt{Roman-Duval2022}). This becomes an intrinsic uncertainty in any constant gas-to-dust ratio assumption. Throughout this thesis,  we make the assumption that dust traces hydrogen and molecular hydrogen gas and do not include helium and heavier elements in our dust-traced gas mass calculations.}

\item{\textbf{Tracing the cold neutral medium with \textsc{HI} emission}} - The atomic gas is directly traced by line emission from hydrogen atoms at the 21 cm wavelength.  The emission line is caused by a transition in the spin of the electron within the hydrogen atom being flipped to oppose the spin of the proton in the atom within the atom's ground state.  The wavelength of the photon released when this transition takes place is $\sim 21$ cm.  In the case of the line being excited,  the intensity of this 21 cm emission is directly proportional to the mass surface density of atomic hydrogen in the observed region of the sky (\citealt{Wilson2009}):
\begin{equation}
\label{eq:intro2}
\Sigma_\mathrm{\textsc{HI}} = 1.8 \times 10^{18}  \times m(\mathrm{H}) \times I_\mathrm{\textsc{HI}}
\end{equation}
where $m(\mathrm{H})$ is the mass of a hydrogen atom and $I_\mathrm{\textsc{HI}}$ is the intensity of emission in K km s$^{-1}$ {\magi and the conversion factor is given in units of cm$^{-2}$ [K km s$^{-1}$]$^{-1}$.  In this thesis,  we use the terms `mass of atomic hydrogen in the ISM' and `atomic gas mass' interchangeably.  Equation \ref{eq:intro2} assumes that the gas is optically thin.  This equation also neglects any self absorption from \textsc{HI} - a phenomenon which occurs due to colder \textsc{HI}-emitting regions in the foreground of warmer \textsc{HI}-emitting regions absorbing \textsc{HI} emission at the same velocity.  This self absorption could affect the atomic gas mass estimates by approximately 30\% in nearby galaxies (e.g.  \citealt{Braun2009},  \citealt{Braun2012}),  forming an intrinsic uncertainty in this method. }
\end{itemize}

{\magi \noindent In this thesis,  we mainly focus on the uncertainty coming from the noise in our observations due to the complicated nature of $X$-factor and HI-conversion factor variations.  However,  we are mindful that the intrinsic uncertainties propagate throughout calculations which use CO or HI intensities to infer gas masses.  This additional uncertainty effectively becomes a scaling factor to our mass estimates.}

\section{The star formation process \& clouds}
\begin{figure}[h]
\centering
\includegraphics[width=10cm]{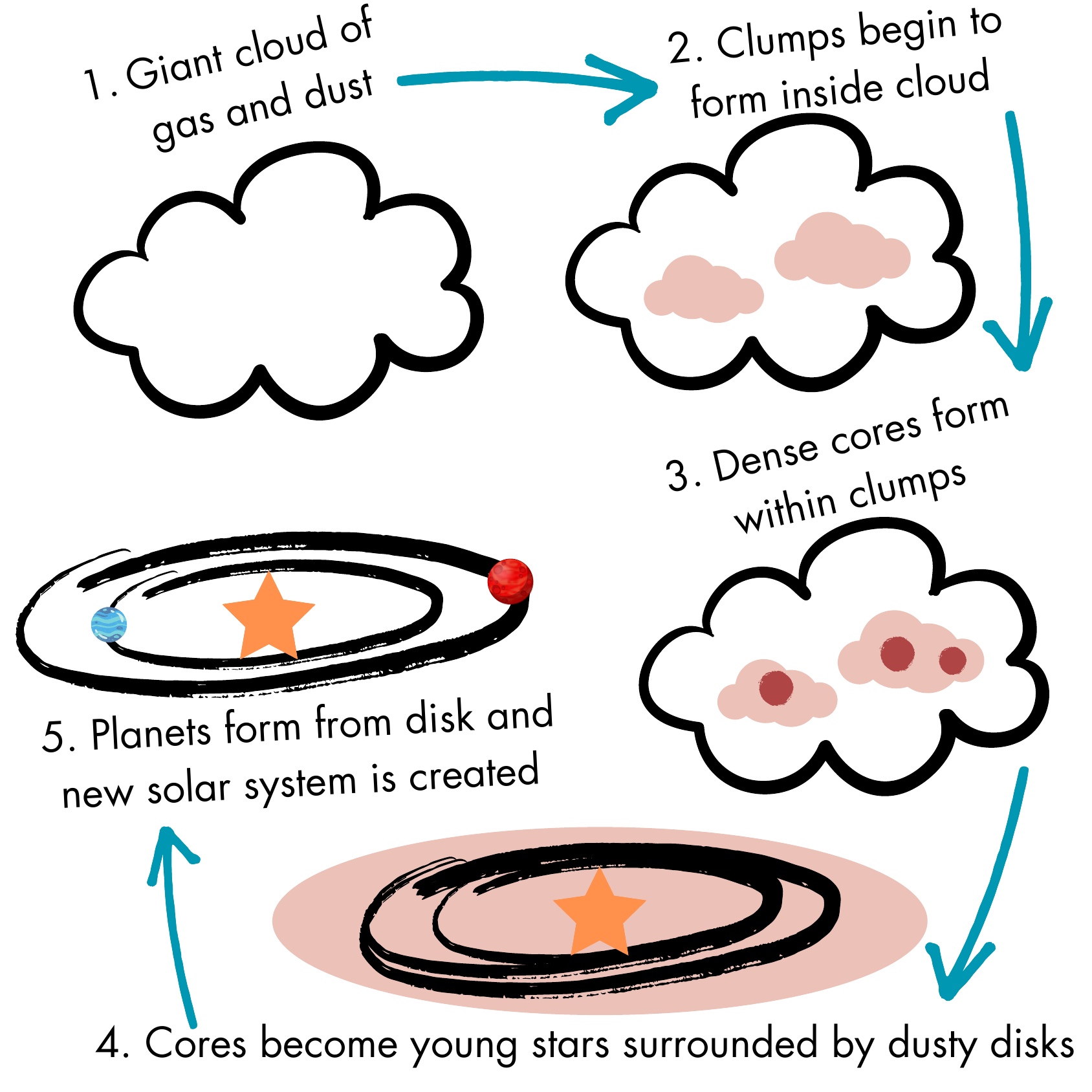} \caption{Cartoon showing a brief overview of the star formation {\magi process} starting with cloud collapse.  The dense cloud of dust and gas is in freefall and begins to collapse.  The cloud fragments into smaller cores.  Accretion of the surrounding gas begins usually in the form of an accretion disk.  Planets form from the dusty accretion disk (the protoplanetary disk) as gravity causes material to collide/stick together and the condensation conditions cause matter to condense into gaseous and solid material. }
\label{fig:sf_cycle}
\end{figure}

\noindent Star formation is the essential astrophysical process which dictates the life we live today,  creating beautiful stars like our Sun and responsible for the planets orbiting around it.  The star formation cycle within galaxies begins with the collapse of a dense cloud of dust and gas, as the inherent battle between the cloud's own gravity and {\magi internal} pressure is won by gravity.  This stellar nursery gives way to form dense clumps as gravity accelerates the accretion of material onto a central core.  Since the accelerating matter is also rotating,  a disk-like structure forms surrounding the core. The inner disk feeds the birth of protostars and the outer disk forms planets.  Figure \ref{fig:sf_cycle} shows a simplistic visual representation of the star formation {\magi process}. 

So how do clouds collapse? The physical threshold above which a cloud collapses is known as the Jeans mass.  Above this mass,  the cloud will begin to contract {\magi under its own gravity} unless some other force is able to stop it.  Within a spherical self-gravitating gas cloud,  {\magi in hydrostatic equilibrium},  any perturbation (e.g.  pressure exerted as collapse begins) moves through the cloud at the speed of sound,  $c_s$.  The time it takes for the perturbation to move across the cloud with a diameter of $L$ is the sound crossing time, $t_{\mathrm{sound}}$:
\begin{equation}
t_{\mathrm{sound}} = \frac{L}{c_s}
\end{equation}
The freefall time,  $t_{\mathrm{ff}}$,  is the time taken for particles to fall from the edge of the cloud all the way to its centre under gravity,  with no pressure:
\begin{equation}
t_{\mathrm{ff}} = \sqrt{\frac{3\pi}{32G\rho}}
\end{equation}
$G$ is the gravitational constant and $\rho$ is the density of the medium,  i.e.  the cloud. 

The cloud will continue to collapse when the freefall time is shorter than sound crossing time,  i.e.  when $ t_{\mathrm{ff}} < t_{\mathrm{sound}}$.  If the sound crossing time is shorter than the freefall time,  pressure will balance the gravitational force and the cloud will remain in hydrostatic equilibrium.  The diameter of the cloud when both timescales are balanced is known as the Jeans length,  defined as:
\begin{equation}
L_{\mathrm{J}} = c_s \sqrt{\frac{3\pi}{32 G\rho}}
\end{equation}
The Jeans mass is then the mass of the cloud with a radius of $R_{\mathrm{J}} = \frac{1}{2} L_{\mathrm{J}}$:
\begin{equation}
M_{{\mathrm{J}}} = \frac{4\pi}{3} R_{\mathrm{J}}^{3} \rho
\end{equation}
{\magi An alternative way of deriving the Jeans mass is by taking the virial theorem into account,  which states that the total kinetic energy of the cloud must be equal to half its potential energy.  This derivation of Jeans mass shows:
\begin{equation}
M_{{\mathrm{J}}} \propto T^{3/2} \rho^{-1/2}
\end{equation}
where $T$ is the temperature of the cloud.  Therefore,  if the temperature of the cloud decreases and the density of the cloud increases enough,  the cloud or parts of the cloud will begin to contract and collapse.  In practice,  of course,  the initial assumptions of an ideal spherical cloud in hydrostatic equilibrium can be challenged by what we see and deduce from observations: clouds are often not in virial equilibrium,  let alone thermostatic equilibrium.  Observed `cloud-like' structures are also not spherical and shocks in the ISM make them highly filamentary (\citealt{Andre2010},  \citealt{Peretto2012}, \citealt{Ragan2014},  \citealt{Tokuda2020}) - a factor which would alter the freefall time. }

{\magi Observationally,  we can identify these clouds with typical sizes between one and a few hundred parsecs and masses of $\sim$10$^{2} - 10^{6}$ M$_{\odot}$ and above.} So far we know that the process of converting the molecular gas mass of clouds into stellar mass is very inefficient.  As such,  a common quantity measured by {\magi G}alactic astronomers to study star formation is the star formation efficiency per freefall time, $\epsilon_{\mathrm{ff}}$:
\begin{equation}
{\magi \epsilon_{\mathrm{ff}} = \frac{\mathrm{SFR} \times t_{\mathrm{ff}}}{M_{\mathrm{cloud}}}}
\end{equation}
{\magi SFR is the overall average star formation rate of the cloud over its lifetime},  and $M_{\mathrm{cloud}}$ is the total mass of the cloud.

In {\magi the G}alaxy,  despite having a total molecular gas mass of $10^9$ M$_{\odot}$,  the rate of star formation is only $\sim$ 2 M$_{\odot}$ yr$^{-1}$ (\citealt{Chomiuk2011}).  If all 10$^{9}$ M$_{\odot}$ of the molecular gas in the {\magi Milky Way} was indeed in freefall,  we would expect the rate of star formation to be $\sim 300$ M$_{\odot}$ yr$^{-1}$.  This discrepancy between the observed star formation rate and expected star formation rate for gas in freefall demonstrates that something must be stopping freefall collapse {\magi and} that the initial assumptions of a gravitationally bound cloud collapsing under a characteristic timescale {\magi may be incorrect}.

{\magi Many studies have tried to understand what may be driving this {\magi star formation} inefficiency.  Studies have investigated whether the entire cloud is in global collapse (\citealt{Goldreich1974}),  or parts of the cloud fragment to collapse (\citealt{Zuckerman1974}),  or if the cloud is supported by some other source like turbulence (\citealt{Krumholz2005},  \citealt{Padoan2016}),  magnetic fields (\citealt{Zuckerman1974}, \citealt{Mouschovias1991},  \citealt{Vazquez-Semadeni2011}),  or stellar feedback (\citealt{Dobbs2011},  \citealt{Chevance2020a}).  What remains to be seen is how much of these factors,  or anything else supporting a cloud against gravitational collapse,  could be driven by environmental factors. }

In this thesis,  we endeavour to understand what environmental factors play a role in driving the star formation inefficiency.  To do this,  we examine the gas depletion timescale within individual molecular clouds in a nearby spiral galaxy,  Andromeda.  This is the amount of time it takes for a cloud to deplete its gas reserve by forming stars and is irrespective of the freefall time:
\begin{equation}
t_{\mathrm{dep}} = \frac{M_{\mathrm{cloud}}}{\mathrm{SFR}}
\end{equation}
where SFR is the star formation rate. 

There are many observational indicators used to estimate the star formation rate within the ISM of galaxies.  Each method has its strengths and limitations but below is a brief overview of how some of the key methods using optical,  submillimetre,  infrared and ultraviolet observations can trace star formation:
\begin{itemize}
\item \textbf{Counting young stellar objects} - The most direct way to obtain a SFR is to count stars.  However,  this method only works if we have a complete dataset and often it is impossible to obtain this due to poor resolution and dust obscuring the light from stars.  Instead,  we can count young stellar objects (YSOs) - protostars at different evolutionary stages - which are embedded within molecular clouds and emit in the infrared regime.  YSOs have typical lifetimes of $\sim 2$ Myr.  The SFR {\magi (given in units of M$_{\odot}$ yr$^{-1}$)} is calculated in this method by:
\begin{equation}
\mathrm{SFR_{\mathrm{YSO}}} = N_{\mathrm{YSO}} \times \overline{M_{\mathrm{YSO}}} \times t^{-1}
\end{equation}
where $t$ is the lifetime of YSOs (which has some uncertainty),  $\overline{M_{\mathrm{YSO}}}$ is the assumed mean mass of YSOs which is somewhat dependent on the assumed initial mass function (IMF),  and $N_{\mathrm{YSO}} $ is the total number of YSOs.  This method is usually used in regions which are resolved,  such as molecular clouds in the Milky Way which are within a distance of 0.5 - 1 kpc from our solar system (\citealt{Calzetti2013}).

\item \textbf{Far-ultraviolet (FUV) continuum emission} -  Young,  massive stars emit FUV radiation at wavelengths between 1250 - 2500 $\mathring{\mathrm{A}}$ (\citealt{Kennicutt1998}).  This can be directly observed and used to probe recent star formation over the past 100 Myr.  {\magi The SFR {\magi (given in units of M$_{\odot}$ yr$^{-1}$)} is calculated in this method by:
\begin{equation}
\mathrm{SFR_{\mathrm{UV}}} = 3.0 \times 10^{-47} \;  \lambda L(\lambda)
\end{equation}
where $\lambda$ is the wavelength in Angstroms and $L(\lambda)$ is the UV luminosity in erg s$^{-1}$ {\magi (\citealt{Calzetti2013})}.  The constant $3.0 \times 10^{-47}$ arises from calibrating using a spectral energy distribution from the stellar population synthesis model by \cite{Leitherer1999}.} The main limitation of this method is that FUV radiation is easily absorbed by dust and re-emitted at infrared wavelengths.  This can be overcome by combining the FUV emission with infrared emission,  probing dust-obscured star formation and unobscured star formation.

\item \textbf{Infrared continuum emission} - Dust grains absorb FUV radiation from stars and re-emit this light at infrared wavelengths.  Dust grains inside star-forming regions are subject to a strong radiation field,  causing them to heat up.  As such,  the emission in these regions coming from warm dust peaks at different wavelengths to the emission from dust in the diffuse ISM.  The difference between emission from warm dust heated by stars and diffuse dust is stark at wavelengths between $20 \lesssim \lambda \lesssim 60$ $\mu$m.  So emission at these wavelengths enable us to successfully trace star-forming regions {\magi where dust is heated}.  In the $5 \lesssim \lambda \lesssim 20$ $\mu$m wavelength regime,  dust emission comes from polycyclic aromatic hydrocarbons (PAHs).  This emission is present in photodissociation regions within molecular clouds and therefore can be linked to star formation.  It is also possible to integrate over the total infrared {\magi (TIR)} luminosity in the wavelength range $5 \leq \lambda \leq 1000$ $\mu$m,  covering the full infrared and submillimetre regime  {\magi (\citealt{Calzetti2013}).  The SFR {\magi (given in units of M$_{\odot}$ yr$^{-1}$)} is calculated in this method by:
\begin{equation}
\mathrm{SFR_{\mathrm{TIR}}} = 2.8 \times 10^{-44} \;  L(\mathrm{TIR})
\end{equation}
where $L(\mathrm{TIR})$ is the TIR luminosity in erg s$^{-1}$.  The constant $2.8 \times 10^{-44}$ assumes the stellar population undergoes constant star formation over 100 Myr and is calibrated using a stellar population synthesis model (\citealt{Calzetti2013}).} This method makes the assumptions that all of the light that is emitted by young stars is {\magi reprocessed} by dust and that dust is only being heated by these young stars.  As such,  a limitation of this method is that it doesn't account for any dust heating by older stars.

\item \textbf{H$\alpha$ line emission} - Hot gas (HII regions) ionised by radiation from massive stars of ages 3-10 Myr can be traced by the H$\alpha$ line emission at a wavelength of $\sim$ 656.3 nm.  This method assumes that all of the photons emitted by massive (O- and B- type) stars ionise the surrounding gas and therefore indirectly allows us to trace the number of massive stars.  Assuming an initial mass function,  we can then extrapolate to the number of low mass stars and calculate a global SFR for a galaxy.  {\magi The SFR in this method is calculated by:
\begin{equation}
\mathrm{SFR_{\mathrm{H\alpha}}} = 7.9 \times 10^{-42} \;  L(\mathrm{H\alpha})
\end{equation}
where $\mathrm{SFR_{\mathrm{H\alpha}}}$ is given in units of M$_{\odot}$ yr$^{-1}$,  $L(\mathrm{H\alpha})$ is the H$\alpha$ luminosity in erg s$^{-1}$,  and the constant $7.9 \times 10^{-42}$ is calibrated using a stellar population synthesis model and assuming a Salpeter initial mass function (\citealt{Kennicutt1998}).} H$\alpha$ emission is less sensitive to absorption by dust than FUV emission.  However,  a limitation of this method is that we need to assume that no ionising photons from the stars will escape.

\item \textbf{Combined tracers} - There are a variety of tracers used to estimate the star formation rate in galaxies.  Most of these methods use a linear combination of different tracers and a suitable calibration factor to help calibrate against each other (\citealt{Calzetti2007}).  For a more detailed review,  we refer the reader to the paper by \cite{Kennicutt2012} and article by \cite{Calzetti2013}.  In this thesis,  we use a combination of FUV + 24 $\mu$m emission to trace young massive stars (FUV) and warm dust emission from star-forming regions where the dust has been heated by young stars (24 $\mu$m).  {\magi A full explanation of how these tracers are used and calibrated is given in Chapter 5,  Section \ref{ssec:sfrobsmethod}.}
\end{itemize}

\section{The role of interstellar dust}
In 1811,  William Herschel suggested that there were vast empty spaces in galaxies inbetween stars (\citealt{Herschel1811}).  Nearly a century later,  Edward Barnard showed osbervational evidence suggesting that these empty spaces might actually be bodies obscuring the light from stars and not just empty spaces (\citealt{Barnard1910},  \citealt{Barnard1919}).  Fast-tracking to 1930s,  the concept of interstellar dust as material which must be absorbing the starlight was first quantitatively evidenced by Robert Trumpler (\citealt{Trumpler1930}) through the discovery of interstellar reddening.  Despite these early discoveries,  it is only in the last 40 years or so that the research in far-infrared astronomy has advanced our understanding of the role that interstellar dust emission plays in the ISM of galaxies. 

\subsection{A snapshot of infrared space missions}
In 1983,  the \textit{Infrared Astronomical Satellite} (IRAS; \citealt{Neugebauer1984}) became the first space-based observatory to study far-infrared emission,  providing observations in the wavelength range 8 -  120 $\mu$m,  with filters centred on 12, 25, 60 and 100 $\mu$m wavelengths.  IRAS revealed around 350,000 infrared objects including point sources,  extended sources, comets and asteroids.  Particularly exciting discoveries were clouds of dust around the star Vega (\citealt{Aumann1984},  \citealt{Harvey1984}) and galaxies which were more than 10$^{11}$ times brighter than our Sun in the infrared (\citealt{Soifer1984}; see \cite{Soifer1985} for a brief overview of first results).  IRAS provided insight into emission from relatively hot dust for the first time and set the stage for future space missions probing infrared emission from galaxies.

Subsequently,  in 1995,  the \textit{Infrared Space Observatory} (ISO; \citealt{Kessler1996}) mission launched to facilitate studies at 2.5 - 240 $\mu$m wavelengths.  ISO observed many different types of galaxies including nearby spiral galaxies like Andromeda and the Triangulum galaxy,  galaxies which have undergone a burst of star formation,  ultra-luminous infrared galaxies and distant dusty star-forming galaxies (redshift ($z$) $\leq$ 1.5).  ISO revealed a colder dust component (typical $T_{\mathrm{dust}} \approx 20$ K) and a warmer dust component ($T_{\mathrm{dust}} \approx 30 - 40$ K) in spiral galaxies by shifting our observational capacity to wavelengths beyond 200 $\mu$m for the first time (see \cite{Genzel2000} for a review of extragalactic science results).

In 2003,  the \textit{Spitzer Space Telescope} launched to observe at wavelengths 3.6 - 160 $\mu$m (\citealt{Werner2004}).  \textit{Spitzer} observed dust debris surrounding stars,  supernovae,  exoplanets and helped form a picture of dust-obscured star-forming regions (where young stars {\magi can heat up} the surrounding dust) in our Milky Way (e.g. \citealt{Werner2006}) and other galaxies (e.g.  \citealt{KennicuttJr.2003}). 

While the aforementioned space observatories provided important observations of emission from dust,  the launch of the \textit{Herschel Space Observatory} (2009 - 2014; \citealt{Pilbratt2010}) laid the foundation for resolved studies of dust,  both in our galaxy (e.g. \citealt{Kirk2013},  \citealt{Molinari2016}) as well as external galaxies (e.g. \citealt{Boselli2010},  \citealt{Kennicutt2011}, \citealt{Fritz2012},  \citealt{Oliver2012}, \citealt{Meixner2013}).  \textit{Herschel} was the first telescope to probe the submillimetre regime with observations not only capturing the emission at shorter wavelengths of 70, 100 and 160 $\mu$m but also longer wavelengths of 250,  350 and 500 $\mu$m.  This allowed astronomers to sample well the wavelengths around the peak of the spectral energy distribution (SED) of the emission from dust at $\sim$ 160 $\mu$m.  Some of the notable \textit{Herschel} surveys of galaxies include the Herschel-ATLAS (H-ATLAS; \citealt{Eales2010}, \citealt{Valiante2016}),  Herschel Reference Survey (HRS; \citealt{Boselli2010}),  Herchel Virgo Cluster Survey (HeViCS; \citealt{Davies2010}),  Herschel Fornax Cluster Survey (HeFOCS; \citealt{Davies2013}),  HERschel Inventory of the Agents of Galaxy Evolution (HERITAGE; \citealt{Meixner2014}),  Key Insights on Nearby Galaxies: A Far-Infrared Survey (KINGFISH; \citealt{Kennicutt2011},  and the Herschel Exploitation of Local Galaxy Andromeda (HELGA; \citealt{Fritz2012}) survey.

\begin{figure}[h]
\centering
\includegraphics[width=16cm]{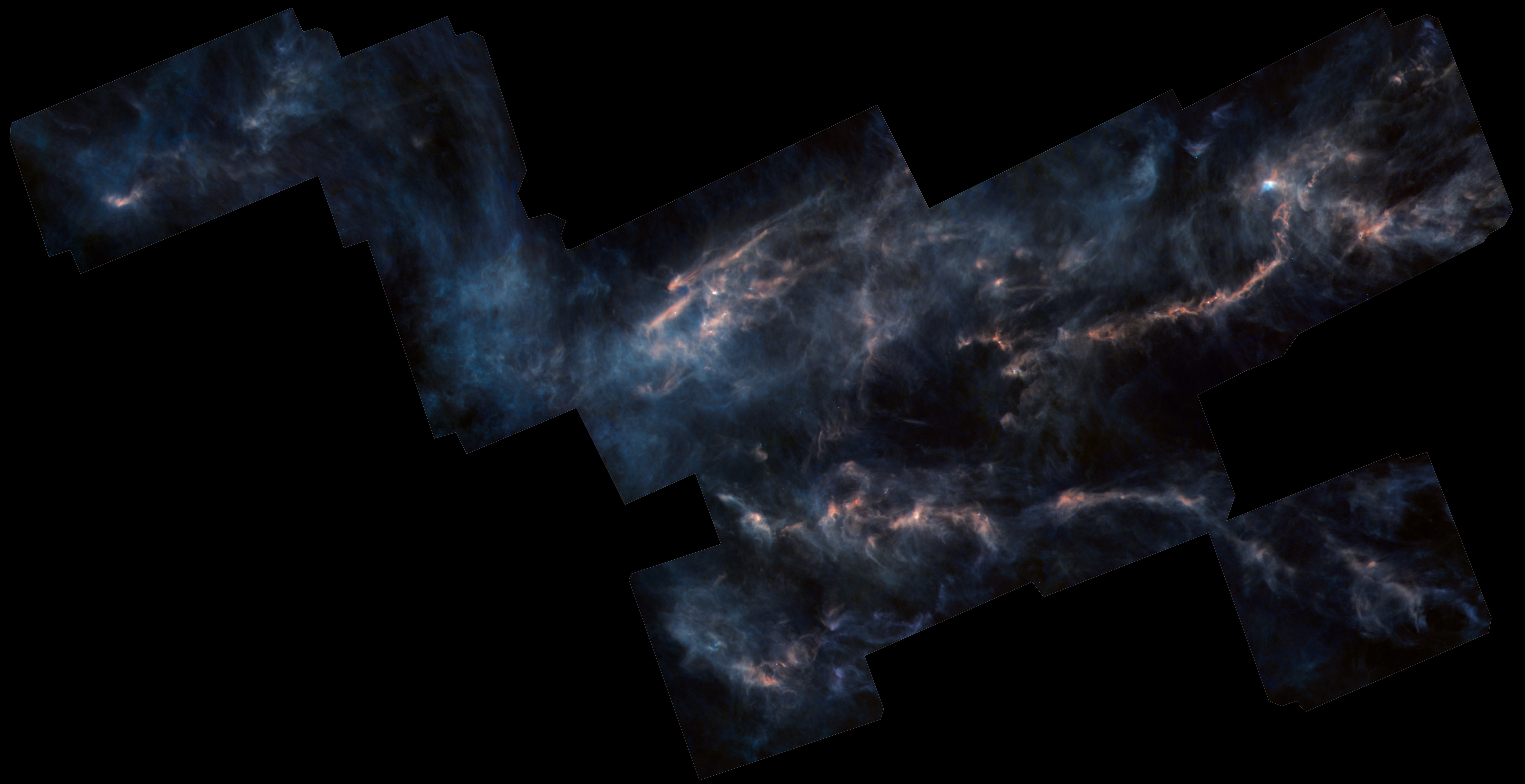}
\caption{A {\magi combined} false colour image of the Taurus Molecular Cloud in the Milky Way {\magi created from observations taken by the} \textit{Herschel Space Observatory} {\magi at 160 $\mu$m (blue),  250 $\mu$m (green),  350 $\mu$m (split between green and red) and 500 $\mu$m (red).} The brighter,  redder colours show the dense regions,  while the darker,  bluer colours show the less dense regions. The dense regions present as a range of filamentary networks linking bright clumps which are sites of star formation.  \copyright{ESA/Herschel/NASA/JPL-Caltech; acknowledgement: R. Hurt (JPL-Caltech).}}
\label{fig:taurus}
\end{figure}

\textit{Herschel} showed the ubiquity of filamentary structures within the ISM of the {\magi Milky Way} (e.g.  {\magi \citealt{Molinari2010},  \citealt{Arzoumanian2011}},  \citealt{Peretto2012},  {\magi \citealt{Hennemann2014}}) and detected over a million galaxies,  including nearby galaxies and distant ones (at $z \geq 2$) which were previously undetected.  \textit{Herschel} also gave us a spectacular view of dust in star-forming regions by observing molecular clouds like the Taurus Molecular Cloud (Figure \ref{fig:taurus}; {\magi \citealt{Kirk2013},  \citealt{Palmeirim2013}}) and the Orion Molecular Cloud {\magi (\citealt{Polychroni2013}, \citealt{Andre2013})} in our own galaxy. 

The majority of this thesis incorporates \textit{Herschel} observations in some manner; see Section \ref{ssec:herschm31} for a detailed review of findings from \textit{Herschel}.

\subsection{What is interstellar dust?}
Dust grains are smaller than the width of a human hair (with a radius of 0.0003 - 0.3 $\mu$m; \citealt{Galliano2018}),  composed of heavy elements such as carbon,  nitrogen,  oxygen,  magnesium,  silicon,  iron.  These grains are incredibly important for a lot of the physics happening within galaxies.  Dust absorbs over 50\% of the starlight in the Universe (\citealt{Dole2006}) despite only making up about 1\% of the ISM of galaxies.  Dust is usually found well mixed with gas in cold,  dense regions.  Dust plays a key role in the formation of stars.  It catalyses the formation of molecular hydrogen which is a necessary molecule for star formation.  {\magi Dust also aids the synthesis of more complex molecules,  like water,  through grain-surface chemistry.  Furthermore,} it shields star-forming regions from radiation that can dissociate molecules and plays a significant role in regulating the temperature of interstellar gas via heating and cooling (\citealt{Draine2011}).

\subsection{Dust creation and destruction}
\begin{figure}[h]
\centering
\includegraphics[width=12cm]{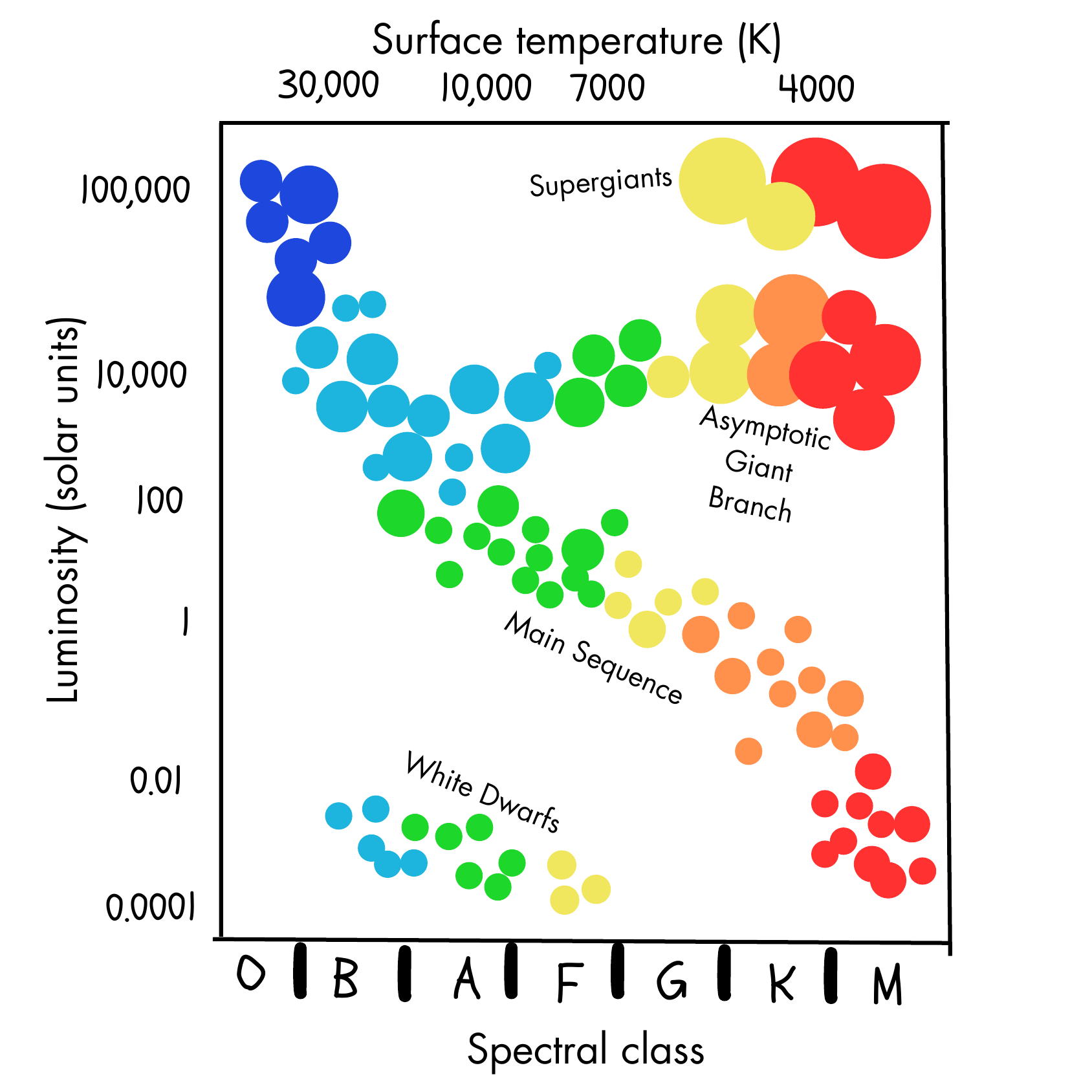}
\caption{A cartoon of the Hertzsprung-Russell diagram showing the spectral class and surface temperature of a star against the luminosity.  The asymptotic giant branch and the main-sequence phase as well as other phases are labelled.}
\label{fig:hr}
\end{figure}

The main mechanism by which dust is created in the ISM is still up for debate.  The three proposed mechanisms are: dust creation in stellar winds coming from evolved stars from the asymptotic giant branch (AGB; e.g.  \citealt{Ferrarotti2006},  \citealt{Olofsson2010}, \citealt{Schneider2014},  \citealt{Hofner2018}),  dust creation in supernova explosions (e.g.  \citealt{Gomez2012} \citealt{Matsuura2015}),  and dust creation and evolution via grain growth in the ISM.  Figure \ref{fig:hr} shows {\magi where the asymptotic giant branch sits on} the Hertzprung-Russell diagram for reference.

AGB stars have low to intermediate main-sequence masses in the range of $\sim$ 1 - 7 M$_{\odot}$.  These stars lose their mass quickly through cool,  expanding winds once they leave the main-sequence phase.  The cores of AGB stars are inert and made of carbon and oxygen.  The core contains most of the mass of the star and is surrounded by an inner shell of helium and outer shell of hydrogen which burn alternately,  as well as an outer envelope.  Every $\sim$ 10$^4$ years,  the helium shell becomes hot and dense enough to explode and then leaves a period of quiescence before the next explosion - a process known as a "thermal pulsation".  This is when many of the different elements and carbon-rich material are created.  This thermal pulsation also causes the hydrogen and helium shells of the star to mix with its outer envelope,  leading to a gradual "dredge-up" of carbon-rich gas and an increase of the C:O ratio,  usually after several pulses {\magi (\citealt{Pols2001},  \citealt{Hofner2018})}.  Intermediate mass stars (> 4 M$_{\odot}$) may bypass this carbon enrichment by converting the carbon directly into nitrogen at the hot base of the outer envelope,  a process known as "hot bottom burning" {\magi (\citealt{Lattanzio1992},  \citealt{Lattanzio1996})}.  With metal enrichment by the star and temperatures low enough for dust to survive (< 2000 K),  the outer envelopes of AGB stars provide a favourable environment for dust formation.  The type of dust produced is dependent on whether the envelope is carbon-dominant or oxygen-dominant.  Stars with an oxygen-rich envelope produce silicates and those with a carbon-rich envelope produce carbonaceous grains (e.g \citealt{Salpeter1974}).

While AGB stars produce much of the dust in galaxies,  there is evidence from studies of nearby galaxies (\citealt{Matsuura2009} \citealt{Mattsson2014}, \citealt{Goldman2022}) and high-redshift ($z \gtrsim 3$) galaxies (\citealt{Morgan2003},  \citealt{Dunne2003},  \citealt{Rowlands2014}, \citealt{Michalowski2015}) that AGB stars alone cannot explain all of the dust found in the ISM.  {\magi In the Large Magellanic Cloud (LMC),  there is evidence that AGB stars could contribute up to 70\% of the dust seen at present time and in the Small Magellanic Cloud (SMC),  up to 15\% (\citealt{Schneider2014}),  if there is no significant dust destruction.  For the Andromeda galaxy,  this contribution has been estimated to be between 0.9 and 36\% (\citealt{Goldman2022}).  At high-redshift,  however,  most stars in galaxies would not have evolved into the AGB phase to produce all of the observed dust.}

This creates a "dust budget crisis".  Where else is the dust produced? Supernovae provide an alternative source of dust manufacturing.  Type Ia supernovae are the endpoint of the evolution of an aged white dwarf star,  with a carbon-oxygen core,  which exists in a binary system with a companion.  As this white dwarf evolves,  it continually accretes material from its stellar companion which can either be a main-sequence star (forming a "single-degenerate system") or another white dwarf (forming a "double-generate system").  The accretion of material creates an envelope of hydrogen and helium around the white dwarf.  The increase in mass causes the white dwarf to contract and heat the material at the boundary between the core and the envelope. This produces a helium layer in the core followed by carbon and oxygen.  When the mass of the accreting white dwarf reaches the Chandrasekhar mass limit (1.44 M$_{\odot}$),  nuclear fusion of carbon in the core ignites the explosion triggering the deflagration and detonation of the white dwarf within a few seconds. The entire mass of the white dwarf is distributed into surrounding regions and the companion star is blown away.  There is little evidence of dust formation in these types of supernova explosions (\citealt{Gomez2014}).

Type II supernovae are the endpoint of the evolution of high-mass stars (stars with main-sequence mass of $\gtrsim 8$ M$_{\odot}$).  High-mass stars contain enough mass for the fusion of heavier and heavier metals (elements heavier than hydrogen and helium).  When they eventually begin the fusion of elements like iron and nickel,  {\magi the fusion uses more energy than it produces. This means that the core is no longer able to produce the thermal pressure from fusion reactions needed to support against gravity and instead begins to collapse under its own gravity. } When the mass of the {\magi core} exceeds the Chandrasekhar mass limit,  the core implodes,  creating a violent explosion.  Following this explosion,  {\magi a supernova remnant rapidly} sweeps up material in the surrounding ISM as it expands outwards.  Shocks trigger the formation of heavy elements within this remnant.  As the mass of the remnant increases,  it cools and the expansion slows down.  This creates a favourable environment for dust formation with low temperatures, high elemental abundance and number densities of particles.  Theoretical modelling suggests that between 0.1 $-$ 1 M$_{\odot}$ of dust can form in Type II supernovae if we ignore any dust destruction (e.g.  \citealt{Todini2001},  \citealt{Bianchi2007}).  There is also a plethora of observational evidence from the last twenty years showing that dust production in Type II supernovae is efficient (e.g.  \citealt{Matsuura2011},  \citealt{Gomez2012},  \citealt{Gomez2014} and references therein).  For a more comprehensive review of dust formation scenarios in supernova remnants,  we refer the reader to \cite{Sarangi2018}.

\begin{figure}[h]
\centering
\includegraphics[width=12cm]{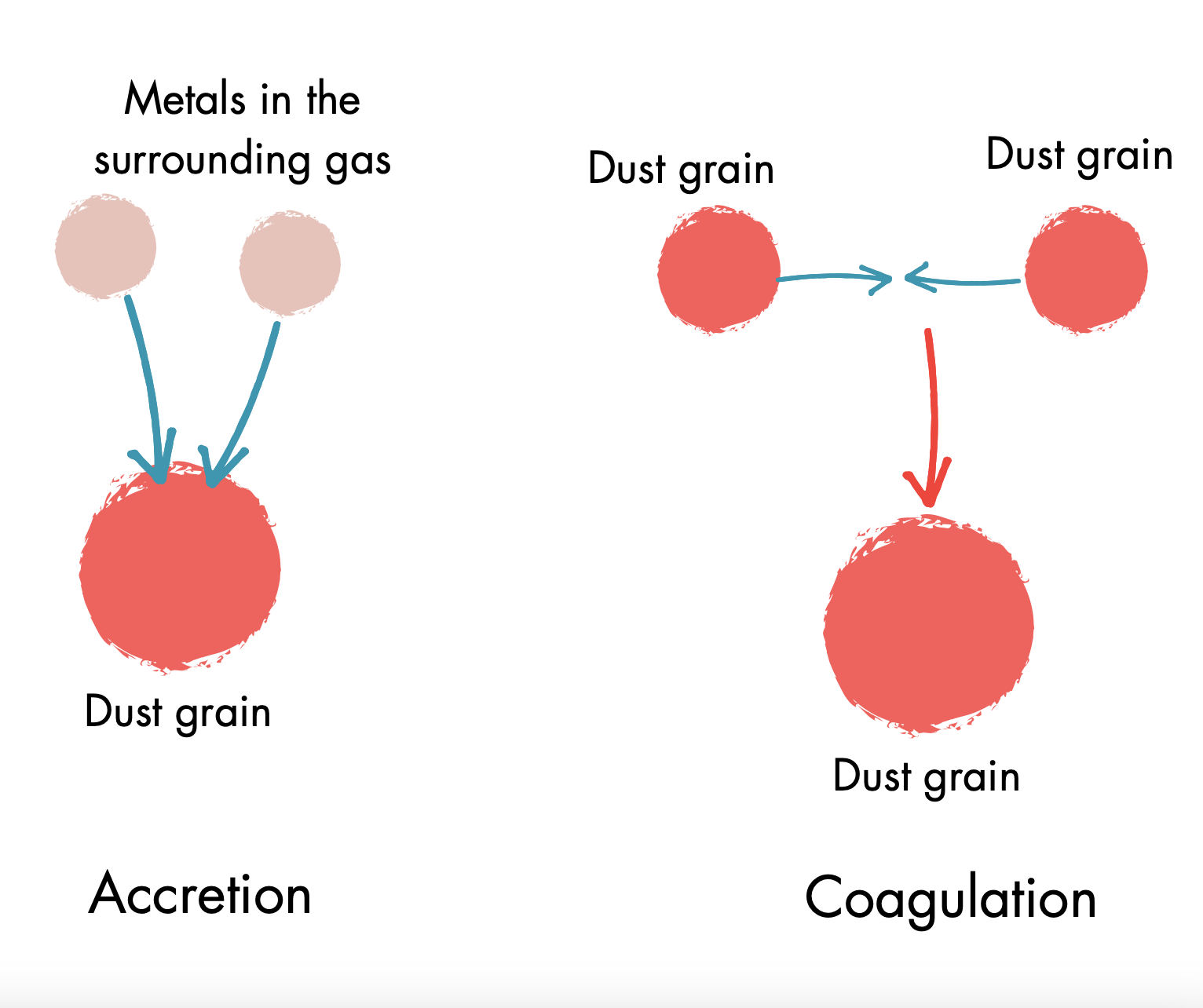} \caption{Cartoon depicting the difference between accretion and coagulation of dust grains in the ISM.}
\label{fig:graingrowth}
\end{figure}

The final mechanism relevant to the formation of dust is grain growth in the ISM itself (e.g. \citealt{Zhukovska2014}).  Grain growth in the ISM can contribute to the overall dust mass budget of galaxies.  In cold,  dense environments,  dust grains can have an icy mantle where molecules from the surrounding gas are accreted onto the mantle and can condense (\citealt{Dwek1998},  \citealt{Ormel2011}).  Dust grains can also grow via coagulation where two colliding grains can stick together (\citealt{Chokshi1993}, \citealt{Stepnik2002}, \citealt{Hirashita2012}).  Figure \ref{fig:graingrowth} shows a simple cartoon of how dust grains grow by these two processes. 

Dust is destroyed by forward and reverse shocks of supernova remnants which heat the ISM gas and destroy grains by various processes (\citealt{Jones2011}).  Dust grains can be broken down when they collide with other grains. This process is known as "shattering". They can also be vapourised,  resulting in material being released back from the dust phase into the gas phase.  Moreover,  collision between dust grains and ions from the gas can be destructive if the impending ion has sufficiently high energy and velocities ($\gtrsim$ 50 km s$^{-1}$) to break down the grain it is colliding with.  This process is known as "sputtering".  The timescale over which dust is destroyed is $\sim 3 \times 10^{8}$ years which is much shorter than the timescale for the creation of dust at $\sim 3 \times 10^{9}$ years,  {\magi making it unclear how the amount of dust that we observe has survived the destructive events,  and adding to the dust budget crisis.}

{\magi \subsection{Radiative transfer and dust emission}
\label{ssec:dustmbb}
\noindent The radiative transfer equation describes the physical processes of absorption,  emission and scattering experienced by a beam of radiation as it traverses a region of space:
\begin{equation}
\frac{dI_{\nu}}{ds} = -\alpha_{\nu}I_{\nu} + j_{\nu}
\end{equation}
where $I_{\nu}$ is the specific intensity of the radiation field at frequency $\nu$,  $dI_{\nu}$ is the change in specific intensity,  $ds$ is the distance that the radiation travels through the medium,  $\alpha_{\nu}$ is the absorption coefficient,  and $j_{\nu}$ is the emission coefficient.  We can rewrite the radiative transfer equation as a differential equation with respect to the optical depth:
\begin{equation}
\label{eq:intro_rt}
\frac{dI_{\nu}}{d\tau_{\nu}} = -I_{\nu} + S_{\nu}
\end{equation}
where $\tau_{\nu}$ is the optical depth,  and $S_{\nu}$ is the source function.  In the case when the medium is in local thermal equilibrium (LTE), and the source function does not vary,  equation \ref{eq:intro_rt} has the solution:
\begin{equation}
\label{eq:intro_rt2}
I_{\nu} = S_{\nu}(1-e^{-\tau_{\nu}}) + I_{\nu,0}e^{-\tau_{\nu}}
\end{equation}
where $I_{\nu,0}$ is specific intensity on the back side of the medium along the chosen line of sight.  Within the optically thin limit,  where the optical depth $\tau_{\nu} << 1$,  we can approximate that $e^{-\tau_{\nu}} \simeq 1-\tau_{\nu}$.  So equation \ref{eq:intro_rt2} becomes:
\begin{equation}
I_{\nu} \simeq S_{\nu}\tau_{\nu}  + I_{\nu,0}
\end{equation}
When $I_{\nu,0} << S_{\nu}\tau_{\nu}$,  we can approximate:
\begin{equation}
I_{\nu} \simeq S_{\nu}\tau_{\nu}
\end{equation}
In LTE,  the source function, $S_{\nu}$,  becomes the Planck function, $B_{\nu}(T)$:
\begin{equation}
B_{\nu}(T) = \frac{2hv^{3}}{c^{2}} \frac{1}{e^{\frac{h\nu}{k_{b}T}}-1}
\end{equation}
where $T$ is the temperature,  $k_{b}$ is the Boltzmann constant,  $h$ is Planck's constant,  and $c$ is the speed of light.  For an observed dusty medium:
\begin{equation}
\tau_{\nu} = \kappa_{\nu} \Sigma_{\mathrm{dust}}
\end{equation}
where $\kappa_{\nu}$ is the specific opacity (given in units of m$^2$ kg$^{-1}$) and $\Sigma_{\mathrm{dust}}$ is the dust mass surface density obtained as the mass of dust in the medium divided by the physical area. Therefore, the specific intensity of radiation from an optically thin dusty medium in LTE can be described by:
\begin{equation}
\label{eq:intro_mbb}
I_{\nu} = B_{\nu}(T_{\mathrm{dust}}) \kappa_{\nu} \Sigma_{\mathrm{dust}} 
\end{equation}
where $T_{\mathrm{dust}}$ is the dust temperature.  $\kappa_{\nu}$ is defined as:
\begin{equation}
    \kappa_{\nu} = \kappa_{0} \Big (\frac{\nu}{\nu_{0}} \Big)^{\beta}
\end{equation}
\noindent where $\kappa_{0}$ is a reference opacity for a reference frequency $\nu_0$.  $\beta$ is the dust emissivity index,  a key parameter which acts as a modifier to the shape of the blackbody spectrum for dust emission. }

\subsection{The far-infrared/sub-mm spectral energy distribution}
\begin{figure}[h]
\centering
\includegraphics[width=12cm]{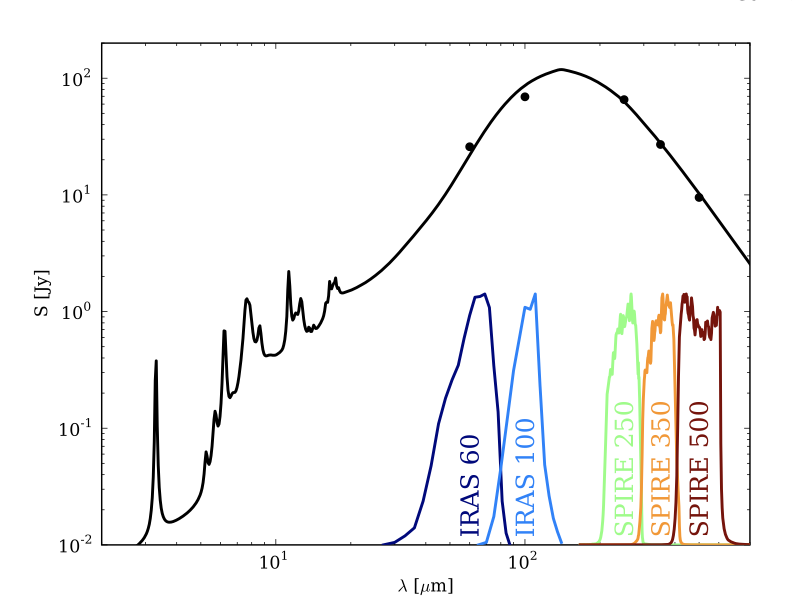} 
\caption{An example spectral energy distribution of a galaxy M100 from the Herschel Reference Survey.  This figure has been taken from \cite{Boselli2012}.  The points show observations taken by different instruments across a range of wavelengths and the dark line is the best fit model (from \citealt{Draine2007}).  \copyright{\cite{Boselli2012}}.}
\label{fig:exampsed}
\end{figure}

\noindent Dust grains absorb ultraviolet light from stars and re-emit this light at infrared and submillimetre wavelengths.  The far-infrared/sub-mm spectral energy distribution of dust continuum emission is modelled with the assumption that it emits like a modified blackbody.  More sophisticated models also exist which use multiple temperature components,  and model different features of the emission spectrum,  taking into account the assumed size and composition of dust grains (e.g. \citealt{Draine2007}, \citealt{Jones2013},  \citealt{Jones2017}).  Figure \ref{fig:exampsed} shows an example best-fit SED {\magi of the galaxy M100}.  In the $60 < \lambda < 100 \mu$m wavelength range,  the emission comes from warm dust grains in equilibrium with the radiation field.  At $\sim$ 160 $\mu$m,  the intensity of dust emission peaks.  The $250 \leq \lambda \leq 500$ $\mu$m wavelength range has contributions from a cooler,  extended diffuse dust component and more compact dusty regions found in cold,  molecular clouds.  At 850 $\mu$m,  we can see emission from cold dust.

\section{CO-dark molecular gas and the variable $X$-factor}
\label{sec:xfactor}
Far-ultraviolet (FUV) photons at wavelengths $911.75 < \lambda < 1117.8$ $\mathring{\mathrm{A}}$ carry enough energy to photodissociate the CO molecule into atomic carbon and oxygen (e.g.  \citealt{VanDishoeck1988}).  Although the abundance of the CO molecule makes it a widely accepted as a tracer of molecular gas,  a major drawback of tracing molecular gas in this way is that CO photodissociates due to its inability to shield from the incoming FUV radiation of stars.

In the Milky Way,  there is evidence that as much as $\sim$ 30 - 50 \% of the gas in molecular clouds is CO-dark (\citealt{Velusamy2010},  \citealt{Abdo2010},  \citealt{PlanckCollaboration2011}, \citealt{Lee2012}). There is {\magi also evidence of} a decrease in the dark gas fraction from 71\% to 43 \% going from the inner {\magi Galaxy} to the outer {\magi Galaxy when excluding the Galactic plane emission where most clouds are located (\citealt{Paradis2012a})}.  In the Magellanic Clouds,  there is evidence from \textit{Herschel} observations that 89\% of H$_2$ in the LMC is CO-dark and and 77\% in the SMC is the same (\citealt{Pineda2017}). 

Other than the CO-dark gas problem,  the CO tracer method also faces the challenge of the variability in the $X$-factor.  {\magi Here,  we are referring to the constant of proportionality which translates the intensity of CO emission into the column density of molecular hydrogen as shown in equation \ref{eq:intro1}.  Adopting a constant $X_\mathrm{CO}$ factor,  while common practice,  assumes a very ideal system where the effects of the environment,  dynamics,  spatial scales and observational assumptions are all incorporated into one constant value. 

Studies have shown that} the $X$-factor varies between galaxies and within galaxies (e.g. \citealt{Sandstrom2013},  \citealt{Brok2023}).  {\magi The value of this $X$-factor can increase with decreasing metallicity} (\citealt{Narayanan2012},  {\magi \citealt{Gong2020}),  probably because for the same amount of H$_2$,  the gas may be CO-dark.} The $X$-factor is correlated with regions of the ISM where there is a low CO optical depth (\citealt{Teng2023}).  {\magi Since the optical depth of CO determines how much CO emission can escape a cloud and reach the observer,  regions of high CO optical depth would naturally decrease the amount of CO emission,  thus increasing the $X$-factor required to maintain the same amount of H$_2$. 

This issue of the variability of the $X$-factor is further complicated by the fact that we can measure it using multiple methods.  For example,  the $X$-factor can be constrained by using a different tracer of H$_2$ such as a higher-order CO transition or a CO isotopologue which has its own conversion factor (e.g.  \citealt{Dickman1978}),  or through the gamma-ray emission in regions where cosmic rays collide with hydrogen molecules (e.g.  \citealt{Abdo2010}). }

{\magi All of these effects mean that the $X$-factor could be up to a factor of two uncertain (\citealt{Bolatto2013}).} For a detailed review of $X$-factor variations, we refer the reader to the paper by \cite{Bolatto2013}.  Throughout this thesis,  we assume a fixed $X_{\mathrm{CO}}$ across the Andromeda galaxy although there is evidence that it may not be the same in every environment.  We are mindful that the choice of $X_{\mathrm{CO}}$ may strongly affect the results of studies of molecular gas traced by CO {\magi and this becomes an intrinsic uncertainty in our calculations.  If measuring the total amount of molecular gas is reliant on the value of the $X$-factor,  then it is possible that we could be over or underestimating the masses and the depletion times of gas in clouds. }

The evidence of CO-dark gas in nearby galaxies and the variability of the $X$-factor suggests that we need a different method to trace molecular gas in galaxies.  Since dust and gas are well mixed within molecular clouds,  dust can be used to trace molecular gas.  The advantage of dust over a molecule like CO is that it does not photodissociate.  Therefore,  we use both CO emission and dust continuum emission to trace molecular clouds in this thesis,  as alternative ways of probing the physics of clouds.

\section{The Andromeda galaxy}
The Andromeda galaxy (M31) is our nearest neighbouring spiral galaxy in the Local Group.  It is more massive than the Milky Way (\citealt{Yin2009}) with double the disk mass of our own galaxy (with a total disk {\magi (gas + stars)} mass of $7 \times 10^{10}$ M$_{\odot}$ compared to $3.7 \times 10^{10}$ M$_{\odot}$ for the Milky Way).  The galaxy is at a distance of approximately 785 kpc (\citealt{McConnachie2005}) meaning that light from M31 takes over 2 million years to reach us.  Due to its proximity to us,  M31 has been observed at almost every wavelength.  Here we introduce M31 through a small selection of these multiwavelength observations.  For reference,  Table \ref{tab:multiwavprops} lists the wavelength ranges corresponding to the emission from different parts of the electromagnetic spectrum. 
\begin{table*}[h!]
\caption{Table showing the wavelength ranges of observations.}
\label{tab:multiwavprops}      
\centering                                      
\begin{tabular}{c c}          
\hline       
Type of observation & Wavelength range \\
\hline
Radio & 100 m $\geq \lambda \geq 1$ cm \\
Millimetre \& submillimetre & 10 mm $\geq \lambda \geq 0.1$ mm\\
Infrared & 100 $\mu$m $\geq \lambda \geq 1$ $\mu$m\\
Optical & 1 $\mu$m $\geq \lambda \geq 300$ nm\\
Ultraviolet & 300 nm $\geq \lambda \geq 10$ nm \\
\hline                      
\end{tabular}
\end{table*}

Mid to far-infrared observations by the \textit{Spitzer Space Telescope} at 24,  70 and 160 $\mu$m (\citealt{Gordon2006}) have shown that while it is difficult to distinguish the spiral arms in M31,  we can see hot dust settled in ring-like structures especially at a galactocentric radius 10 kpc (\citealt{Habing1984}) but also in an inner and outer ring at 5 and 15 kpc respectively.  There is also a very old stellar population in the central bulge of M31,  traceable by 3.6 $\mu$m emission (\citealt{Barmby2006}).   Overall, \textit{Spitzer} has revealed that M31 is estimated to have $\sim 1$ trillion stars.
\begin{figure}[h!]
\centering
\includegraphics[width=16cm]{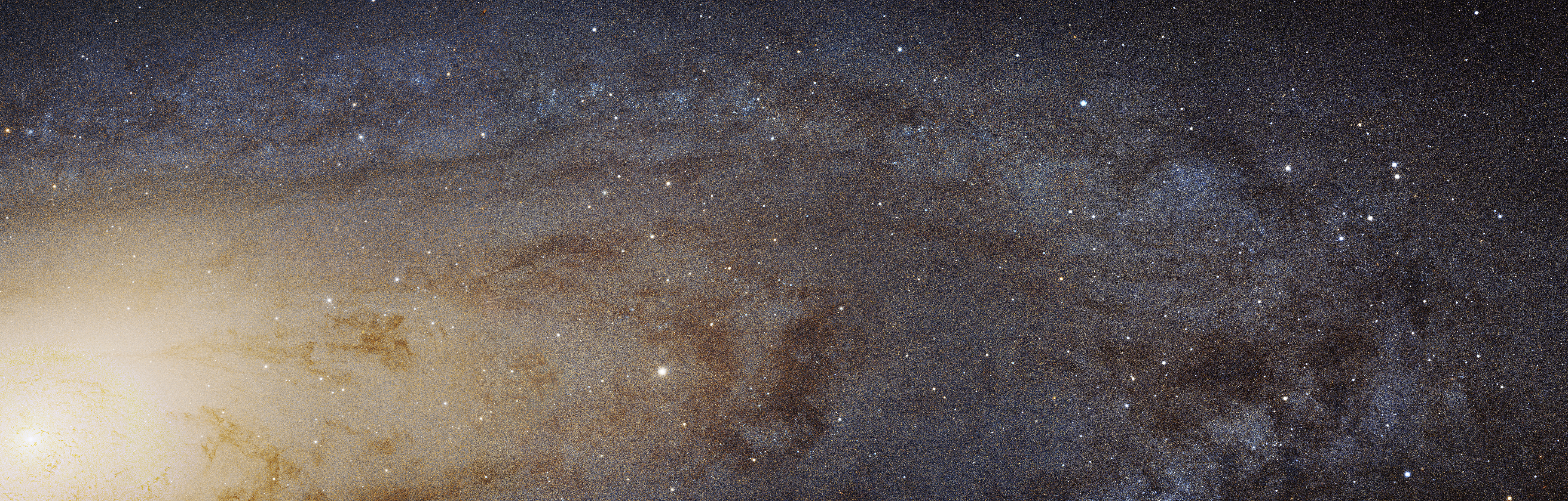} \caption{M31 as seen by the \textit{Hubble Space Telescope's} blue and red filters.  \copyright{NASA, ESA, J. Dalcanton (University of Washington), the PHAT team, and R. Gendler}.}
\label{fig:optihub_m31}
\end{figure}

\begin{figure}[h!]
\centering
\includegraphics[width=12cm]{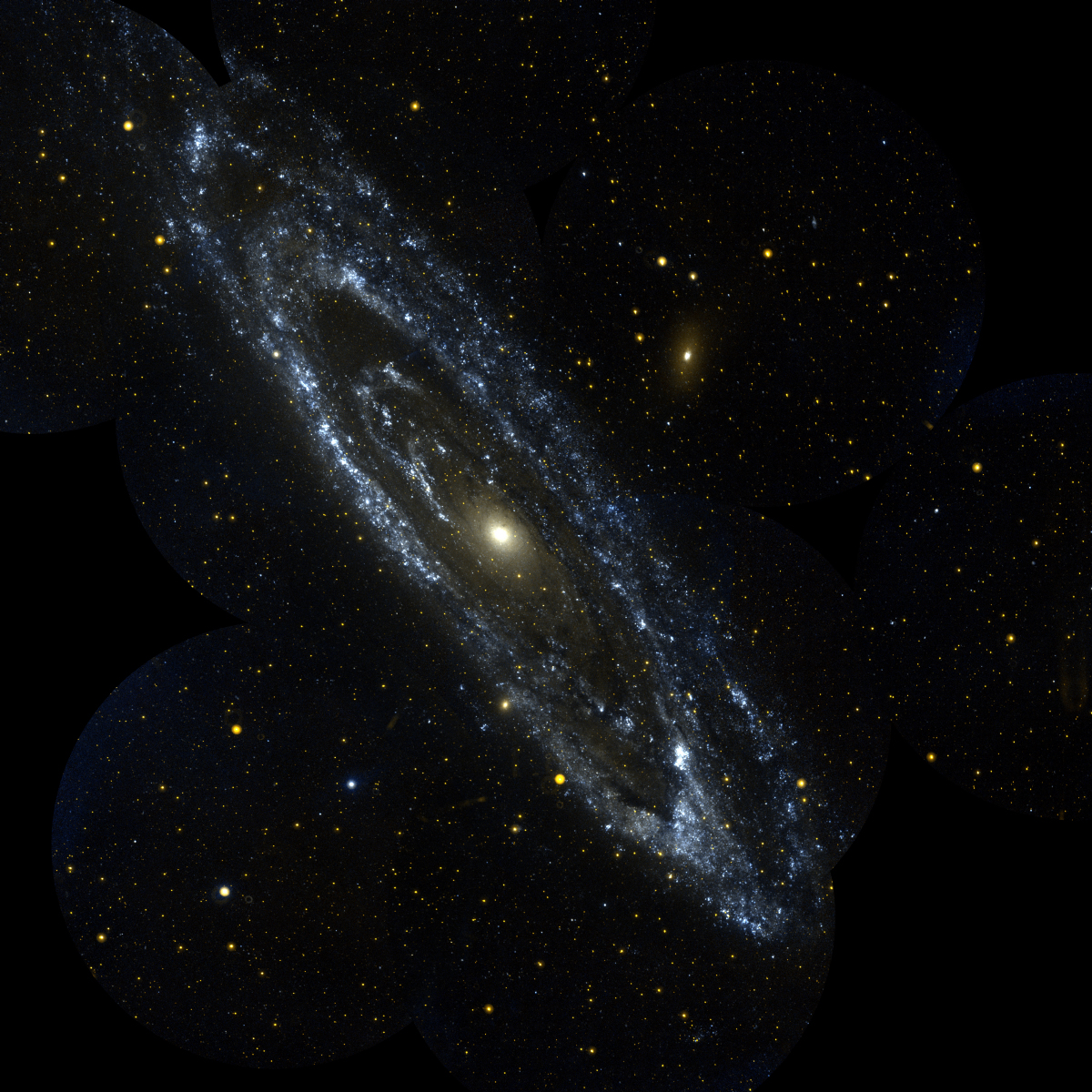} \caption{Two-colour composite image of M31 taken by the Galaxy Evolution Explorer (GALEX) showing the far-ultraviolet (135–175 nm) in blue and near-ultraviolet (175–275 nm) in orange.  The ultraviolet emission comes from young,  massive stars in the rings of M31 (in blue) and an older stellar population in the central bulge (ball of light in the centre).  The regions which are especially bright in both bands are shown white.  \copyright{NASA/JPL-Caltech}.}
\label{fig:uv_m31}
\end{figure}

Approximately one-third of M31 has been mapped by the \textit{Hubble Space Telescope} (Figure \ref{fig:optihub_m31}) for the Panchromatic Hubble Andromeda Treasurey (PHAT) survey (\citealt{Dalcanton2012}).  This survey detected $\sim 117$ million stars (\citealt{Williams2014}) within the galaxy and found that star formation must have been active in the 10 kpc ring for at least the last 500 Myr (\citealt{Lewis2015}).  The PHAT team found an average SFR of $0.28 \pm{0.03}$ M$_{\odot}$ yr$^{-1}$,  in the one-third of M31 mapped by \textit{Hubble}.  Additionally,  \cite{Williams2017} have shown that M31 is a more quiescent galaxy compared to the {\magi Milky Way},  with majority of the star formation taking place over 8 Gyr ago,  and with recent star-formation activity kickstarting at $\sim 2$ Gyr ago and then returning to a state of relative quiescence. 

Radio observations at the 21 cm wavelength from the Westerbork Synthesis Radio Telescope (WSRT; \citealt{Braun2009}) and Extended Very Large Array (VLA; \citealt{Koch2021}) show large amounts of atomic gas distributed in the outer rings of the galaxy.  The cold molecular medium in M31 has also been mapped by using $^{12}$CO(J=1-0) observations (\citealt{Nieten2006},  \citealt{Caldu-Primo2016}) made using the 30-m telescope at the Institut de radioastronomie millimétrique (IRAM) and the Combined Array for Research in Millimeter-wave Astronomy (CARMA) interferometer,  showing that M31 contains $\sim 3.6 \times 10^8 M_{\odot}$ of molecular gas. 

The Galaxy Evolution Explorer (GALEX; \citealt{Thilker2005}) took near and far-ultraviolet observations of M31 and found that intermediate-mass stars emitting UV radiation can be seen all the way out to a galactocentric distance of 27 kpc (Figure \ref{fig:uv_m31}).  There is also catalogue of massive stars recorded using H$\alpha$ emission and \textit{UBVRI} photometry of observations taken at the Kitt-Peak Observatory (\citealt{Massey2006}). 

\noindent In this thesis,  we study far-infrared and submillimetre observations of M31.

\subsection{The Herschel view of the Andromeda galaxy} 
\label{ssec:herschm31}
\begin{figure}[h!]
\centering
\includegraphics[width=16cm]{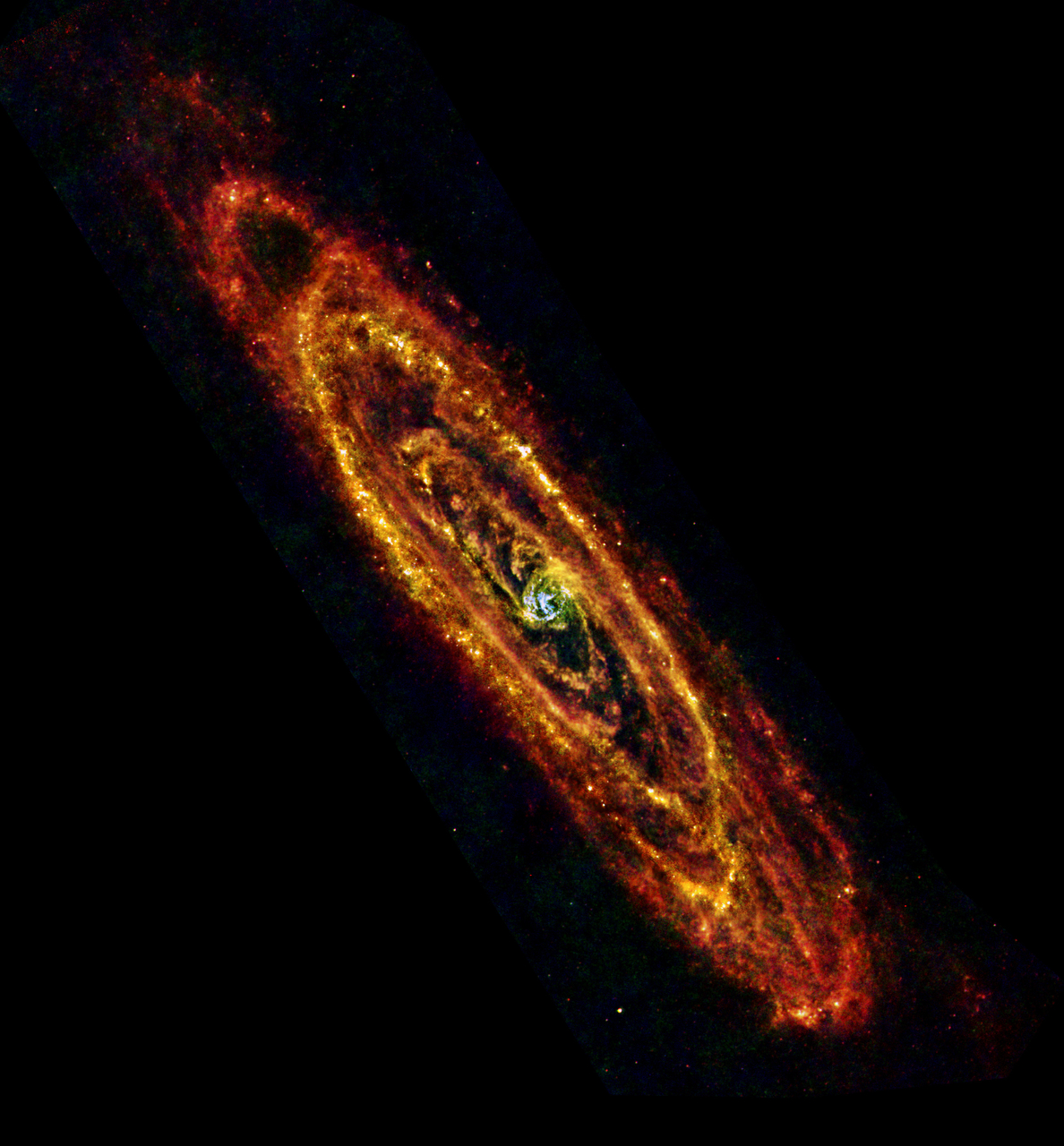} \caption{The \textit{Herschel} view of M31.  The blue colour shows 70 $\mu$m emission and the green colour shows 100 $\mu$m emission,  both coming from warmer dust likely heated by the older stellar population in the centre of the galaxy.  The red colour shows combined emission at 160 and 250 $\mu$m coming from cooler dust in regions where it is mixed in with the gas.  \copyright{ESA/PACS \& SPIRE Consortium/O. Krause/H. Linz.}}
\label{fig:hersch_m31combo}
\end{figure}
\textit{Herschel} completed two sets of observations of M31,  one proposed by the HELGA team (\citealt{Fritz2012}) and one proposed by the \cite{Draine2014} team (\citealt{Groves2012}).  Both sets of observations were taken independently,  at 70, 100,  160,  250, 350 and 500 $\mu$m wavelengths.  Studies of these observations have set the scene for a significant portion of the analysis in this thesis.  Figure \ref{fig:hersch_m31combo} shows a composite image of \textit{Herschel} observations.

\begin{figure}[h!]
\centering
\includegraphics[width=16cm]{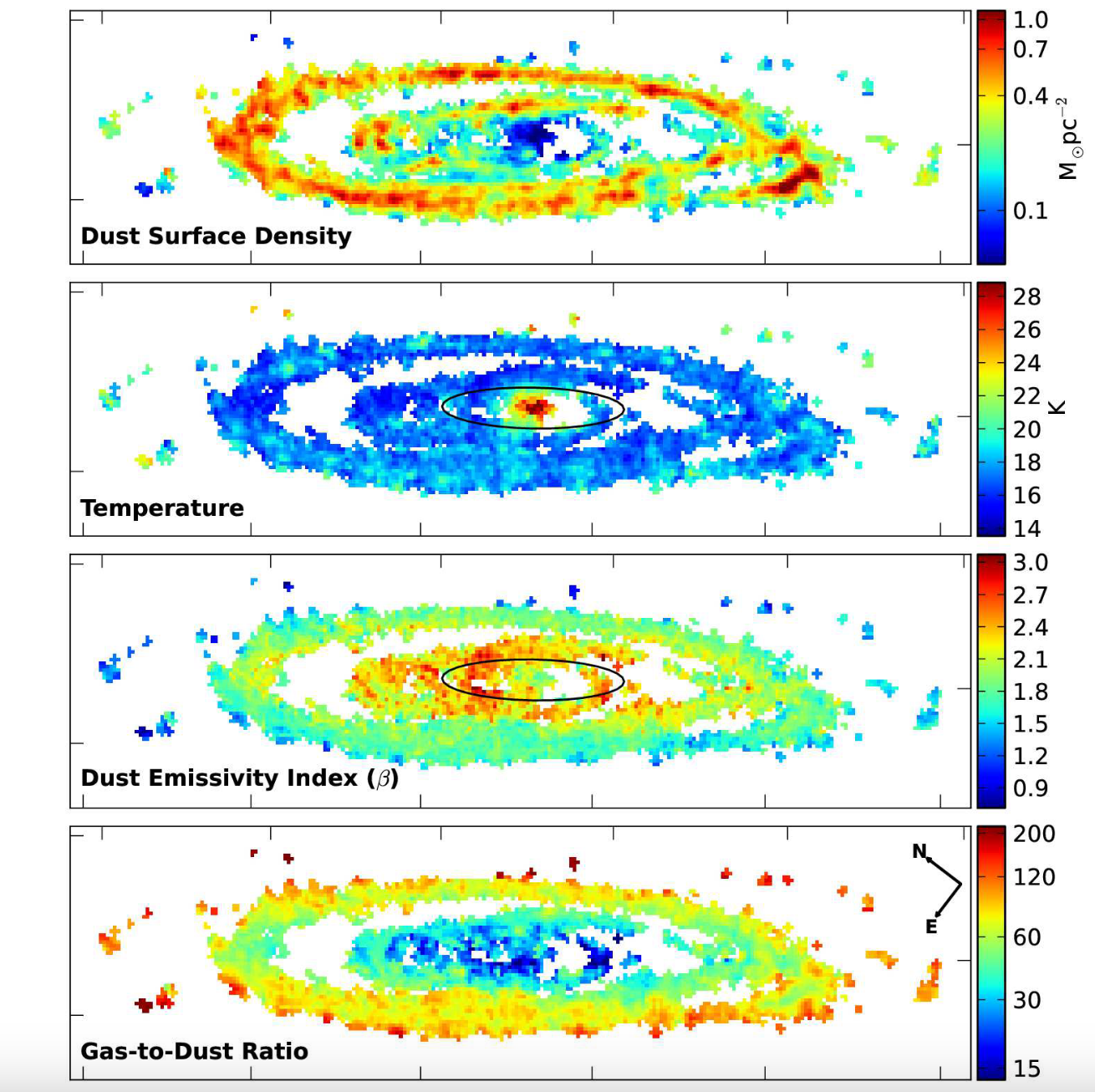} \caption{Figure taken from \cite{Smith2012} showing the spatial distribution of dust mass surface density,  dust temperature,  dust emissivity index and the gas-to-dust ratio across M31.  The colorbar represents the values of these dust properties.  The dust mass surface density is the highest in the inner 5 kpc and outer 10 kpc ring.  The GDR increases from the centre of the galaxy to the outskirts. The solid black circle outlines a transition galactocentric radius of 3.1 kpc.  At radii smaller than this transition radius,  the temperature decreases and $\beta$ increases with radius.  At radii larger than this transition radius,  the temperature increases and $\beta$ decreases with radius.  \copyright{HELGA Collaboration.}}
\label{fig:helgares}
\end{figure}

The HELGA team (\citealt{Fritz2012}) found that M31's 10 and 15 kpc rings contained $\sim$78\% of the total dust in the galaxy.  The 10 kpc ring contained a higher peak dust mass surface density than the 15 kpc ring.  In 2012,  \cite{Smith2012} showed for the first time that there are radial variations in observational dust properties at kpc spatial scales.  The authors found that the dust emissivity index,  $\beta$,  decreases as you move from the inner galaxy to the outskirts.  They also found that the dust temperature increases as you move from the inner galaxy to the outskirts.  {\magi The authors do not attribute this radial change to the well-known $\beta$-temperature degeneracy.  Within their study,} the central bulge region (inner 3 kpc) showed an increase in dust temperature to $\sim 30$ K,  correlated with emission from an older stellar population. This suggested that dust in the central bulge is being heated by this older stellar population.  Along with radial variations in $\beta$ and temperature,  the authors found that the total (atomic + molecular) gas-to-dust ratio (GDR) also increased as you moved outwards through the galaxy when using \textsc{HI} emission to detect atomic gas and CO emission to trace molecular gas. This is consistent with the radially decreasing metallicity gradient previously found in M31 (\citealt{Galarza1999}) assuming that a constant fraction of the metals in the ISM are locked up in dust grains {\magi (}\citealt{Edmunds2001}).  Figure \ref{fig:helgares} shows the distribution of these gas and dust properties across M31.

\cite{Draine2014} also found radial variations in $\beta$ using a different method and different \textit{Herschel} observations of M31.  \cite{Whitworth2019} derived dust temperature, emissivity index and integrated column density maps from the \textit{Herschel} M31 observations with significantly greater resolution through the application of a more sophisticated Bayesian algorithm (\citealt{Marsh2015}).  Resolved analysis of $\beta$ in the central region of M31 (\citealt{Marsh2018}) using these maps has also confirmed the radial variations as noted by previous studies. 

The HELGA team also searched for molecular gas within M31 which might be `CO-dark'.  Using a similar approach to \cite{PlanckCollaboration2011},  the HELGA team (\citealt{Smith2012}) adopted dust as a tracer for gas and compared the \textsc{HI} + CO gas column density with the dust-traced gas column density to look for any differences.  The team did not find any evidence for CO-dark gas in M31 but proposed that the poor angular resolution and source confusion along the line of sight might be a limiting factor in detecting any dark gas.

Furthermore,  there has been a hunt for any excess emission at 500 $\mu$m and beyond,  in galaxies like M31,  owing to such an excess previously detected in the Milky Way (\citealt{Paradis2012b}) and nearby galaxies (\citealt{Galametz2011}).  \citealt{Smith2012} found that the far-infrared spectral energy distribution of M31 could be fit with a single temperature modified blackbody model but found no evidence of there being very cold dust ($\leq$ 10 K).  Since cold dust radiates less efficiently than warmer dust,  there may still be quite a lot of cold dust in M31 without having a significant effect on the spectral energy distribution found by \cite{Smith2012}.

Aside from dust properties,  molecular clouds and star formation in M31 have also been studied with \textit{Herschel} observations.  With the assumption that dust and molecular gas are well mixed in molecular clouds,  \cite{Kirk2015} used dust continuum emission {\magi from \textit{Herschel} observations} as a tracer to identify 326 giant molecular clouds (clouds with masses ranging between $10^4$ and $10^7$ M$_{\odot}$) {\magi at $\approx 95$ pc spatial resolution}.  The clouds found within the 10 kpc ring of M31 formed 25\% of the molecular gas within the galaxy.  The rate of star formation,  as traced by a combination of FUV emission from stars and the dust-obscured star-formation measured by 24 $\mu$m emission,  across the whole galaxy has been measured as 0.25 M$_{\odot}$ yr$^{-1}$ (\citealt{Ford2013}).  

\subsection{New James Clerk Maxwell Telescope observations}
The missing but crucial piece of the puzzle thus far has been observing M31 at wavelengths  $> 500 \mu$m.  Observations at wavelengths $> 500 \mu$m would be more sensitive to {\magi the} emission from very cold dust or to a submillimetre excess similar to what has been detected in nearby galaxies (\citealt{Gordon2014}, \citealt{Chang2020}).  Hence,  a new set of observations of M31 were taken between 2017 and 2022 by the world's largest single-dish submillimetre telescope: \textit{James Clerk Maxwell Telescope} (JCMT).  These observations were carried out in the HARP And SCUBA-2 High-resolution Terahertz Andromeda Galaxy (HASHTAG) large programme (UK PI: Dr Matthew Smith) with $\sim 275$ hours of observing time,  and covered the entire galaxy (\citealt{Smith2021}).  The HASHTAG observations were taken at two submillimetre wavelengths: 450 and 850 $\mu$m.  The HASHTAG survey will allow us to look at dust in M31 for the first time at nearing 30 pc spatial scales,  nearly four times better resolution than \textit{Herschel} studies and below the size of a typical giant molecular cloud.  JCMT observations at 850 $\mu$m in particular allow us to study the tail-end of the far-infrared/sub-mm spectral energy distribution and help constrain the amount of cold dust in the galaxy. 

\begin{landscape}
\begin{figure}[h]
\centering
\includegraphics[width=22cm]{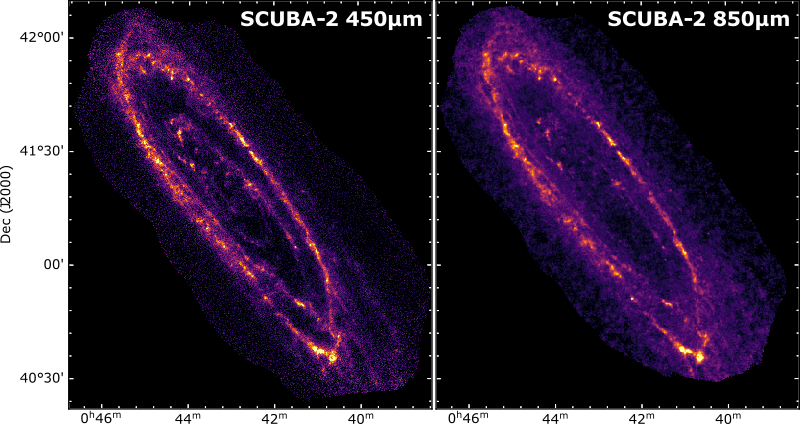}
\caption{JCMT observations of M31 at 450 and 850 $\mu$m (Figure 1 from \citealt{Smith2021}),  taken by the SCUBA-2 camera aboard the telescope.  {\magi The 450 $\mu$m and 850 $\mu$m images have an angular resolution of 7.9" and 13",  respectively.  \copyright{HASHTAG Collaboration.}}}
\label{fig:hashtag_m31}
\end{figure}
\end{landscape}

At these long wavelengths,  absorption of incoming radiation by the Earth's atmosphere before reaching the telescope and variations in the atmosphere is a problem.  To account for this,  the JCMT image at 450 $\mu$m has been combined with space-based data from \textit{Herschel} at 500 $\mu$m and the 850 $\mu$m image has been combined with \textit{Planck} observations at 353 GHz,  adding in the large scale structure obtained by the two space-based observatories and keeping the high resolution level obtained by the ground-based telescope. The technical details of the data reduction process and simulation techniques used to provide the final images can be found in the paper by \cite{Smith2021}.  Figure \ref{fig:hashtag_m31} shows the JCMT images made with 70\% complete observation run.

Following the recent HASHTAG survey data release and with the wealth of ancillary observations in mind,  it is timely that we explore the variations in dust properties and cold dust emission in dense molecular clouds as captured by the HASHTAG survey.  Hence,  nearly half of this thesis is spent analysing the HASHTAG {\magi SCUBA-2} observations of M31.  {\magi The HASHTAG HARP instrument observations of $^{12}$CO(J=3-2) line emission (\citealt{Li2020}) is not used in this work.}

\newpage
\newpage

\subsection{Why study M31?}
Despite \textit{Herschel} revolutionising our view of dust in galaxies like M31,  previous studies have left some unanswered questions.  In M31,  why does $\beta$ vary with radius? Existing studies fail to find the cause of radial variations in dust temperature and $\beta$.  While some thought has been given to the influence of the environment on dust grains through modelling (\citealt{Ysard2012}, \citealt{Jones2013}),  existing observational work does not test the effect of dense molecular gas on the radial variations in $\beta$.  While there is no evidence of a submillimetre excess at the wavelength of 500 $\mu$m,  another unknown is whether there is any excess emission at longer wavelengths like 850 $\mu$m.  Answering this question would help us to constrain the amount of very cold dust or give us clues about any different dust types in M31.

Moreover,  while studies have measured the star formation rate for M31,  we do not know if the star formation efficiency depends on position within the galaxy.  We also do not know to what extent observational dust properties like dust temperature and $\beta$ affect star formation in this galaxy.  These unsolved mysteries suggest that we do not yet have a complete picture of the effect of ISM interactions with star formation in our big neighbour. 

Since CO-dark gas is yet to be detected in M31 (\citealt{Smith2012}),  we are also left with the question of how much dark gas is there,  if any,  and where is it? Finding any CO-dark gas would have implications for the amount of molecular gas we are actually detecting using the CO tracer method and would raise questions about whether the CO is missing due to the photodissociation of this molecule.

Previous \textit{Herschel} studies of M31 were limited to the resolution of the lowest resolution image at 500 $\mu$m wavelength meaning that they resolved dust emission at $\sim$ 140 pc spatial scales.  With a mixture of M31 observations from HELGA team (\citealt{Smith2012}, \citealt{Fritz2012}) and \cite{Draine2014} and the {\magi HASHTAG survey from JCMT},  we endeavour to understand the environmental factors which may be influencing the more local variations in dust properties detected by previous studies.  We study the interplay of dust with gas properties and star formation in M31 at the scale of individual molecular clouds (sub-kpc).

\section{The central research questions of this work}
\noindent In this thesis,  we investigate the following research questions:
\begin{itemize}
    \item \textbf{Are the radial variations in the dust emissivity index within M31 caused by an increase of this parameter in dense gas regions?} In Chapter \ref{chapter:betavar},  we use $^{12}$CO(J=1-0) observations taken by the Combined Array for Research in Millimeter Astronomy (CARMA) and dust maps derived from \textit{Herschel} images to study variations in $\beta$ across an area of $\approx$ one-third of M31 at 30 pc spatial scales.  We produce two molecular cloud catalogues,  one traced by $^{12}$CO emission and one traced by dust continuum emission. We investigate whether there is evidence that $\beta$ is different inside and outside molecular clouds.  As a secondary goal,  we search for any evidence for CO-dark gas in M31.  Chapter \ref{chapter:betavar} is based on derived measurements from \textit{Herschel} data from before the HASHTAG data became available and before the simulated galaxy data were available for the work in Chapter \ref{chapter:ppmaptest}.
    \item \textbf{Is there an excess in observed dust emission compared to the modified blackbody model prediction at long wavelengths (450 \& 850 $\mu$m) in M31?} In Chapter \ref{chapter:sedfit},  we fit the spectral energy distribution of 100,  160,  250 $\mu$m observations of M31 from \textit{Herschel} and 450 and 850 $\mu$m observations from the new HASHTAG survey with a single temperature modified blackbody model.  We show the distribution of dust mass surface density,  dust temperature and $\beta$ from this survey for the first time.  We search for a submillimetre excess in M31 at 450 and 850 $\mu$m wavelengths.
    \item \textbf{Is dust a good tracer of the total gas content in M31?} In Chapter \ref{chapter:dustmass},  we present a new dust-selected cloud catalogue containing clouds extracted from the dust mass surface density map produced in Chapter \ref{chapter:sedfit} using observations from \textit{Herschel} and the HASHTAG survey observations from JCMT.  We examine how much CO-traced molecular gas and atomic gas is in our dust-traced clouds and search for any CO-dark gas.
    \item \textbf{Does star formation efficiency in M31 depend on position within the galaxy and do dust properties influence the star formation efficiency of molecular clouds in M31?} In Chapter \ref{chapter:sfe},  we use the dust-selected cloud catalogue from Chapter \ref{chapter:dustmass} and a star formation rate surface density map created using \textit{Herschel} observations to determine the star formation efficiency of individual clouds in our dust-selected catalogue.  We search for any correlations between the average dust temperature,  average $\beta$ and the gas depletion times of each cloud.
      \item \textbf{Can we accurately estimate dust emission in external galaxies using a more sophisticated Bayesian algorithm and reproduce sensible dust properties for extragalactic scales?} In Chapter \ref{chapter:ppmaptest},  we use a simulated galaxy from the Auriga simulation suite to test a Bayesian algorithm previously developed at Cardiff University called \textsc{PPMAP}.  The procedure shows powerful promise in removing the need to degrade observations to the lowest resolution image when capturing the dust content of a galaxy.  We test for the first time the capability of \textsc{PPMAP} to reproduce dust mass surface density estimates at extragalactic scales by examining how it behaves with an increasing number of iterations,  changes in the signal-to-noise levels of input images and changes in the inclination angle of the galaxy in the input image. 
\end{itemize}
\noindent Chapter \ref{chapter:Conclusion} summarises our key conclusions and discusses future work.

%% file: Chapters/chapter_betavar.tex
\chapquote{``There’s power in allowing yourself to be known and heard,  in owning your unique story,  in using your authentic voice.  And there’s grace in being willing to know and hear others. This,  for me,  is how we become.'' }{Michelle Obama}{}

\noindent This chapter is based on work presented in \cite{Athikkat-Eknath2022}. The collaborative work from co-authors came mainly in the form of reviewing.  All of the analysis presented in the chapter is original work.

\section{Introduction}
\label{sec:intro_beta}
Although primarily known for its obscuration effects on starlight, interstellar dust plays many other important roles in the interstellar medium (ISM): heating the ISM through the photoelectric effect (\citealt{Draine1978}), providing a cooling mechanism via its infrared emission in dense regions, shielding molecules from dissociating radiation, and even providing a surface for catalysing the formation of hydrogen molecules (\citealt{Gould1963}, \citealt{Hollenbach1971}). Many authors have suggested that the continuum emission from interstellar dust can be used to trace the mass of gas in galaxies (e.g. \citealt{Eales2012}, \citealt{Magdis2012}, \citealt{Liang2018}, \citealt{Groves2015}, \citealt{Scoville2016}, \citealt{Scoville2017}, \citealt{Tacconi2018}, \citealt{Janowiecki2018}), as an alternative to traditional tracers like carbon monoxide (CO). However, this method can only work if the properties of dust are universal, or if their variation with environment or epoch is known.

The dust emissivity index ($\beta$) is an important property that acts as a modifier to the shape of the blackbody spectrum which describes the emission from dust. In the optically thin limit (where the optical depth $\tau \ll$ 1), the specific intensity of dust emission is given by:
\begin{equation}
\label{eq:1}
  I_{\nu} \propto B_{\nu}(T)\nu^{\beta}
\end{equation}
where $B_{\nu}(T)$ is the Planck function and $\nu$ is the frequency. \cite{Smith2012} have used {\it Herschel} observations to investigate the properties of the dust
in the Andromeda galaxy (M31),
discovering that $\beta$ varies radially within the galaxy's disk. This general trend has been confirmed by \cite{Draine2014} using independent \textit{Herschel} data for M31 but with a different method. Variations in $\beta$ have also been reported in two other large spiral galaxies within the Local Group, the Milky Way (MW) and the Triangulum galaxy (M33). In M33, there is evidence for radial variations in $\beta$ and dust temperature, both of which decrease with galactocentric radius (\citealt{Tabatabaei2014}). So far, however, we have very little understanding of what is causing such radial variations. \cite{Tabatabaei2014} have found that $\beta$ is higher in regions where there is molecular gas traced by $^{12}$CO(J=2-1) or strong H$\alpha$ emission in M33 but the authors did not investigate whether there is any difference in $\beta$ between low-density and high-density environments at the same radius. In the MW, the \textit{Planck} team have shown that $\beta$ decreases from smaller Galactic longitudes to larger Galactic longitudes (\cite{PlanckCollaboration2014b}; see their Fig. 9). They have also found an increase in $\beta$ by $\sim$0.23 in the regions dominated by molecular gas along the line of sight when compared to the more diffuse atomic medium (\cite{PlanckCollaboration2014a}; see their Fig. 12); although as in M33, it is not clear whether the increase of $\beta$ in dense environments is the explanation of any radial variations in $\beta$.

Radial variations in $\beta$ have also been seen in galaxies outside of the Local Group. In a sample of 61 galaxies from Key Insights into Nearby Galaxies: Far-Infrared Survey with \textit{Herschel} (KINGFISH), \cite{Hunt2015} have found that the radial effects of $\beta$ can vary from one galaxy to another (see their Fig. 10). For example, some galaxies show negative radial gradients (e.g. NGC0337, NGC3049, NGC3077, NGC4559, NGC4725) whereas others show positive radial gradients (e.g. NGC1482, NGC3773, NGC4321, NGC4594). Evidence for  variation in the global values
of $\beta$ has come from a study of the gas and dust in 192 galaxies
in the JCMT dust and gas In Nearby Galaxies Legacy Exploration (JINGLE) project (\citealt{Lamperti2019}, \citealt{Smith2019}). The authors find  correlations between $\beta$ and properties such as stellar mass, stellar mass surface density, metallicity, \textsc{HI} mass fraction, star formation rate (SFR), specific SFR, SFR surface density, and the ratio of SFR and dust mass for these galaxies. The strongest positive correlation is found between $\beta$ and stellar mass surface density.

Our study focuses on variations in dust properties,  in particular the dust emissivity index ($\beta$), within M31.  Due to its proximity to us \citep[at a distance of $\approx$ 785 kpc;][]{McConnachie2005}, we can learn about properties of dust and the ISM in M31 at the scale of individual molecular clouds.  M31 also provides a unique perspective as the biggest spiral galaxy in the Local Group, with the added incentive that we can observe the galaxy from the outside, unlike observing the Milky Way from within which limits us from getting a global view of our galaxy and has problems such as superimposed sources at different distances along the line-of-sight.

There are many archival datasets containing observations of M31 in different wavebands (\citealt{Thilker2005}, \citealt{Braun2009}, \citealt{Dalcanton2012}, \citealt{PlanckCollaboration2015}), including key far-infrared and sub-mm observations from the Herschel Exploitation of Local Galaxy Andromeda (HELGA) survey (\citealt{Smith2012}, \citealt{Fritz2012}) and $^{12}$CO(J=1-0) observations (\citealt{Nieten2006}) made using the 30-m telescope at the Institut de Radioastronomie Millim\'etrique (IRAM) to trace the distribution of molecular gas over the whole galaxy. We take advantage of the high resolution observations covering part of M31 obtained with the Combined Array for Research in Millimeter-wave Astronomy (CARMA) inteferometer to trace molecular gas (A.~Schruba et al, in preparation). The central research question this work attempts to address is: `Are the radial variations in the dust emissivity index ($\beta$) in the Andromeda galaxy caused by an increase of $\beta$ in dense molecular gas regions?' As such, we measure and compare $\beta$ in dense molecular gas regions with $\beta$ in non-dense regions.

Given the evidence from {\it Fermi} (\citealt{Abdo2010}), {\it Planck} (\citealt{PlanckCollaboration2011}) and {\it Herschel} (\citealt{Pineda2013}) that there is CO-dark molecular gas in the Milky Way, probably because of photodissociation of the CO molecule \citep{Hollenbach1997}, we trace clouds in two different ways, using CO emission and dust continuum emission. This allows us to carry out the additional interesting project of comparing
the cloud catalogues produced by the two different methods.

This chapter is structured as follows: we first describe the observational data that are used in this work (Section \ref{sec:obs}). Next, we outline our source extraction methodology (Section \ref{sec:se}), followed by our results (Section \ref{sec:res}) and discussion (Section \ref{sec:disc}). Finally, we summarise our conclusions in Section \ref{sec:conc}.

\section{Observations}
\label{sec:obs}
\subsection{CARMA survey of Andromeda}
\label{ssec:carmaobs}
\begin{figure*}
\centering
\includegraphics[width=17cm]{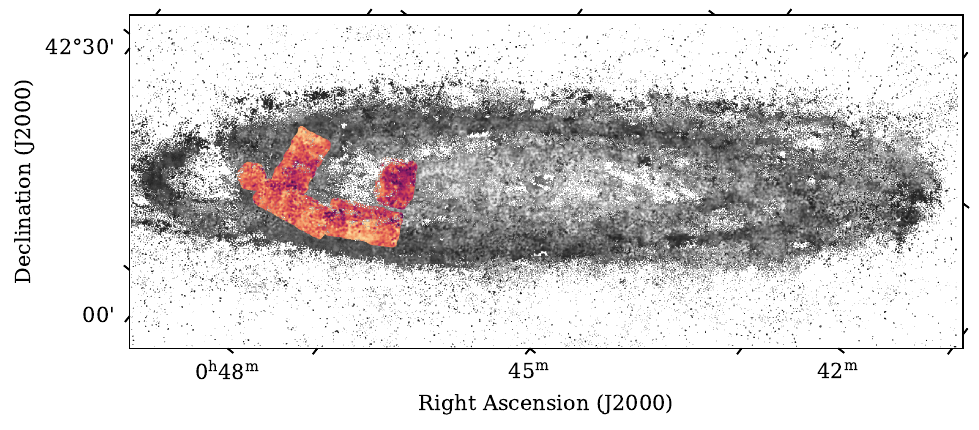} \caption{Map of dust emissivity index ($\beta$) made from PPMAP. The maroon-yellow part shows the region of the $\beta$ map overlapping the $^{12}$CO(J=1-0) map made with CARMA, partially covering the inner ring at 5 kpc and the dusty, star-forming ring at 10 kpc.}
\label{fig:carmareg} 
 \end{figure*}

We use a map of $^{12}$CO(J=1-0) integrated intensity obtained from observations of M31 made using the CARMA interferometer. The data were taken as part of the `CARMA Survey of Andromeda' (A.~Schruba et al., in preparation) across an angular area of 365 arcmin$^2$ and on-sky physical area covering 18.6 kpc$^2$, which includes parts of M31's inner 5 kpc gas ring and 10 kpc dusty, star-forming ring (\citealt{Habing1984}). This corresponds to a deprojected physical area of $\approx$ 84.6 kpc$^2$ at {\magi the distance to M31 of 785 kpc (\citealt{McConnachie2005}) and inclination of 77$^{\circ}$ (\citealt{Fritz2012})}. The region covered in our work is highlighted in Figure \ref{fig:carmareg}. The high-resolution CARMA data were combined with observations from the IRAM 30 m telescope (\citealt{Nieten2006}), to capture the emission at large angular scales missed by the interferometric
data (\citealt{Caldu-Primo2016}, A.~Schruba et al. in preparation). Without this correction, 43\% of the CO flux would have been lost \citep{Caldu-Primo2016}. The data were merged using the \texttt{immerge} task from the data reduction software \texttt{miriad} (\citealt{Sault1995}),  which performs a linear combination of the low resolution and high resolution data cubes in Fourier space. The data
were merged using unit weights for the single-dish data at all spatial frequencies, which leaves the CARMA beam unchanged (A.~Schruba et al. in preparation).  The merged CARMA + IRAM data have a pixel scale size of 2" and the beam width is approximately 5.5" or 20 pc. For our analysis, we convolve the 5.5" $^{12}$CO(J=1-0) map with a Gaussian kernel of full width half maximum (FWHM) $\theta_\mathrm{kernel}$ = $\sqrt{\theta^{2}_{8"} - \theta^{2}_{5.5"}}$ = 5.8" to obtain a resulting map with a FWHM of 8" - the effective resolution of our dust observations (see Section \ref{ssec:ppmapobs}). The map is then reprojected to match the 4" pixel scale size of our dust maps in order to ensure consistent pixel-by-pixel analysis across the CO and dust data.

\cite{Caldu-Primo2016} give a 1$\sigma$ error in the molecular gas mass surface density derived from their CO observations,  on the assumption of a line width of 10 km s$^{-1}$,  of 0.83 M$_{\odot}$ pc$^{-2}$.  We have estimated the error in the molecular column density using the robust empirical technique of calculating the standard deviation in groups of pixels. This is a conservative technique since the error will also include a contribution from the variance in the distribution of the molecular gas. We find no evidence for a variation in sensitivity across the image and estimate that the 1$\sigma$ sensitivity of our CO integrated intensity map is 1.20 $\mathrm{K \; km \; s}^{-1}$. If we adopt a constant conversion factor, $X_{\mathrm{CO}}$ = 1.9 $\times$ 10$^{20}$ cm$^{-2}$ [K km s$^{-1}$]$^{-1}$ (\citealt{Strong1996}), this corresponds to a molecular gas mass surface density of 3.63 M$_{\odot}$ pc$^{-2}$. We do not account for helium in our molecular gas mass surface density ($\Sigma_{\mathrm{H_2}}$) calculations. The molecular gas mass surface density including helium {\magi and heavy elements} can be calculated using $\Sigma_{\mathrm{gas, mol}} = 1.36 \times \Sigma_{\mathrm{H_2}}$, {\magi as accounting for these elements require a roughly 36\% correction (\citealt{Bolatto2013}).}

\subsection{PPMAP algorithm applied to Herschel observations}
\label{ssec:ppmapobs}
\cite{Smith2012} have produced dust temperature ($T_{\mathrm{dust}}$), dust emissivity index ($\beta$) and dust mass surface density ($\Sigma_{\mathrm{dust}}$) maps of M31 by applying {\magi single-temperature} spectral energy distribution (SED) fitting to continuum emission observations made by \textit{Herschel} centred at wavelengths of 70, 100, 160 and 250, 350, 500 $\mu$m. In their work, \cite{Smith2012} convolved all \textit{Herschel} images to the resolution of the lowest resolution image and made the assumption that all dust was at a single temperature along the line of sight.

Since then, \cite{Marsh2015} have created a way of increasing the spatial resolution of maps of the dust properties through a procedure called point process mapping (PPMAP). When applied to the \textit{Herschel} data, PPMAP provides data products with an angular resolution of 8",  roughly corresponding to the resolution of the shortest \textit{Herschel} wavelength (70 $\mu$m). By removing the requirement
to convolve the images to \textit{Herschel}'s lowest angular resolution of 36" (at 500 $\mu$m), PPMAP allows us to probe spatial scales of $\approx$ 30 pc.

PPMAP discards the assumption that the $T_{\mathrm{dust}}$ and $\beta$ along the line of sight is uniform, with the only key assumption being that dust emission is in local thermal equilibrium and is optically thin. Through the application of PPMAP to \textit{Herschel} observations, \cite{Whitworth2019} have produced $T_{\mathrm{dust}}$, $\beta$ and $\Sigma_{\mathrm{dust}}$ maps of M31 (see their Fig. 2 (b) and (c)) which we use in our analysis in this chapter.  PPMAP produces estimates of the dust mass surface density in each pixel at a set of discrete dust temperatures and emissivity values. In creating their dust maps, \cite{Whitworth2019} have used twelve values of dust temperature ($T_{\mathrm{dust}}$ = 10.0 K, 11.6 K, 13.4 K, 15.5 K, 18.0 K, 20.8 K, 24.1 K, 27.8 K, 32.2 K, 37.3 K, 43.2 K, 50.0 K) and four values of dust emissivity index ($\beta$ = 1.5, 2.0, 2.5 and 3.0). We have {\magi summed} over these intervals to create a map of $\Sigma_{\mathrm{dust}}$ and maps of the mass-weighted $T_{\mathrm{dust}}$ and $\beta$ values.  We choose to perform our analysis on the average $T_{\mathrm{dust}}$, $\beta$ and $\Sigma_{\mathrm{dust}}$ maps because these are what we need for comparison with the work of \cite{Smith2012} and \cite{Draine2014}. This is also a more conservative approach than using the data in each PPMAP slice as, although \cite{Marsh2015} have tested the PPMAP slice results using synthetic observations of the Milky Way, there have been no such tests done using synthetic observations on extragalactic scales.  Further details about the PPMAP algorithm can be found in the work of \cite{Marsh2015}, with the full method for creating the M31 maps described in \cite{Whitworth2019}. We have removed a scaling factor from the $\Sigma_{\mathrm{dust}}$ map for our analysis\footnote{PPMAP produces a map of the total mass surface density of interstellar matter (gas + dust) using a dust mass opacity coefficient of $\kappa_{300, \; \mathrm{PPMAP}} = 0.010 \; \mathrm{m^2 \; kg^{-1}}$. Here we convert the PPMAP results to the surface density of dust alone using $\kappa_{350}$ = 0.192 m$^2$ kg$^{-1}$ (\citealt{Draine2003}) and a gas-to-dust ratio of 100.}.

The minimum value in our map of dust mass surface density is $\Sigma_{\mathrm{dust}} \simeq 0.05$  M$_{\odot}$ pc$^{-2}$, which corresponds roughly to a 5$\sigma$ detection.  We focus on the CARMA-observed region within the PPMAP dust maps (see Figure \ref{fig:carmareg}) for ease of comparison of dust properties with observations of $^{12}$CO(J=1-0). The rise in $\beta$ with galactocentric radius in the central 3 kpc of M31 followed by a fall in $\beta$ beyond this radius, first seen by \cite{Smith2012}, has been confirmed by \cite{Whitworth2019} in their reanalysis of the original \textit{Herschel} data using PPMAP.

\subsection{\textsc{HI} ancillary data}
\label{ssec:HIobs}
We use the \textsc{HI}  column density map of M31 obtained by \cite{Braun2009} using the Westerbork Synthesis Radio Telescope (WSRT) at an angular resolution of 30" and spatial resolution of $\approx$ 110 pc. The observed \textsc{HI} in M31 has a smooth distribution with a `clumping factor' of $\approx$ 1.3  (\citealt{Leroy2013}). Therefore, although these data do not match the resolution of our CO and dust maps, the \textsc{HI} map provides an estimate of the contribution of
the smooth HI distribution at the position of a molecular cloud. We emphasise that our estimate of the \textsc{HI} contribution at the scale of an individual cloud is at best a rough estimate. The \textsc{HI} map has not been corrected for opacity effects as the best method of doing this is still uncertain. Localised opacity corrections can lead to an increase of the inferred \textsc{HI} gas mass by $\approx$ 30\% or more (e.g. \citealt{Braun2009}, \citealt{Koch2021}), and so this is an additional uncertainty in our estimates of the contribution of HI at the position of a cloud. The data have a pixel scale size of 10" which we reproject into 4" $\times$ 4" pixels to match the PPMAP map projection. We focus on the CARMA-observed region within this reprojected map. The 1$\sigma$ sensitivity of the \textsc{HI} column density map is 4.05 $\times$ 10$^{19}$ cm$^{-2}$, corresponding to an atomic gas mass surface density value of 0.32 M$_{\odot}$ pc$^{-2}$. We do not account for helium in our atomic gas mass surface density ($\Sigma_{\mathrm{\textsc{HI}}}$) calculations.

\subsection{Astrometric offsets}
We estimate the astrometric accuracy of the instruments as FWHM of the beam divided by the signal-to-noise ratio of the pointing source observations. Typical pointing calibrator observations reach a signal-to-noise level of $\approx$ 10. Therefore, we estimate that the CARMA object positions are accurate to within 0.5”. The IRAM astrometric offset should not be more than 2.3”. We expect that the positions of objects from the raw \textit{Herschel} maps are accurate to 2”\footnote{SPIRE astrometry-corrected maps readme: \url{http://archives.esac.esa.int/hsa/legacy/HPDP/SPIRE/SPIRE-P/ASTROMETRY/README.html}}. We do not think these astrometric offsets are likely to be significant because the smallest area of a cloud in our source extraction has been taken as 10 pixels (see Table \ref{tab:param}),  making a cloud {\magi bigger} than the beam of either the CO map and the effective beam of the dust map.

\section{Source extraction}
\label{sec:se}
\subsection{Dendrogram}

\begin{table*}
\caption{Input dendrogram parameters for source extraction.}              
\label{tab:param}      
\centering                                      
\begin{tabular}{c c c c}          
\hline\hline                        
Data & Minimum value  & Minimum structure significance value & Minimum no. of pixels \\    
& \texttt{min\_value} & \texttt{min\_delta} & \texttt{min\_npix} \\
\hline                                   
     CO & 4.57 K\,km s$^{-1}$ & 2.40 K\,km s$^{-1}$ & 10 \\      
     Dust & 0.44 M$_{\odot}$ pc$^{-2}$ & 0.296 M$_{\odot}$ pc$^{-2}$ & 10 \\
\hline                                             
\end{tabular}
\end{table*}

A common method for accessing the hierarchical structure of molecular clouds is to use a dendrogram. A dendrogram allows us to segregate the denser regions from the more diffuse regions and access any nested sub-structure. We compute dendrograms for the CO and PPMAP maps using the Python package \texttt{astrodendro 0.2.0} (\citealt{Rosolowsky2008}). This package allows us to construct an empirically motivated segmentation of `clouds' within our data.  If we adopt the analogy within the \texttt{astrodendro} documentation\footnote{\texttt{astrodendro} documentation: \\ \url{https://dendrograms.readthedocs.io/en/stable/}},  a dendrogram can be represented as a "tree" with a trunk, and the nested structures of this trunk are called "branches" which contain sub-structures in the form of "leaves".  The resulting sources (molecular clouds) extracted by our computed dendrograms are analogous to leaves on the tree.

The dendrogram computation requires the specification of three parameters:
\begin{enumerate}
  \item The minimum intensity value of a pixel (\texttt{min\_value}): the dendrogram will discard any pixels fainter than this threshold.
  \item The minimum significance value for a leaf (nested structure) to be identified as an independent object (\texttt{min\_delta}): if the difference between a new local maximum pixel value (peak of prospective structure) and the last pixel value examined in an existing structure (the point at which the new structure may be merged onto the existing one) is greater than this parameter, a structure is considered to be significant enough to be independent.
  \item The minimum size (defined in number of pixels) required to identify a structure as an independent object (\texttt{min\_npix}): if the number of pixels in a structure does not match or exceed this value,  the structure is merged with an existing structure,  {\magi or discarded if it has no parent structure}.
\end{enumerate}

For our work, we select these parameter values carefully as described in Sections \ref{ssec:gmc_co} and \ref{ssec:gmc_ppmap}. Further details about the dendrogram algorithm can be found in the documentation of the \texttt{astrodendro} package and in the work of \cite{Rosolowsky2008}.

\subsection{Identifying molecular clouds with CO}
\label{ssec:gmc_co}
We choose to run the dendrogram on the CO integrated intensity map rather than trying to find clouds in the original CO data cube. Our reasoning for this is as follows: firstly, although the CO linewidth information would allow us to distinguish whether a cloud found by the dendrogram is a single source or multiple clouds along the line-of-sight, knowledge of the 3D structure of the gas does not provide any benefit in understanding the cause of the spatial variation in the dust emissivity index as we do not have access to matching information on our dust emissivity index map. Moreover, the CO method could still suffer from the problem that approximately one-third of the molecular gas may not contain any CO (see {\magi Section \ref{sec:intro_beta}}) and so would not give us a perfect benchmark catalogue of clouds to compare all other catalogues with. In order to make as direct a comparison as possible of the two tracers (CO and dust) for finding clouds, we compute the dendrogram on the CO integrated intensity map.

The input parameters used for our source extraction are listed in Table \ref{tab:param}. To calculate the 1$\sigma$ noise threshold, we select a region within our smoothed and reprojected map where we see no obvious sources and calculate the standard deviation ($\sigma_{\mathrm{std}}$) of pixel values. Our minimum detection threshold (\texttt{min\_value}) is 3$\sigma$, which we add to a mean background level of 0.97 $\mathrm{K \; km \; s}^{-1}$ to account for a diffuse constant level of CO present beneath denser structures. For our minimum structure significance threshold (\texttt{min\_delta}), we choose a value of 2$\sigma$. This means that clouds will have a peak CO intensity of at least 5$\sigma$ above the mean background level. The final key requirement for our dendrogram extraction is that an independent structure should have a minimum size (\texttt{min\_npix}) of 10 pixels. This size threshold is larger than the beam area (calculated using the diameter as the FWHM = 8") divided by the area of one pixel.

Clouds are defined as the objects at the highest level of the dendrogram hierarchy (i.e. dendrogram `leaves'), containing no nested sub-structures. We find 140 sources from our CO observations (see Figure \ref{fig:carma_cons_sources}) which we assume are molecular clouds traced by CO.

\subsubsection{Determining the gas and dust properties of clouds}
\label{sssec:gdpropmethod}
The molecular gas mass surface density in a pixel,  $\Sigma_{\mathrm{H_2}}$, is related to the CO intensity in the pixel, $I_{\mathrm{CO}}$,  by:
\begin{equation}
\label{eq:2}
\Sigma_\mathrm{H_2} = I_\mathrm{CO} \times X_\mathrm{CO} \times m(\mathrm{H_2})
\end{equation}
where $m(\mathrm{H_2})$ is the mass of a hydrogen molecule. We adopt a constant $X_{\mathrm{CO}}$ = 1.9 $\times$ 10$^{20}$ cm$^{-2}$ [K km s$^{-1}$]$^{-1}$ from \cite{Strong1996}. We calculate the total CO-traced molecular gas mass in each cloud by multiplying the gas mass surface density values by the area of a pixel and summing over all gas mass values in the dendrogram leaf.

We also calculate the dust-mass-weighted mean $T_{\mathrm{dust}}$ and mean $\beta$ for our clouds using the pixels corresponding to our dendrogram leaves in these maps.  We calculate the total dust mass in a cloud by multiplying the $\Sigma_{\mathrm{dust}}$ values from the PPMAP map by the area of a pixel and summing over all dust mass values in the dendrogram leaf. The CO-traced molecular gas-to-dust ratio (GDR) is obtained for each cloud by:
\begin{equation}
  \label{eq:3}
    \mathrm{ CO\mbox{-}traced \; molecular \; GDR} = \frac{M_{\mathrm{H_2}}}{M_{\mathrm{dust}}}
\end{equation}
where $M_{\mathrm{H_2}}$ is the total CO-traced molecular gas mass in the cloud, and $M_{\mathrm{dust}}$ is the total mass of dust in the cloud.

The gas mass surface density of atomic hydrogen in a pixel is given by:
\begin{equation}
\label{eq:4}
  \Sigma_{\mathrm{\textsc{HI}}} = N(\mathrm{H}) \times m(\mathrm{H})
\end{equation}
where $N$(H) is the column density of atomic hydrogen obtained from the \textsc{HI} map and $m(\mathrm{H})$ is the mass of a hydrogen atom. The total atomic gas mass in the cloud, $M_{\mathrm{HI}}$, is calculated by multiplying $\Sigma_{\mathrm{\textsc{HI}}}$ with the area of a pixel and summing over all atomic gas mass values in the dendrogram leaf. The total (molecular + atomic) GDR is obtained by:
\begin{equation}
  \label{eq:5}
  \mathrm{ Total \; GDR} = \frac{(M_{\mathrm{\textsc{HI}}} + M_{\mathrm{H_2}})}{M_{\mathrm{dust}}}
\end{equation}

To propagate the error in molecular GDR values, we make the approximation that the noise from the CO map dominates over error contributions from PPMAP dust measurements. To propagate the error in total GDR values, we add the noise from both the CO and \textsc{HI} maps in quadrature. We neglect systematic errors from PPMAP. The error in the CO intensity and \textsc{HI} column density within each cloud are calculated using $\mathrm{N}_{\mathrm{pix, beam}} \times  \sqrt{\mathrm{N}_{\mathrm{pix, cloud}}/\mathrm{N}_{\mathrm{pix, beam}}} \times \sigma_{\mathrm{std}}$, where $\sigma_{\mathrm{std}}$ is the standard deviation of pixel values within a region of each map where we see no obvious sources. $\mathrm{N}_{\mathrm{pix, beam}}$ is number of pixels in the beam and $\mathrm{N}_{\mathrm{pix, cloud}}$ is the number of pixels in a cloud.

We calculate the radial distance of a cloud from the centre of M31 (RA: 00$^h$ 42$^m$ 44.33$^s$, Dec: 41$^{\circ}$ 16' 7.5" (\citealt{Skrutskie2006})) assuming that the pixel with the peak intensity marks the centre of the cloud. 

\subsubsection{Determining dust properties in non-dense regions}
We compare the values of $\beta$ and $T_{\mathrm{dust}}$ in the pixels within our clouds with the values of the pixels that fall outside the clouds (non-dense regions). We identify all the pixels which fall within the non-dense regions by masking out pixels within our dendrogram leaves. We split the pixels into two radial bins (5-7.5 kpc and 9-15 kpc) and create histograms of $T_{\mathrm{dust}}$ and $\beta$ values at these radii, both inside and outside dense regions (see Section \ref{ssec:radvarres}).

\subsection{Identifying molecular clouds with dust}
\label{ssec:gmc_ppmap}

\begin{figure}
\centering
\hspace{-2cm}
\includegraphics[width=12.25cm]{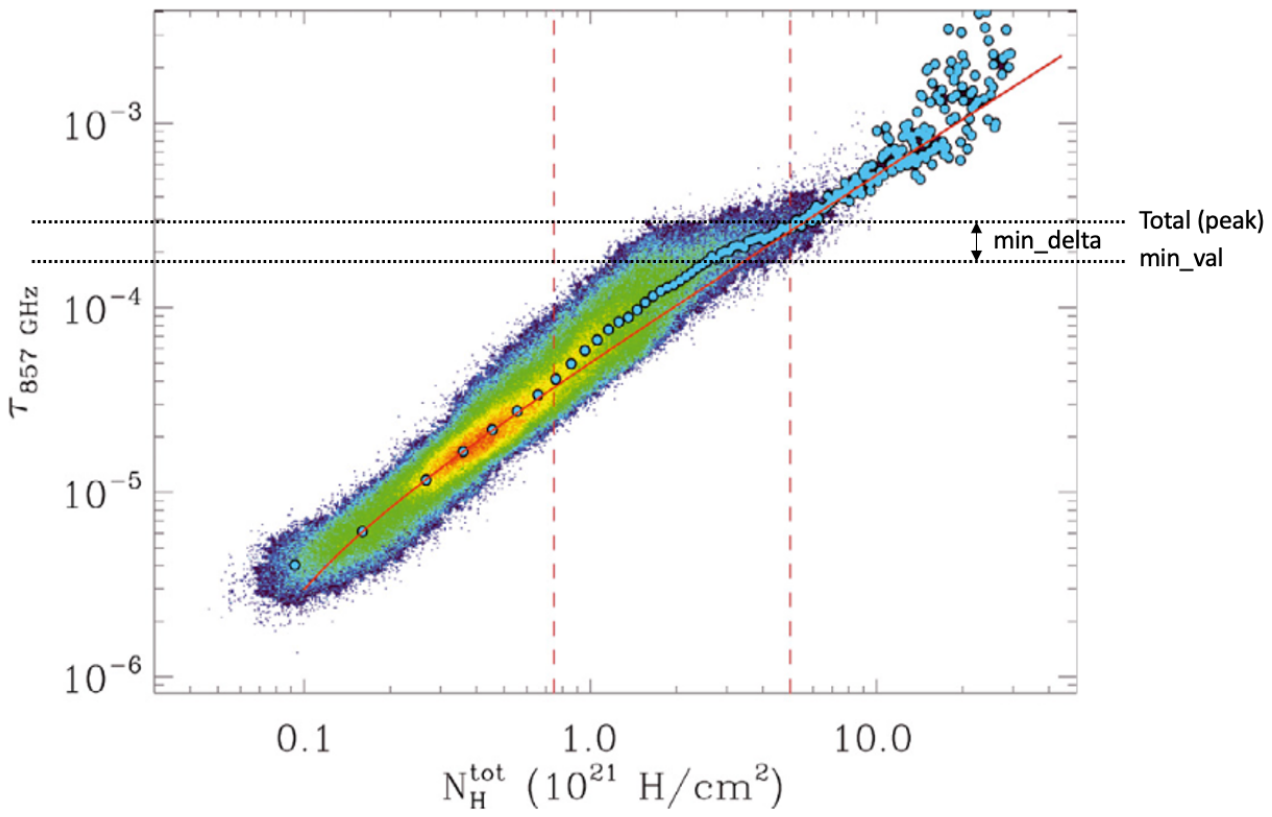}
\vspace{-1.75cm} \caption{Adaptation of Fig. 6 from \citet{PlanckCollaboration2011} showing the correlation between the dust optical depth $\tau$ at 857 GHz and the total gas column density. The dashed vertical red line at $\mathrm{N}_{\mathrm{H}}^{\mathrm{tot}} \approx 8.0 \times 10^{20}$ H cm $^{-2}$ shows the threshold above which excess thermal emission from dust traces CO-dark molecular gas. The dashed vertical red line at N$_{\mathrm{H}}^{\mathrm{tot}} \approx 5.0 \times 10^{21}$ H cm$^{-2}$ shows the threshold at which the gas column density becomes dominated by the molecular gas traced by CO emission. The lower black dotted horizontal line shows the $\tau_{857 \; \mathrm{GHz}}$ value used to calculate the \texttt{min\_value} dendrogram parameter for our $\Sigma_{\mathrm{dust}}$ map. The difference between the black dotted lines shows the $\tau_{857 \; \mathrm{GHz}}$ value used to calculate the \texttt{min\_delta} dendrogram parameter. All of our clouds have a peak dust mass surface density of $\Sigma_{\mathrm{dust}} \geq 0.74 \; \mathrm{M}_{\odot}$ $\mathrm{pc}^{-2}$ (Equation \ref{eq:chap2_6}), equivalent to a peak optical depth of $\tau_{857 \; \mathrm{GHz}} \geq 3.0 \; \times 10^{-4}$ (upper black dotted line).  \copyright{\cite{PlanckCollaboration2011}.  Reproduced with permission from ESO.}}
\label{fig:planck_cartoon}
\end{figure}

\begin{figure*}
\centering
\includegraphics[width=15cm]{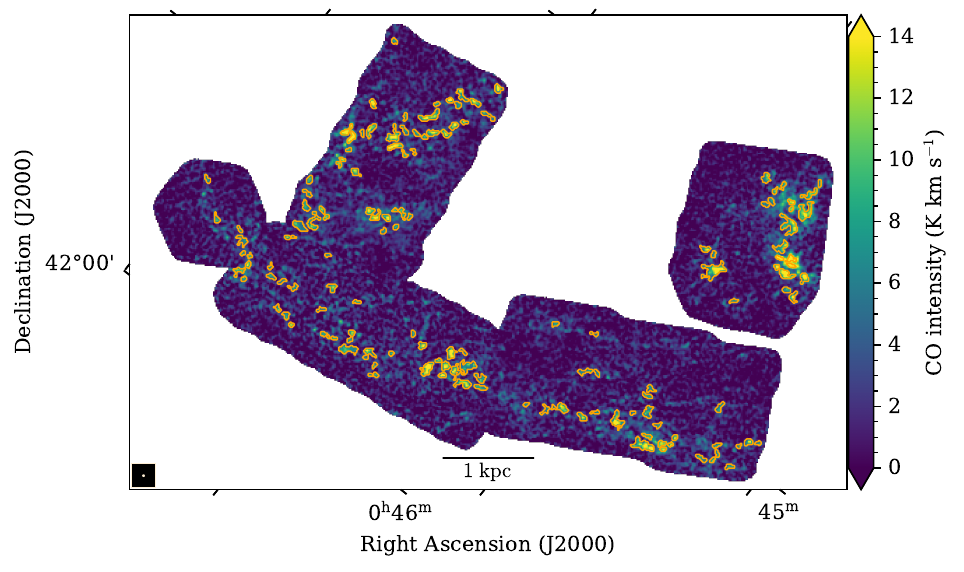}
\caption{Map of $^{12}$CO(J=1-0) intensity from CARMA + IRAM, convolved to 8" resolution and reprojected to 4" pixel size. Orange contours show the 140 sources extracted using a dendrogram.}
\smallskip
\label{fig:carma_cons_sources}
\end{figure*}

\begin{figure*}
\centering
\includegraphics[width=15cm]{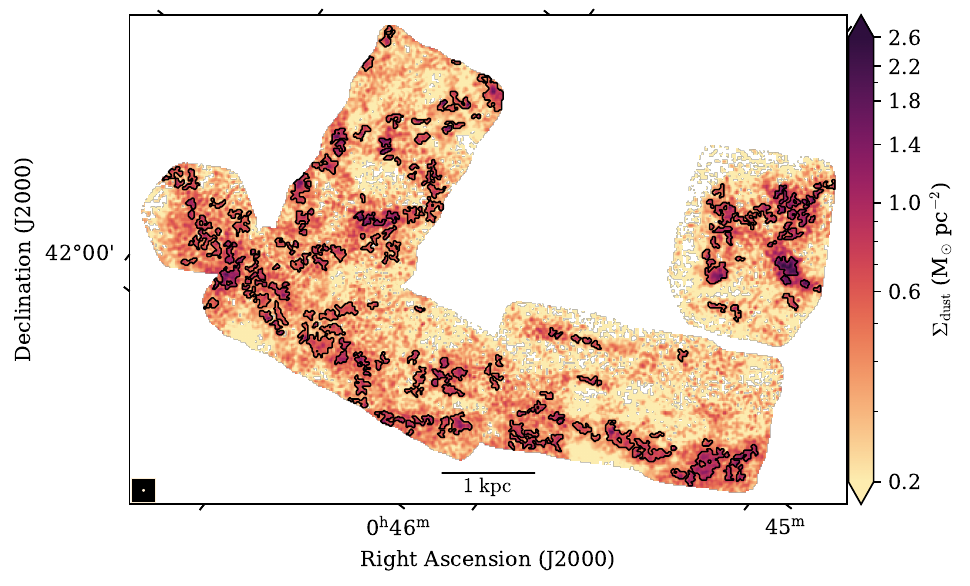}
\caption{Map of dust mass surface density produced by PPMAP at 8" resolution and 4" pixel size. Black contours show the 196 sources extracted using a dendrogram.}
\smallskip
\label{fig:dust_cons_sources}
\end{figure*}

We also apply the dendrogram algorithm to our dust mass surface density map and identify sources (dendrogram `leaves') which we assume are molecular clouds traced by dust. Our motivation for finding dust-selected clouds stems from the existence of molecular gas in the MW which is not traced using the traditional CO method (see {\magi Section \ref{sec:intro_beta}}).

Our dendrogram parameter selection is based on the work of the \textit{Planck} team (\citealt{PlanckCollaboration2011}) and allows us to extract a similar number of clouds as in our CO-selected catalogue. Figure \ref{fig:planck_cartoon} is an adaptation of Fig. 6 from \cite{PlanckCollaboration2011} showing the correlation between the optical depth of dust at a frequency of 857 GHz ($\tau_{857 \; \mathrm{GHz}}$) vs the combined atomic and molecular gas column density (N$_{\mathrm{H}}^{\mathrm{tot}}$) in the MW. The colour scale shows the number density of pixels on a logarithmic scale. At  N$_{\mathrm{H}}^{\mathrm{tot}}$ below $0.8 \times 10^{21}$ H cm$^{-2}$, the observed correlation between $\tau_{857 \; \mathrm{GHz}}$ and N$_{\mathrm{H}}^{\mathrm{tot}}$ (blue circles) follows a linear relationship in linear space (modelled by the red line). Above N$_{\mathrm{H}}^{\mathrm{tot}}$ $\approx$ 5 $\times$ 10$^{21}$ H cm$^{-2}$, N$_{\mathrm{H}}^{\mathrm{tot}}$ is dominated by contributions from CO, and $\tau_{857 \; \mathrm{GHz}}$ is visibly consistent with the observed linear correlation. An excess of $\tau_{857 \; \mathrm{GHz}}$ is seen between these two column density values (marked by the vertical red dashed lines) as the blue circles deviate from the linear relationship. This deviation from the modelled linear relationship is attributed by the \textit{Planck} team to CO-dark molecular gas.

We select the optical depth of dust $\tau_{857 \; \mathrm{GHz}} = 3.0 \times 10^{-4}$ (see highest black dotted line in Figure \ref{fig:planck_cartoon}) to derive our input dendrogram parameters because it roughly corresponds to the upper limit of the deviation seen between the blue circles and the red line. This allows us to select a  \texttt{min\_value} below this (see lowest black dotted line in Figure \ref{fig:planck_cartoon}) and possibly pick up regions of CO-dark molecular gas. A frequency of 857 GHz roughly corresponds to a wavelength of 350 $\mu$m. We substitute our chosen $\tau_{857 \; \mathrm{GHz}} = 3.0 \; \times  10^{-4}$ and the dust mass absorption coefficient, $\kappa_{350, \; \mathrm{Draine}}$ = 0.192 m$^2$ kg$^{-1}$ (from \citealt{Draine2003}), into Equation \ref{eq:chap2_6} to find the corresponding $\Sigma_{\mathrm{dust}}$:
 \begin{equation}
 \label{eq:chap2_6}
 \begin{split}
 \Sigma_{\mathrm{dust}} = \frac{\tau_{857 \; \mathrm{GHz}}}{\kappa_{\mathrm{350, \; Draine}}} = \frac{3.0 \times 10^{-4}}{0.192 \; \mathrm{m^2} \; \mathrm{kg^{-1}}}  = 0.74 \; \mathrm{M_{\odot}} \; \mathrm{pc^{-2}}
 \end{split}
 \end{equation}

For consistency with the 3:2 ratio of \texttt{min\_value} : \texttt{min\_delta} used to create our CO-selected catalogue, we split this $\Sigma_{\mathrm{dust}} = 0.74 \; \mathrm{M_{\odot}} \; \mathrm{pc}^{-2}$ value using the same ratio and obtain the \texttt{min\_value} and \texttt{min\_delta} for our dust-selected source extraction. The input dendrogram parameter values are listed in Table \ref{tab:param}.  Using these parameters, we find 196 clouds. Figure \ref{fig:dust_cons_sources} shows our dust-selected clouds. The spatial resolution of the \textit{Planck} observations ($\approx$ 0.3 pc; \citealt {PlanckCollaboration2011}) is, of course, much finer than our resolution. Therefore, our observations will not be as sensitive to CO-dark molecular gas as the \textit{Planck} observations of the MW.

We calculate the total CO-traced molecular gas mass within our dust-selected clouds by finding the corresponding pixel locations in the CO map, and following the methodology described in Section \ref{ssec:gmc_co}. We also calculate the total dust mass of each cloud, CO-traced molecular GDR and total GDR as described in Section \ref{ssec:gmc_co}.

\section{Results}
\label{sec:res}
\subsection{Properties of extracted clouds}
\label{sec:cloud_mass}
We obtain two molecular cloud catalogues: one of clouds traced by CO and one of clouds traced by dust (Appendix \ref{ap:cloud_catscarmappmap}). In this section, we provide our analysis of the cloud properties.

\begin{figure}
\centering
\includegraphics[width=12cm]{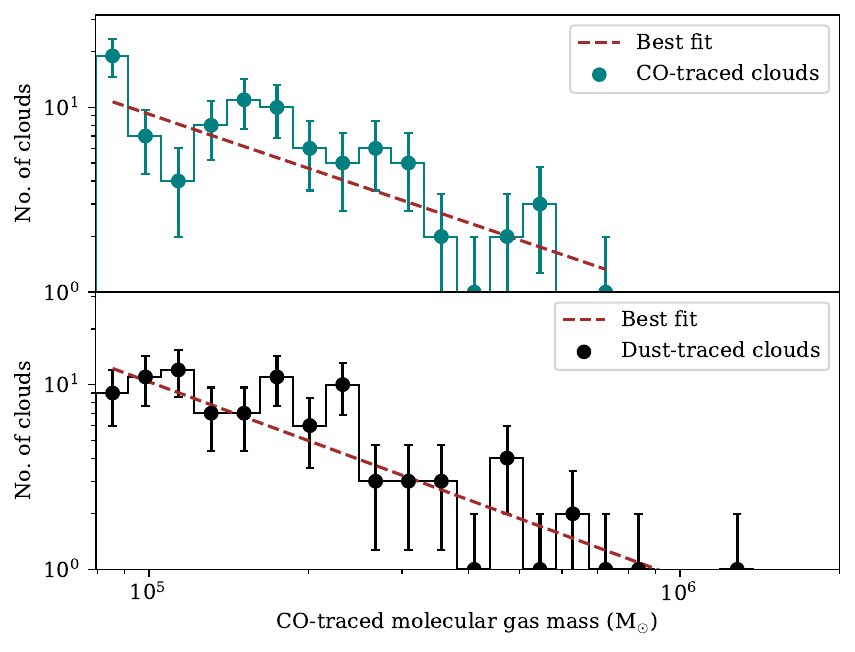} \caption{Cloud mass functions for CO-selected (teal) and dust-selected (black) catalogues, for cloud masses greater than 10$^{4.9}$ M$_{\odot}$. The histograms show the number of clouds per mass interval. The scatter points show the central mass value in each bin. The brown dashed line shows the best fit power law to the cloud mass function.}
\label{fig:cmf_all}
\end{figure}

Figure \ref{fig:cmf_all} shows the distribution of molecular cloud masses for our CO-selected and dust-selected catalogues. We fit a power law model to our cloud masses using the \texttt{SciPy} Levenberg-Marquardt least-squares minimisation package \texttt{lmfit}. We fit a power law of the form:
\begin{equation}
\label{eq:7}
  \mathrm{dN}_{\mathrm{cloud}} = \mathrm{k}M_{\mathrm{H_2}}^{-\alpha} \; dM_{\mathrm{H_2}}
\end{equation}
\noindent where dN$_{\mathrm{cloud}}$ is the number of molecular clouds in the mass interval ($M_{\mathrm{H_2}}$, $M_{\mathrm{H_2}}$ + $dM_{\mathrm{H_2}}$),  k is a normalisation factor, $M_{\mathrm{H_2}}$ is the total CO-traced molecular gas mass of the cloud, and $\alpha$ is the power law exponent.

Since the CO map is used to calculate the {\magi CO-bright} molecular gas mass of all dust-selected clouds,  only the clouds with at least a 3$\sigma$ detection in the CO map have been included in our fits.  177 of 196 dust-traced clouds have a CO detection above this threshold. The molecular gas masses of the CO-selected clouds range from 3.9 $\times$ 10$^4$ M$_{\odot}$ $\le$ M$_{\mathrm{cloud}}$ $\le$ 7.1 $\times$ 10$^5$ M$_{\odot}$. The molecular gas masses of the dust-selected clouds with a 3$\sigma$ CO detection range from 1.8 $\times$ 10$^4$ M$_{\odot}$ $\le$ M$_{\mathrm{cloud}}$ $\le$ 1.3 $\times$ 10$^6$ M$_{\odot}$.
We only include clouds with a molecular gas mass $>10^{4.9}\ M_{\odot}= 7.9 \times 10^{4}\ M_{\odot}$ in our fitting because the decrease in the number of clouds at lower masses suggests that our mass functions are increasingly incomplete at lower masses. We bin data from both catalogues into 22 bins which are equidistant in logarithimic space between the mass limits 10$^{4.9}$ M$_{\odot}$ and 10$^{6.2}$ M$_{\odot}$. Our fits are performed in linear space and we assume Poisson errors.

\begin{table}[h]
\caption{Cloud mass function best fit results and reduced $\chi^2$.}              
\label{tab:cmf_params}      
\centering                                      
\begin{tabular}{c c c c}         
\hline\hline                       
Data & Best fit $\alpha$ exponent & Reduced $\chi^2$ \\    
\hline                                  
     CO & 1.98 $\pm$ 0.24 & 1.38   \\      
     Dust & 2.06 $\pm$ 0.14 & 0.74   \\
\hline                                             
\end{tabular}
\end{table}

The best fit $\alpha$ values for clouds from each catalogue and the reduced $\chi^2$ parameter for our fits are given in Table \ref{tab:cmf_params}.
The reduced $\chi^2$ parameter for both catalogues is close to one, indicative of a good fit. The slope values of $\approx$ 2 are similar to measurements of the slope of the cloud mass function in the MW and nearby galaxies (e.g. \citealt{Rice2016}, \citealt{Rosolowsky2021}). They are also similar to the value for M31 by \cite{Kirk2015}: $\alpha$ = 2.34 $\pm$ 0.12.

Table \ref{tab:cprops} shows the 16th, 50th and 84th percentiles of cloud properties for both the CO-selected and the dust-selected clouds.

\begin{table*}
\caption{Statistical properties of clouds extracted from the $^{12}$CO(J=1-0) map and dust mass surface density map. Values have been rounded to 1 decimal place.}
\label{tab:cprops}      
\centering                                      
\begin{tabular}{l l l}          
\hline\hline                  
Properties & CO-selected & Dust-selected \\
\hline
No. of sources extracted & 140 & 196 \\
& & \\
50$^{84\mathrm{th}}_{16\mathrm{th}}$ percentile total CO-traced molecular gas mass in cloud & 9.4$^{23.0}_{6.0}$ $\times$ 10$^{4}$ M$_{\odot}$ & 7.1$^{20.6}_{2.6}$ $\times$ 10$^{4}$ M$_{\odot}$ \\
& & \\
50$^{84\mathrm{th}}_{16\mathrm{th}}$ percentile total dust mass in cloud & 2.8$^{7.0}_{1.8}$ $\times$ 10$^{3}$ M$_{\odot}$ & 5.5$^{10.4}_{2.4}$ $\times$ 10$^{3}$ M$_{\odot}$ \\
& & \\
50$^{84\mathrm{th}}_{16\mathrm{th}}$ percentile density weighted average $T_{\mathrm{dust}}$ in cloud & 15.6$^{17.1}_{14.3}$ K & 14.6$^{15.9}_{13.8}$ K \\
& & \\
50$^{84\mathrm{th}}_{16\mathrm{th}}$ percentile density weighted average $\beta$ in cloud & 2.1$^{2.3}_{2.0}$ & 2.2$^{2.5}_{2.0}$ \\
& & \\
50$^{84\mathrm{th}}_{16\mathrm{th}}$ percentile molecular GDR of cloud & 34.5$^{48.3}_{25.8}$ & 14.1$^{25.8}_{7.2}$ \\
& & \\
50$^{84\mathrm{th}}_{16\mathrm{th}}$ percentile total GDR of cloud & 80.0$^{107.6}_{54.5}$ & 55.1$^{70.1}_{34.9}$ \\
& & \\
50$^{84\mathrm{th}}_{16\mathrm{th}}$ percentile equivalent radius of cloud & 36.9$^{50.8}_{29.7}$ pc & 47.8$^{68.7}_{33.3}$ pc \\
& & \\
\hline                                   
\end{tabular}
\end{table*}

\subsection{Radial variations in dust properties of clouds traced by CO}
\label{ssec:radvarres}
\begin{figure*}
\centering
\includegraphics[width=16cm]{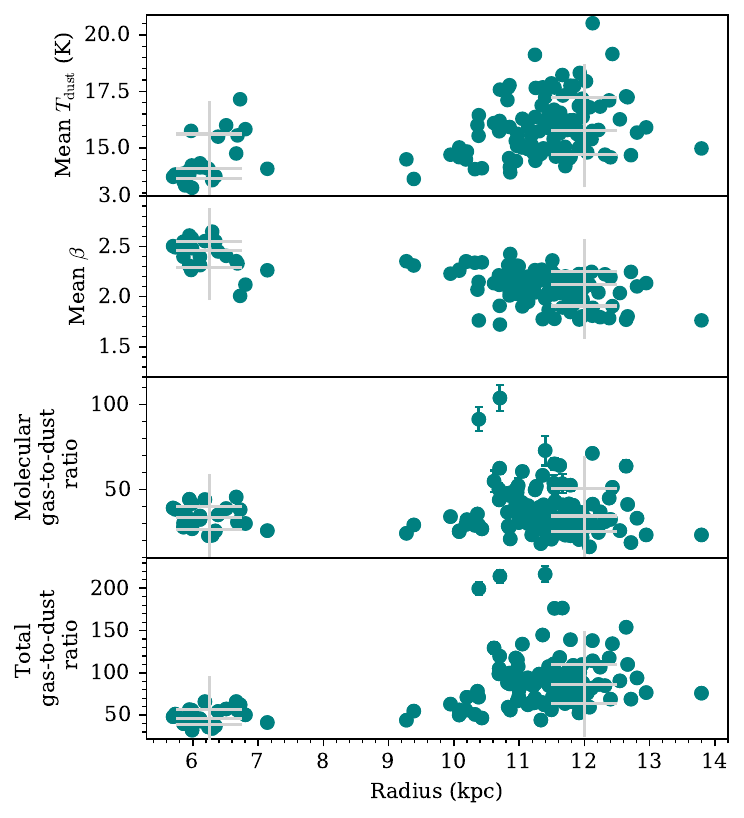} \caption{Radial distribution of gas and dust properties of clouds from the CO-selected catalogue. Each scatter point represents a dendrogram leaf (molecular cloud). The light grey crosses show the 16th, 50th and 84th percentiles of each cloud property in the 5-7.5 kpc radial bin and the 9-15 kpc radial bin. \textit{First row:} Radial distribution of mean $T_{\mathrm{dust}}$ in each cloud. \textit{Second row:} Radial distribution of mean $\beta$ in each cloud. \textit{Third row:} Radial distribution of CO-traced molecular GDR of each cloud. \textit{Fourth row:} Radial distribution of total (\textsc{HI} + H$_2$) GDR of each cloud. The GDR errorbars have been calculated using the method described in Section \ref{sssec:gdpropmethod}.}
\label{fig:properties}
\end{figure*}

Figure \ref{fig:properties} shows the mean $T_{\mathrm{dust}}$, mean $\beta$, CO-traced molecular GDR, and the total GDR as a function of radius for clouds from the CO-selected catalogue. We find that our extracted clouds are located within two
ranges of radius: between 5-7.5 kpc and between 9-15 kpc, reflecting the
inner ring at $\approx$ 5 kpc and the dusty star-forming ring at $\approx$ 10 kpc.

\begin{landscape}
\begin{figure*}
\centering
\includegraphics[width=23cm]{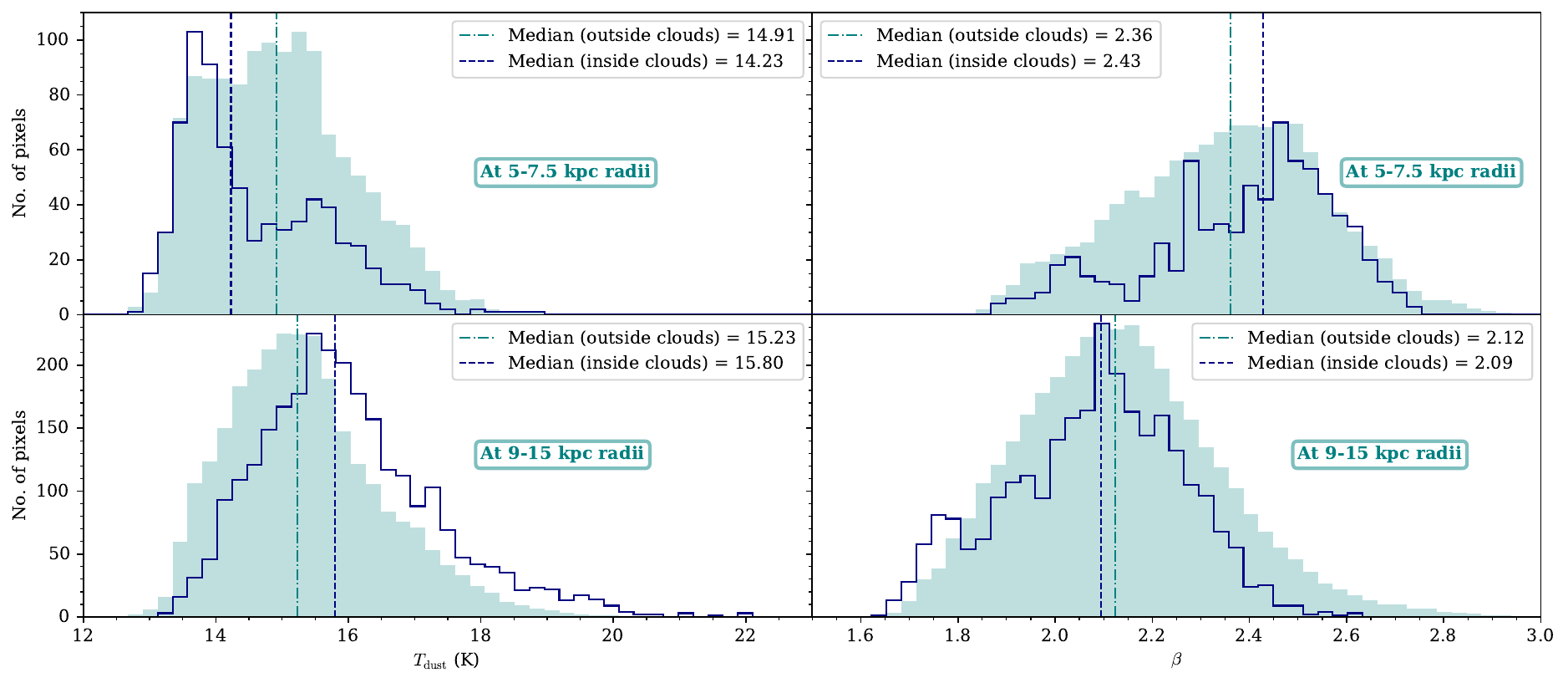} \caption{Distribution of $T_{\mathrm{dust}}$ and $\beta$ inside (navy) and outside molecular clouds (shaded teal) from the CO-selected catalogue. \textit{Top left:} $T_{\mathrm{dust}}$ distribution for pixels in the radius range 5-7.5 kpc. \textit{Top right:} $\beta$ distribution for pixels in the radius range 5-7.5 kpc. \textit{Bottom left:} $T_{\mathrm{dust}}$ distribution for pixels in the radius range 9-15 kpc. \textit{Bottom right:} $\beta$ distribution for pixels in the radius range 9-15 kpc. In all four plots, the histogram of pixels outside clouds has been rescaled to have the same height as the histogram of pixels inside clouds. The navy dashed vertical line shows the median value of $T_{\mathrm{dust}}$ and $\beta$ inside clouds. The teal dashed-dotted vertical line shows the median value outside clouds.}
\label{fig:tempbetaCARMAvar}
\end{figure*}
\end{landscape}

Figure \ref{fig:tempbetaCARMAvar} shows the mean $T_{\mathrm{dust}}$ and
mean $\beta$ in these two radial bins, both for the pixels inside and outside clouds.
The left-hand column of Figure \ref{fig:tempbetaCARMAvar} shows histograms of the $T_{\mathrm{dust}}$. We see that the median $T_{\mathrm{dust}}$ increases with galactocentric radius. The distributions of $T_{\mathrm{dust}}$ for pixels inside clouds but at different radii are significantly different (two-sided Mann-Whitney U test, p-value $\ll$ 0.01; see Table \ref{tab:mannwhitneyu} for full results), as are the distributions of $T_{\mathrm{dust}}$ for the pixels outside clouds (p-value $\ll$ 0.01). However, the temperature difference is fairly small. The average $T_{\mathrm{dust}}$ is 1.57 K higher for the clouds in the 10 kpc ring than for the clouds in the 5 kpc ring, and the average $T_{\mathrm{dust}}$ is 0.32 K higher for the pixels outside the clouds in the 10 kpc ring than for the pixels outside the clouds in the 5 kpc ring.
In both radial bins, the distributions of $T_{\mathrm{dust}}$ inside and outside molecular clouds are significantly different (p-value $\ll$ 0.01), although again the difference in temperature is actually quite small, $\simeq 0.6-0.7$ K.

The right-hand column of Figure \ref{fig:tempbetaCARMAvar} shows our most interesting result, with histograms of $\beta$ in pixels inside and outside molecular clouds, split into the two radial bins. The distributions of $\beta$ for pixels inside clouds at different radii are significantly different (two-sided Mann-Whitney U test, p-value $\ll$ 0.01; see Table \ref{tab:mannwhitneyu} for full results), as are the distributions of $\beta$ for the pixels outside clouds (p-value $\ll$ 0.01). The average $\beta$ is 0.34 lower for the clouds in the 10 kpc ring than for the clouds in the 5 kpc ring, and the average $\beta$ is 0.24 lower for the pixels outside the clouds in the 10 kpc ring than in the 5 kpc ring. Our result is consistent with the decreasing trend in $\beta$ with increased galactocentric radius (going from $\beta \approx$ 2.5 to $\beta \approx$ 1.9) discovered by \cite{Smith2012} beyond a radius of 3.1 kpc. Our result is also in agreement with the results from
other \textit{Herschel} observations of M31 by \cite{Draine2014}, which found larger $\beta$ in the central $\approx$ 7 kpc than in the outer disk (see their Fig. 13). Although \cite{Draine2014} find a shallower decrease in $\beta$ beyond $\gtrsim$ 7 kpc than \cite{Smith2012}, our result agrees with the general trend found by both studies of a radial decrease in $\beta$ between the radii of 5-15 kpc.

\newpage
\newpage

In both radial bins, the distributions of $\beta$ inside and outside molecular clouds are significantly different (p-value $\ll$ 0.01). However, the more interesting aspect of our result is that there is a much smaller difference between the average $\beta$ inside molecular clouds compared to outside molecular clouds in both radial bins. The median value of $\beta$ is only greater by 0.07 inside clouds in the 5 kpc ring and is actually less by 0.03 inside the clouds in the 10 kpc ring.  Therefore, we find no evidence for the radial variations in $\beta$ in M31 being caused by a large change in $\beta$ in regions of dense gas.  {\magi Furthermore,  the change in the shape of the distributions of $\beta$ affects both inside and outside clouds almost identically,  implying that the change in $\beta$ is most affected by the galactic position rather than being inside or outside molecular clouds. }

\subsection{The gas-to-dust ratio in clouds found using the two methods}
\label{ssec:gdrres}

\begin{landscape}
\begin{figure*}[h]
\centering
\hspace{-0.4cm}
\includegraphics[width=23cm]{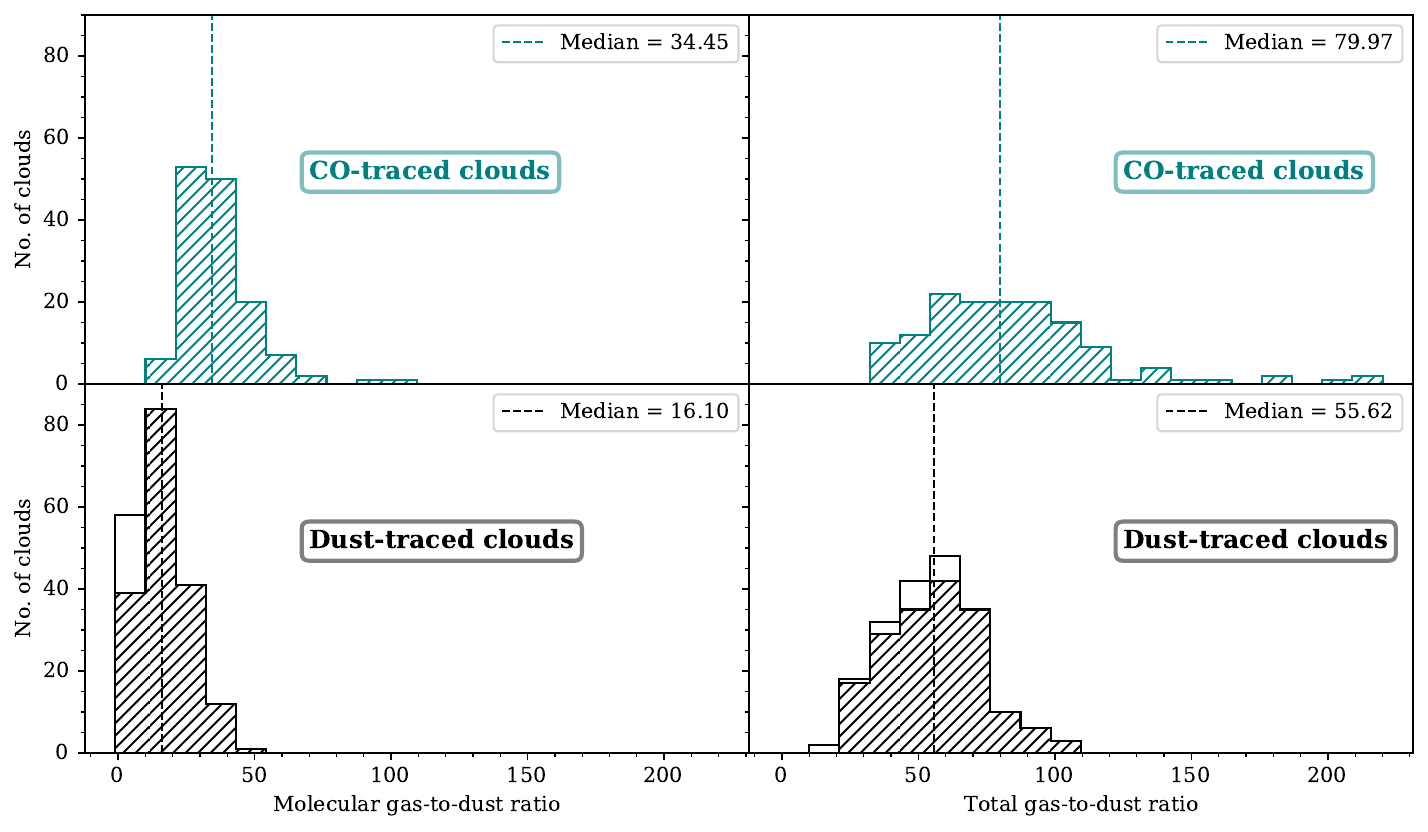} \caption{Histograms showing the number of clouds from the CO-selected and dust-selected catalogues and their CO-traced molecular GDR and total GDR. \textit{Top:} Clouds from the CO-selected catalogue. \textit{Bottom:} Clouds from the dust-selected catalogue. The hatched regions indicate clouds with a GDR measurement and the non-hatched regions represent an upper limit. The vertical dashed lines indicate the median GDR values for each hatched histogram.}
\label{fig:gdr_mass_cat}
\end{figure*}
\end{landscape}

The left-hand column of Figure \ref{fig:gdr_mass_cat} shows histograms of the CO-traced molecular GDR of clouds from both of our cloud catalogues. The median molecular GDR of clouds from the dust-selected catalogue is roughly half the value for the CO-selected
catalogue. We compare the two distributions using a Mann-Whitney U test,  finding that the distributions are significantly different (p-value $\ll$ 0.01; see Table \ref{tab:mannwhitneyu} for full results). The right-hand column of Figure \ref{fig:gdr_mass_cat} shows histograms of the total (\textsc{HI} + H$_2$) GDR of clouds from both of our cloud catalogues. We compare the two distributions using the Mann-Whitney U test, again finding a significant difference (p-value $\ll$ 0.01), with the GDR for the dust-selected clouds being lower than for the CO-selected clouds.

 \newpage
 \begin{landscape}
 \thispagestyle{empty}
 
\begin{table*}
 \caption{Two-sided Mann-Whitney U test results for $T_{\mathrm{dust}}$ and $\beta$ inside and outside molecular clouds at the inner and outer ring; and distributions of molecular GDR and total GDR for CO-traced and dust-traced clouds with at least a 3$\sigma$ detection.}              
 \label{tab:mannwhitneyu}      
 \centering                                      
 \begin{tabular}{c c c c}          
 \hline\hline                        
 Samples being compared & Sample sizes (no. of pixels/clouds) & {U-statistic} & {p-value} \\    
\hline
      $T_{\mathrm{dust}}$ inside clouds in inner ring vs outer ring & Inner ring = 734; Outer ring = 2672  & 396546.0 & 1.3 $\times 10^{-31}$ \\                               
      $T_{\mathrm{dust}}$ outside clouds in inner ring vs outer ring & Inner ring = 9807; Outer ring = 50826 & 205272680.0 & 8.0 $\times 10^{-169}$ \\      
      $T_{\mathrm{dust}}$ inside vs outside clouds in the inner ring & Inside clouds =734 ; Outside clouds = 9807 & 4415974.5 & 9.5 $\times 10^{-25}$ \\
      $T_{\mathrm{dust}}$ inside vs outside clouds in the outer ring & Inside clouds = 2672; Outside clouds = 50826 & 48974193.5 & 1.0 $\times 10^{-130}$ \\
      \hline
      $\beta$ inside clouds in inner ring vs outer ring & Inner ring = 734; Outer ring = 2672 & 1713391.0 & 1.0 $\times 10^{-211}$ \\
      $\beta$ outside clouds in inner ring vs outer ring & Inner ring = 9807; Outer ring = 50826 & 384633211.5 & {$\ll$ 0.01} \\
      $\beta$ inside vs outside clouds in the inner ring & Inside clouds = 734 ; Outside clouds = 9807 & 3111901.5 & 8.9 $\times 10^{-10}$ \\
      $\beta$ inside vs outside clouds in the outer ring & Inside clouds = 2672; Outside clouds = 50826 & 77007902.5 & 1.3 $\times 10^{-31}$ \\
 \hline                                             
      Molecular GDR for CO-traced vs dust-traced clouds & CO-traced clouds = 140; Dust-traced clouds = 177 & 22854.0 & 3.8 $\times 10^{-38}$ \\
      Total GDR for CO-traced vs dust-traced clouds & CO-traced clouds = 140; Dust-traced clouds = 177 & 19146.0 & 7.7 $\times 10^{-17}$ \\
 \hline
 \end{tabular}
 \end{table*}
 
\end{landscape}

\newpage
\newpage

\section{Discussion}
\label{sec:disc}

\subsection{Dust emissivity index inside and outside molecular clouds}
\label{ssec:betavarydisc}
\cite{Smith2012} have found radial variations in $\beta$ in M31, with the value of $\beta$ decreasing from $\simeq$ 2.5 at a radius of 3 kpc to $\simeq$ 1.8 at 12 kpc. \cite{Draine2014} have also found, using a different \textit{Herschel} dataset and a different method, the same general trend; with $\beta$ decreasing from a value of $\simeq 2.35$ at a radius of 3 kpc to a value of $\simeq$ 2.0 at a radius of 12 kpc. Some possible causes of these variations are large grain coagulation or the accretion of a mantle in denser environments since some dust models (e.g. \citealt{Kohler2015}) predict that this will lead to an increase in the value of $\beta$ by $0.3-0.5$.

In our study, we find a much smaller difference (of order 0.03 to 0.07) between the median values of $\beta$ in low-density and high-density environments at the same radius than the much larger radial change. Our results are similar to the findings of \cite{Roman-Duval2017} in the Magellenic Clouds, who find no correlation between $\beta$ and gas mass surface density at spatial scales of 75 pc in the Large Magellenic Cloud (LMC) and 90 pc in the Small Magellenic Cloud (SMC). Our result contrasts with the increase in $\beta$, found along sight lines dominated by molecular gas, in the MW by \cite{PlanckCollaboration2014a}. The \textit{Planck} team find that $\beta$ increases from 1.75 in the atomic medium of the Galactic plane to 1.98 in the molecular medium.  Our result also contrasts with what has been found in M33 where $\beta$ (from SED fitting of \textit{Herschel} observations at $\approx$ 150 pc spatial scales) is strongly and positively correlated with molecular gas traced by $^{12}$CO(J=2-1) emission (\citealt{Tabatabaei2014}). The authors do not attempt to separate effects of radius on $\beta$ from the effects of a dense environment in their study. Therefore, the apparent increase in $\beta$ in dense environments in M33 might be caused by a radial gradient that is unconnected to gas density but caused by there being more clouds at small radii in the galaxy.  We note that our study is the first one that has tried to separate the effect of gas density and radius on $\beta$ in M31, although we examine a smaller dynamic range in galactocentric radii.

To explore the possibility that our results are the consequence of our choice of dendrogram parameters, we have run the dendrogram with a set of parameters that resulted in a peak CO threshold of 3$\sigma$ rather than 5$\sigma$. We found that our results
were very similar.

Our strong conclusion therefore is that in M31, molecular gas surface density is not the driver of radial variations in $\beta$ at 30 pc spatial scales. This suggests that, at these spatial scales, large grain coagulation in dense environments is not having a big effect on $\beta$. The only alternative is that there is some genuine radial change in the composition or the structure of the dust grains as we move out through the galaxy. What these changes are is still a mystery, although one speculative possibility is that there is a radial change in the ratio of carbonaceous and silicate dust grains, perhaps caused by a changing C/Si abundance ratio.

\subsection{CO-dark gas?}
\label{ssec:dustyismdisc}

\begin{figure*}[h!]
\centering
\includegraphics[width=14cm]{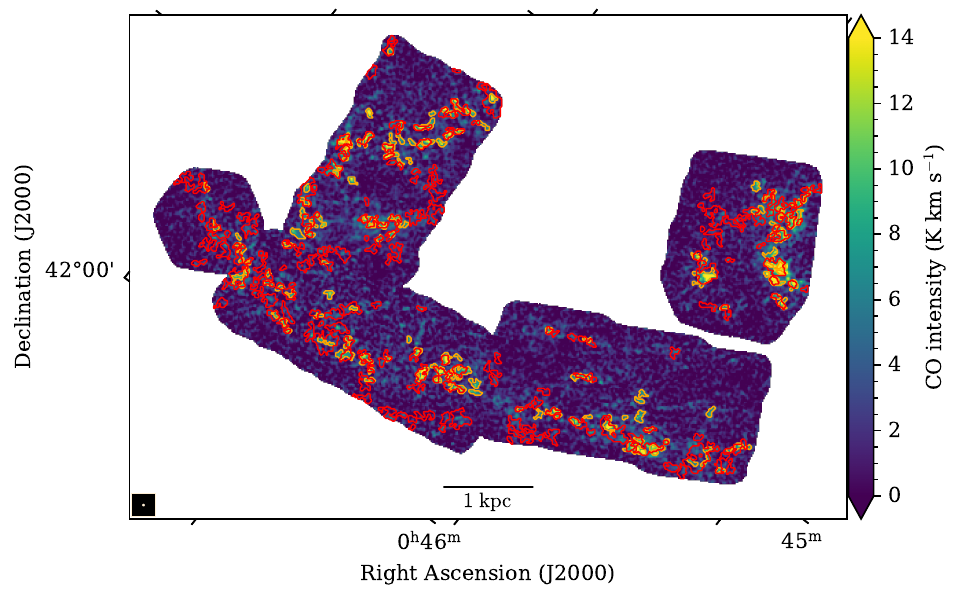}
\caption{Map of $^{12}$CO(J=1-0) intensity taken from CARMA + IRAM, convolved to 8" resolution and reprojected to 4" pixel size. Orange contours show the 140 sources from the CO-selected catalogue. The red contours show the sources from the dust-selected catalogue.}
\smallskip
\label{fig:comap_overlap}
\end{figure*}

\begin{figure*}[h!]
\centering
\includegraphics[width=14cm]{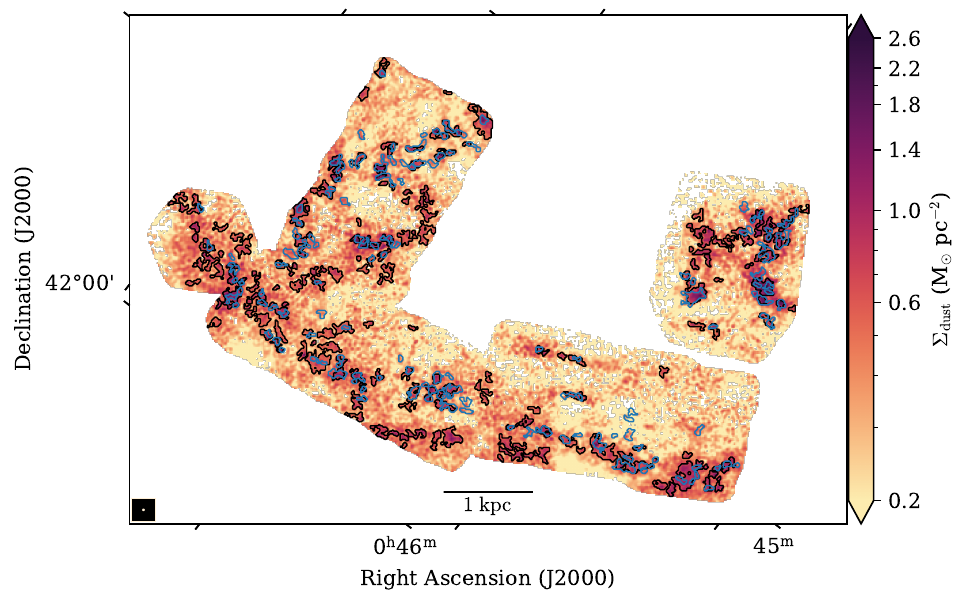}
\caption{Map of dust mass surface density produced by PPMAP at 8" resolution and 4" pixel size. Black contours show the 196 sources from the dust-selected catalogue. The blue contours show the sources from the CO-selected catalogue.}
\smallskip
\label{fig:dustmap_overlap}
\end{figure*}

In Section \ref{ssec:gdrres}, we have compared the GDR distributions of clouds from the two catalogues,
finding that the GDR for the dust-selected clouds is significantly lower than for the CO-selected clouds. Figure \ref{fig:gdr_mass_cat} shows that there are clouds in the dust-selected catalogue with molecular GDR below $\approx$ 16, which are not found in the CO-selected catalogue. Figures \ref{fig:comap_overlap} and \ref{fig:dustmap_overlap} show the overlap of our CO-selected and dust-selected clouds.  Alongside dust-selected clouds with low CO emission, we serendipitously find some CO-selected clouds with low dust emission. We are uncertain of the astrophysical meaning of these clouds.

One possible explanation for clouds in the dust-selected catalogue with low molecular GDR is that they may simply be emphemeral structures in atomic gas. We have investigated this by adding in the atomic gas mass and finding the total (\textsc{HI} + H$_2$) GDR of clouds (see right-hand column of Figure \ref{fig:gdr_mass_cat}). However, even after adding in the atomic gas mass, we find some dust-selected clouds with lower total GDR than any of the CO-selected clouds.

Are the clouds showing lower levels of CO simply because they are smaller, which might suggest that they are not genuine molecular cloud structures? To answer this question, we have compared the physical sizes and total masses of the clouds in our dust-selected catalogue.  We have found the total ISM mass of each cloud by multiplying the dust mass surface density of our cloud by a constant GDR of 100 (\citealt{Hildebrand1983}) and calculating the total dust mass of the cloud as described in Section \ref{ssec:gmc_ppmap}.  We have used the dust mass rather than the CO to estimate the total ISM mass because we want to examine the possibility that there are molecular clouds made up of a large proportion of CO-dark gas.  As for the physical size, we have simply taken the area of each cloud in square parsec.

\begin{figure*}[h]
\centering
\includegraphics[width=13cm]{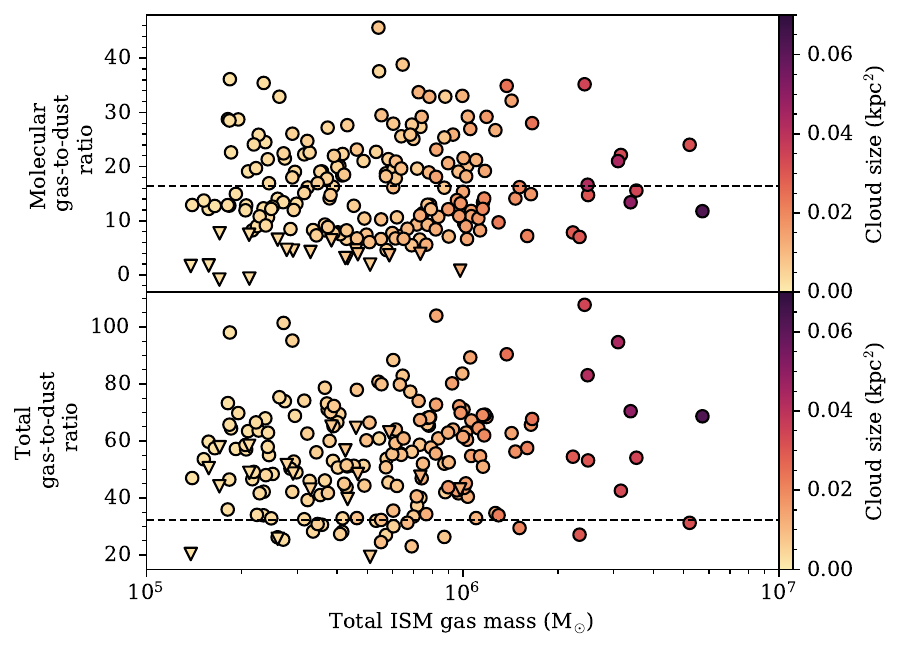} \caption{\textit{Top row:} Total ISM mass (inferred from the dust) versus CO-traced molecular GDR of dust-selected clouds. \textit{Bottom row:} Total ISM mass (inferred from dust) versus total GDR of dust-selected clouds. The points are coloured by cloud size given in terms of its physical area in kpc$^2$. The circles represent dust-selected clouds with at least a 3$\sigma$ detection in the CO map. The triangles represent dust-selected clouds with less than a 3$\sigma$ detection in the CO map. The dashed horizontal black line shows the lowest molecular GDR obtained (GDR $\approx$ 16) and lowest total GDR obtained (GDR $\approx$ 32) in clouds from the CO-selected catalogue.}
\label{fig:gdr_mass_cloudsize}
\end{figure*}

The top panel of Figure \ref{fig:gdr_mass_cloudsize} shows the total ISM mass versus the molecular GDR in the dust-selected clouds coloured by cloud size. The bottom panel shows the total ISM mass vs total GDR of clouds. We find that there is still a population of large clouds ($\geq$ 0.03 kpc$^2$) with ISM mass greater than 10$^6$ M$_{\odot}$ and low total GDR ($\lesssim$ 50). There are some clouds with total GDR below $\approx$ 32 which are not found in the CO-selected catalogue.

One possibility is that these are real molecular clouds with low levels of CO, i.e. clouds that are largely made up of CO-dark gas. The alternative explanation remains that these structures are largely made up of atomic gas, perhaps the result of source confusion along the line of sight. Although we have tried to correct for the contribution of atomic gas,  we are limited by the resolution of the available radio data. The HARP and SCUBA-2 High-Resolution Terahertz Andromeda Galaxy (HASHTAG) survey \textit{James Clerk Maxwell Telescope} observations will give a higher resolution view of the dust,  and a new \textit{Extended Very Large Array} (EVLA) survey will give a higher resolution view of the atomic gas,  helping to distinguish between these two possibilities.

\section{Summary}
\label{sec:conc}

M31 forms an excellent testbed for understanding the variations in dust properties and the interplay of dust and gas in a spiral galaxy similar to our own. We investigate whether radial variations in the dust emissivity index ($\beta$) are caused by an increase of $\beta$ in dense molecular gas regions in M31. We probe the ISM of M31 at significantly improved spatial resolution ($\approx$ 30 pc) compared to previous studies, using combined CARMA + IRAM $^{12}$CO(J=1-0) observations and {\it Herschel} observations of the dust which have been reanalysed using the PPMAP algorithm.
We use a dendrogram to create molecular cloud catalogues in two ways: using CO and dust as a tracer. Our key findings are:

 \begin{enumerate}
    \item We see a radial variation in $\beta$ in agreement with previous studies (e.g. \citealt{Smith2012}, \citealt{Draine2014}, \citealt{Whitworth2019}), with a decrease in $\beta$ going from the inner ring to the outer dusty, star-forming ring.
    \item We find no evidence for radial variations in $\beta$ being caused by an increase of $\beta$ in dense molecular gas regions at radii between $5-7.5$ kpc and $9-15$ kpc.
    \item We find a population of clouds in our dust-selected catalogue with lower median CO-traced molecular GDR than in our CO-selected catalogue. These may be
    clouds containing CO-dark molecular gas, although we are unable to rule
    out the possibility that these structures are confused with features
    in the atomic phase of the ISM.
 \end{enumerate}

 We conclude that an increase of $\beta$ in dense molecular gas regions is not the prominent driver of the radial variations in $\beta$ in M31.

%% file: Chapters/chapter_sed.tex
\chapquote{Humans have a tendency to look for things in the places where it is easiest to search for them rather than in the places where the truth is more likely to be found.}{Esther Perel}{}

\section{Introduction}
\label{sec:intro}
An excess of emission from dust grains at longer wavelengths ($\geq$ 500 $\mu$m) has been observed in low-metallicity galaxies (e.g. \citealt{Israel2010}, \citealt{Bot2010}, \citealt{Galametz2011}, \citealt{Remy-Ruyer2013}, \citealt{Gordon2014}) and our own galaxy (e.g. \citealt{Paradis2012b}) when compared to the amount of emission predicted by models.  Excess emission at long wavelengths has also been observed in nearby spirals like M33 (\citealt{Relano2018}), NGC 0337, NGC 0628, NGC 1512 and NGC 7793 (\citealt{Galametz2014}).  A recent study by \cite{Chang2020} found this excess in multiple spiral galaxies and suggested that the excess becomes more obvious in galaxies with lower stellar mass and cold dust temperatures between 20 and 22 K (see their Fig. 3).  \cite{Kirkpatrick2013} have found that an excess is not necessarily correlated with metallicity.  Although the precise origin of such an excess is still a mystery,  finding an excess (coined "sub-mm excess") might imply that there is a lot of very cold dust in the galaxy (\citealt{Galliano2005}) or that the dust grains have unusual emission properties (\citealt{Draine2007}, \citealt{Draine2012}).  In this chapter,  we describe a search for a sub-mm excess in the Andromeda galaxy. 

The Andromeda galaxy (M31) is a spiral galaxy within our Local Group, with a known metallicity gradient (\citealt{Gregersen2015}),  which exhibits radial variations in its dust properties (\citealt{Smith2012}, \citealt{Draine2014}, \citealt{Whitworth2019}, \citealt{Athikkat-Eknath2022}).  \cite{Smith2012} have reasonably fit the far-infrared emission from dust shown in \textit{Herschel} observations of M31 with a single temperature modified blackbody (MBB) model at $\approx$ 140 pc spatial scales.   In their paper,  the authors searched for any excess emission at 500 $\mu$m which deviates from the single temperature MBB model but did not find any evidence for this.  Due to the lack of high-resolution observations at longer wavelengths,  the authors were unable to put constraints on the existence of any dust at temperatures below 10 K.  CO observations of M31 imply there must be some very cold gas present (\citealt{Loinard1995}),  but it has been hard to determine whether there is any dust at a similar temperature because the emission from very cold dust is generally swamped by the emission from the warmer dust (\citealt{Eales1989}).

Since dust continuum data are now available from the HARP and SCUBA-2 High-Resolution Terahertz Andromeda Galaxy (HASHTAG) survey enabling us to study M31 at much better than sub-kpc spatial resolution and at long wavelengths,  we exploit these new JCMT observations at 450 and 850 $\mu$m to search for any excess emission.  Throughout this chapter,  we define an `excess' as emission above what is predicted by the underlying modified blackbody model at 450 $\mu$m and 850 $\mu$m.

This chapter is structured as follows: Section \ref{sec:observe_sed} provides the details of the observational data that we use, Section \ref{sec:sedfit} explains how we obtain our spectral energy distributions,  Section \ref{sec:results_sed} shows our results,  including a first look at the spatial distribution of dust properties in M31 captured by the HASHTAG survey.  In Section \ref{sec:disc_sed},  we discuss the implications of our results,  followed by our key conclusions in Section \ref{sec:summary_sed}.

\section{Observations}
\label{sec:observe_sed}
For this work, we make use of observations of M31 taken by the \textit{Herschel Space Observatory} and the \textit{James Clerk Maxwell Telescope} (JCMT).  We use 100 and 160 $\mu$m observations from the work of \cite{Draine2014} and 250 $\mu$m observations from the work of The Herschel Exploitation of Local Galaxy  Andromeda (HELGA) collaboration (\citealt{Fritz2012}, \citealt{Smith2012}). Our long wavelength observations,  at 450 and 850 $\mu$m,  come from HASHTAG survey (\citealt{Smith2021}). The 450 $\mu$m image has been combined with the \textit{Herschel} 500 $\mu$m map (\citealt{Smith2012}) and the 850 $\mu$m has been combined with the \textit{Planck} 353 GHz map (\citealt{PlanckCollaboration2015}) in order to restore large scale structure (\citealt{Smith2021}).

We smooth all datasets to the resolution of the lowest resolution image using a Gaussian convolution kernel. Hence, all five maps are effectively at the resolution of the 250 $\mu$m map at 18". We also reproject the data to match the pixel scale size of the 250 $\mu$m map at 6".

\section{Spectral energy distribution of each pixel}
\label{sec:sedfit}
The spectral energy distribution (SED) of dust emission seen at far-infrared wavelengths can be fit with a modified black blackbody model for the optically thin limit {\magi (see Chapter 1,  equation \ref{eq:intro_mbb})}.  We fit a single temperature modified black blackbody model to each pixel in our datasets.  We assume a distance to M31 of 785 kpc (\citealt{McConnachie2005}),  and a $\kappa_{0}$ of 0.192 m$^{2}$ kg$^{-1}$ (reference opacity for 350 $\mu$m from \citealt{Draine2003}). 

\subsection{Uncertainty in the observations \& colour correction}
\label{ssec:uncertainties}
\begin{table*}
\caption{Table showing background noise values, the fractional calibration uncertainty,  colour correction factors and the mean subtracted signal for the 100,  160, 250, 450 and 850 $\mu$m observations.  The 100 and 160 $\mu$m colour correction factors have been taken from Table 3 of the PACS handbook for modified blackbody emission with $\beta=2$ and $T_{\mathrm{dust}}=20$ K.  The colour correction factor for 250 $\mu$m has been taken from Table 5.8 of the SPIRE handbook for modified blackbody emission with $\beta=2$ and $T_{\mathrm{dust}}=20$ K.  The colour correction factor for 450 and 850 $\mu$m has been taken from the work of \citealt{Smith2021} (private communication).}
\label{tab:uncert}      
\centering                                      
\begin{tabular}{c c c c c}          
\hline       
\makecell{Wavelength \\ ($\mu$m)} & \makecell{Background noise \\ (Jy/pix)} & \makecell{Fractional calibration \\ uncertainty} & \makecell{Colour correction \\ factor} & \makecell{Mean signal \\ subtracted (Jy/pix)}  \\
\hline
100 & 0.00198296  & 0.070 & 1/0.974 & -0.00115763 \\
160 & 0.00160938 & 0.070 & 1/0.971 & 0.0000656883 \\
250 & 0.000990815 & 0.055 & 0.9998 & 0.000261958 \\
450 & 0.000919114 & 0.100 & 0.9944 & 0.000135209 \\
850 & 0.000157034 & 0.050 & 1.0216 & 0.000694068 \\
\hline                                   
\end{tabular}
\end{table*}

To ensure that we can meaningfully interpret the error margins on our observations,  we account for three primary sources of uncertainty within our analysis: instrument calibration uncertainty,  colour calibration uncertainty and background noise.

The flux density of an astronomical source is calibrated against the known flux density of a calibrator source using an appropriate flux conversion factor.  The calibration corrects the flux density to what a monochromatic detector would have measured at a reference wavelength.  The calibrator source is usually one which has been observed over a range of atmospheric conditions and has accurate and known flux properties from modelling or observation (\citealt{Dempsey2013}).  For \textit{Herschel},  the planets Neptune (\citealt{Bendo2013}) and Uranus were used to calibrate flux measurements for the SPIRE instrument.  Five fiducial stellar sources were used for the PACS instrument (\citealt{Nielbock2013}, \citealt{Balog2014}).  For the JCMT measurements,  observations of Mars and Uranus were used for calibration (\citealt{Dempsey2013}).  The calibration uncertainty arises from uncertainties in the model spectrum of the calibrator source and measurement uncertainties.  We incorporate the calibration uncertainty in our uncertainties for SED fitting by finding the fractional uncertainty in flux due to the calibration uncertainty {\magi (Table \ref{tab:uncert},  third column).}

The passbands of instruments can be quite broad in order to maximise sensitivity.  This means that although the filters used may be centred around a particular wavelength,  the instrument may be sensitive to photons from a much wider wavelength range.  The measured flux density of the source depends on the convolution of the source spectrum with the spectral passband.  This results in an increase or decrease when measuring the flux density of the source due to the instrument being more or less sensitive to certain wavelengths within the passband.  If the standard calibration source and the observed astronomical source of interest have different spectral shapes,  then a correction needs to be applied to give the true value for the brightness of the astronomical source.  The multiplicative colour correction factor accounts for the different spectral shapes of the calibrator source and the astronomical source of interest (in our case,  M31).  Prior to smoothing and reprojecting,  we colour-correct each dataset using the appropriate multiplicative factor {\magi (Table \ref{tab:uncert},  fourth column)} taken from the PACS colour correction manual\footnote{\url{https://www.cosmos.esa.int/documents/12133/996891/PACS+Photometer+Passbands+and+Colour+Correction+Factors+for+Various+Source+SEDs} - Table 3,  modified blackbody emission assuming $\beta=2$, $T_{\mathrm{dust}}=20$ K ($\nu^2 B_{\nu}$($T_{\mathrm{dust}}$))},  SPIRE handbook\footnote{\url{http://herschel.esac.esa.int/Docs/SPIRE/spire_handbook.pdf} - Table 5.8,  modified blackbody emission assuming $\beta=2$, $T_{\mathrm{dust}}=20$ K ($\nu^2 B_{\nu}$($T_{\mathrm{dust}}$)),  extended source.} and HASHTAG data pipeline (\citealt{Smith2021},  private communication). 

To calculate the background noise,  we select a small boxed region in all five observations where we see no obvious sources and calculate the standard deviation of the pixel values within this box in our smoothed and reprojected data {\magi (Table \ref{tab:uncert},  second column)}.  We only fit pixels where the signal is at least 5 times above the noise level within the 250 $\mu$m observations. 

For each pixel across all five observations,  we add the background noise value in quadrature with the uncertainty in flux due to the calibration uncertainty to get the total uncertainty in flux.  This combined value is used as the error on each data point in our SED fitting process.  For all of the data,  we also subtract contributions from a non-zero background by finding the mean signal in a region with no obvious sources and subtracting this value from all pixels.  The mean background signal subtracted for each of our five observations is listed in {\magi the final column of} Table \ref{tab:uncert}.

\subsection{Fitting the MBB model to the data}
We use \texttt{lmfit 1.0.3} Python package to fit the model described by equation \ref{eq:1} to each pixel across our convolved and regridded data.  \texttt{lmfit} performs Levenberg-Marquardt least-squares minimisation taking into account the intensity of emission ($I_{\nu, \mathrm{data}}$) from the data, the model intensity ($I_{\nu, \mathrm{model}}$) and the total uncertainty ($\sigma_{\mathrm{tot}}$; see Section \ref{ssec:uncertainties}) in each pixel within our data.  The $\chi^{2}$ parameter is calculated as:
\begin{equation}
\chi^2 = \sum_i \frac{(I_{\nu, \mathrm{data};\; i} - I_{\nu,  \mathrm{model};\; i})^2}{\sigma_{\mathrm{tot},  i}^2}
\end{equation}
We consider values where $\chi^2$ is a minimum as our best fit values for these dust properties.  The number of degrees of freedom is the number of data points minus the number of free parameters minus one.  We allow $M_{\mathrm{dust}}$, $T_{\mathrm{dust}}$ and $\beta$ parameters to vary. 

A common practice in observational and simulation work is to assume an average $\beta$ parameter across the whole galaxy.  However,  \cite{Smith2012} and \cite{Draine2014} have found radial variations in the dust emissivity index across M31,  with a decrease in $\beta$ with increased galactocentric radius,  beyond the central bulge.  \cite{Whitworth2019} have confirmed these radial variations using a more sophisticated Bayesian analysis.  In light of these findings,  we also perform our SED fitting with a fixed $\beta$ (at $\beta = 1.5, 1.8, 2.0, 2.2 \;  \mathrm{and} \; 2.5$) to compare and confirm whether a variable $\beta$ MBB model applies to the new data.

\section{Results}
\label{sec:results_sed}

\subsection{The spatial distribution of dust properties}

\begin{figure}[h!]
\centering
\includegraphics[width=17cm]{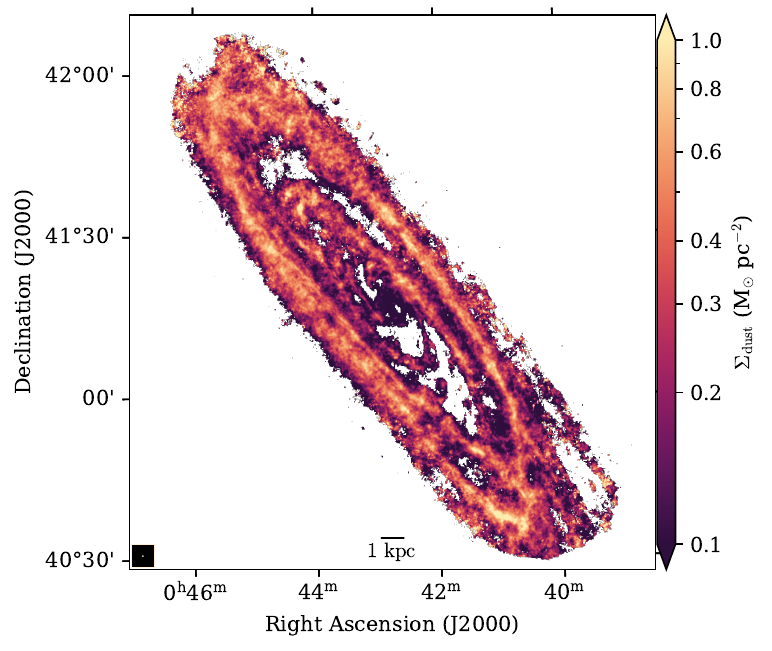} \caption{Dust mass surface density map obtained from SED fitting of 100,  160,  250,  450 and 850 $\mu$m observations.}
\label{fig:sed_surfreswith850}
\end{figure}

\begin{figure}[h!]
\centering
\includegraphics[width=17cm]{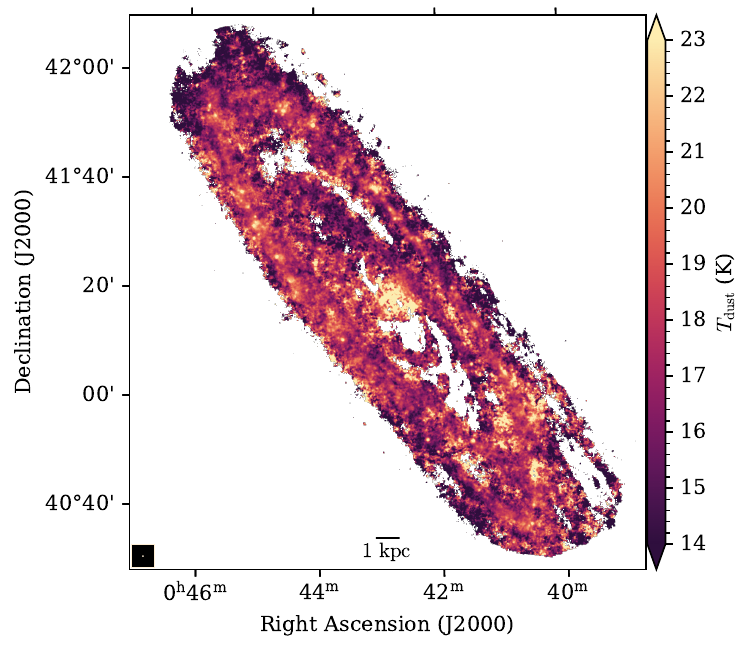} \caption{Dust temperature map obtained from SED fitting of 100,  160,  250,  450 and 850 $\mu$m observations.}
\label{fig:sed_tempreswith850}
\end{figure}

\begin{figure}[h!]
\centering
\includegraphics[width=17cm]{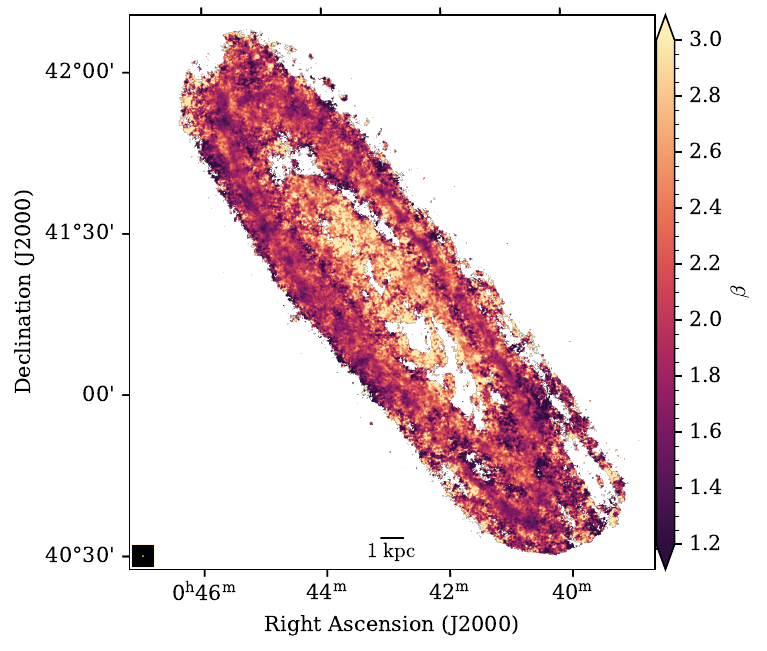} \caption{Dust emissivity index map obtained from SED fitting of 100,  160,  250,  450 and 850 $\mu$m observations.}
\label{fig:sed_betareswith850}
\end{figure}

We fit the SED of M31 with a single temperature modified blackbody (MBB) model,  allowing $M_{\mathrm{dust}}$,  $T_{\mathrm{dust}}$ and $\beta$ to vary.  We convert $M_{\mathrm{dust}}$ to the dust mass surface density $\Sigma_{\mathrm{dust}}$ by dividing each pixel in the $M_{\mathrm{dust}}$ map by the area of one pixel.  Table \ref{tab:MAPprops} shows the median, 16th and 84th percentile values of our best fit $\Sigma_{\mathrm{dust}}$, $T_{\mathrm{dust}}$ and $\beta$.  Figures \ref{fig:sed_surfreswith850} - \ref{fig:sed_betareswith850} show the spatial distribution of these dust properties in M31.

\begin{table}[h!]
\caption{{\magi The median,  16th and 84th percentile values} of $M_{\mathrm{dust}}$,  $T_{\mathrm{dust}}$ and $\beta$ maps {\magi of the entire galaxy} extracted from fitting a single temperature MBB model to observations of M31.  The $\Sigma_{\mathrm{dust}}$ values are also given.  Values have been rounded to 1 decimal place.}
\label{tab:MAPprops}      
\centering                                      
\begin{tabular}{cc}          
\hline \\           
Dust properties & 50th$^{\mathrm{84th}}_{\mathrm{16th}}$ percentile \\
\\
\hline \\
\\
$\Sigma_{\mathrm{dust}}$ & 0.2$_{0.1}^{0.5}$ M$_{\odot}$ pc$^{-2}$ \\
\\
$M_{\mathrm{dust}}$ & 128.9$_{56.9}^{251.8}$ M$_{\odot}$ \\
\\
$T_{\mathrm{dust}}$ & 17.1$_{14.4}^{19.5}$ K \\
\\
$\beta$ & 2.0$_{1.6}^{2.6}$ \
\\
\hline                                   
\end{tabular}
\end{table}

\newpage

\newpage

\newpage

\begin{figure}
\centering
\includegraphics[width=16cm]{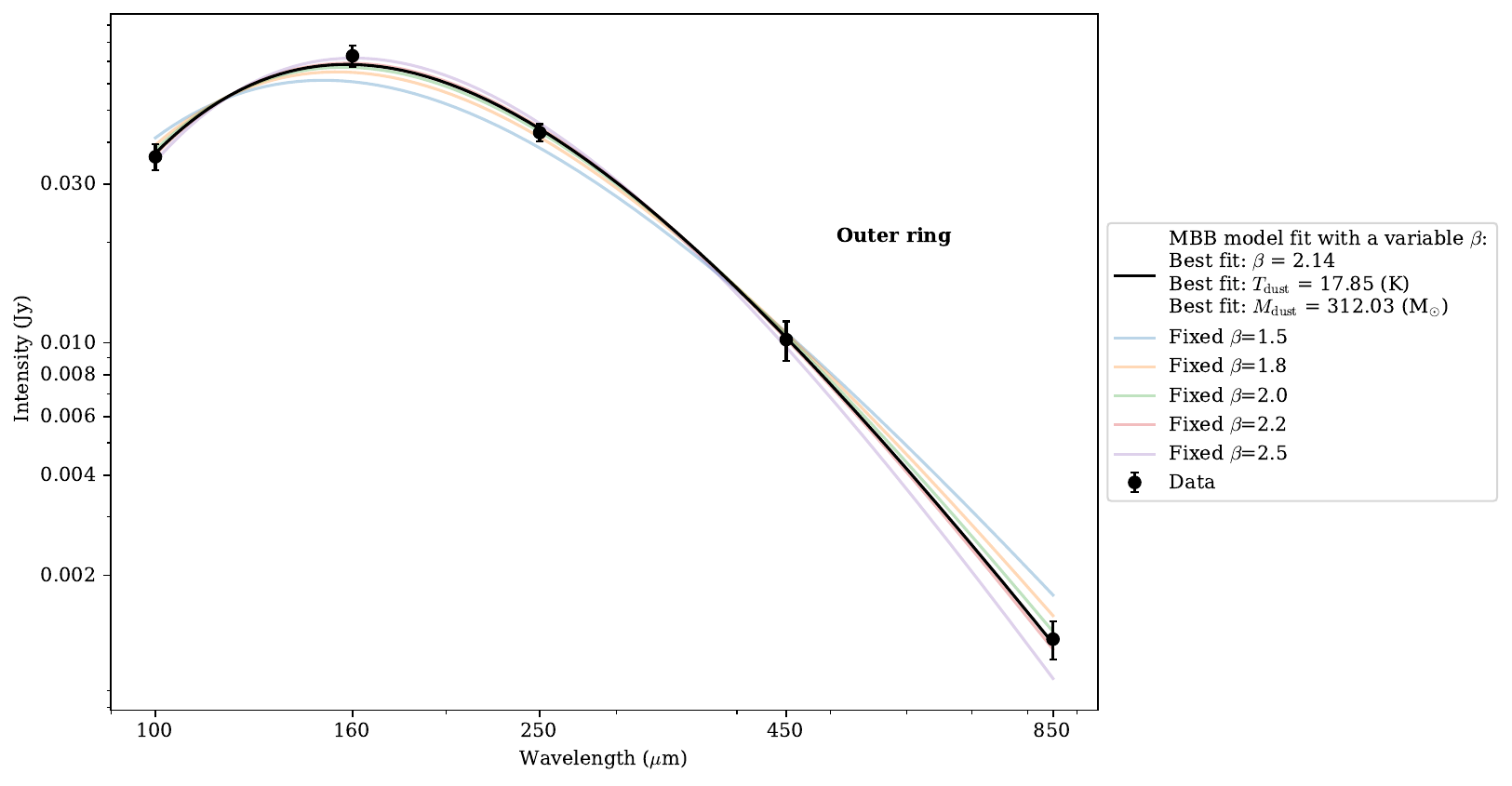} \caption{Single temperature MBB fit to the intensity in one pixel in the outer star-forming ring of M31 (black line).  The coloured lines show fits with a fixed $\beta$ while allowing $M_{\mathrm{dust}}$ and $T_{\mathrm{dust}}$ to vary. }
\label{fig:sed_onepix_outer}
\end{figure}

\begin{figure}
\centering
\includegraphics[width=16cm]{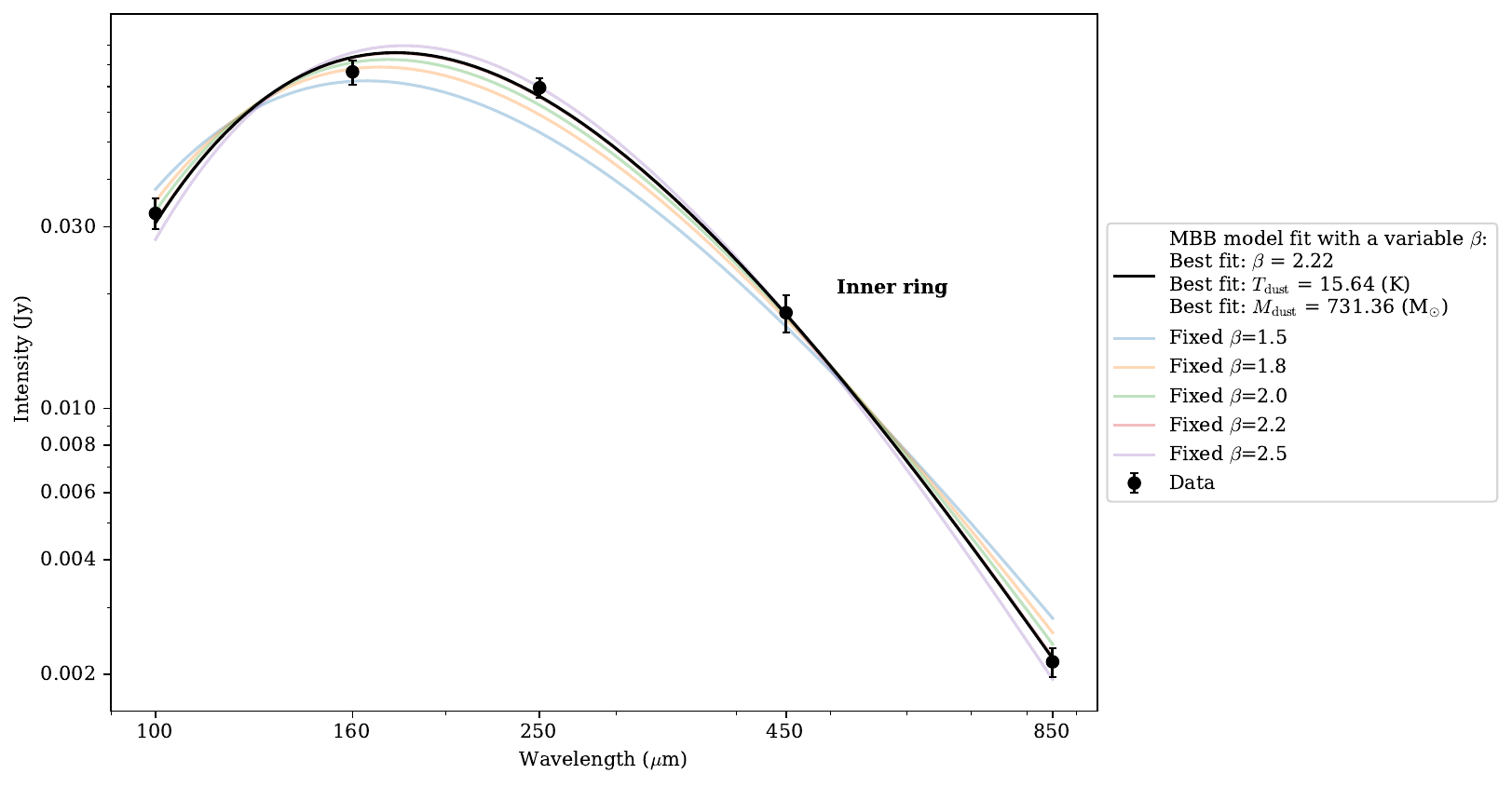} \caption{Single temperature MBB fit to the intensity in one pixel in the inner ring of M31 (black line). The coloured lines show fits with a fixed $\beta$ while allowing $M_{\mathrm{dust}}$ and $T_{\mathrm{dust}}$ to vary. }
\label{fig:sed_onepix_inner}
\end{figure}

\begin{figure}
\centering
\includegraphics[width=16cm]{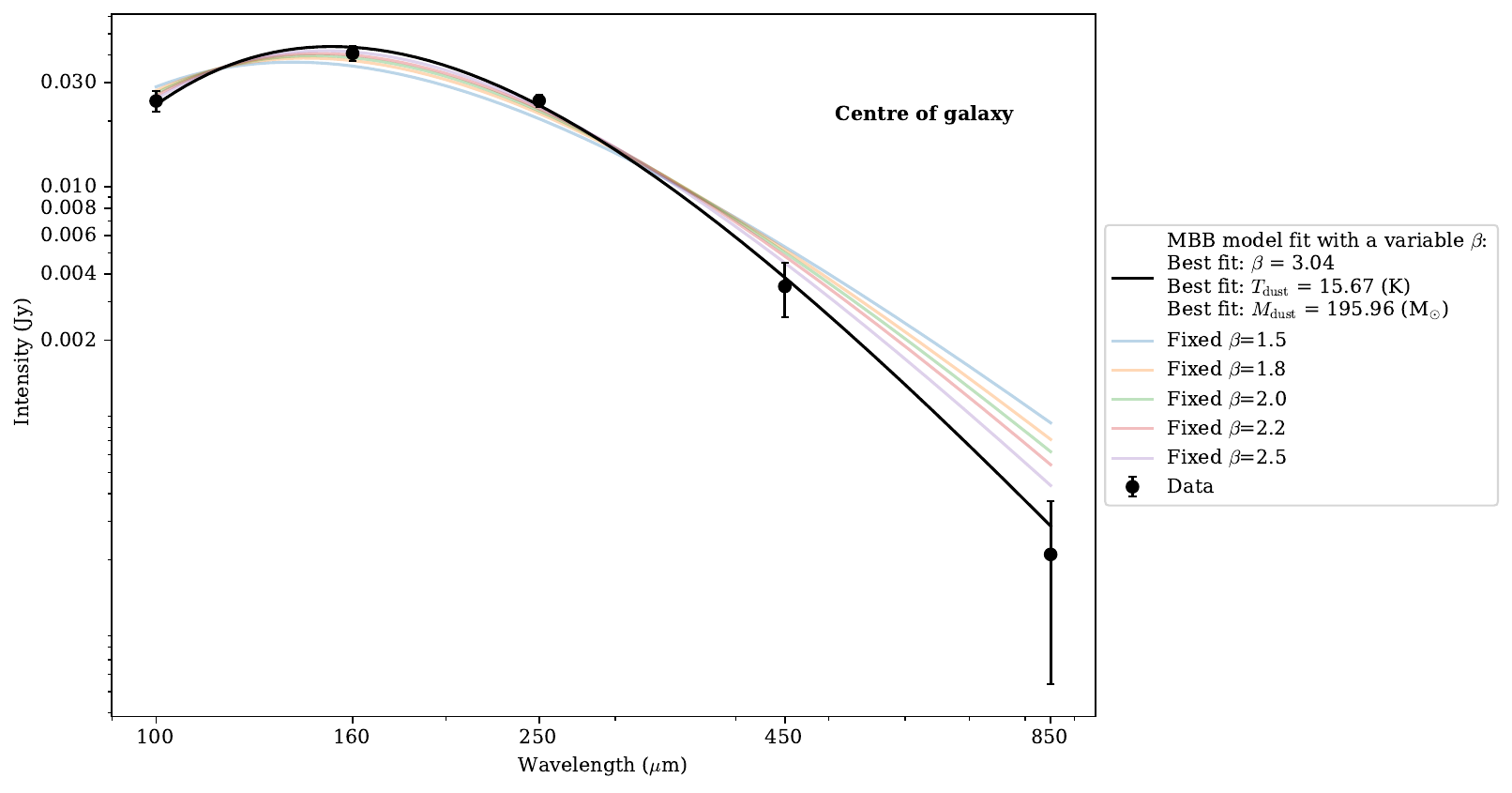} \caption{Single temperature MBB fit to the intensity in one pixel in the galactic centre (black line). The coloured lines show fits with a fixed $\beta$ while allowing $M_{\mathrm{dust}}$ and $T_{\mathrm{dust}}$ to vary. }
\label{fig:sed_onepix_centre}
\end{figure}

\begin{figure}
\centering
\includegraphics[width=16cm]{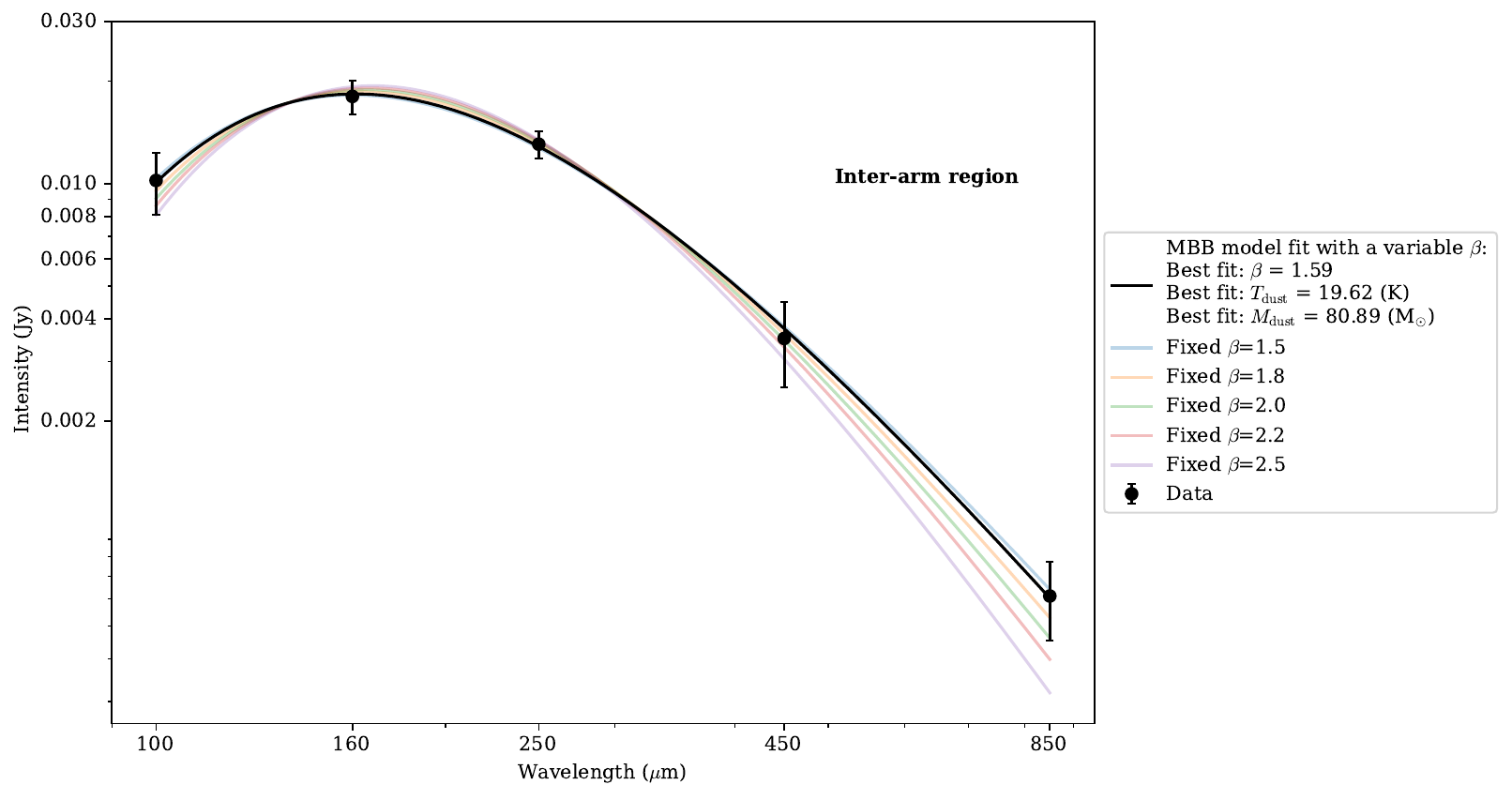} \caption{Single temperature MBB fit to the intensity in one pixel in the interarm region (black line).  The coloured lines show fits with a fixed $\beta$ while allowing $M_{\mathrm{dust}}$ and $T_{\mathrm{dust}}$ to vary. }
\label{fig:sed_onepix_interarm}
\end{figure}

Figures \ref{fig:sed_onepix_outer}-\ref{fig:sed_onepix_interarm} show our best fitting model to the data in one pixel at the outer ring,  inner ring,  galactic centre and interarm regions of the galaxy.   These figures also show our fits for a range of fixed $\beta$ values while still allowing $T_{\mathrm{dust}}$ and $M_{\mathrm{dust}}$ to vary.  We fit 297203 pixels in total.

\begin{figure*}
\centering
\includegraphics[width=16cm]{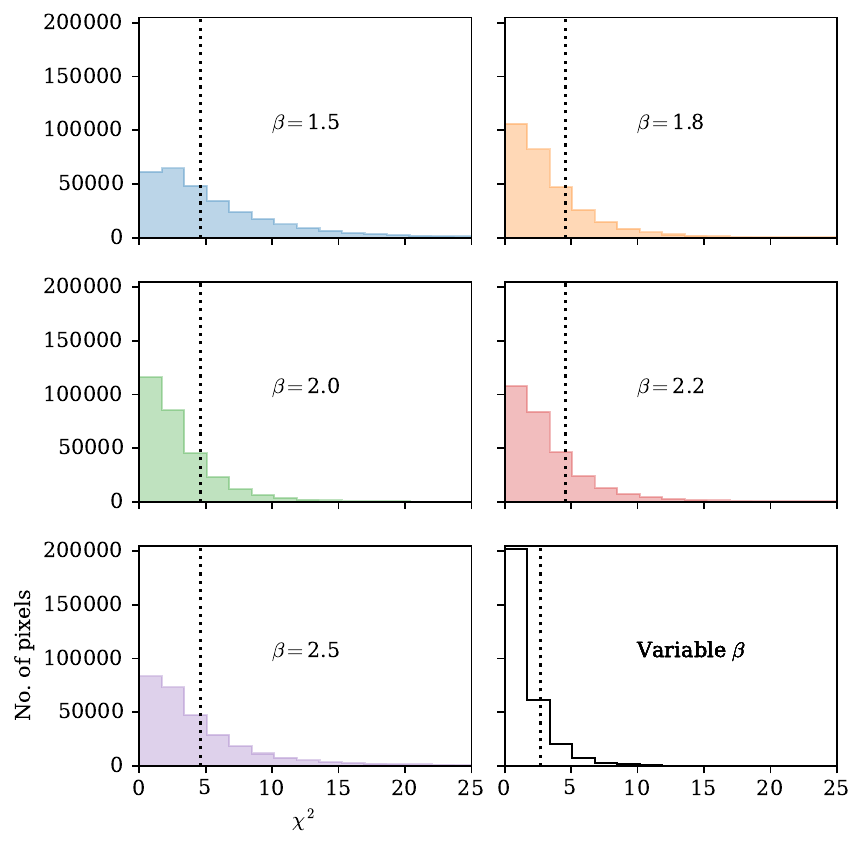} \caption{$\chi^2$ distribution for each model used during SED fitting.  For the bottom right panel,  the dotted black line shows the 10\% significance value,  $\chi^2 = 2.706$,  for 1 degree of freedom taken from the table critical $\chi^2$ of values.  For the remaining panels,  the dotted black line shows the 10\% significance value, $\chi^2 = 4.605$,  for 2 degrees of freedom taken from the table of critical $\chi^2$ values.}
\label{fig:chi_with850}
\end{figure*}

Figure \ref{fig:chi_with850} shows the $\chi^2$ distributions of SED fitting with a fixed $\beta$ model with $\beta$ fixed between 1.5 and 2.5, and a variable $\beta$ model where we leave $\beta$ as a free parameter.  Table \ref{tab:sumchi} shows the sum of $\chi^2$ values for each model.  The sum of $\chi^2$ values is lowest for the variable $\beta$ model,  followed by a fixed $\beta$ of 2.  This shows that statistically,  the MBB model with a variable $\beta$ provides a better fit to the data.  This is also consistent with radial variations in $\beta$ found through previous studies in M31 (\citealt{Smith2012}, \citealt{Draine2014}, \citealt{Whitworth2019}, \citealt{Athikkat-Eknath2022}) which suggest that a fixed $\beta$ does not properly describe emission from dust grains in M31. 

\begin{table}
\caption{Sum of $\chi^2$ values for each MBB model,  with fixed $\beta$ and a variable $\beta$.  Values have been rounded to 1 decimal place.}
\label{tab:sumchi}      
\centering                                      
\begin{tabular}{cc}          
\hline \\           
$\beta$ value & Sum of $\chi^2$ \\
\\
\hline \\
1.5 &  1.7 $\times 10^6$\\
\\
1.8 & 1.0 $\times 10^6$\\
\\
2.0 &  0.9 $\times 10^6$\\
\\
2.2 &  1.0 $\times 10^6$\\
\\
2.5 & 1.5 $\times 10^6$\\
\\
Variable & 0.5 $\times 10^6$\\
\\
\hline                                   
\end{tabular}
\end{table}

\newpage
\subsection{The radial distribution of dust properties}
\begin{figure}
\centering
\includegraphics[width=14cm]{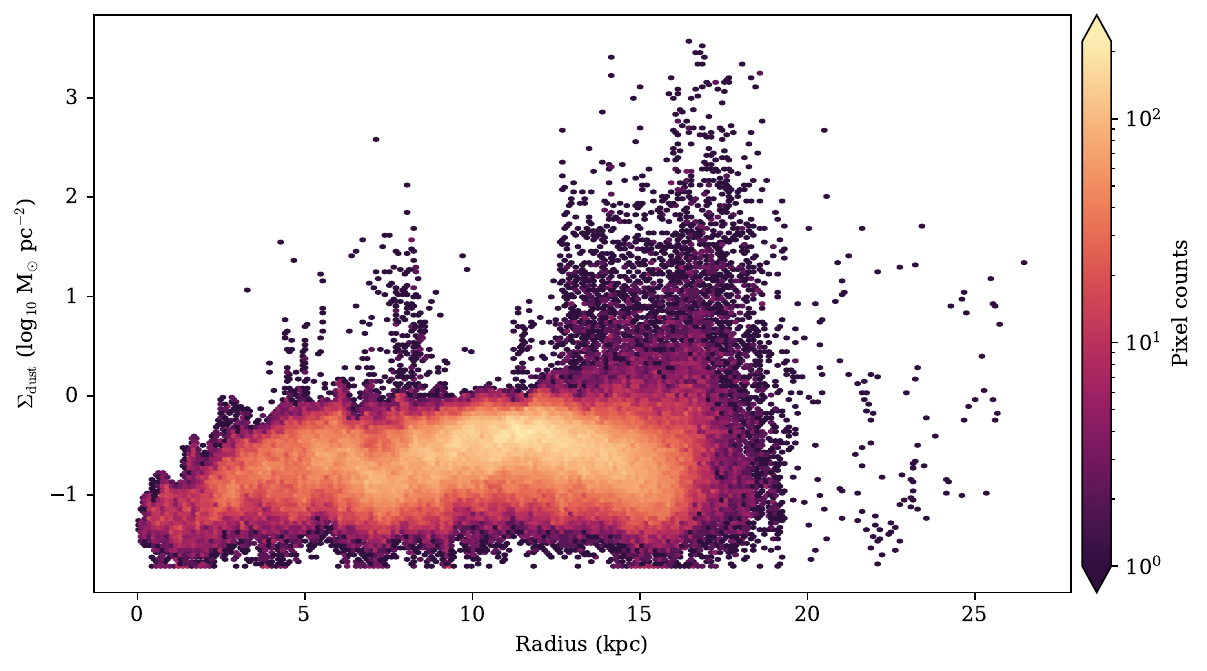} \caption{A 2D histogram binned hexagonnally showing the radial distribution of dust mass surface density.  There are 200 bins.  Each point is a hexagonal bin in the radius-$\Sigma_{\mathrm{dust}}$ parameter space,  coloured by the number of pixels in that bin. }
\label{fig:dustsurfradvar}
\end{figure}

\begin{figure}
\centering
\includegraphics[width=14cm]{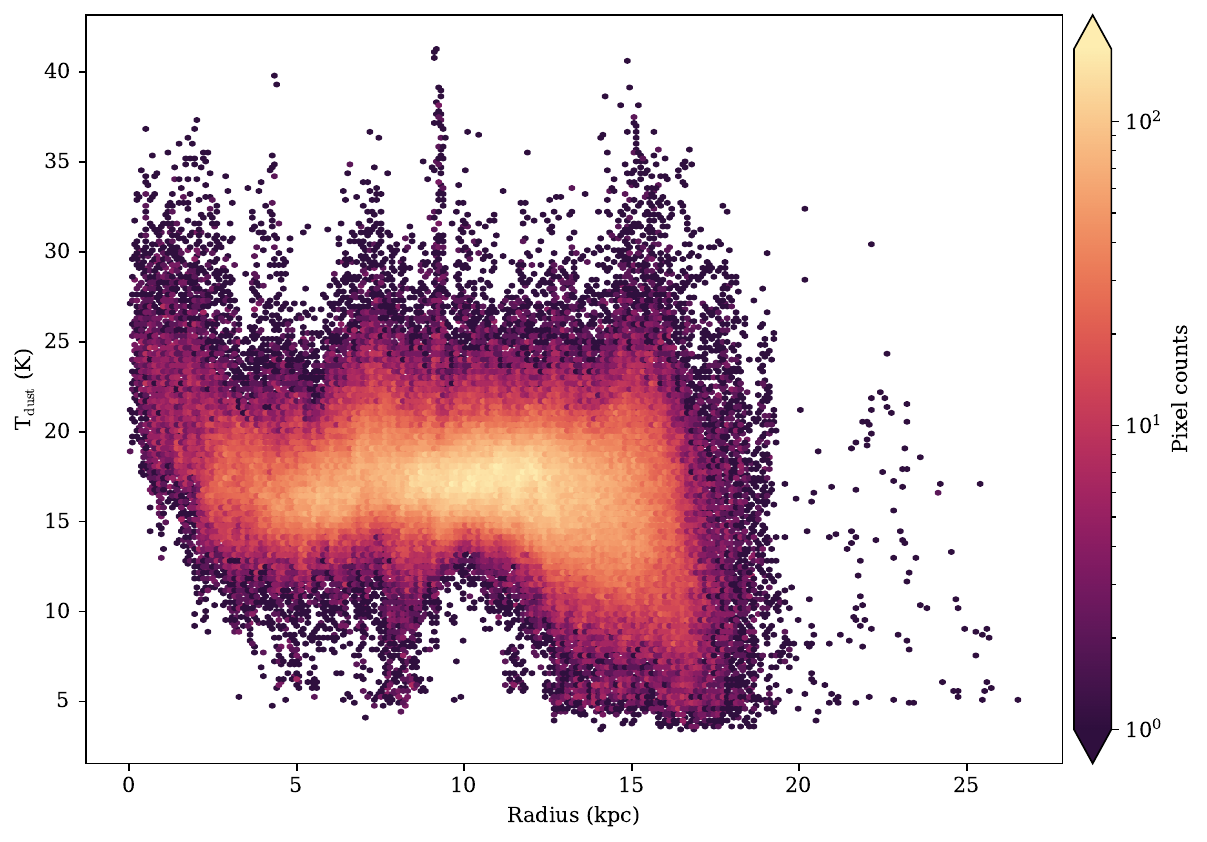} \caption{A 2D histogram binned hexagonnally showing the radial distribution of dust temperature.  There are 200 bins.  Each point is a hexagonal bin in the radius-$T_{\mathrm{dust}}$ parameter space,  coloured by the number of pixels in that bin. }
\label{fig:tempradvar}
\end{figure}

\begin{figure}
\centering
\includegraphics[width=14cm]{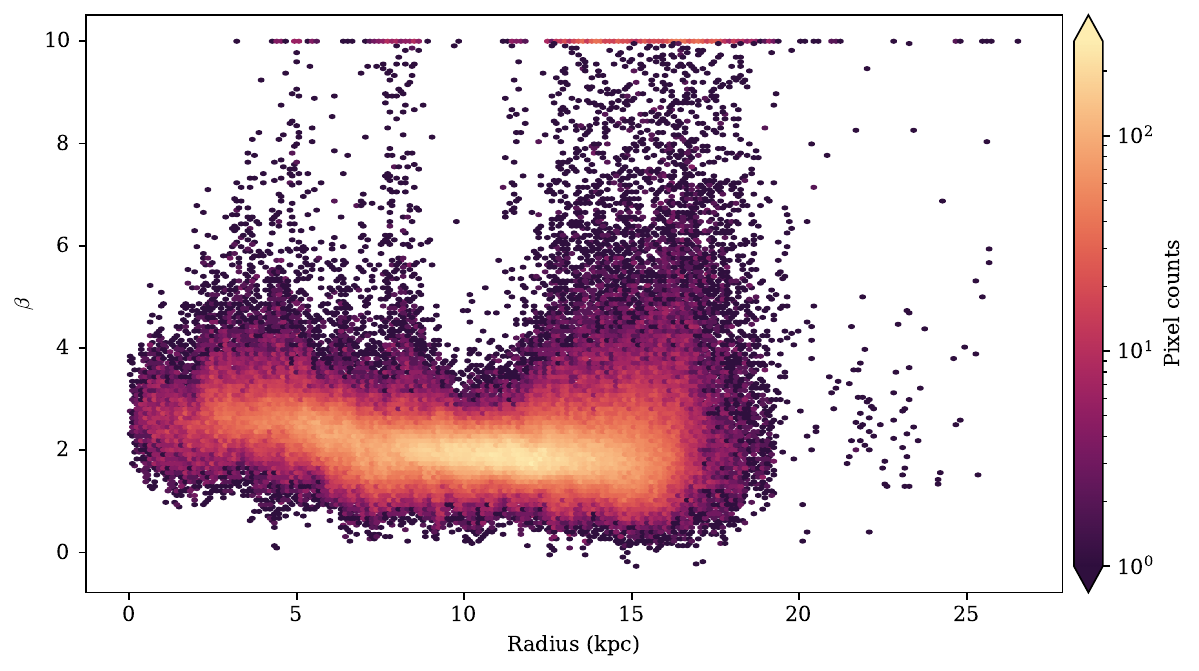} \caption{A 2D histogram binned hexagonnally showing the radial distribution of dust emissivity index.  There are 200 bins.  Each point is a hexagonal bin in the radius-$\beta$ parameter space,  coloured by the number of pixels in that bin.}
\label{fig:betaradvar}
\end{figure}

We examine the radial distribution of the best fit dust mass surface density (Figure \ref{fig:dustsurfradvar}),  dust temperature (Figure \ref{fig:tempradvar}),  dust emissivity index (Figure \ref{fig:betaradvar}) from SED fitting.  We find an increase in dust mass surface density from $\sim 0.02$ to $\sim 1$ {\magi M}$_{\odot}$ pc$^{-2}$ as we move out radially from the galactic centre to a radius of $\sim 5$ kpc.  Between the inner ring at 5 kpc and the 10 kpc ring, we see a slight dip in $\Sigma_{\mathrm{dust}}$,  before seeing the values increase.   Between the 10 and 15 kpc rings,  we largely see the $\Sigma_{\mathrm{dust}}$ values plateauing,  with hundreds of pixels $\sim 0.3$ $M_{\odot}$ pc$^{-2}$.  In Figure \ref{fig:tempradvar},  we see the dust temperature decrease between the galactic centre and the inner ring at 5 kpc.  The temperature values plateau between 5 kpc and 13 kpc,  at which point there is some evidence for a decrease in temperature values.  In Figure \ref{fig:betaradvar},  we see $\beta$ gradually decrease as we move out radially from the galactic centre towards the 15 kpc ring.

\newpage
\subsection{Mapping the sub-mm excess}
\begin{figure}[h!]
\centering
\includegraphics[width=17cm]{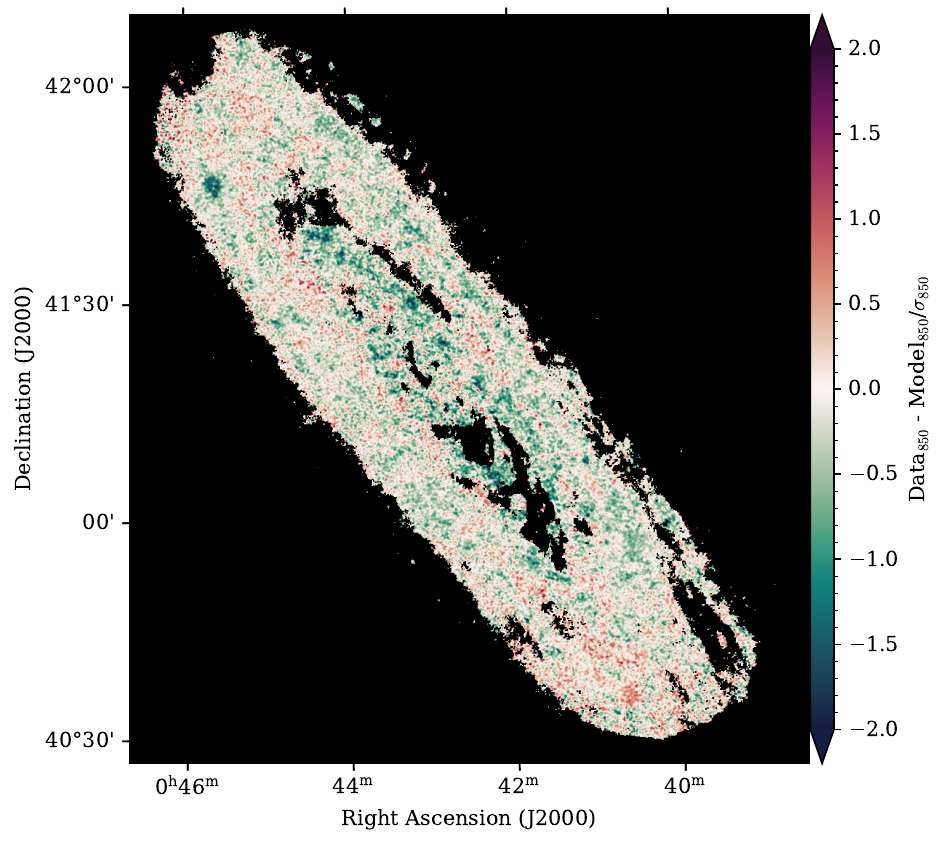} \caption{Distribution of pixels fit with a single temperature MBB model.  The colour map represents excess signal as a fraction of the background noise at 850 $\mu$m.  Pink areas show excess emission and green areas show a deficit in the observed emission compared to the model.}
\label{fig:excess850map}
\end{figure}

\begin{figure}[h!]
\centering
\includegraphics[width=17cm]{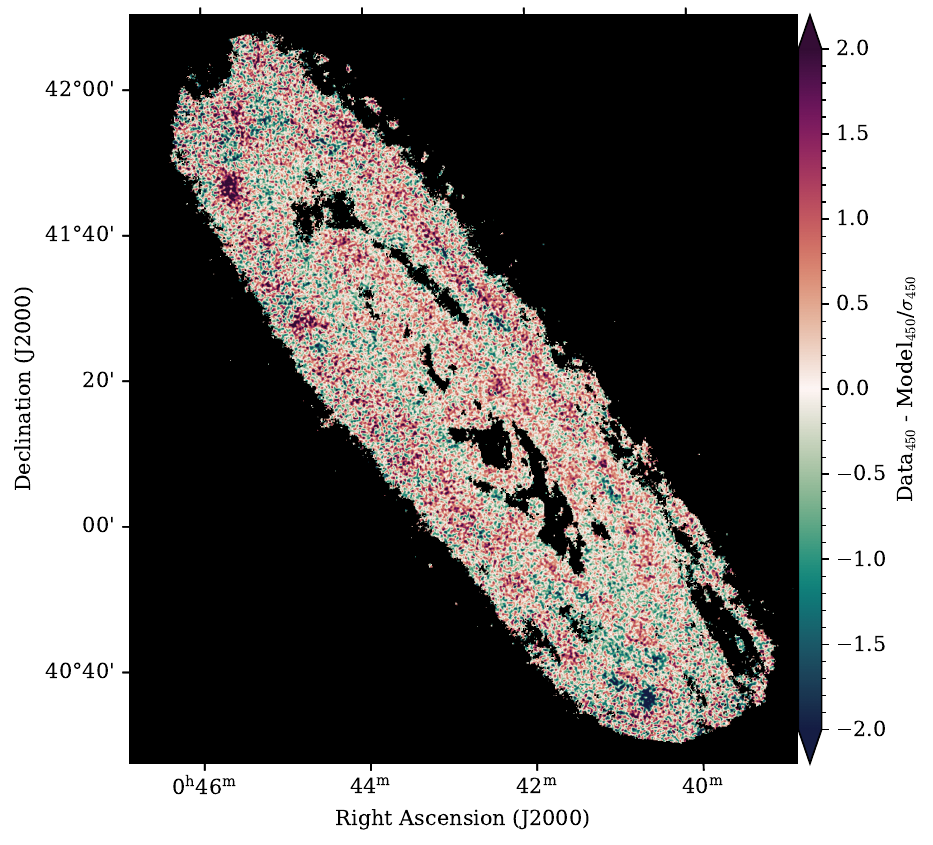} \caption{Distribution of pixels fit with a single temperature MBB model.  The colour map represents excess signal as a fraction of the background noise at 450 $\mu$m. Pink areas show excess emission and green areas show a deficit in the observed emission compared to the model.}
\label{fig:excess450map}
\end{figure}

To distinguish any excess or deficit in emission from the single temperature modified blackbody model,  we studied the residual emission ($\mathrm{Data}_{\lambda} - \mathrm{Model}_{\lambda}$) as a fraction of the background noise ($\sigma_{\lambda}$) in 850 and 450 $\mu$m observations:
\begin{equation}
\mathrm{Fractional \; excess} = \frac{\mathrm{Data}_{\lambda} - \mathrm{Model}_{\lambda}}{\sigma_{\lambda}}
\end{equation}
Figures \ref{fig:excess850map} and \ref{fig:excess450map} show our result.  We do not see any strong evidence for an excess at the long wavelength tail at 850 $\mu$m or even at 450$\mu$m.  There is a concentration of pixels in the outer ring (at $\approx$ RA: 00$^h$ 45$^m$ 37.97$^s$,  Dec: 41$^{\circ}$ 46' 17.09") which show a deficit  at 850 $\mu$m and an excess at 450$\mu$m.  This is also a region where we have higher $\chi^2$ values (median $\chi^2 \simeq 6$) and our model does not fit the data as well.

\begin{figure}[h!]
\centering
\includegraphics[width=18cm]{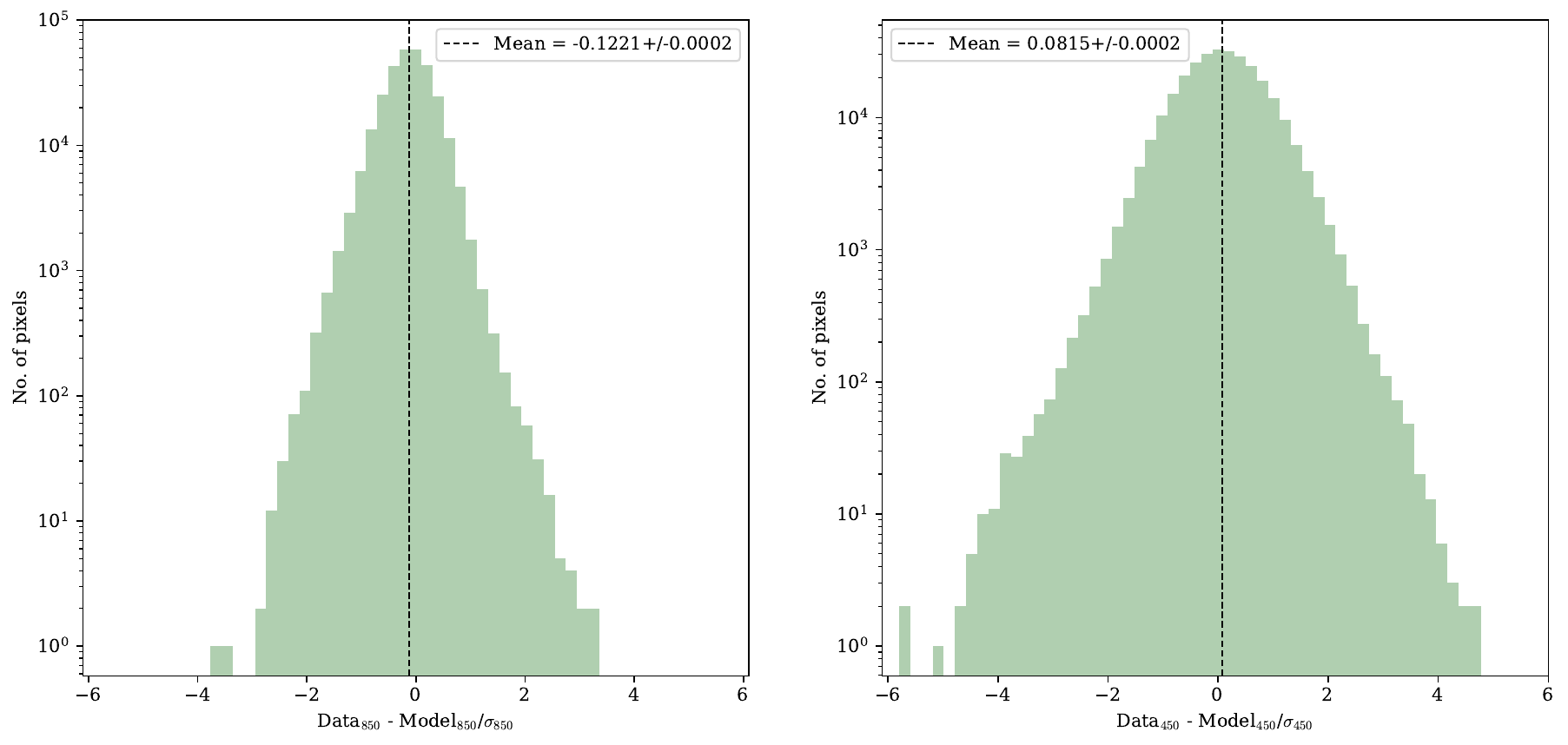} \caption{Histogram of pixel distribution showing the residual at 850 $\mu$m and 450 $\mu$m as a fraction of the background noise in the observational data. The vertical line shows the mean value of this fraction. }
\label{fig:excess850_overerr}
\end{figure}

In Figure \ref{fig:excess850_overerr},  we depict the fractional residual as a histogram.  Any deviation from the zero value suggests that we may be seeing an excess or deficit in emission at 850 and 450 $\mu$m compared to the model.  We find that the mean value is at -0.1 at 850 $\mu$m and 0.08 at 450 $\mu$m,  indicating that there is not much evidence for a sub-mm excess at both of these wavelengths.  Although certain groups of pixels in the maps in Figure \ref{fig:excess850map} \& \ref{fig:excess450map} show an excess,  only 4 pixels out of the total 297203 present with excess signal which is 3 times greater than the noise at 850 $\mu$m.  At 450 $\mu$m,  this total increases to $\approx$ 276 pixels.  {\magi To explore the possibility that the use of 850 $\mu$m observations in our fitting is skewing the result,  we have also tried fitting a single temperature MBB model excluding the 850 $\mu$m observations and then extrapolating the model intensity to this wavelength.  We found that our results were similar.}

\newpage

\newpage
\section{Discussion}
\label{sec:disc_sed}
\subsection{Radial variations in $T_{\mathrm{dust}}$ and $\beta$}
Previously,  \cite{Smith2012} have found radial variations in $\beta$ in M31,  with the value of $\beta$ decreasing from $\simeq$ 2.5 at a radius of 3 kpc to $\simeq$ 1.8 at 12 kpc. \cite{Draine2014} have also found,  using a different \textit{Herschel} dataset and a different method,  the same general trend; with $\beta$ decreasing from a value of $\simeq 2.35$ at a radius of 3 kpc to a value of $\simeq$ 2.0 at a radius of 12 kpc.  In the inner 3 kpc radii,  these authors find $\beta$ increasing from $\simeq 2$ to 2.5 with galactocentric radius.  Our result is different to the work of \cite{Smith2012} and \cite{Draine2014} in the central 3 kpc,  where we find no evidence for a sharp increase in $\beta$ with increasing radius but instead see a high initial $\beta$ values ($\beta \sim 2.8$) near the galactic centre and a steady decrease in $\beta$ with increasing radius.  This difference could be because by adding the 850 $\mu$m observations,  we are now able to better constrain $\beta$ values than these previous studies.  The temperature variations that we find across the galaxy are consistent with \cite{Smith2012} with majority of the pixels showing a decrease of $T_{\mathrm{dust}}$ from $\sim$ 24 K to $\sim$ 16 K with increasing radius in the inner 5 kpc and then a gradual increase from 16 to 20 K with radius beyond 5 kpc. 

The $\beta$-$T_{\mathrm{dust}}$ degeneracy is an effect of measurement errors for $\beta$ and temperature being correlated (\citealt{Shetty2009},  \citealt{Juvela2013}).  Since $\beta$ is high across many beam areas in our resulting maps (Figure \ref{fig:sed_betareswith850}),  and noise is not correlated over such a large area (only across neighbouring pixels in smaller areas),  we do not think the $\beta$-$T_{\mathrm{dust}}$ degeneracy is having an impact on our analysis.  We apply the Spearman-rank coefficient test to our $\beta$ and $T_{\mathrm{dust}}$ values and find a coefficient value $\approx$ -0.7 with a p-value $\ll$ 0.01. Statistically, this suggests a strong anti-correlation between the two properties with > 99\% confidence.  However,  by examining the $T_{\mathrm{dust}}$ and $\beta$ maps in Figures \ref{fig:sed_tempreswith850} \& \ref{fig:sed_betareswith850},  we can see that there is no obvious co-related trend between variations in $T_{\mathrm{dust}}$ and $\beta$ across the whole galaxy.  In practice,  what we statistically find as an anti-correlation might simply be a result of {\magi a real} decrease in $\beta$ with galactocentric radius and an increase in $T_{\mathrm{dust}}$ with galactocentric radius (\citealt{Smith2012}),  independently of each other. 

\subsection{Is there a sub-mm excess?}
The primary motivation behind our study was to search for any submilimetre excess at 450$\mu$m and 850 $\mu$m which may be indicative of cold dust dominance within M31 or might provide clues towards the intrinsic make-up of dust grains and, hence,  their emission properties.  We do not find any clear evidence for such an excess at either wavelength.  We use a simple single temperature modified blackbody model to describe the intensity of emission from dust grains at these wavelengths and find that this model sensibly fits our data.  Our results are complementary to the study by \cite{Smith2012} who do not find an excess at 500 $\mu$m, with the additional advantage that we now probe the longer wavelength emission at 850$\mu$m and examine higher resolution images.  As a first approximation,  our study is also consistent with the work of \cite{Draine2014} who use more sophisticated models from \cite{Draine2007} and find that their models slightly overpredict emission at 500 $\mu$m,  i.e.  without finding an excess.

It is possible to argue that that M31 does not contain very cold dust contributing to any excess emission at long wavelengths since we have no clear evidence for this.  The question then is,  how much dust is required to produce a significant sub-mm excess in M31? We provide a crude estimate of this.  We define excess emission as:
\begin{equation}
\mathrm{Excess} = \mathrm{Data} - \mathrm{Model} = I_{\nu, \; \mathrm{data}} - I_{\nu, \; \mathrm{model}}
\end{equation}
where $I_{\nu}$ is the flux density in Jy.  How much $I_{\nu, \; \mathrm{data}}$ is needed to see real excess emission 5 times above the background noise? That is,  can we calculate the $I_{\nu, \; \mathrm{data}}$ such that:
\begin{equation}
I_{\nu, \; \mathrm{data}} - I_{\nu, \; \mathrm{model}} = 5\sigma
\end{equation}
where $\sigma$ is the background noise at that particular wavelength.  Rearranging this equation,  we get:
\begin{equation}
I_{\nu, \; \mathrm{data}} =  5\sigma \; + \; I_{\nu, \; \mathrm{model}}
\end{equation}
For the pixel in the outer ring (Figure \ref{fig:sed_onepix_outer}),  this means that we would need to see emission of 0.002 Jy.  So in order to detect excess signal at 850 $\mu$m which is 5 times above the noise, and assuming that the dust is at a temperature of 10 K with a $\beta = 2.14$,  the required dust mass is:
\begin{equation}
M_{\mathrm{dust}} = \frac{I_{\nu, \; \mathrm{data}} \; D^{2}}{B_{\nu}(T_{\mathrm{dust}}=10 \; \mathrm{K}) \; \kappa_{\nu}(\beta=2.14)}
\end{equation}
We convert the dust mass to dust mass surface density by dividing by the area of one pixel (521 pc$^2$).
Therefore,  we find that to see real excess emission within our data which is 5 times the noise value at 850 $\mu$m in this pixel,  $1.4 \times 10^3$ M$_{\odot}$ of dust would be required.  This amounts to a dust mass surface density of $\sim 2.7$ M$_{\odot}$ pc$^{-2}$.  For even colder dust at 5 K,  we would require $\sim 9.2 \times 10^3$ $M_{\odot}$ of dust to produce excess emission 5 times above the background noise.  This amounts to a dust mass surface density of 17.6 M$_{\odot}$ pc$^{-2}$.  Our observations provide very little constraint on the amount of very cold dust in M31.  There would need to be $\frac{9.2 \times 10^3 \; \mathrm{M}_{\odot}}{312 \; \mathrm{M}_{\odot}} = 29$ times as much dust at 5 K as the dust we actually detect to be able to be sure that it exists.  This could be because cold dust is less efficient at radiating (e.g.  \citealt{Eales1989}). 

\section{Summary}
\label{sec:summary_sed}
In this chapter,  we have used 100, 160 and 250 $\mu$m observations from the \textit{Herschel Space Observatory} and new JCMT observations at 450 and 850 $\mu$m from the HASHTAG survey to hunt for a submillimetre excess in M31 at 68 pc spatial scales.  We define an excess as a significant observed deviation in emission from the predicted emission by a single temperature modified blackbody model at 450 and 850 $\mu$m.  We also examine the effect of using a fixed vs variable dust emissivity index ($\beta$) on constraining the dust mass and temperature across M31.  Our key results and conclusions are as follows:
\begin{enumerate}
\item We have produced maps showing the spatial distribution of $\Sigma_{\mathrm{dust}}$,  $T_{\mathrm{dust}}$,  $\beta$ from SED fitting,  for the first time incorporating new observations at 450 and 850 $\mu$m.
\item We find that a variable dust emissivity index ($\beta$) provides a statistically better fit of the MBB model to our data,  consistent with past studies showing radial variations in this parameter across M31. The next best model is a single temperature MBB model with a fixed $\beta$ of 2.
\item We find no strong evidence for excess emission at long wavelengths,  in particular at 850 $\mu$m suggesting that our observations are not sensitive to cold dust.  To see a real astrophysical excess that is 5 times above the noise level,  we would require dust mass at least one order of magnitude higher.  Therefore,  we cannot rule out the possibility that is still a large amount of cold dust in M31 masked by the emission from warmer dust.
\end{enumerate}








%% file: Chapters/chapter_dustmass.tex
\chapquote{``We open doors so others can walk through them."}{Alexandria Ocasio-Cortez}{}

\section{Introduction}
Over the past decade,  a method that has been increasingly used for estimating the mass of gas in a galaxy has involved determining the mass of dust from far-infrared observations and using it as a tracer for gas mass (\citealt{Eales2012}, \citealt{Tacconi2018}, \citealt{Liang2018}).  This is an independent method from the use of traditional tracers like carbon monoxide (CO) and direct emission from \textsc{HI}.  For looking at molecular gas in particular,  many studies (e.g. \citealt{Hughes2014},  \citealt{Rice2016},  \citealt{Rosolowsky2021}) have traditionally used CO emission because this molecule is abundant in regions of H$_2$ and can be excited at low temperatures to instigate emission.  However,  a problem with using CO as a tracer is that FUV radiation from stars can photodissociate CO into atomic carbon and oxygen.  This means that by using this tracer,  we may be missing regions of H$_2$ in which CO has been photodissociated.

In the Milky Way (MW),  it has been increasingly shown that CO-dark molecular gas exists by studies using a variety of observational tracers.  For example,  \cite{PlanckCollaboration2011} have used all sky maps to show that the correlation between the optical depth of dust at a frequency of 857 GHz and the combined atomic and molecular gas column density inferred from CO deviate from a linear relationship in a certain column density range (see Section \ref{ssec:gmc_ppmap} in Chapter 2 for details).  The authors attribute this deviation from the linear relationship to CO-dark molecular gas.  {\magi \cite{Paradis2012a}} have {\magi applied a similar methodology to} the \textit{Planck} maps {\magi but using extinction data} to show a decrease in the dark gas fraction from 71\% to 43 \% going from the inner {\magi Galaxy} to the outer {\magi Galaxy.  Their study excludes the Galactic plane where most of the molecular clouds are located.}

{\magi However, } studies of molecular clouds in the MW have also shown that the fraction of CO-dark gas within individual clouds is not insignificant.  For example,  \cite{Velusamy2010} have used [CII] emission in 53 clouds to probe the transition stage at which a cloud goes from being predominantly atomic to molecular gas,  finding that $\sim$ 25 \% of the total molecular hydrogen gas is CO-dark.  \cite{Lee2012} have studied the Perseus molecular cloud using dust,  CO and HI observations to find that the dark gas fraction amounts to as much as 30\% in Perseus.  CO-dark gas has also been detected in the Magellanic Clouds.  \cite{Pineda2017} used deep \textit{Herschel} observations across 54 sight lines to show that 89\% of H$_2$ in the LMC is CO-dark and 77\% in the SMC is the same.  \cite{Chevance2020} have found that $\sim$ 75\% of H$_2$ in 30 Doradus in the LMC is CO-dark.

In M31,  CO-dark gas is yet to be detected (\citealt{Smith2012}).  Our previous work (Chapter 2,  \citealt{Athikkat-Eknath2022}) using \textsc{PPMAP} estimates of dust surface density showed some clouds traced by dust with low CO emission.  However,  due to low resolution \textsc{HI} data,  we were unable to rule out whether these clouds were dominated by atomic hydrogen and if they were truly CO-dark.

A possible solution to the CO-dark gas problem is to use an alternative tracer for identifying molecular clouds.  Interstellar dust is an apt tracer of molecular gas regions as it is usually found shielding cold,  dense regions and is not photodissociated,  unlike CO.  Furthermore,  while many studies have used CO emission to trace molecular gas and compare cloud properties,  very few authors (e.g.  \citealt{Kirk2015}, \citealt{Williams2019},  \citealt{Athikkat-Eknath2022}) have created dust-selected molecular cloud catalogues for nearby galaxies.  Motivated by the unreliability of the CO molecule in tracing all of the molecular hydrogen,  and with the availability of the new HASHTAG data at 450 $\mu$m and 850 $\mu$m and the dust mass surface density map produced in the previous chapter from SED fitting,  the goal of this study is to create a catalogue of clouds in an alternative way,  using dust to trace the molecular gas,  and then to examine how much CO-traced molecular gas and atomic hydrogen is in our dust-traced clouds.

This chapter is structured as follows: we first describe the observational data that are used in this work (Section \ref{sec:observe}). Next, we outline our source extraction methodology and how we obtain cloud properties (Section \ref{sec:dustmass_methods}),  followed by our results (Section \ref{sec:results}) and discussion (Section \ref{sec:dustmass_disc}).  Finally,  we summarise our conclusions in Section \ref{sec:dustmass_summary}.

\section{Observations}
\label{sec:observe}
We use the dust mass surface density map obtained from SED fitting of M31 observations from Chapter {\magi 3} (Figure \ref{fig:sed_surfreswith850}) to extract molecular clouds.  To examine the CO-traced molecular gas content,  we use a map of  $^{12}$CO(J=1-0) integrated intensity obtained from observations of M31 made using the IRAM 30 m telescope (\citealt{Nieten2006}) at an angular resolution of 23".  The data have a pixel scale size of 2.4" which we reproject into 6" to match the pixel scale of the 250 $\mu$m \textit{Herschel} observations. The resolution of this CO map is lower than the resolution of the 250 $\mu$m observations; however,  we use this map to capture a rough estimate of the CO contribution in a cloud across regions where high resolution data are unavailable. 

{\magi We use the same map of $^{12}$CO(J=1-0) integrated intensity observed using the CARMA interferometer as described in Chapter 2,  Section \ref{ssec:carmaobs} (\citealt{Caldu-Primo2016}). These observations cover approximately one-third of M31. } For our analysis, we smooth 5.5" $^{12}$CO(J=1-0) map with a Gaussian kernel to obtain a resulting map with a FWHM of 18" - the resolution of the 250 $\mu$m observations. The map is then reprojected to match the 6" pixel scale size of the 250 $\mu$m map for consistent pixel-by-pixel analysis. 

We use the \textsc{HI} column density map of M31 obtained by \cite{Braun2009} using the Westerbork Synthesis Radio Telescope (WSRT) at an angular resolution of 30" and spatial resolution of $\approx$ 110 pc.  The data have a pixel scale size of 10" which we reproject into 6" to match the pixel scale of the 250 $\mu$m observations. The resolution of this \textsc{HI} map is lower than the resolution of the 250 $\mu$m observations; however,  we use this map to capture a rough estimate of the \textsc{HI} contribution in regions where high resolution data are unavailable.  Further details of this map are in Section \ref{ssec:HIobs} and \cite{Braun2009}. 

A smaller portion of M31,  overlapping with the CARMA and PHAT observations,  was very recently observed using the \textit{Extended Very Large Array} (EVLA; \citealt{Koch2021}) to capture the \textsc{HI} intensity.  The interferometric data were combined with single-dish observations of the MW and the Local Volume taken by the Effelsberg-Bonn \textsc{HI} Survey (EBHIS; \citealt{Winkel2016}). The original data, with an angular resolution of 10",  have been convolved with a Gaussian kernel to match the resolution of the 250 $\mu$m map at 18".  The data have also been reprojected from a pixel scale size of 0.9" to 6".  We do not apply a signal-to-noise cut for our CO and HI maps.

\section{Methods}
\label{sec:dustmass_methods}

\subsection{Source extraction}
\label{sec:dustmass_sourceextract}

We use a dendrogram to identify the denser regions nested within the $\Sigma_{\mathrm{dust}}$ map.  We compute a dendrogram by providing the three required parameters for the Python package \texttt{astrodendro} (\citealt{Rosolowsky2008}): the minimum dust mass surface density value of a pixel (\texttt{min\_value}),  the minimum significance value for a nested dense structure to be identified as an independent object (\texttt{min\_delta}) and the minimum size threshold (defined in number of pixels) for the nested object (\texttt{min\_npix}).  We choose the same \texttt{min\_value} as that used in our \textsc{PPMAP} surface density source extraction (Section \ref{sec:se}) but adjust the \texttt{min\_delta} by visually checking whether the dendrogram's segregation attempt closely matches what we see by eye.  Our \texttt{min\_npix} again is identical to that used in Section \ref{sec:se} as this ensures that the smallest cloud size is greater than the beam size ($\approx$ 7 pixels).  The dendrogram parameter values are listed in Table \ref{tab:param_sf}.  Following a similar framework as Section \ref{sec:se},  we classify the `leaves' of the dendrogram as our molecular clouds. 

\begin{table}[h!]
\caption{Input dendrogram parameters for source extraction.}              
\label{tab:param_sf}      
\centering                                      
\begin{tabular}{c c c c}          
\hline\hline                        
Data & Minimum value  & Minimum structure significance value & Minimum no. of pixels \\    
& \texttt{min\_value} & \texttt{min\_delta} & \texttt{min\_npix} \\
\hline                                   
     $\Sigma_{\mathrm{dust}}$ map & 0.44 M$_{\odot}$ pc$^{-2}$ & 0.05 M$_{\odot}$ pc$^{-2}$ & 10 \\
\hline                                             
\end{tabular}
\end{table}

\subsection{Obtaining dust and gas properties for each cloud}
\label{ssec:hashtag_gasmethods}
We calculate the total dust mass of each cloud as the sum of the dust masses, obtained from SED fitting,  in the pixels of each cloud.  We calculate the dust-mass-weighted mean $T_{\mathrm{dust}}$ and mean $\beta$ for our clouds using the pixels corresponding to our dendrogram leaves in the $T_{\mathrm{dust}}$ and $\beta$ maps obtained from SED fitting (see Figures \ref{fig:sed_tempreswith850} \& \ref{fig:sed_betareswith850} from Chapter 3).

We determine the total gas mass of each cloud in two ways.  In the first method,  we use dust as a tracer of the total gas,  hence obtaining the total ISM gas mass of each cloud.  We multiply the dust mass in each pixel within a cloud by a constant GDR of 100 (\citealt{Hildebrand1983},  \citealt{Marsh2015}) to give the total gas mass of the cloud:
\begin{equation}
M_{\mathrm{dust-traced-gas}} =  100 \times M_{\mathrm{dust}}
\end{equation}
\noindent Initially,  we use the constant GDR to match the one used by \cite{Marsh2015}.  We assess whether this is a sensible value of GDR to assume in Section \ref{ssec:offsetgdr}.

In our second method,  we use $^{12}$CO(J=1-0) to emission to trace molecular gas and \textsc{HI} emission to identify atomic gas.  We first identify the regions corresponding to our dusty cloud in the high and low resolution $^{12}$CO(J=1-0) and \textsc{HI} maps. We convert the intensity of  $^{12}$CO(J=1-0) shown in both our IRAM and CARMA maps and \textsc{HI} emission from the VLA map into the mass of molecular hydrogen and atomic hydrogen as follows:
\begin{equation}
M_{\mathrm{H_2}} =  X_\mathrm{CO}  \times I_\mathrm{CO} \times m(\mathrm{H_2}) \times A_{\mathrm{pix}}
\end{equation}
where $m(\mathrm{H_2})$ is the mass of a hydrogen molecule and $I_\mathrm{CO}$ is the intensity of emission in K km s$^{-1}$,   $X_{\mathrm{CO}}$ is a constant of proportionality,  $1.9 \times 10^{20}$ cm$^{-2}$ [K km s$^{-1}$]$^{-1}$ (\citealt{Strong1996}).  $A_{\mathrm{pix}}$ is the area of one pixel.
\begin{equation}
M_{\mathrm{\textsc{HI}}} = 1.8 \times 10^{18} \times I_\mathrm{\textsc{HI}} \times m(\mathrm{H}) \times A_{\mathrm{pix}}
\end{equation}
where $I_\mathrm{\textsc{HI}}$ is the intensity of emission in K km s$^{-1}$ and $m(\mathrm{H})$ is the mass of a hydrogen atom.  Since the Westerbork map is in units of \textsc{HI} column density,  we simply multiply this by mass of hydrogen. 

\noindent We add up the mass of molecular and atomic hydrogen in all of the pixels within the cloud region to obtain the total gas mass of the cloud.  Where we have high resolution $^{12}$CO(J=1-0) and \textsc{HI} intensity values available for all of the cloud,  these values are selected for calculating the gas mass.  Otherwise,  we use the values from the lower resolution $^{12}$CO(J=1-0) and \textsc{HI} maps as a best estimate of the total gas mass.  We note that the values from these maps are an approximation of the contribution due to the larger beam size of the lower resolution maps.

As such,  our two methods give us the total dust-traced gas mass and total molecular gas + atomic gas mass of each cloud.  We do not account for helium in our gas mass calculations.  We calculate the radial distance of a cloud from the centre of M31 {\magi in the same way as in Chapter 2 (Section \ref{ssec:gmc_co})},  and assume an inclination angle of 77$^{\circ}$ (\citealt{Fritz2012}). 

\section{Results}
\label{sec:results}

\subsection{Molecular cloud catalogue}
\label{ssec:cloudcat_hashtag}
\begin{figure}[h!]
\centering
\includegraphics[width=16cm]{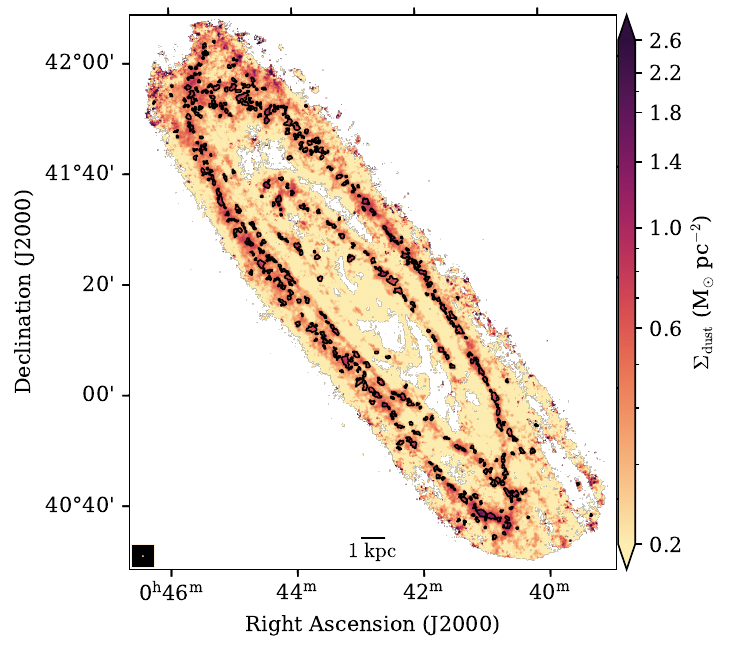} \caption{Map of dust mass surface density obtained from SED fitting of 100, 160, 250, 450 and 850 $\mu$m observations.  Black contours show the 422 extracted sources using a dendrogram.}
\label{fig:cloud_extract}
\end{figure}

We have produced a new catalogue of dust-traced molecular clouds from \textit{Herschel} and JCMT observations (Appendix \ref{ap:hashtagcloudcat}).  In this section,  we provide our analysis of the dust and gas properties of these clouds.  Out of the initial 893 clouds extracted,  we select clouds which have a peak dust mass surface density with at least 5 times the uncertainty value produced during SED fitting.  This gives us 422 clouds for our catalogue.  Figure \ref{fig:cloud_extract} shows the boundaries for these 422 sources as contours on our dust mass surface density map.

\subsection{Cloud mass function}
\begin{figure*}[h]
\centering
\includegraphics[width=16cm]{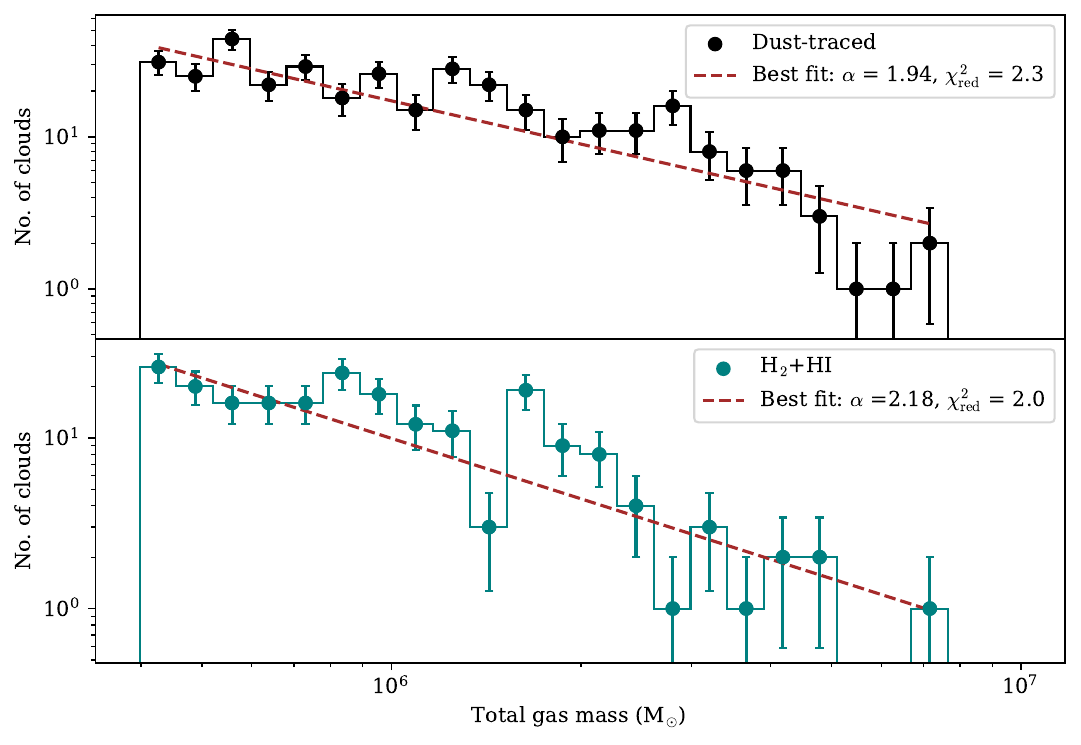} \caption{Cloud mass functions for total gas mass traced by dust (black) and H$_2$+\textsc{HI} (teal), for cloud masses greater than 10$^{5.6}$ M$_{\odot}$. The histograms show the number of clouds per mass interval. The scatter points show the central mass value in each bin. The brown dashed line shows the best fit power law to the cloud mass function.}
\label{fig:cmf_hashtag}
\end{figure*}

Figure \ref{fig:cmf_hashtag} shows the distribution of total gas mass in our extracted clouds obtained using the methods described in Section \ref{ssec:hashtag_gasmethods}.  We fit a power law model to our distribution of cloud masses.  We use the \texttt{SciPy} Levenberg-Marquardt least-squares minimisation package \texttt{lmfit} to fit the power law:
\begin{equation}
\label{eq:7}
  \mathrm{dN}_{\mathrm{cloud}} = \mathrm{k}M_{\mathrm{gas}}^{-\alpha} \; dM_{\mathrm{gas}}
\end{equation}
where dN$_{\mathrm{cloud}}$ is the number of molecular clouds in the mass interval ($M_{\mathrm{gas}}$, $M_{\mathrm{gas}}$ + $dM_{\mathrm{gas}}$),  k is a normalisation factor, $M_{\mathrm{gas}}$ is the total gas mass of the cloud,  and $\alpha$ is the power law exponent.

The total gas mass of the clouds as traced by dust ranges from 2.4 $\times$ 10$^5$ M$_{\odot}$ $\le$ M$_{\mathrm{dust-traced \; gas}}$ $\le$ 1.1 $\times$ 10$^7$ M$_{\odot}$.  The total molecular + atomic gas mass of the clouds ranges from 6.2 $\times$ 10$^4$ M$_{\odot}$ $\le$ M$_{\mathrm{H}_2}$ +  M$_{\mathrm{HI}}$ $\le$ 7.6 $\times$ 10$^6$ M$_{\odot}$.  We only include clouds with a total gas mass $>10^{5.6}$ M$_{\odot}$ (which is $\simeq 4 \times 10^{5}$ M$_{\odot}$) in our fitting because the decrease in the number of clouds at lower masses suggests that our mass functions are increasingly incomplete at lower masses.  This threshold also surpasses the median value of cloud masses found using both dust and molecular+atomic gas.  We bin data from both catalogues into 25 bins which are equidistant in logarithimic space between the mass limits 10$^{5.6}$ M$_{\odot}$ and 10$^{7}$ M$_{\odot}$.  Our fits are performed in linear space and we assume Poisson errors ($\sqrt{\mathrm{dN}_{\mathrm{cloud}}}$).

The best fit $\alpha$ values for dust-traced cloud masses is $1.94 \pm 0.11$ and for molecular + atomic cloud masses is $2.18 \pm 0.16$.  The reduced $\chi^2$ parameter for our fits are given Figure \ref{fig:cmf_hashtag}.  For comparison,  in the MW,  \cite{Rice2016} have found a steeper slope for clouds in the outer galaxy ($\alpha \gtrsim 2$) and a shallower slope for the inner galaxy ($\alpha \sim 1.8$) by using CO to trace clouds (see their Table 6).  In nearby galaxies,  the PHANGS team have found a steeper slope ($\alpha \gtrsim 2.1$; {\magi \citealt{Rosolowsky2021}} - see their Table 5),  again for clouds traced by CO emission.  Our values of $\alpha$ are lower than the value for M31 obtained by \cite{Kirk2015} using \textit{Herschel} dust continuum observations  ($\alpha$ = 2.34 $\pm$ 0.12).  Our values are similar to $\alpha = 1.98 \pm 0.24$ for CO-traced clouds found in our own previous work in Chapter 2 (Section \ref{sec:cloud_mass}).  Both of our $\alpha$ values from this work are also similar to $\alpha = 2.06 \pm 0.14$ which we found in Chapter 2 for dust-traced clouds.

\subsection{Comparing different tracers of gas}
\begin{figure}[h!]
\centering
\includegraphics[width=16cm]{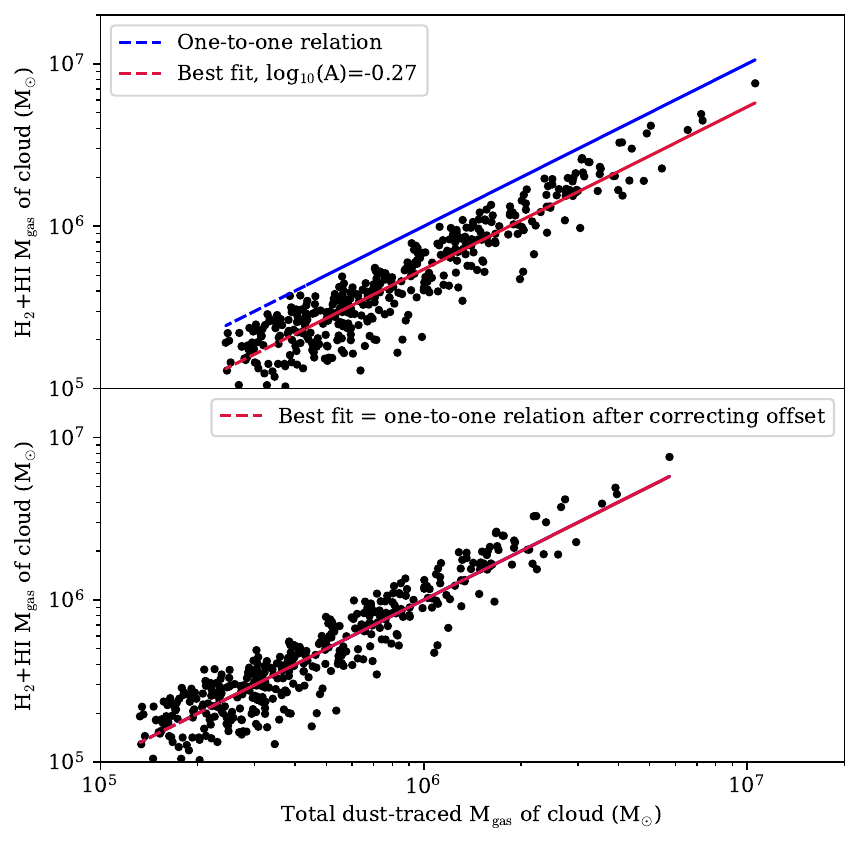} \caption{\textit{Top row:} Dust-traced gas vs CO+\textsc{HI}-traced gas mass of clouds (black).  The red dashed line shows our best fit to the data in logarithmic space assuming a fixed slope of unity.  The blue dashed line shows the one-to-one relation line between the two properties. \textit{Bottom row:} Dust-traced gas vs CO+\textsc{HI}-traced gas mass of clouds after applying multiplicative correction factor. The best fit line is identical to the one-to-one relation.}
\label{fig:dustvsgas_cloudmass}
\end{figure}

We compare the dust-traced total gas mass with the molecular + atomic gas mass of each cloud.  Due to missing values in the CO maps,  we remove three clouds from our fits and fit over 419 cloud masses.  Figure \ref{fig:dustvsgas_cloudmass} shows our result.  We find a strong correlation between the gas traced by dust and the molecular + atomic gas in each cloud.  We perform a Spearman rank correlation test and find a Spearman rank coefficient of 0.92 and a p-value of $1.8 \times 10^{-172}$,  i.e.  p-value $\ll 0.01$. This indicates that there is a very strong positive correlation between the total dust-traced gas mass and the total molecular + atomic gas mass of the clouds,  with statistical significance of greater than 99\%.  This suggests that dust is a good tracer of the total gas content of these clouds. 

We fit a power law in logarithmic space to the distribution of clouds (Figure \ref{fig:dustvsgas_cloudmass}) using the least squares minimisation Python package \texttt{lmfit}:
\begin{equation}
\mathrm{log}_{10}(M_{\mathrm{H_2}} + M_{\mathrm{\textsc{HI}}}) = \mathrm{log}_{10}(\mathrm{A}) + \mathrm{N} \; \mathrm{log}_{10}(M_{\mathrm{dust-traced \ gas}})
\end{equation}
We allow the intercept parameter,  A,  to vary and fix the slope,  N,  to 1.  Since the {\magi observational noise} in the dust-traced gas mass and the atomic + molecular gas mass {\magi is} much smaller than the intrinsic scatter between our cloud masses,  we do not weight by errors during the fitting.  We calculate the dispersion in the total molecular + atomic gas mass distribution as the standard deviation:
\begin{equation}
\label{eq:dustmass1}
\mathrm{Dispersion} = \sqrt{\frac{\Sigma (\mathrm{log}_{10} \; M_{\mathrm{gas,  \; data}} - \mathrm{log}_{10} \; M_{\mathrm{gas,  \; model}})^2}{n_{\mathrm{data}}}}
\end{equation}
where $M_{\mathrm{gas,  \; data}}$ is the total molecular + atomic gas mass obtained from the data (i.e.  $M_{\mathrm{H_2}} + M_{\mathrm{\textsc{HI}}}$),  $M_{\mathrm{gas,  \; model}}$ is the total molecular + atomic gas mass obtained from the best fit model,  and $n_{\mathrm{data}}$ is the number of data points.  We find a dispersion of 0.13 dex. 

We find a best fit intercept of $\mathrm{log}_{10}\mathrm{A} = -0.27 \pm 0.01$.  This means that there is an offset of $\sim 0.5$ in linear space from the one-to-one relation.

\section{Discussion}
\label{sec:dustmass_disc}

Our catalogue is the third dust-traced cloud catalogue for M31 known to date.  Previously,  \cite{Kirk2015} have produced a dust-traced molecular cloud catalogue from \textit{Herschel} observations and our own study (Chapter 2; \citealt{Athikkat-Eknath2022}) has produced a catalogue of clouds extracted from PPMAP's dust surface density measurements which were derived from \textit{Herschel} observations.  Our data are more {\magi complete} than previous work as they include direct observations of emission at wavelengths above 500 $\mu$m and have not been convolved to the lowest resolution \textit{Herschel} image, at 36".  Instead,  we are able to probe 68 pc spatial scales and for the first time,  colder dust than previous studies. 

The total CO-traced molecular cloud mass across our 419 clouds ranges from $7.2 \times 10^2 M_{\odot} \le M_{\mathrm{H}_2} \le 2.8 \times 10^6 M_{\odot}$ {\magi (before offset correction)}.  In comparison to the dust-selected {\magi catalogue} produced in Chapter 2,  where the molecular cloud mass ranged from $1.8 \times 10^4 M_{\odot} \leq M_{\mathrm{H}_2} \leq 1.3 \times 10^6 M_{\odot}$,  we see a wider range of molecular gas masses with this work.

\subsection{Offset in the gas-to-dust ratio}
\label{ssec:offsetgdr}
In Figure \ref{fig:dustvsgas_cloudmass},  we showed  the dust-traced gas mass vs the molecular + atomic gas mass.  There is an offset in our best fit relation (in red) and the expected one-to-one relationship line (in blue).  Since the slope of our best fit line is fixed at unity,  this offset could imply two things: either our assumed constant GDR value is wrong or there is some missing gas mass untraced by one or more of the observational tracers.  If we assume that our initial assumption of GDR = 100 is incorrect,  the offset of $0.54$ in linear space becomes a multiplicative factor to find the true GDR.  Hence,  we would need a GDR of $\simeq 54$ to reproduce a one-to-one relation between the dust-traced gas mass and the molecular + atomic gas mass.  The best empirical measurements of the GDR for the Milky Way to date have been published in the work of \cite{Roman-Duval2022}. The authors find a GDR of 167.2 (calculated from the integrated dust-to-gas ratio for the MW of $5.98 \times 10^{-3}$ provided in their Table 5) using depletion measurements.  Their value is even more contrasting to the value that we obtain for M31 and would make our offset larger if we assume this instead is the correct GDR for M31.

\begin{figure}[h!]
\centering
\includegraphics[width=12cm]{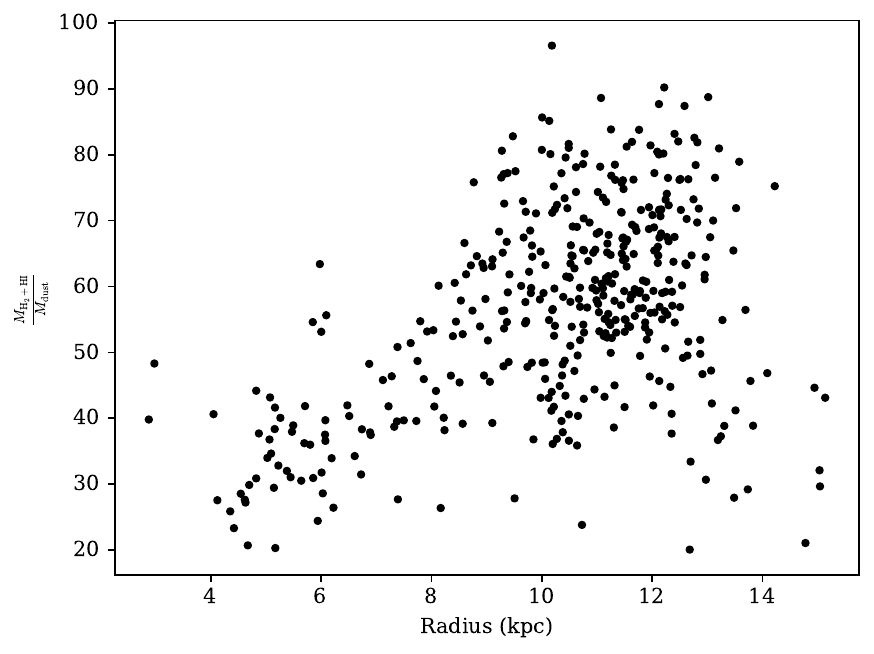} \caption{The galactocentric radius vs gas-to-dust ratio of each cloud for 419 clouds.}
\label{fig:gdrrad}
\end{figure}

Studies have also shown that the gas-to-dust ratio in M31 varies radially and with ISM surface density (e.g. \citealt{Walterbos1988},  \citealt{Smith2012},  \citealt{Clark2023}).  We examine our measured gas-to-dust ratio vs radius in Figure \ref{fig:gdrrad}.  We see a general increase in GDR values from 20 to 90 between 4 and 10 kpc,  followed by a lot of scatter in the GDR in the outer ring at 10 kpc.  Beyond 12 kpc,  the GDR values fall below 50.  The radial increase in the GDR up to a galactocentric distance of 10 kpc is consistent with the steady increase in GDR values with radius seen by \cite{Smith2012}.  However,  the large scatter (up to a factor of 2 difference) at the 10 kpc radius and decrease in the GDR beyond 12 kpc is not consistent with their pixel-by-pixel analysis.  The most likely explanation for the general increase in GDR with increasing radius is the negative metallicity gradient seen in M31 (\citealt{Galarza1999}) as GDR is linearly anti-correlated with metallicity (e.g.  \citealt{Sandstrom2013}). The clouds with low values of GDR are interesting because these could be ones with CO-dark gas\footnote{Scaling by the GDR value from \cite{Roman-Duval2022} would keep the general trend found in Figure \ref{fig:gdrrad} the same but would change the true values of our measured GDR (y-axis of Figure \ref{fig:gdrrad}). We would still have some clouds with low values of GDR when compared to the vast majority of the clouds.}.

\begin{figure}[h!]
\centering
\includegraphics[width=12cm]{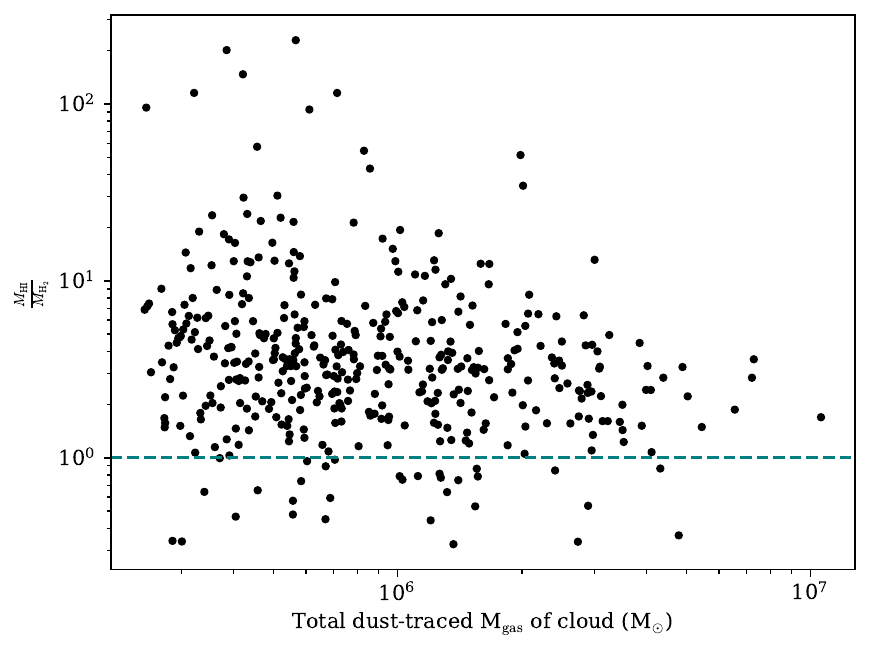} \caption{The ratio of atomic gas to CO-traced molecular gas as a function of the total dust-traced gas mass in each cloud for 419 clouds.}
\label{fig:cohigasmass}
\end{figure}

Alternatively,  since there is evidence that approximately one-third of the molecular gas in the MW is CO-dark (\citealt{Pineda2013}),  our assumption that CO is tracing all of the molecular gas in the clouds of M31 may be wrong.  Our 54\% offset value is also roughly in agreement with \cite{Abdo2010} who find that CO-dark gas, which is traced by an excess in dust and $\gamma$-ray emission in the Gould Belt,  contributes to 50\% of all molecular gas traced by CO.  Therefore,  here we examine if the offset could be a result of a fraction of missing molecular gas.  In this case:
\begin{equation}
\mathrm{GDR} = \frac{M_{\mathrm{H_2}} + M_{\mathrm{H_2},  \; \mathrm{CO-dark}} + M_{\mathrm{\textsc{HI}}}}{M_{\mathrm{dust}}}
\end{equation}
Since we assume a constant GDR of 100:
\begin{equation}
100 = \frac{M_{\mathrm{H_2}} + M_{\mathrm{H_2},  \; \mathrm{CO-dark}} + M_{\mathrm{\textsc{HI}}}}{M_{\mathrm{dust}}}
\end{equation}
\begin{equation}
M_{\mathrm{H_2},  \; \mathrm{CO-dark}} = (100 \times M_{\mathrm{dust}}) - M_{\mathrm{H_2}} - M_{\mathrm{\textsc{HI}}}
\end{equation}
We find a 50th$^{\mathrm{84th}}_{\mathrm{16th}}$ {\magi percentile} CO-dark gas mass value of $3.3^{8.0}_{1.6} \times 10^5$ M$_{\odot}$. The total CO-dark gas mass of all clouds is $2.0 \times 10^8$ M$_{\odot}$.  This means that CO-dark gas forms $41.7 \%$ of the total dust-traced gas mass ($4.9 \times 10^8$ M$_{\odot}$) across all clouds when assuming a GDR of 100.  If we assume that the GDR of M31 matches the GDR of the MW obtained from the best empirical depletion measurements to date from \cite{Roman-Duval2022},  this percentage would need to be 108\%,  suggesting that a very large unfeasible amount of the molecular gas would need to be CO-dark. Therefore,  the most likely explanation is that the GDR value which we have assumed is incorrect.

To examine in another way whether we are seeing any molecular gas clouds which are CO-dark or if our clouds are largely \textsc{HI} dominated,  we study the ratio of atomic gas to CO-traced molecular gas in each cloud.   Figure \ref{fig:cohigasmass} shows the ratio of atomic gas to CO-traced molecular gas as a function of the total dust-traced gas mass in each cloud (assuming a GDR of 100).  The teal dashed line in this figure shows the threshold at which a cloud is composed of equal amounts of CO-traced molecular gas and atomic gas.  Any clouds above this line are \textsc{HI} dominated.  In our sample of clouds,  we find that majority of the clouds fall above this threshold.  This implies that majority of our clouds are \textsc{HI} dominated.  Once again it is clear that if our clouds are largely \textsc{HI} dominated,  then the existence of CO-dark gas does not explain the 54\% offset between the dust tracer method and the \textsc{HI}+H$_2$ gas tracer method.  However,  we cannot rule out the possibility that a fraction of the molecular gas within these clouds is still CO-dark. 

{\magi \subsubsection{Intrinsic uncertainties}
The offset is affected by how well we can obtain our gas measurements and dust measurements.  Below we list the possible issues which could impact the offset that we are seeing between the dust-traced gas masses and the CO+HI traced gas masses:
\begin{itemize}
\item The conversion factor applied to the observed CO intensity (e.g. \citealt{Bolatto2013}) might vary by a factor of 2.  The aforementioned variations in metallicity seen in M31 could be correlated with the $X_{\mathrm{CO}}$ factor (e.g.  \citealt{Chiang2023}).  For example,  a decrease in metallicity will mean less CO compared to H$_2$ and an increase in $X_{\mathrm{CO}}$ to make up the same amount of H$_2$.  If we use a different $X_{\mathrm{CO}}$,  this becomes a scaling factor in our molecular gas measurements. 
\item The $\kappa$ value in galaxies is very uncertain and negatively correlated with ISM surface density (e.g.  \citealt{Clark2019}).  $\kappa$ in external galaxies could very up to a factor of $\approx 5$ (\citealt{Clark2019}).  $\kappa$ can be affected by dust grain size,  morphology,  and chemical composition - properties which may be degenerate with each other.  While there are dust models which prescribe these properties,  they are hard to constrain observationally (\citealt{Whittet1992}).  $\kappa$ is inversely proportional to the GDR and using a different $\kappa$ will affect the GDR as such. 
\end{itemize}}

Therefore, a number of factors exist which could offset the one-to-one relation between dust-traced gas mass and molecular + atomic gas mass in this work and the 54\% offset might very well be caused by a complex combination of these factors but is probably mostly driven by the incorrect assumption of GDR = 100.

\section{Summary}
\label{sec:dustmass_summary}
We have used the dust mass surface density measurements obtained from SED fitting of \textit{Herschel} and JCMT observations of M31 to trace molecular clouds and produce a cloud catalogue.  We have compared the use of dust as a tracer of interstellar gas against combined CO-traced molecular and atomic gas,  particularly motivated by the existence of CO-dark molecular gas in the MW and nearby galaxies.  Our key results are as follows:
\begin{enumerate}
\item We produce a catalogue of 422 dust-traced clouds, probing a spatial resolution of 68 pc.
\item We find a very strong positive correlation between the total dust-traced gas mass and the total molecular + atomic gas mass of the clouds,  with statistical significance of greater than 99\%,  suggesting that dust is a good tracer of the ISM gas mass at the scale of individual molecular clouds.
\item We show that there is a discrepancy between the amount of gas mass in the clouds in M31 which are traced by dust vs combined CO and \textsc{HI}.  We propose that this discrepancy could be due to an incorrect assumption of the constant gas-to-dust ratio or due to contributions from CO-dark gas.  From our CO-dark gas fraction calculation,  an unfeasible amount of molecular gas in our clouds would need to be CO-dark could to rectify the discrepancy.  Moreover, a large number of our clouds are \textsc{HI}-dominated.  Therefore,  variations in the GDR value is a more likely explanation for the offset.
\item We find a large variation in the GDR of clouds with galactocentric radius and a general trend of increase in GDR with increasing radius,  which could be due to M31's negative metallicity gradient.
\end{enumerate}

%% file: Chapters/chapter_starformation.tex
\chapquote{``If not me, who? If not now,  when?"}{Emma Watson}{}

\section{Introduction}
{\magi Molecular clouds are the birthplace of stars. } While we know that stars form within these clouds of dust and gas,  the process of converting cloud mass into stellar mass is highly inefficient.  Within the Milky Way,  observations show a star formation rate of $\approx 2$ M$_{\odot}$ yr$^{-1}$ (\citealt{Chomiuk2011}).  If the efficiency of star formation was 100\%,  and assuming stars are forming at a constant rate,  we should be seeing stars forming at a rate of $\approx 300$ M$_{\odot}$ yr$^{-1}$ if all of the gas is in freefall.  This inefficiency is also reflected in nearby galaxies (\citealt{Leroy2008},  \citealt{Utomo2018}), implying that dense,  molecular gas regions in galaxies must have something interfering with the process of freefall collapse.  

The star formation efficiency (SFE) is a measure of how quickly stars begin to form while eating up the surrounding gas.  In extragalactic terms,  this can be defined as the ratio of star formation rate (SFR) to the total gas mass:
\begin{equation}
\mathrm{SFE} = \frac{\mathrm{SFR}}{M_{\mathrm{gas}}}
\end{equation}
\noindent The inverse of this equation gives the depletion timescale,  $t_{\mathrm{dep}}$,  which is the amount of time it takes for a cloud to deplete its gas reserve by forming stars
\begin{equation}
\label{sfeq:totdep}
t_{\mathrm{dep}} = \frac{M_{\mathrm{gas}}}{\mathrm{SFR}}
\end{equation}

One big challenge with calculating the SFE is the lack of a direct SFR tracer which fully captures emission from stars deeply embedded within clouds of dust and gas.  There are also over ten methods of determining a star formation rate for galaxies (\citealt{Kennicutt2012}, \citealt{Davies2016}) with no consensus on which one is the best.  For example,  some studies of nearby galaxies have traced star formation using FUV + 24 $\mu$m emission to estimate the SFR.  FUV emission comes directly from massive stars.  24 $\mu$m emission is used as a tracer of dust-enshrouded star formation and captures dust heating by ultraviolet photons of young stars (e.g. \citealt{Leroy2008},   \citealt{Williams2018}).  Others have used the total infrared luminosity (TIR) + H$\alpha$ emission to find the star formation rate (e.g.  \citealt{Kennicutt2009}).  The total far-IR emission is used to measure the total star formation rate on the assumption that all the OB stars are completely hidden by dust.  H$\alpha$ emission comes directly from gas ionised by massive stars.  Other tracer combinations include H$\alpha$ + UV and H$\alpha$ + 24$\mu$m emission (e.g. \citealt{Leroy2012},  \citealt{Liu2011}). 

In this chapter,  we measure the star formation efficiency of individual clouds in the nearby galaxy Andromeda (M31) to see if SFE varies across different parts of M31.  If such variations exist,  then this could imply that the physics of star formation in clouds might not be the same everywhere in M31.  While such a variety of star formation tracers exist,  in this study,  we examine the SFR of individual molecular clouds using FUV + 24$\mu$m emission to capture both the massive star formation and dust-obscured star formation.

M31 is interesting because it is a spiral galaxy similar to our own with a dusty,  star forming ring (\citealt{Habing1984}).  M31 has a very old stellar population in the central bulge, traceable by 3.6 $\mu$m emission (\citealt{Barmby2006}).  It is difficult to distinguish the spiral arms in this galaxy but we can see ring-like structures (\citealt{Gordon2006}) at galactocentric radii of 5, 10 and 15 kpc.  {The PHAT survey has detected $\sim 117$ million stars (\citealt{Dalcanton2012},  \citealt{Williams2014}) in one-third of M31 and active star formation in the 10 kpc ring lasting at least over the past 500 Myr (\citealt{Lewis2015}). } The PHAT team have found an average SFR of $0.28 \pm{0.03} \; \mathrm{M}_{\odot}$ yr$^{-1}$,  by counting individual stars (\citealt{Lewis2015}).  Additionally,  \cite{Williams2017} have shown that M31 is a more quiescent galaxy compared to the MW,  with majority of the star formation taking place over 8 Gyr ago,  and with recent star-formation activity kickstarting at $\sim 2$ Gyr ago and then returning to a state of quiescence.  {\magi The \textit{Herschel} study of M31 by} \cite{Ford2013} found a global value for SFR of 0.25 M$_{\odot}$ yr$^{-1}$.

For our clouds in the dust-selected cloud catalogue from Chapter {\magi 4} (Section \ref{ssec:cloudcat_hashtag}),  we determine the SFE of each individual cloud using the SFR map created by \cite{Ford2013}.  The central research questions that we aim to address in this chapter are: `Does star formation efficiency in M31 depend on position within the galaxy?' and `Do observational dust properties influence the star formation efficiency of molecular clouds in M31?' This work is novel in that we are able to compare the dust properties and star formation properties at 68 pc spatial scales,  for the first time incorporating emission from dust at the 850 $\mu$m wavelength.  This is also the first time that a study has looked at the SFE of individual clouds in this galaxy.

This chapter is structured as follows: Section \ref{sec:sfeobs} provides the details of the observational data that we use and how we obtain our SFR map and Section \ref{sec:sferes} gives our results.  In Section \ref{sec:sfediscussion},  we discuss the implications of our results,  followed by our key conclusions in Section \ref{sec:sfesummary}.

\section{Observations \& methods}
\label{sec:sfeobs}

\subsection{Dust, CO and \textsc{HI} observations}
We use the same map of  dust mass surface density,  $^{12}$CO(J=1-0) integrated intensity and \textsc{HI} intensity of M31 for this work as described in Chapter {\magi 4} (Section \ref{sec:observe}).  The dust and gas properties of each cloud are obtained using the method described in Chapter {\magi 4},  Section \ref{ssec:hashtag_gasmethods}.  The outline of our extracted clouds from the dust mass surface density map and corresponding regions in the SFR,  IRAM CO and WSRT \textsc{HI} maps are in Figures \ref{fig:postage_stamp_top}-\ref{fig:postage_stamp_bottom}.

\subsection{Star formation rate map}
\label{ssec:sfrobsmethod}
We use the SFR surface density map created by \cite{Ford2013} using 24 $\mu$m emission from the \textit{Spitzer Space Telescope} (\citealt{Gordon2006}) and FUV emission at $\sim 0.153$ $\mu$m from GALEX (\citealt{Thilker2005}) and the prescription following \cite{Leroy2008} (their Appendix D) and \cite{Calzetti2007}.  The older stellar population of M31 has been removed from the SFR map using the 3.6 $\mu$m emission map from \textit{Spitzer} (\citealt{Barmby2006}).  The foreground stars have also been removed. The SFR map was created by combining linearly the intensity of 24 $\mu$m and FUV emission and calibrating sources against expected H$\alpha$ emission.  We give a brief description of the method behind the making of the map below. 

H$\alpha$ emission (with a characteristic wavelength of 0.656 $\mu$m) is the second most direct tracer for massive stars,  preceded only by the direct counting of stars in the optical regime.  It is used to trace gas which is ionised by photons from high-mass stars.  Unfortunately,  H$\alpha$ emission can be obscured by dust.  Instead,  it is possible to use the Paschen $\alpha$ (Pa$\alpha$) line emission to trace the ionised gas as this has very little dust obscuration effects. The Pa$\alpha$ line is detected at a longer wavelength of 1.875 $\mu$m. The intrinsic ratio of Pa$\alpha$/H$\alpha$ is determined by atomic physics which means that it is a fundamentally fixed value.  Using the Pa$\alpha$ detection and the fixed ratio of the two lines,  any missing dust-absorbed H$\alpha$ emission can be accounted for. To calculate the SFR traced by H$\alpha$ emission,  the aim is to work out how many high-mass stars would need to be present to produce the observed H$\alpha$ values. This process involves some basic assumptions like assuming a particular initial mass function and the lifetime of a star. The initial mass function is used to estimate the number of low-mass stars that do not produce any H$\alpha$ emission. The number of years that a star lives is used as a normalisation factor. We then have the dust-corrected H$\alpha$ emission from the source.

Other tracers can also be used to estimate the SFR but need to be calibrated against the H$\alpha$ measurement.  \cite{Leroy2008} use 24$\mu$m and FUV emission to trace star formation.  FUV emission comes from young,  high-mass stars.  24 $\mu$m emission traces hot dust which has been heated by embedded stars {\magi or protostars}.  Since both of these tracers are affected by dust, there are a number of unknowns when using them to trace SFR (e.g. the clumpiness of dust, the size of dust grains,  etc). We can use the dust-corrected H$\alpha$ measurements to calibrate these tracers and obtain a SFR.  These corrections manifest as coefficients within the equation below.  This equation forms the basis of obtaining {\magi the SFR map that we use}:
\begin{equation}
\mathrm{SFR} = (a \times I_{\mathrm{24 \mu m}}) + (b \times I_{\mathrm{FUV}})
\end{equation}
where $I_{\mathrm{24 \mu m}}$ is the intensity of 24 $\mu$m emission and $I_{\mathrm{FUV}}$ is the intensity of FUV emission.
We can look at the SFR in many regions of a galaxy or different galaxies and solve for the best calibration coefficients simultaneously by searching for the combination of coefficients which best produce the SFR values found from using the dust-corrected H$\alpha$ emission as a tracer.  Once we have these calibration coefficients,  we can use 24 $\mu$m and FUV emission to obtain the SFR.  Within the MW,  it is possible to observe FUV emission unaffected by dust,  in which case there is no need to calibrate the intensity of this emission.  However,  for M31,  \cite{Ford2013} use the following formula:
\begin{equation}
\Sigma_{\mathrm{SFR}} = (3.2^{+1.2}_{-0.7} \times 10^{-3} I_{\mathrm{24 \mu m}}) + (8.1 \times 10^{-2} I_{\mathrm{FUV}})
\end{equation}
where $\Sigma_{\mathrm{SFR}}$ is the star formation rate surface density.  The coefficient $a$ in this equation is the H$\alpha$ + 24 $\mu$m calibration found by \cite{Calzetti2007},  scaled up by 30\% to account for FUV emission being absorbed more than H$\alpha$ emission.  The coefficient $b$ in this equation is the FUV-to-SFR calibration estimated by \cite{Salim2007}.  We correct the map for inclination effects of M31 by dividing by $cos(incl)$ where $incl$ is M31's inclination angle.  We multiply the SFR surface density map by the area of a pixel to obtain a map of the SFR.  The original SFR map has an angular resolution of 6" and a pixel scale size of 1.5".  We smooth and reproject the SFR map to match the resolution of the \textit{Herschel} 250 $\mu$m image at 18" and the pixel scale size of 6". 

We obtain a total SFR for each cloud by going to the corresponding region of our dendrogram-extracted clouds from our catalogue (Chapter {\magi 4},  Section \ref{ssec:cloudcat_hashtag}) and adding up the SFR values in each pixel of the cloud.

\newpage

\begin{landscape}

\begin{figure*}[h]
\centering
\includegraphics[width=23cm]{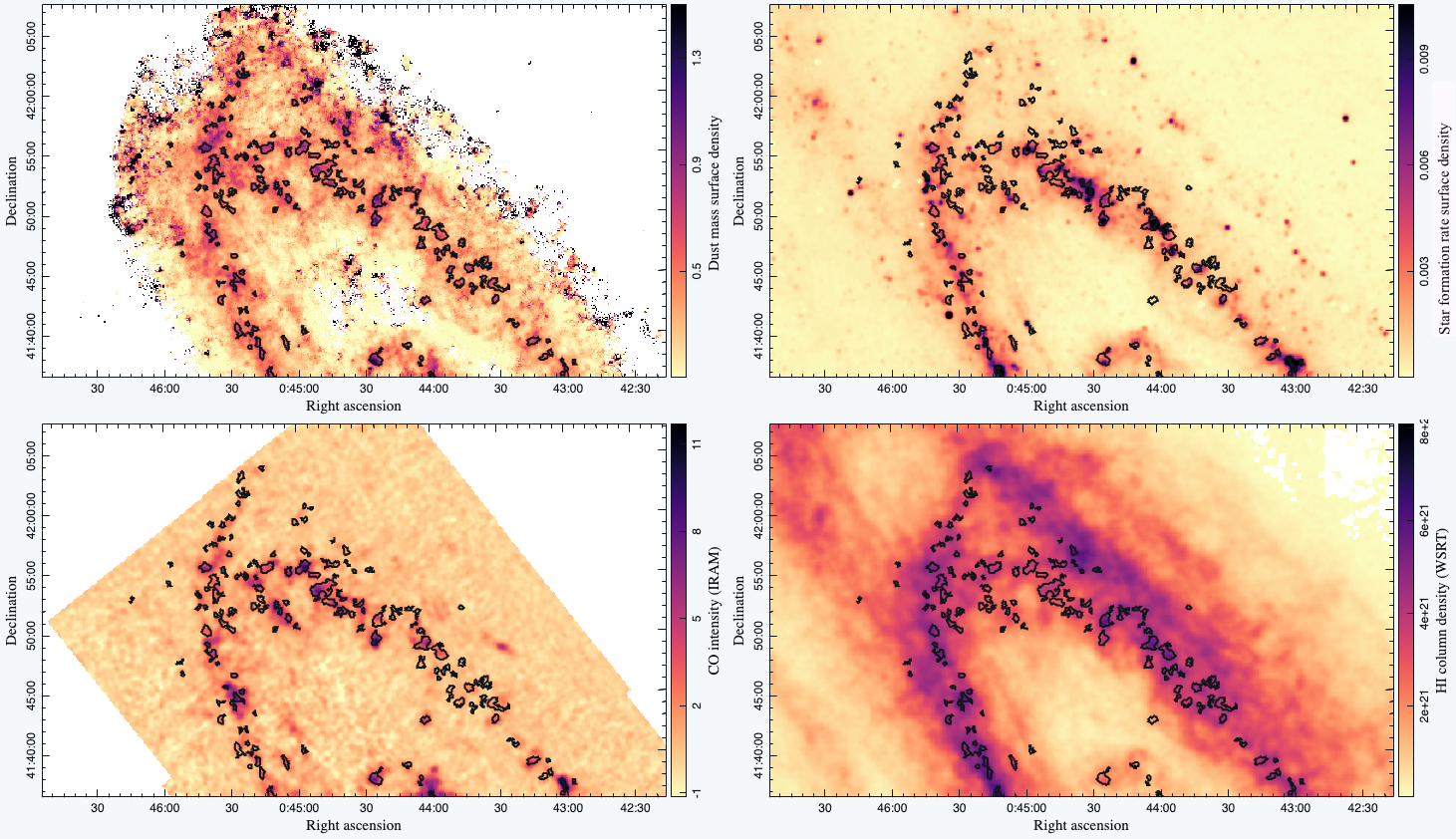} \caption{Distribution of clouds in the top third of M31.  Contours show cloud masks in the smoothed and reprojected SFR map and reprojected IRAM CO and WSRT \textsc{HI} maps.  \textit{Top left:} Contours of clouds on our dust mass surface density map in units of $M_{\odot}$ pc$^{-2}$. \textit{Top right:} Map of star formation rate surface density traced by 24 $\mu$m and FUV emission in units of $M_{\odot}$ pc$^{-2}$ yr$^{-1}$.  \textit{Bottom left:} CO intensity measurements from IRAM single-dish telescope in units of K km s$^{-1}$.  \textit{Bottom right:} HI column density map from WSRT in units of cm$^{-2}$. }
\label{fig:postage_stamp_top}
\end{figure*}

\newpage

\begin{figure*}[h]
\centering
\includegraphics[width=23cm]{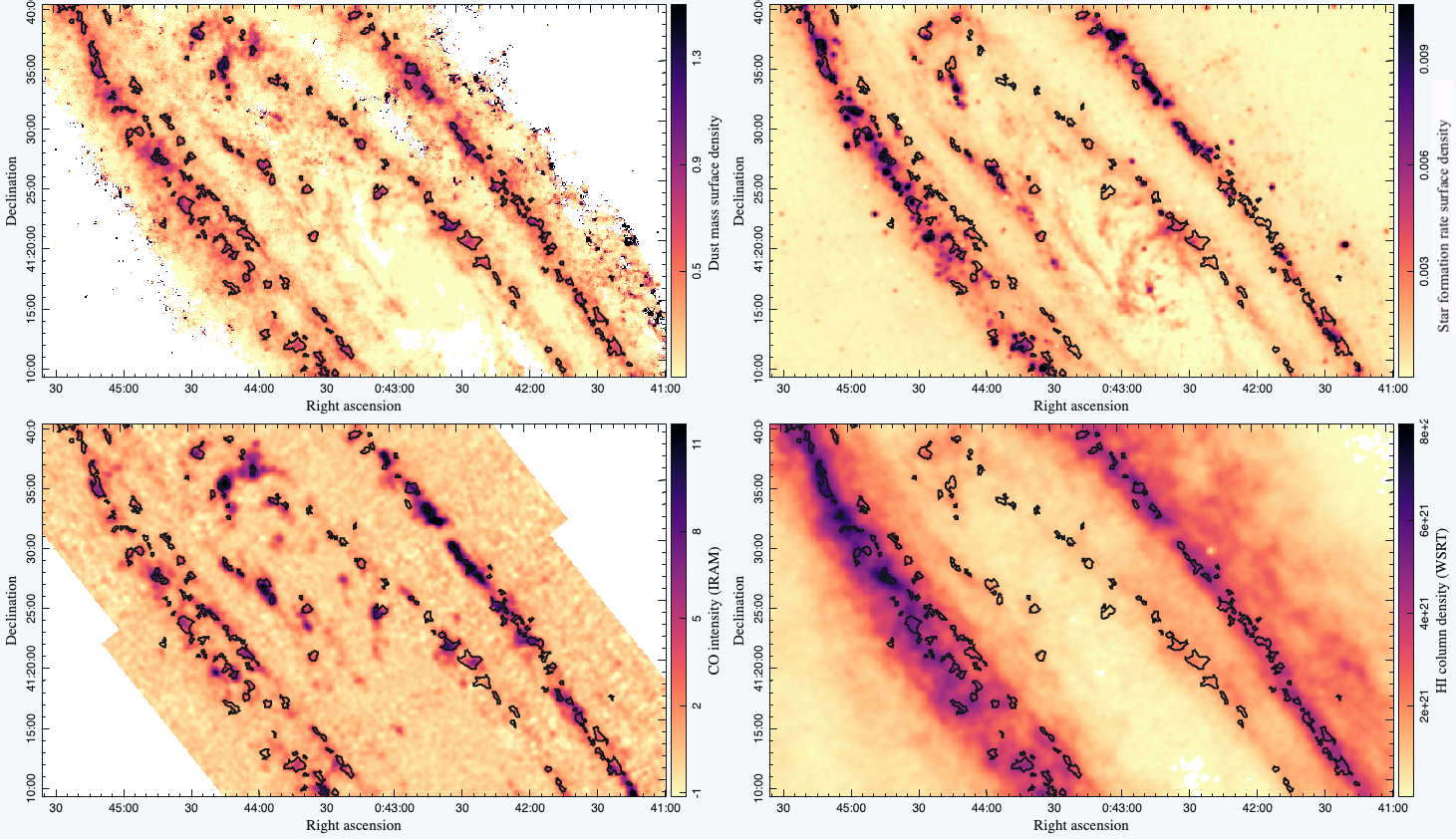} \caption{Distribution of clouds in the middle third of M31.  Contours show cloud masks in the smoothed and reprojected SFR map and reprojected IRAM CO and WSRT \textsc{HI} maps.  \textit{Top left:} Contours of clouds on our dust mass surface density map in units of $M_{\odot}$ pc$^{-2}$. \textit{Top right:} Map of star formation rate surface density traced by 24 $\mu$m and FUV emission in units of $M_{\odot}$ pc$^{-2}$ yr$^{-1}$.  \textit{Bottom left:} CO intensity measurements from IRAM single-dish telescope in units of K km s$^{-1}$.  \textit{Bottom right:} HI column density map from WSRT in units of cm$^{-2}$. }
\label{fig:postage_stamp_middle}
\end{figure*}

\newpage

\begin{figure*}[h]
\centering
\includegraphics[width=23cm]{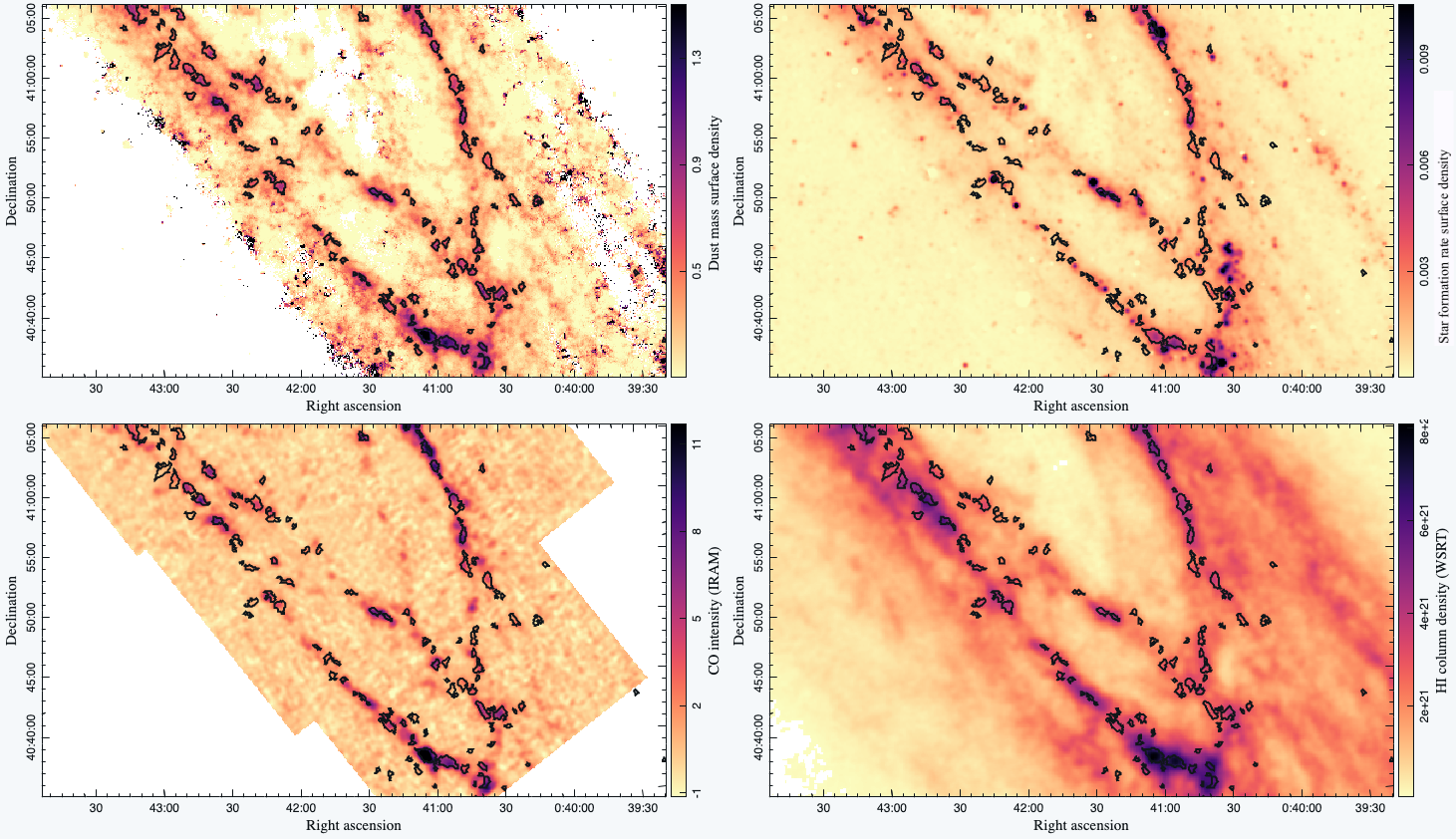} \caption{Distribution of clouds in the bottom third of M31.  Contours show cloud masks in the smoothed and reprojected SFR map and reprojected IRAM CO and WSRT \textsc{HI} maps.  \textit{Top left:} Contours of clouds on our dust mass surface density map in units of $M_{\odot}$ pc$^{-2}$. \textit{Top right:} Map of star formation rate surface density traced by 24 $\mu$m and FUV emission in units of $M_{\odot}$ pc$^{-2}$ yr$^{-1}$.  \textit{Bottom left:} CO intensity measurements from IRAM single-dish telescope in units of K km s$^{-1}$.  \textit{Bottom right:} HI column density map from WSRT in units of cm$^{-2}$. }
\label{fig:postage_stamp_bottom}
\end{figure*}

\end{landscape}
\section{Results}
\label{sec:sferes}

\subsection{The total gas mass of clouds and their star formation rate}
\begin{figure}[h!]
\centering
\includegraphics[width=16cm]{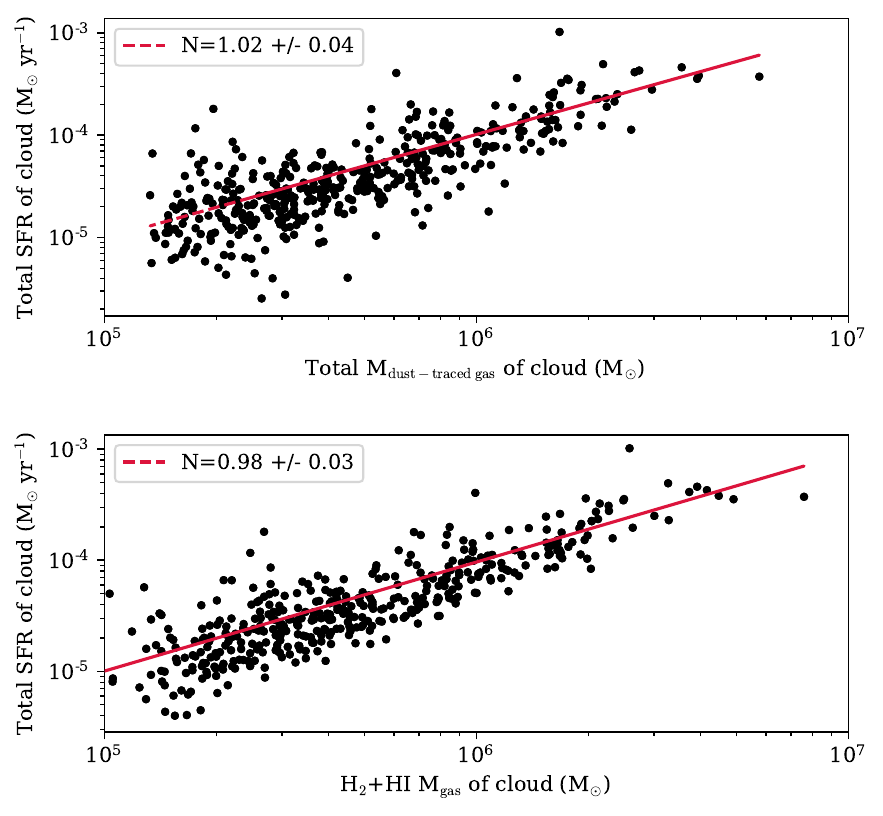} \caption{\textit{Top panel:} The total offset-corrected dust-traced gas mass vs total SFR for 422 clouds in M31.  \textit{Bottom panel:} The total \textsc{HI} + CO-traced H$_2$ gas mass vs total SFR for 419 clouds. The red dashed line in each panel shows the best fit model to the data points.  The errorbars for SFR values are present in this figure but not visible due to the noise being very small.}
\label{fig:ksplot}
\end{figure}

{\magi The} top panel of Figure \ref{fig:ksplot} shows the total dust-traced gas mass within each cloud vs the total star formation rate as traced by a combination of FUV + 24 $\mu$m emission.  The bottom panel of Figure \ref{fig:ksplot} shows the relationship between the total \textsc{HI} + CO-traced H$_2$ gas of clouds vs the total SFR.  For the total dust-traced gas mass,  we apply the offset correction factor calculated in Chapter 4.  We use the least squares minimisation Python package \texttt{lmfit} to fit the following model to the data:
\begin{equation}
\mathrm{SFR} = \mathrm{A} \; M_{\mathrm{gas}}^{\mathrm{N}}
\end{equation}
where A is the intercept,  $M_{\mathrm{gas}}$ is the total gas mass in each cloud as traced by dust,  assuming a constant GDR of $\simeq$ 54 (top panel of Figure \ref{fig:ksplot}),  or combined atomic and CO-traced molecular gas (bottom panel of Figure \ref{fig:ksplot}).  N is the best-fit gradient value.  We weight by the error in SFR calculated as the standard deviation in a region of the SFR map where we see no obvious emission.  We find that both of the best-fit slopes are very close to 1,  suggesting that the total gas mass and SFR of our clouds mostly follow a linear relation.  As the total cloud gas mass increases,  so does the total SFR for both panels in Figure \ref{fig:ksplot}.  The median error of the offset-corrected dust-traced gas mass of clouds is $1.1 \times 10^{4}$ M$_{\odot}$.  The median error on the SFR of clouds is $3.0 \times 10^{-7}$ M$_{\odot}$ yr$^{-1}$. The errors in our total $M_{\mathrm{gas}}$ and total SFR values cannot explain the dispersion in the relationship.  We find a dispersion in SFR of 0.3 dex for the top panel (calculated using a similar method to equation \ref{eq:dustmass1}).  We find a dispersion in SFR of 0.2 dex for the bottom panel.  The average percentage change in the SFR dispersion is 69\% for the top panel and and 54\% for the bottom panel in linear space.

\subsection{Is SFR correlated with molecular or atomic gas?}
\begin{figure}[h!]
\centering
\includegraphics[width=16cm]{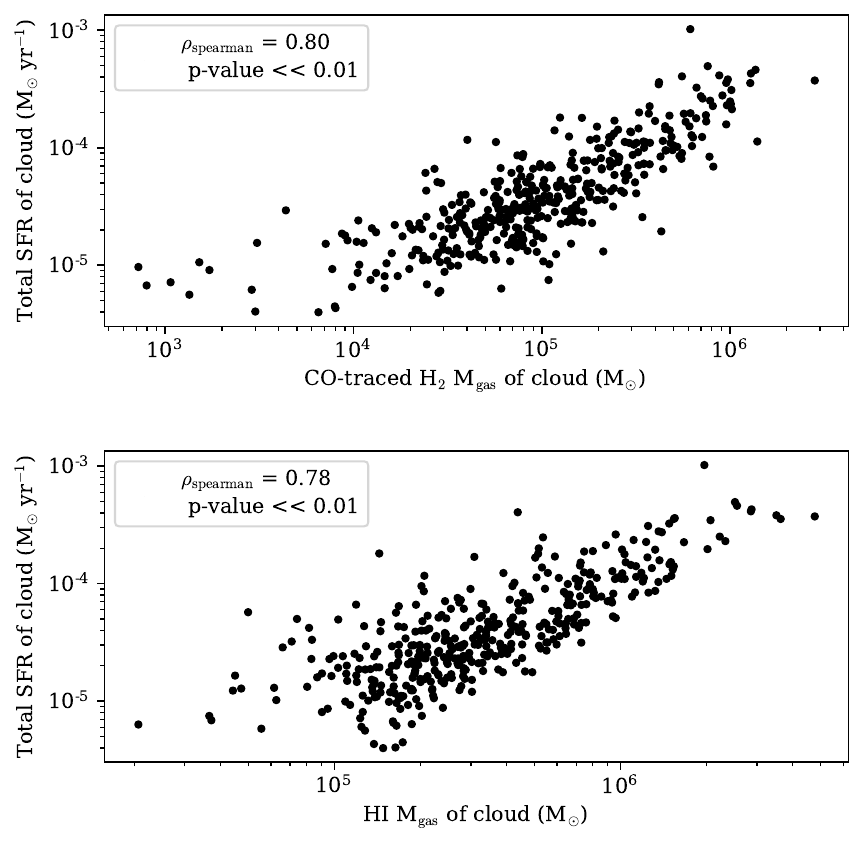} \caption{\textit{Top panel:} Total CO-traced H$_2$ gas mass vs total SFR of 419 clouds. \textit{Bottom panel:} Total \textsc{HI} gas mass vs total SFR of 419 clouds. }
\label{fig:cohi_sfr}
\end{figure}
Since star formation is typically associated with molecular gas,  we examine whether the dispersion in SFR found in Figure \ref{fig:ksplot} could be due to clouds having different amounts of \textsc{HI}.  We separate the CO-traced molecular gas and atomic gas phases of the clouds,  and compare these with the total SFR of each cloud.  Figure \ref{fig:cohi_sfr} shows our result.  We find a very strong positive correlation between the total CO-traced molecular gas mass and the total SFR of clouds (Spearman rank coefficient = 0.8,  p-value $\ll 0.01$).  This is similar to the linear relation found for dense gas and SFR for galaxies (\citealt{Kennicutt2012}).  Surprisingly,  we also find a very strong positive correlation between the total \textsc{HI} gas and the total SFR of clouds (Spearman rank coefficient = 0.78,  p-value $\ll 0.01$).  Our result is contrasting to the weak correlation found between HI surface density and SFR surface density for a sample of 33 nearby spiral galaxies by \cite{Schruba2011}.  While we established in Chapter 4 that a lot of our clouds are HI-dominated,  here we are finding that both H$_2$ and HI are strongly correlated with the star formation rate and, therefore,  must be important for the physics of star formation in some way.  A model by \cite{Krumholz2011} has suggested that at very low metallicities,  the carbon in the ISM is able to cool gas on shorter timescales than what would be required by dust grains to convert the \textsc{HI} to H$_2$,  and star formation could occur in the atomic gas phase.  But this is difficult to test as low metallicity environment naturally implies low CO abundance.

\newpage

\subsection{Radial variations in the cloud depletion times}
\begin{table}
\caption{Table listing the total {\magi s}tar formation rate and gas depletion times for extracted clouds.}
\label{tab:sfeprops}      
\centering                                      
\begin{tabular}{cc}          
\hline \\           
Property & 50th$^{\mathrm{84th}}_{\mathrm{16th}}$ percentile \\
\\
\hline
SFR (M$_{\odot}$ yr$^{-1}$) & $3.4^{11.0}_{1.5} \times 10^{-5}$ \\
Offset-corrected total dust-traced gas depletion time (Gyr) & 12.2$_{7.0}^{20.4}$ \\
\textsc{HI} + CO-traced H$_2$ depletion time (Gyr) & 12.5$^{19.1}_{7.6}$ \\
CO-traced molecular gas depletion time (Gyr) & 2.6$^{4.9}_{1.3}$ \\
\hline
\end{tabular}
\end{table}

\begin{figure}[h!]
\centering
\includegraphics[width=16cm]{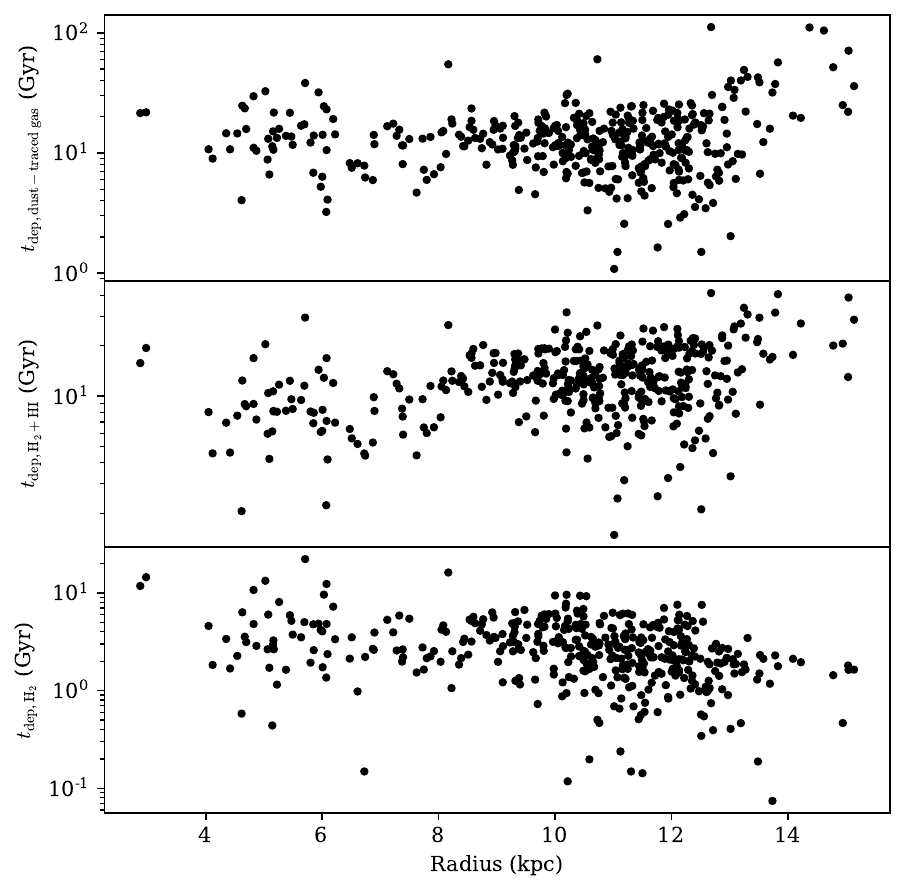} \caption{\textit{Top panel:} Galactocentric radius vs offset-corrected total dust-traced gas depletion time of clouds.  \textit{Middle panel:} Galactocentric radius vs combined \textsc{HI} + CO-traced H$_2$ gas depletion time of couds. \textit{Bottom panel:} Galactocentric radius vs CO-traced molecular gas depletion time of clouds.}
\label{fig:rad_dep}
\end{figure}

The gas depletion time provides a measure of how quickly the gas reservoir of a cloud is being converted into stars.  We examine whether the distance of the cloud from the galactic centre has any influence on the gas depletion time.  The top panel of Figure \ref{fig:rad_dep} shows the galactocentric distance of the cloud against the offset-corrected total dust-traced gas depletion time of the cloud.  We see a wide range of depletion times and no other prominent trends except perhaps a slight increase in depletion time beyond 12 kpc. The middle panel of Figure \ref{fig:rad_dep} shows the radius vs the combined \textsc{HI} + CO-traced H$_2$ gas depletion time.  We see a slight increase in depletion time (10 Gyr to 13 Gyr) between the clouds in the inner ring ($\sim$ 5 kpc) and the clouds in the outer ring ($\gtrsim$ 10 kpc). 

The bottom panel of Figure \ref{fig:rad_dep} shows the radius vs the CO-traced H$_2$ gas depletion time.  Without the \textsc{HI},  we now see a slight decrease in gas depletion time beyond 12 kpc.  We see some interesting sources between the galactic centre and a radius of 9 kpc with high depletion times ($t_{\mathrm{dep,  H_2}} > 10$ Gyr). We also see some other sources with low depletion times ($t_{\mathrm{dep,  H_2}} < 0.3$ Gyr) above a radius of 6 kpc.  Some clouds seen in the middle panel at radii $\gtrsim$ 12 kpc with high gas depletion times ($\gtrsim$ 10 Gyr) are no longer seen with such high gas depletion times in the bottom panel.  This suggests that these clouds had a larger combined H$_2$ + \textsc{HI} depletion time due to the presence of atomic gas.  Since the molecular gas depletion times of the clouds on average are smaller than the total gas depletion times traced by both dust and combined CO + \textsc{HI},  either the molecular gas in our clouds is being used up much quicker than the HI or dust-traced gas, or there is generally less molecular gas present in these clouds.  Since we established in Chapter 4 that a lot of our clouds are HI-dominated,  the latter explanation is more likely.  However,  since we earlier found a strong correlation between the amount of \textsc{HI} and the total SFR,  the longer total gas depletion times do not rule out any star formation activity within the clouds dominated by HI. 

The 16th,  50th and 84th percentile values of the total star formation rate and gas depletion times of clouds are listed in Table \ref{tab:sfeprops}.  While our large gas depletion times suggest that the star formation in M31 may be quite inefficient,  the interesting result here is that firstly,  we see a broad mix of cloud depletion times and secondly,  that we see this across different positions in the galaxy.  We discuss this result further in Section \ref{sec:sfediscussion}. 

\newpage
\subsection{The depletion time and cloud dust properties}
\begin{figure}[h!]
\centering
\includegraphics[width=16cm]{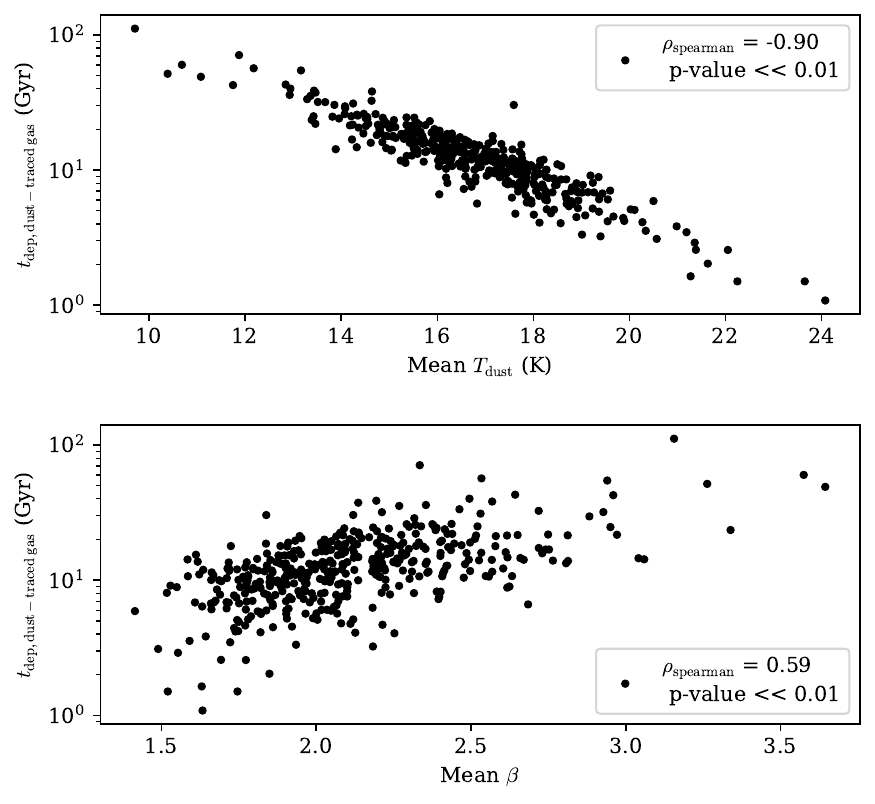} \caption{\textit{Top panel:} Density weighted mean dust temperature vs offset corrected total dust-traced gas depletion time of clouds.  \textit{Bottom panel:} Density weighted mean $\beta$ vs offset corrected total dust-traced gas depletion time of clouds.}
\label{fig:sfetempbeta}
\end{figure}

How do observational dust properties fare with the total dust-traced gas depletion time of each cloud? The top panel of Figure \ref{fig:sfetempbeta} shows the dust mass surface density weighted mean $T_{\mathrm{dust}}$ of each cloud against its dust-traced gas depletion time.  We see a very strong anti-correlation between the mean $T_{\mathrm{dust}}$ and depletion time (Spearman rank coefficient = -0.9,  p-value $\ll 0.01$).  This is a very interesting result as it indicates that how quickly a cloud depletes its gas reservoir is directly related to the temperature of dust within that cloud. 

The bottom panel of Figure \ref{fig:sfetempbeta} shows the dust mass surface density weighted mean $\beta$ of each cloud against the cloud's dust-traced gas depletion time.  We find a moderate positive correlation between $\beta$ and the depletion time of each cloud (Spearman rank coefficient $\sim$ 0.6,  p-value $\ll 0.01$).  Once again,  this indicates that the observational properties of dust do have some connection to how efficiently stars are formed in clouds.  We discuss this further in Section \ref{sec:sfediscussion}.

\section{Discussion}
\label{sec:sfediscussion}

\subsection{A mix of star formation efficiencies}
Following the offset correction in Chapter 4,  we find a long median total dust-traced gas depletion time of 12.2 Gyr for our clouds.  This is most likely because the total gas as traced by dust in these clouds contains gas in both the atomic and molecular phase.  Our median CO-traced molecular gas depletion time is much lower at 2.6 Gyr.  However,  this is still longer than the depletion times found in nearby galaxies.  In NGC628,  \cite{Kreckel2018} have found a molecular gas depletion time of 1.3 Gyr at comparable (50 pc) spatial scales.  Their value is similar to the value of $\approx$ 1.5 Gyr found in M51 by \cite{Leroy2017}.  Our $t_{\mathrm{dep,  H_2}}$ values are larger than the values found for these galaxies and in MW studies of clouds showing $t_{\mathrm{dep,  H_2}} \leq 1$ Gyr (e.g. \citealt{Vutisalchavakul2016}).  Our median $t_{\mathrm{dep,  H_2}}$ is also larger than the 1.1 Gyr found previously in M31 without incorporating the high resolution CARMA CO data (\citealt{Tabatabaei2010}). 

\cite{Querejeta2021} find median $^{12}$CO(J=2-1)-traced molecular gas depletion times of $1 < t_{\mathrm{dep,  H_2}} < 2.3$ Gyr across different regions of a sample of 74 galaxies from the Physics at High Angular resolution in Nearby GalaxieS (PHANGS) survey (see their Table 3 for specific values for the galactic centre,  bar,  spiral arm,  interarm and disk regions).  Their findings suggest that the centres of galaxies have a lower depletion time but other regions do not seem to show any systematic trends despite having a huge spread in values.  Our CO-traced molecular gas depletion times are consistent with their findings.  We do not see any systematic trends with radius.  This suggests that the location of the cloud is not having a huge affect on the molecular gas depletion or total gas depletion in M31.

Finding a broad mix of depletion times is analogous to clouds having a variety of star formation efficiencies.  What is driving the variation in SFE is unclear.  However,  the  broad range that we find suggest that either the clouds genuinely have a variety of SFEs or that we are catching the clouds at different instances during their collapse,  i.e.  at different evolutionary stages.  A cloud evolution framework has been previously suggested in observational studies by \cite{Kawamura2009} and \cite{Miura2012} for the LMC and M33,  whereby clouds evolve from showing no signatures of massive star formation to showing signatures of embedded star formation traced by compact \textsc{HII} regions and then finally,  more active star formation with stellar clusters.  Future work with a high resolution H$\alpha$ emission map and an optical map of M31 to classify the clouds according to this framework may help distinguish any trends in cloud properties with evolutionary stage. 

\subsection{Do dust properties influence SFE?}
We find a strong correlation between the dust emissivity index,  $\beta$,  and the total dust-traced gas depletion time of clouds.  This suggests that as the star formation efficiency of a cloud increases,  the value of $\beta$ decreases.  A possible reason for this could be changes in composition of dust grains within these clouds.  For example,  $\beta$ is expected to be lower for carbonaceous grains compared to silicate grains (\citealt{Jones2013}).  Therefore,  a speculative possibility is that more silicate grains could be present in clouds with low SFE compared to clouds with with high SFE.

We find a neat anti-correlation between dust temperature and total dust-traced gas depletion time.  This suggests that as stars begin to form and the total gas reserve depletes,  dust grains in these clouds are getting increasingly heated and their temperature is raised.  The dust could be being heated by OB-type stars a few Myr in age.  Our result is somewhat similar to the positive trend identified between dust temperature and SFE  calculated using only dust mass by \cite{Davies2019} for spiral galaxies (see their Figure 9).  Our result is also consistent with the work of \cite{Clemens2013} who find that dust emission at $\sim 20$ K may be powered by ongoing star formation across a sample of 234 star-forming galaxies.  There is also evidence for a strong correlation between dust temperature and SFR normalised by dust mass within a sample of 192 galaxies from the JINGLE survey (\citealt{Lamperti2019}),  which our result agrees with.

\begin{figure}[h]
\centering
\includegraphics[width=16cm]{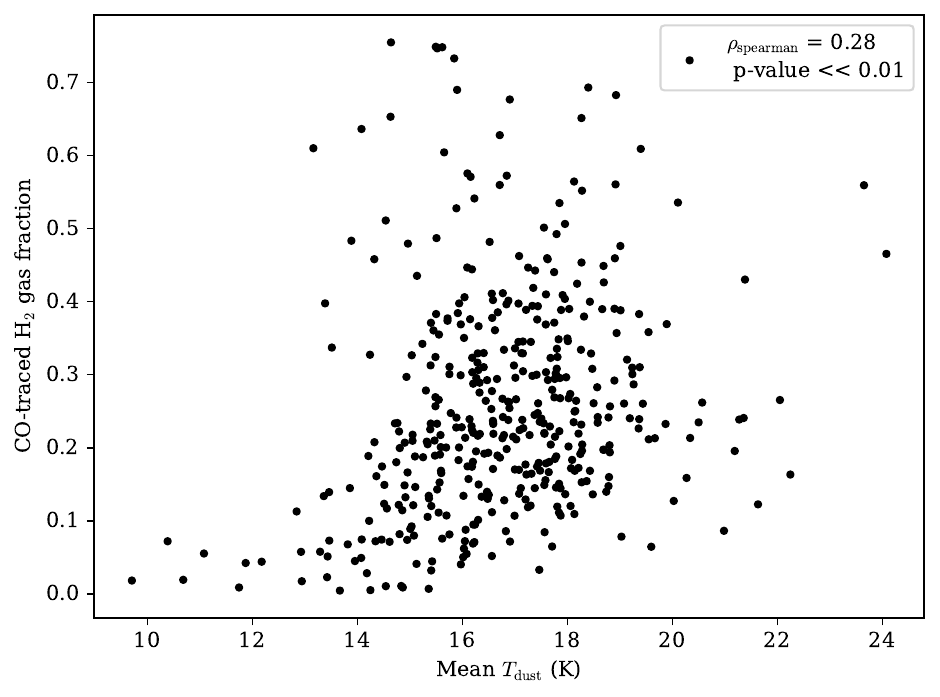} \caption{Density weighted mean dust temperature vs CO-traced molecular gas fraction in each cloud. }
\label{fig:sfe_tempco}
\end{figure}

Are there clouds with cold dust (< 15 K) and a longer gas depletion time (> 10 Gyr) because these clouds are in the atomic gas phases rather than being dominated by CO-traced molecular gas? To find the answer to this question,  we check for a correlation between the CO-traced molecular gas fraction in each cloud and mean dust temperature.  Figure  \ref{fig:sfe_tempco} shows our result.  We find a weak correlation between dust temperature and the CO fraction,  with more than 99\% statistical significance.  This suggests that the temperature of the dust is not strongly affected by whether the cloud is predominantly made of CO-traced molecular gas.  Since dust temperature is not strongly correlated with the CO-traced gas fraction but is strongly correlated with star formation efficiency (Figure \ref{fig:sfetempbeta}),  it is possible that both HI and CO-traced gas are driving star formation efficiency in M31.  This argument is also consistent with our result in Figure \ref{fig:cohi_sfr} where both gas phases are strongly correlated with the rate of star formation. 

Overall,  we find that dust temperature is strongly correlated with SFE,  and $\beta$ is strongly anti-correlated with SFE,  suggesting that a cloud's SFE has some connection to the observational properties of dust within that cloud.  However,  to what extent these dust properties influence the ongoing star formation remains unclear.

\section{Summary}
\label{sec:sfesummary}
The main goal of this study is to calculate the star formation efficiency of individual clouds within M31 and investigate whether the efficiency of clouds is dependent on its position within the galaxy or observational dust properties.  We do this by using the cloud catalogue from Chapter 4,  probing 68 pc spatial scales,  and the star formation rate measurements produced by \cite{Ford2013} using \textit{Herschel} observations.  Our star formation tracers of FUV + 24 $\mu$m emission are sensitive to dust-heated star-forming regions and direct emission from massive stars.  Our key results and conclusions are as follows:
\begin{enumerate}
\item We find long total dust-traced gas depletion times across our clouds with a median value of 12.2 Gyr. 
\item We find an average CO-traced molecular gas depletion time for our clouds of 2.6 Gyr,  larger than values found for the MW and nearby galaxies. 
\item We find that the total star formation rate {\magi of} clouds is strongly correlated with both CO-traced molecular gas and \textsc{HI} gas.
\item We do not see any evidence of total or molecular gas depletion time being associated with distance from the galactic centre.
\item We find a broad range of total dust-traced and CO-traced gas depletion for our clouds suggesting that we may be catching clouds at different evolutionary stages.
\item  We find a strong anti-correlation between dust temperature and total dust-traced gas depletion time and a moderate correlation between $\beta$ and depletion time,  suggesting that observational dust properties have a significant connection to star formation. 
\item We find a weak correlation between the CO-traced molecular gas fraction of clouds and their average dust temperature,  suggesting that dust heating by stars is not dependent on the gas phase of the surrounding ISM. 
\end{enumerate}

%% file: Chapters/chapter_ppmaptest.tex
\chapquote{``Do the best you can until you know better. Then when you know better,  do better."}{Maya Angelou}{}

\section{Introduction}
\label{sec:intro}
Submillimetre observations of galaxies from telescopes like the \textit{Herschel Space Observatory} and the \textit{James Clerk Maxwell Telescope} have been instrumental in detecting variations in the properties of dust grains in nearby galaxies (\citealt{Smith2012}, \citealt{Draine2014}, \citealt{Gordon2014}) and studying the nature of dust in star-forming regions (\citealt{Tabatabaei2014}, \citealt{Kirk2015}). However, standard modelling practices of spectral energy distribution (SED) fitting to dust continuum data, across multiple wavelengths,  have the well-known issue of needing to convolve all observations to the resolution of the lowest resolution image when fitting pixel-by-pixel.  This results in a huge loss of information from the high-resolution shorter wavelength datasets.  One promising modelling advancement in tackling this problem is the introduction of the algorithm \textsc{PPMAP} (\citealt{Marsh2015}) which implements point process mapping to estimate the dust content from submilllimetre observations of galaxies without compromising in resolution.  The second problem with the standard method is that it requires us to assume a single dust temperature or at the most two dust temperatures.  \textsc{PPMAP} doesn't require this assumption and makes an attempt to estimate how much dust there is at different temperatures,  and if desired,  different values of the dust emissivity index ($\beta$).  A summary of how the \textsc{PPMAP} algorithm works is provided in Section \ref{sec:ppmapinfo}.

Since its initial development, \textsc{PPMAP} has been applied to a few observations of filamentary and molecular cloud structures within our own galaxy.  For example, \cite{Marsh2017} have reanalysed observations of the Galactic plane from the Hi-GAL survey using \textsc{PPMAP} to show that dust temperature decreases with increasing galactocentric distance. \cite{Howard2019} have applied the algorithm to observations of the star-forming filament L1495, in the Taurus molecular cloud, finding cooler dust inside the filament compared to the outer regions. The authors have also found variations in the dust emissivity index, a key property which influences the shape of the blackbody spectrum followed by dust emission, with lower values of the parameter found inside the filament compared to the outside. Observations of the filamentary structures L1688 and L1689 in the Ophiuchus molecular cloud have also been reanalysed using \textsc{PPMAP} (\citealt{Howard2021}) with the study finding hot dust due to heating by B-type stars in L1689, and colder dust on average in L1688 compared to L1689. The changes in dust properties discovered by these works imply that the environment influences the properties of dust grains and vice versa. Therefore, it is also apt to apply \textsc{PPMAP} to extragalactic observations to understand the cause behind these variations.

Outside of the Milky Way, \textsc{PPMAP} has been applied to observations of the Andromeda galaxy (henceforth M31) by \cite{Marsh2018} for the central regions and by \cite{Whitworth2019} for the entire galaxy. The authors confirm the existence of warm dust in the centre of M31 and {\magi an} increase in the dust emissivity index within the inner 3 kpc.  They confirm a decrease of $\beta$ with increasing radius for radii greater than 3 kpc,  corroborating the results of previous studies (e.g. \citealt{Smith2012}, \citealt{Draine2014}) which have used the standard SED fitting procedure.  By further analysing \textsc{PPMAP} dust column density and emissivity index maps for a small portion of the galaxy, \cite{Athikkat-Eknath2022} have studied the influence of gas density on dust properties and found that density is not the prominent driver of radial variations in the dust emissivity index within M31.

In order to validate these findings and to use \textsc{PPMAP} for extragalactic studies where signal-to-noise ratios are typically lower and there is more mixing of dust temperatures along the line-of-sight,  it is important to test the capability of \textsc{PPMAP} to produce reliable dust column density estimates both on local and global scales.  Previous testing has only been done on {\magi models of} Galactic objects. 

\cite{Marsh2015} have tested \textsc{PPMAP} with a simulation of a prestellar core. The authors generated artificial images at five \textit{Herschel} wavelengths (70, 160, 250, 350 and 500 $\mu$m) and ran \textsc{PPMAP} on these images.  The properties of the original simulation were then compared to the output of \textsc{PPMAP}.  The authors found that \textsc{PPMAP} successfully recovers the true mass and integrated column density of a model prestellar core object.  A similar test was performed using a simulated turbulent cloud.  The authors found that \textsc{PPMAP} also estimates the true mass distribution, integrated column density and spatial structure of the model turbulent cloud well (\citealt{Marsh2015}), with a larger error margin only seen for dust estimates in the individual temperature slices.  While these results attest to the capability of \textsc{PPMAP} to be applied to nearby small-scale structures,  does the algorithm produce reasonable estimates of the dust content of entire external galaxies?

The aim of this chapter is to test the \textsc{PPMAP} algorithm with data of a simulated galaxy to see if the algorithm reproduces the same dust mass surface density values as what went into it.  Since we know how the simulated dust mass surface density values were created,  this test allows us to check how well the algorithm deals with physics on extragalactic scales.  Our study uses existing multiwavelength simulated galaxy images from the Auriga simulation suite (\citealt{Grand2017}) which have been post-processed with the \textsc{SKIRT} radiative transfer code (\citealt{Camps2015}, \citealt{Camps2020}) to incorporate emission from dust grains (\citealt{Kapoor2021}). 

This chapter is structured as follows: we first summarise the \textsc{PPMAP} algorithm (Section \ref{sec:ppmapinfo}) and describe the post-processed simulated data from the Auriga simulation suite that are used in this work (Section \ref{sec:ppmapdata}). Next, we outline our tests to see how well \textsc{PPMAP} recovers the Auriga galaxy's dust content (Section \ref{sec:ppmapmethods}), followed by our results (Section \ref{sec:ppmapresults}). We discuss the implications of our results for future use of \textsc{PPMAP} and summarise our conclusions in Section \ref{sec:ppmapsummary}.

\section{Using the PPMAP algorithm}
\label{sec:ppmapinfo}
PPMAP estimates the dust column density by taking in observational data of astrophysical systems across different wavelengths and the corresponding point spread function and applying an algorithm based on Bayes theorem to find the expectation value of the best-fit model.

We provide \textsc{PPMAP} with the following inputs: a temperature grid (in logarithmic intervals), the point spread function of the \textit{Herschel} instruments at each wavelength and the data (emission map of Auriga galaxy) at each wavelength (70, 100, 160, 250, 350 and 500 $\mu$m). An example set of input parameters used for one of our tests can be found in Table \ref{tab:ppmap_params}.

\textsc{PPMAP} contains a \textsc{PREMAP} component which ensures that all of the input datasets are projected in the same way before starting the iteration process.  Using these inputs, the algorithm firstly creates a model grid of cells with each cell defined by a set of parameters. For example, for a model grid of dust column density, each cell is parametrised by the position on the sky in pixel coordinates (x,y), dust temperature ($T_{\mathrm{dust}}$) and dust emissivity index ($\beta$). The model assumes that dust emission is optically thin at all input wavelengths.  We assume a constant $\beta$ of 1.8 across all cells.  We choose this value as it is one which approximately matches the \textsc{SKIRT} post-processing,  but we cannot be exact as the \textsc{SKIRT} post-processing dust recipe uses multi-grain properties (e.g.  for carbon and silicate grains).

The algorithm iteratively tries to solve for the model column density distribution, \textbf{$\Gamma$}, given the data, \textbf{d}. The expectation value for $\Gamma$ is given as:
\begin{equation}
E(\Gamma_n|\mathbf{d}) = \sum \Gamma_n \; P(\mathbf{\Gamma}|\mathbf{d})
\end{equation}
where n represents the n$^{th}$ cell and $P(\mathbf{\Gamma}|\mathbf{d})$ is given using Bayes' theorem:
\begin{equation}
P(\mathbf{\Gamma} |\mathbf{d}) = \frac{P(\mathbf{d}|\mathbf{\Gamma}) \; P(\mathbf{\Gamma})}{P(\mathbf{d})}
\end{equation}

$P(\mathbf{d}|\mathbf{\Gamma}$) is a probability distribution which takes into account Gaussian noise and the system response incorporating the convolution of the point spread function with the data. $P(\mathbf{\Gamma})$ is the prior distribution\footnote{The prior consists of the product of Gaussian distributions. More information about the prior are given in equations 5 and 6 in \cite{Marsh2015}.} which includes a key parameter (\texttt{dilution} in Table \ref{tab:ppmap_params}) that has the role of urging the algorithm to try to fit the data with the fewest possible number of components, depending on how small a value it is set at. $P(\mathbf{d})$ is a normalisation factor given by equation 9 of \cite{Marsh2015}.

To begin with, the procedure attempts to fit a model to the observations as if the observations have a high noise level. This means that initially the observations are regarded as so uncertain that they do not need to be fit very closely. The noise level is then gradually reduced and the procedure updates the expectation value of the model, given the data for each iteration. This process is repeated until the noise level matches the actual noise level in the data. The reduced $\chi^2$ is calculated for each iteration. \textsc{PPMAP} continues iterating until it is able to get a reduced $\chi^2$ of the order of unity without the need for inflating the observational uncertainties any longer. The procedure stops when the reduced $\chi^2$ value becomes greater than a reference reduced $\chi^2$ value produced by the algorithm. At that point we have, in principle, a solution that satisfies the data within the true observational errors and which also satisfies the \textit{a priori} statistical model that states that the model with the fewest components is likely to be the correct one. For the sake of minimising computational expense, the iterative process is done by splitting the input emission maps into multiple "tiles", with each tile containing 40 $\times$ 40 pixels. Further details about the PPMAP algorithm can be found in the work of \cite{Marsh2015}.

The final raw data products of \textsc{PPMAP} are a column density map {\magi of interstellar matter (gas + dust)},  scaled using a dust mass opacity coefficient {\magi $\kappa_{300, \; \mathrm{PPMAP}} = 0.010 \; \mathrm{m^2 \; kg^{-1}}$},  and a $T_{\mathrm{dust}}$ map in each of the temperature slices provided as an input.  {\magi This raw map includes an implicit} assumption that the gas-to-dust ratio is 100.  We multiply by the ratio of $\kappa_{300, \; \mathrm{PPMAP}}$ and $\kappa_{350}$ = 0.192 m$^2$ kg$^{-1}$ (\citealt{Draine2003}),  and divide by the gas-to-dust ratio of 100 to turn the \textsc{PPMAP} output into a {\magi dust mass surface density} map with the same value of $\kappa$ as in the work of \cite{Draine2003}.

\begin{table*}
\centering
\caption{Input parameters provided for \textsc{PPMAP} for the 8000 iteration run. The \texttt{maxiterat} parameter was changed to edit the number of iterations for other runs.}
\label{tab:ppmap_params}
\begin{tabular}{ccc}
\hline
Name of variable & Description & Parameter value \\
\hline
\texttt{gloncent} & Right Ascension at centre ($^{\circ}$) & 180.0 \\
\texttt{glatcent} & Declination at centre ($^{\circ}$) & 0.0 \\
\texttt{fieldsize} & Field of view dimensions ($^{\circ}$) & 0.48, 0.48 \\
\texttt{pixel} & Output sampling interval (arcsec) & 3.0 \\
\texttt{dilution} & \textit{a priori} dilution & 0.3 \\
\texttt{maxiterat} & Maximum no. of integration steps & 8000 \\
\texttt{distance} & Distance to the galaxy (Mpc) & {\magi 8.56} \\
\texttt{kappa300} & Reference opacity (cm$^2$ g$^{-1}$) & 0.1 \\
\texttt{nbeta} & Number of dust emissivity index values & 1 \\
\texttt{betagrid} & Dust emissivity index values & 1.8 \\
\texttt{betaprior} & \textit{a priori} mean and sigma of beta & 1.8, 0.35 \\
\texttt{ncells} & Nominal size of subfield & 40 \\
\texttt{noverlap} & Size of subfield overlap & 20 \\
\texttt{Nt} & Number of temperatures & 8 \\
\texttt{temprange} & Range of temperatures (K) & 10.0 50.0 \\
\texttt{nbands} & Number of wavebands & 6 \\
\texttt{wavelen} & Wavelengths ($\mu$m) & 70, 100, 160, 250, 350, 500 \\
\texttt{sigobs} & Noise levels of the input images (MJy sr$^{-1}$) & 4.3293, 3.0475, 1.5046,  \\ & & 0.6526, 0.3050, 0.1336 \\
\hline
\end{tabular}
\end{table*}

\section{Data setup}
\label{sec:ppmapdata}
\subsection{Auriga simulation suite}
The Auriga simulation suite comprises of 30 Milky Way-type galaxies created from performing magneto-hydrodynamical cosmological simulations. The galaxies have evolved from host dark matter halos selected from the EAGLE simulations (\citealt{Schaye2015}) from redshift (\textit{z}) 127 to present day (\textit{z}=0). The interstellar medium of these galaxies contains star-forming gas in a cold, dense phase inside a hot phase (\citealt{Grand2017}), modelled by an equation of state from \cite{Springel2003}. The star formation process is initiated if the gas density in a cell is high enough and below a temperature threshold provided by the equation of state. The Auriga galaxies do not contain any interstellar dust. For more details about the physical processes modelled within the Auriga galaxies, we refer the reader to \cite{Grand2017}.

\cite{Kapoor2021} have produced synthetic observations of the Auriga galaxies by post-processing using the \textsc{SKIRT} radiative transfer code (\citealt{Camps2015}, \citealt{Camps2020}). In our work, we use the post-processed image of one of these halos, halo 3 (henceforth referred to as Au 3). The properties of Au 3 before the inclusion of dust emission are provided in Table \ref{tab:au3_tab}.

\begin{table*}
\centering
\caption{Properties of Auriga galaxy halo 3 (Au 3) taken from \citealt{Grand2017}}
\label{tab:au3_tab}
\begin{tabular}{cc}
\hline
Stellar mass ($\mathrm{M_{\odot}}$) & 7.75 $\times 10^{10}$  \\
Radial scalelength (kpc) & 7.26 \\
Virial radius (kpc) & 239.02 \\
Virial mass ($\mathrm{M_{\odot}}$) & 145.78 $\times 10^{10}$ \\
\hline
\end{tabular}
\end{table*}

\subsection{Adding dust into the Auriga galaxy}
\label{ssec:dustgrid}
Since dust is not included in the Auriga simulations, \cite{Kapoor2021} have added dust to Au 3 by assuming that a fixed fraction of metals are locked inside dust grains.  The dust-to-metal ratio ($f_{\mathrm{dust}}$) is left as a free parameter in \textsc{SKIRT} and the value which best fits is calibrated against real galaxy observations from a selection of 45 galaxies from the the DustPedia sample (\citealt{Davies2017}).  A more detailed description of this is provided in Section 2.3.2 (Table 2) and Section 3.2 of \cite{Kapoor2021}. The authors multiply the gas density values from Auriga by a metallicity taken from the Auriga simulation and a fixed dust-to-metal ratio ($f_{\mathrm{dust}}$=0.14) to create a dust-containing interstellar medium (DISM) component given in terms of a dust mass surface density distribution on a spatial grid.  Diffuse dust can be composed of amorphous silicates or amorphous hydrocarbons,  modelled using The Heterogeneous dust Evolution Model for Interstellar Solids (\textsc{THEMIS};  \citealt{Jones2017}). 

\begin{figure}
\centering
\includegraphics[width=18cm]{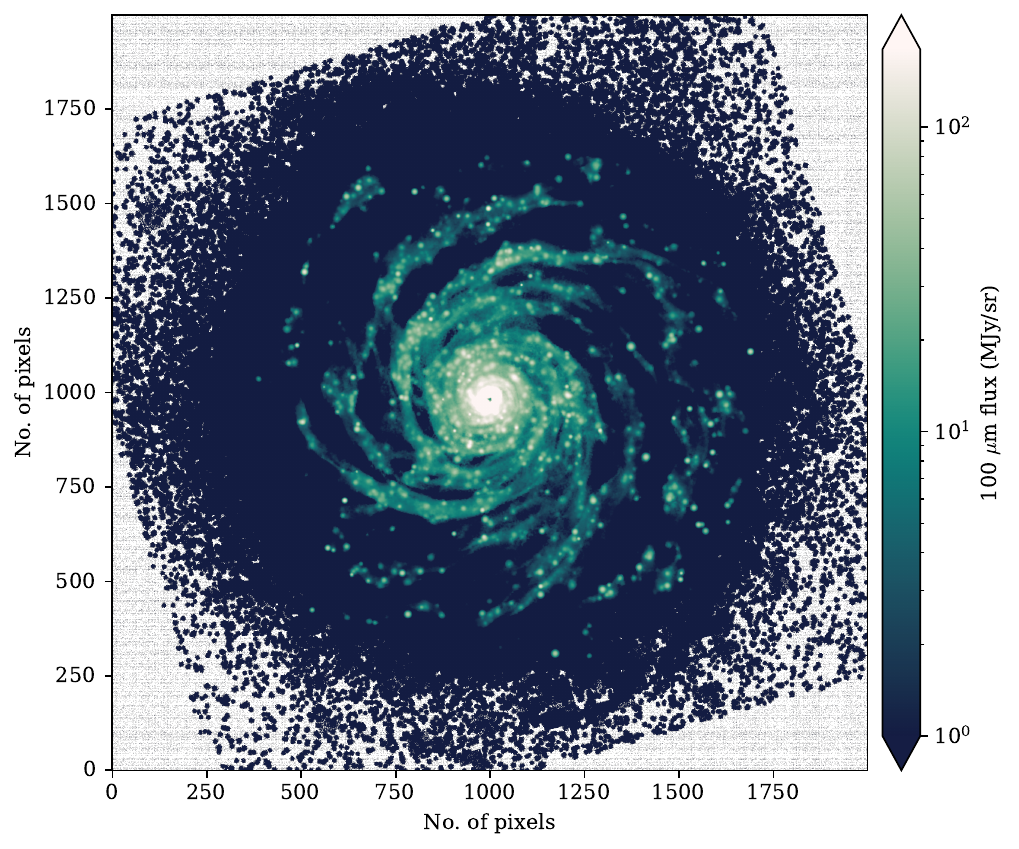}
\caption{Post-processed image of Au 3 (\citealt{Kapoor2021}) showing 100 $\mu$m emission at zero degree inclination.}
\label{fig:au3raw}
\end{figure}

Figure \ref{fig:au3raw} shows an image of Au 3 following post-processing with \textsc{SKIRT} (\citealt{Kapoor2021}). Stellar emission is the primary source of emission in Au 3. \cite{Kapoor2021} have categorised star formation in two ways. Stellar particles from the Auriga simulations with an age below 10 Myr are considered to be star-forming regions containing dust.  Particles with an age above 10 Myr are called stars and follow a Chabrier initial mass function (\citealt{Chabrier2003}).  The star-forming regions follow the spectral energy distribution (SED) from \textsc{MAPPINGS III} (\citealt{Groves2008}) which models SEDs incorporating HII regions and photodissociation regions (PDRs).  The SED of stars is modelled using the model from \cite{Bruzual2003}.  For the purpose of our work,  we remove any emission from stars and star-forming regions to solely focus on the dust emission {\magi (see Section \ref{ssec:synth})}.  This is because the stars and star-forming regions follow templates where the underlying dust mass surface density is not specified so it is difficult to retrieve this information.  Our study examines diffuse dust in the galaxy only and not localised dust at the scale of individual star-forming clouds.

We obtain the grid of dust mass surface density values taken in by \textsc{SKIRT} through the use of the \texttt{ProjectedMediaDensityProbe} class in the radiative transfer code which allows us to project the dust grid at different inclinations.  We obtain four dust grids: at 0.0,  45.57,  72.54,  and 90.0$^{\circ}$ inclination.

\subsection{Producing synthetic observations}
\label{ssec:skirtconf}
\begin{figure*}
\centering
\includegraphics[width=16cm]{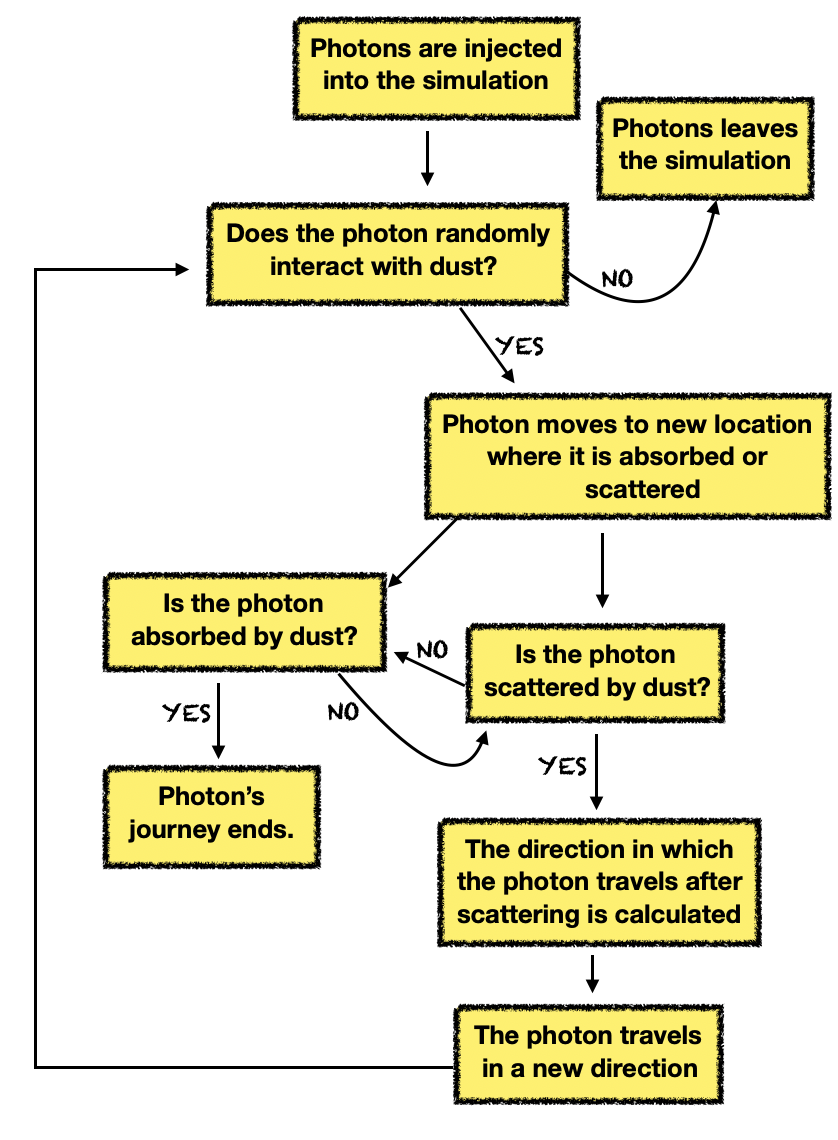}
\caption{Flow chart depicting a photon's journey in the \textsc{SKIRT} radiative transfer code.}
\label{fig:skirtcartoon}
\end{figure*}

\cite{Kapoor2021} have run \textsc{SKIRT} as a panchromatic simulation with 50 different wavelengths between 0.1 and 1244 $\mu$m.  Amongst these are the six \textit{Herschel} wavelengths relevant for this work,  at 70, 100,  160,  250, 350 to 500 $\mu$m.  The simulation has been run with $2 \times 10^4$ photon packets and the dust grid as described in Section \ref{ssec:dustgrid}.  A flow chart depicting the journey of a photon within the radiative transfer code is shown in Figure \ref{fig:skirtcartoon}.  The final post-processed images of Au 3 serve as the six input galaxy images taken in by \textsc{PPMAP}, following some modifications (see Section \ref{sec:ppmapmethods}).

\subsection{Herschel observations}
\label{ssec:hatlashersch}
We use two sets of existing observations from \textit{Herschel} to add noise to the synthetic observations.  Firstly,  we use observations of M31 produced by \cite{Draine2014} for 70, 100 and 160 $\mu$m wavelengths and observations by the HELGA collaboration (\citealt{Fritz2012}, \citealt{Smith2012}) for 250, 350 and 500 $\mu$m. These are used for our signal-to-noise analysis further explained in Section \ref{ssec:test_ppmap}.  We make use of large-field observations of the sky from the Herschel Astrophysical Terahertz Large Area Survey (H-ATLAS), in particular from the GAMA-12 region (\citealt{Valiante2016}). These images are used to superficially add noise to the input images of Au 3 which are fed into \textsc{PPMAP} (see Section \ref{ssec:synth} for further details).

\section{Methods}
\label{sec:ppmapmethods}
\subsection{Modifying synthetic observations}
\label{ssec:synth}
\begin{figure*}
\centering
\includegraphics[width=18cm]{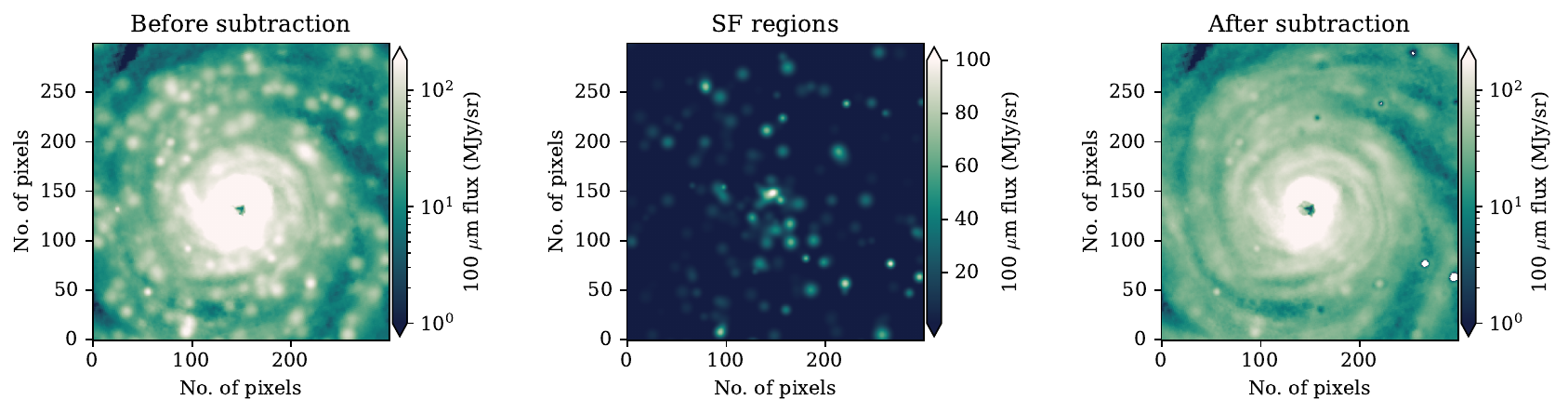} \caption{{\magi Emission maps s}howing the effect of removing star-forming regions from Au 3.  {\magi The colorbar shows 100  $\mu$m emission in units of MJy/sr.}}
\label{fig:sfremoveffect}
\end{figure*}

We extract the synthetic observations of Au 3 at 70, 100, 250, 350 and 500 $\mu$m from the original data cube which showed emission at 50 wavelengths (\citealt{Kapoor2021}). We run the \textsc{SKIRT} radiative transfer code without shooting any photon packets to reproduce the emission coming from stars or star-forming regions at the six wavelengths.  We then subtract the emission from these regions from the galaxy's emission.  This is not a perfect subtraction because \textsc{SKIRT} has a scatter as it includes some stochastic nature when recreating the regions (i.e.  it is randomly drawing from a distribution for each run; see Appendix \ref{sec:skirtscatter}).  This means that we retain some artefacts on the images where there is negative flux. Figure \ref{fig:sfremoveffect} shows the effect of subtracting the emission from star-forming regions on the Au 3 image at 100 $\mu$m.

Following the removal of star-forming regions,  we then convolve the synthetic observations with the point spread function of \textit{Herschel} at the corresponding wavelength.  This gives them the effective angular resolution of $\approx$ 8,  12.5,  14,  18,  25,  36". The data are reprojected to match the pixel scale size of the \textit{Herschel} images at each wavelength: $\approx$ 2, 3, 4, 6, 8, 12".  Since our initial analysis focuses on the face-on galaxy image, we set the distance to the galaxy to match an observed face-on galaxy,  M51,  at 8.56 Mpc (\citealt{McQuinn2016}).

\begin{figure*}
\centering
\includegraphics[width=16cm]{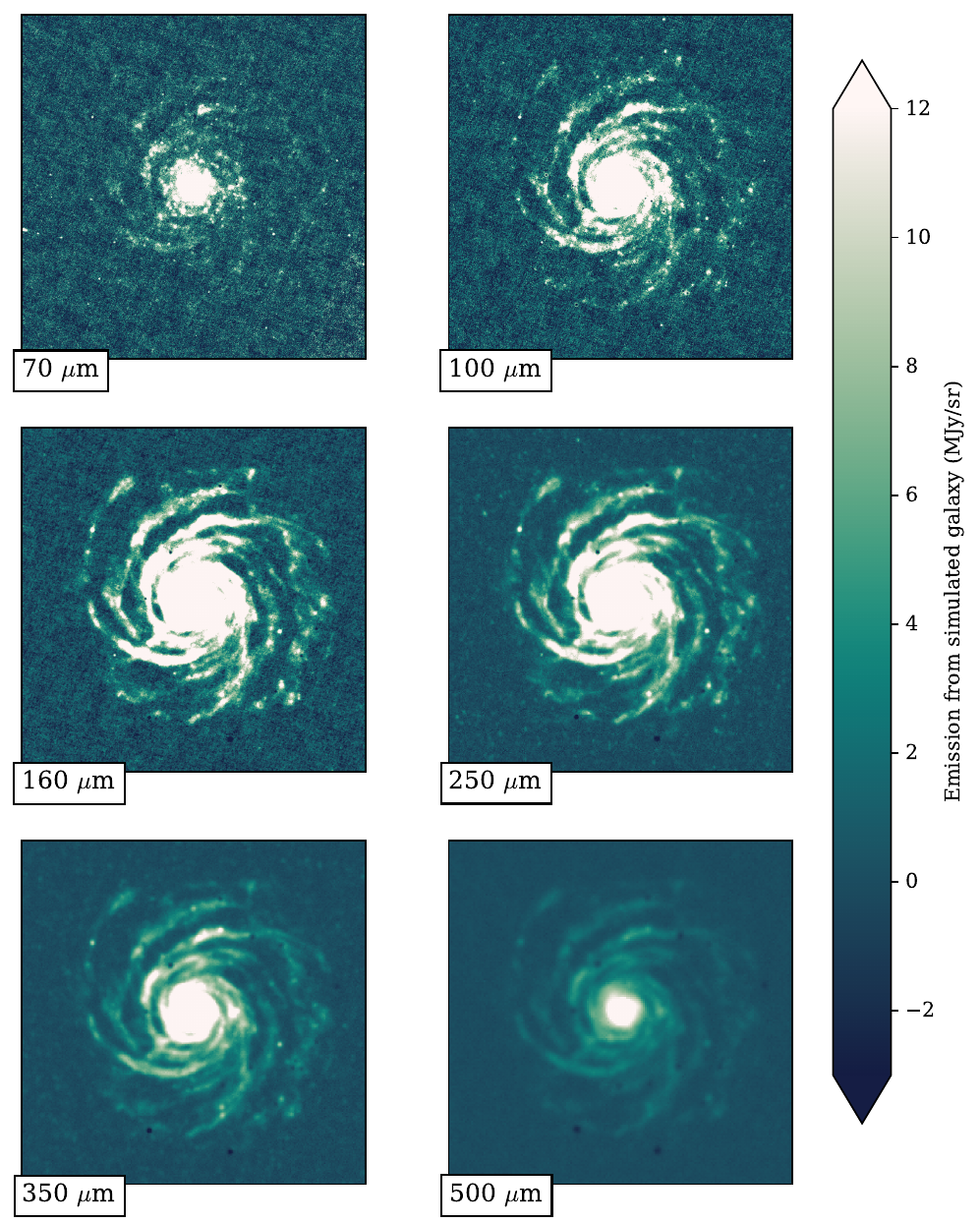}
\caption{Post-processed image of Au 3 (\citealt{Kapoor2021}) after subtraction of star-forming region,  convolution with \textit{Herschel} point spread function,  reprojection and noise-matched to M31 \textit{Herschel} observations (\citealt{Smith2012}, \citealt{Draine2014}). The colorbar shows the emission in units of MJy/sr.  The images at these six wavebands are the input images which are fed into \textsc{PPMAP}.}
\label{fig:au3postadj}
\end{figure*}

To make the synthetic observations more realistic as if they were observed by \textit{Herschel},  we add noise by using vast regions from the H-ATLAS large-field images between 100 to 500 $\mu$m as "noisy fields".  Since the H-ATLAS survey did not observe at 70 $\mu$m,  we use a different region of the 100 $\mu$m H-ATLAS observations to add as a noisy field to the 70 $\mu$m Au 3 image.  After artificially adding noise,  we match the signal-to-noise ratio (SNR) of Au 3 to that of the M31 \textit{Herschel} observations.  We select M31 images to match SNR to because this sets a robust case for the application of \textsc{PPMAP} to new observations of M31 in the HASHTAG survey (\citealt{Smith2021}). We measure the noise levels by finding a background region with no obvious emission and calculating the standard deviation of the intensity values within that region. Our signal is the mean value of the intensity from a region in the spiral arm of the galaxy.  A depiction of our chosen signal and noise regions can be found in Appendix \ref{sec:snrreg}.  {\magi We take the ratio of the signal and the noise from the noise-added images of the simulated galaxy and scale this value to match the SNR of the M31 \textit{Herschel} image at the corresponding waveband.} Figure \ref{fig:au3postadj} shows an example of our final simulated galaxy images ready to be used as an input for the \textsc{PPMAP} algorithm.

\subsection{Testing and optimising PPMAP}
\label{ssec:test_ppmap}
\noindent We test \textsc{PPMAP}'s peformance in three ways:
\begin{enumerate}[label=(\roman*)]
    \item How many iterations are best suited to recover spatial features of extragalactic observations accurately?
    \item How does \textsc{PPMAP} perform when applied to extragalactic observational datatsets with varying signal-to-noise levels?
    \item How does \textsc{PPMAP} perform when applied to galaxies at different inclinations?
\end{enumerate}
For test (i), we apply \textsc{PPMAP} six times, each time with a different number of iterations: 1000, 3000, 8000, 10000, 30000 and 50000 iterations. We  move towards larger intervals between the iterations as running the algorithm with a high number of iterations is more computationally intensive. We apply \textsc{PPMAP} on the zero-inclination Au 3 image with the signal-to-noise ratio (SNR) of our synthetic observations matching the SNR of \textit{Herschel} observations of M31 (\citealt{Draine2014}, \citealt{Fritz2012}, \citealt{Smith2012}).

For test (ii), we run \textsc{PPMAP} on the zero-inclination Au 3 image varying the signal-to-noise levels.  Test (i) already provides us with the result of matching SNR to \textit{Herschel} observations of M31.  For test (ii),  we run the \textsc{PPMAP} algorithm twice more; once with five times less noise than M31 observations and once with two times greater noise than M31 observations. The SNR values for each run can be seen in Table \ref{tab:ppmapobs_snr}.  Both the high and low SNR runs are perfomed with the number of iterations set to 3000,  8000 and 30000 to check how the SNR levels relate with our choice of iterations.

\begin{table*}
\centering
\caption{Signal-to-noise ratio (SNR) of the input images given to \textsc{PPMAP} after adjusting using regions from the \textit{Herschel} M31 observations and H-ATLAS observations.  The values have been rounded to the nearest whole number. These values apply for the 0$^{\circ}$ inclination image.  The three other inclinations have been SNR-matched to M31 observations.}
\label{tab:ppmapobs_snr}
\begin{tabular}{c c cl}
\hline
Wavelength ($\mu$m) & SNR \\
\hline
70 & 2 & \hspace{-0.5em}\rdelim\}{6}{*}[M31-matched]  \\
100 & 7 \\
160 & 23 \\
 250 & 35 \\
 350 & 37 \\
 500 & 37 \\
 \hline
 70 &  1 & \hspace{-1em}\rdelim\}{6}{*}[Low SNR]  \\
100 & 4 \\
160 & 12 \\
 250 & 17 \\
 350 & 19 \\
 500 & 19 \\
 \hline
 70 &  11 & \hspace{-1em}\rdelim\}{6}{*}[High SNR]  \\
100 & 36 \\
160 & 114 \\
 250 & 173 \\
 350 & 184 \\
 500 & 186 \\
 \\
\hline
\end{tabular}
\end{table*}

Using the `best choice' iteration number of 3000 inferred from test (i) and test (ii),  for test (iii), we provide the \textsc{PPMAP} algorithm with Au 3 at four different inclinations: 0$^{\circ}$ (from test (i)),  45.57$^{\circ}$, 72.54$^{\circ}$, 90$^{\circ}$,  and SNR matching the M31 observations.

\section{Results}
\label{sec:ppmapresults}
In this section, we present maps of dust mass surface density for Au 3 estimated by \textsc{PPMAP}. We look at the impact of varying the number of iterations, the signal-to-noise level in the image of the galaxy and observing the galaxy at different inclinations on \textsc{PPMAP}'s performance.

\subsection{PPMAP performance with varying number of iterations}
\label{ssec:itera_vary}
Figure \ref{fig:resultiter} shows the input dust mass surface density values used by \textsc{SKIRT} and the \textsc{PPMAP} output for a varying set of iterations. Visually, the outputs from using 8000 iterations or below produce the best match to the input dust grid for regions in the outskirts of the galaxy. The lower number of iterations best reproduce the spiral arm structure whereas the higher number of iterations ($\geq$ 30000 iterations) better depict the central bright source within the galaxy.  At a high number of iterations,  \textsc{PPMAP} overiterates and produces spurious point-source-like artefacts in its output images.  

\subsection{Scaling the PPMAP output}
In practice,  most extragalactic observational astronomy analysis is done on pixels where there is a confirmed detection of at least 3$\sigma$ if not 5$\sigma$ signal-to-noise level.  Motivated by this,  we produce scatter plots showing the dust mass surface density values of the simulated galaxy that went into \textsc{PPMAP} against the \textsc{PPMAP} output for the different iteration runs after masking out all of the pixels which are below the 5$\sigma$ threshold in the 250 $\mu$m input image.  Figure \ref{fig:scatter_5sig} shows our result.  We fit each iteration result with a linear relation of the form:
\begin{equation}
\label{eq:scatmodel}
\Sigma_{\mathrm{dust,  PPMAP}} = \mathrm{N} \; \Sigma_{\mathrm{dust,  SKIRT}} + \mathrm{A}
\end{equation}
where N is the gradient and A is the intercept.  We allow both N and A to vary.  The idea is to use the this best fit to see if there is a linear relation between the input values and the output values produced by the algorithm.  Any change in N gives insight into how much the output deviates from a one-to-one relation from the input.  The intercept A gives the starting value of \textsc{PPMAP}'s dust mass surface density output.

\begin{landscape}
\begin{figure}
\centering
\includegraphics[width=22cm]{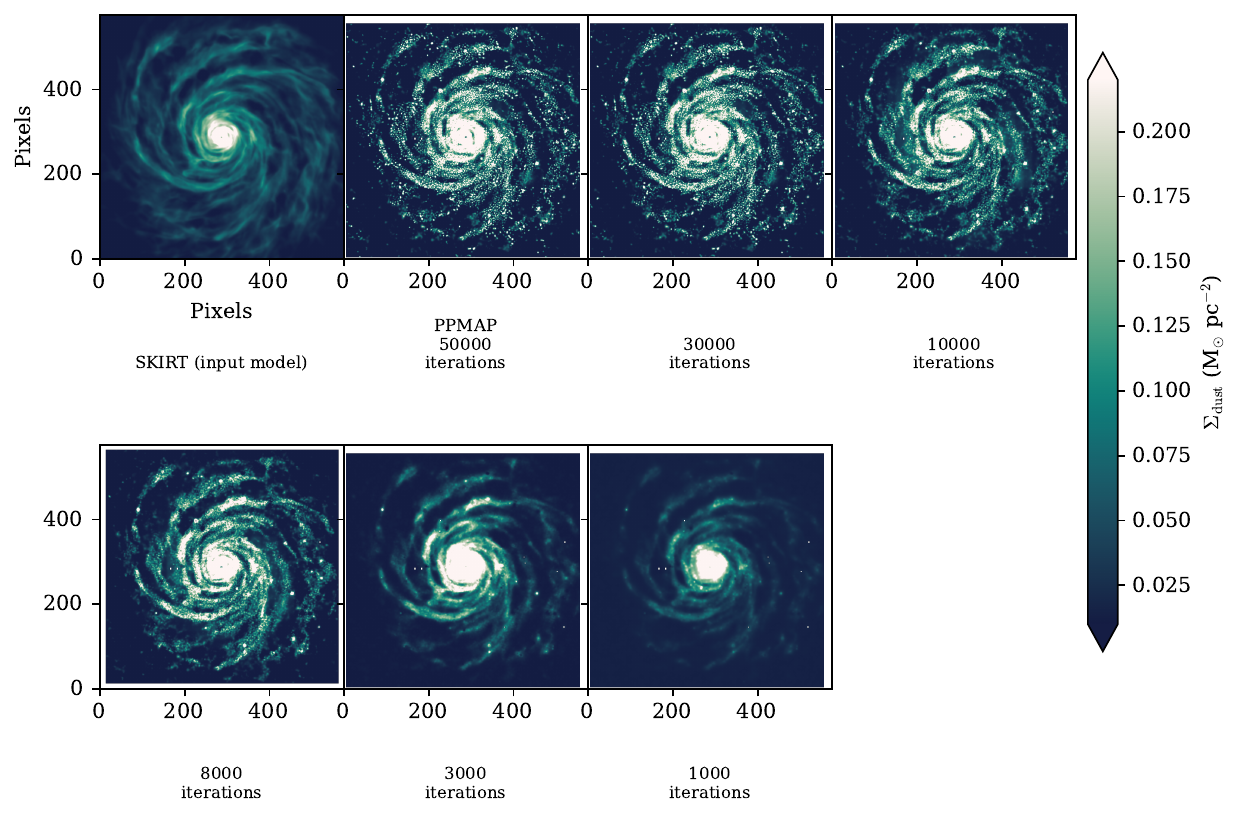} \caption{Comparison of input SKIRT dust grid with \textsc{PPMAP}'s dust mass surface density maps for different iterations numbers.}
\label{fig:resultiter}
\end{figure}
\end{landscape}

\textsc{PPMAP} generally underestimates the dust mass surface density in regions with 5$\sigma$ signal.  For 1000 iterations,  the linear model doesn't fully describe the lower $\Sigma_{\mathrm{dust}}$ values,  possibly because we are catching the algorithm before convergence.  For all runs,  except the 3000 iteration run,  there is a large dispersion around the best-fit line for input dust mass surface densities below 0.5 M$_{\odot}$ pc$^{-2}$.  Overall,  $\sim$ 3000 iterations best visually reproduces the input dust distribution (Figure \ref{fig:resultiter}) and the tightest linear relation between the input and the output values.  Table \ref{tab:spear_iter} shows the Spearman rank correlation coefficient and corresponding p-value for all of our runs.  We find that the input and output dust mass surface density values are very strongly correlated for the 3000 iteration run and all other runs.
 
The central region of the input images has a bright source with a hole in the middle.  \cite{Kapoor2021} attribute this to Active Galactic Nucleus (AGN) feedback,  \textit{"which causes low gas density holes to appear after a local dump of thermal energy."} The higher number of iterations do not handle this central bright source very well and produce square-like artefacts and spurious point-sources while over iterating. 

\interfootnotelinepenalty=10000

For all iteration runs,  there is an offset between the best-fit slope and the one-to-one relation line which shows where the scatter points should lie if \textsc{PPMAP} reproduced exactly what went into it. The offset is most likely caused due to the dust mass absorption coefficient,  $\kappa$,  used in \textsc{SKIRT} from the THEMIS dust model being different from the $\kappa$ assumed by \textsc{PPMAP}\footnote{{\magi The difference between the SKIRT dust model and the PPMAP dust model can be either due to assuming a particular dust type (with certain properties,  composition and emissivity index) or due to assuming a particular temperature distribution. If there is no difference in $\beta$ between the two models,  changing the properties or composition will affect $\kappa$ and hence the slope of our best fit lines on a linear scale.  The temperature distribution is harder to predict and a more complicated correction.  Given PPMAP uses multiple temperature bins while fitting,  and since there is no easy way to retrieve temperature distribution of the input model,  we make an assumption that PPMAP is reproducing the temperature distribution of the input model reasonably well.}}.  Since there is no easy way to retrieve the $\kappa$ used in \textsc{SKIRT} and under the assumption that the offset is only caused by this,  we apply an offset correction to \textsc{PPMAP}'s dust mass surface density output values.  Since the 3000 iteration run produces the output which best matches the input,  we use the slope of this fit (N = 2.1) as the offset correction factor and simply divide all future \textsc{PPMAP} output values by this value before fitting.  Figure \ref{fig:scatter_5sigoffcorr} shows the offset-corrected scatter plots.

\begin{figure}[h!]
\centering
\includegraphics[width=18cm]{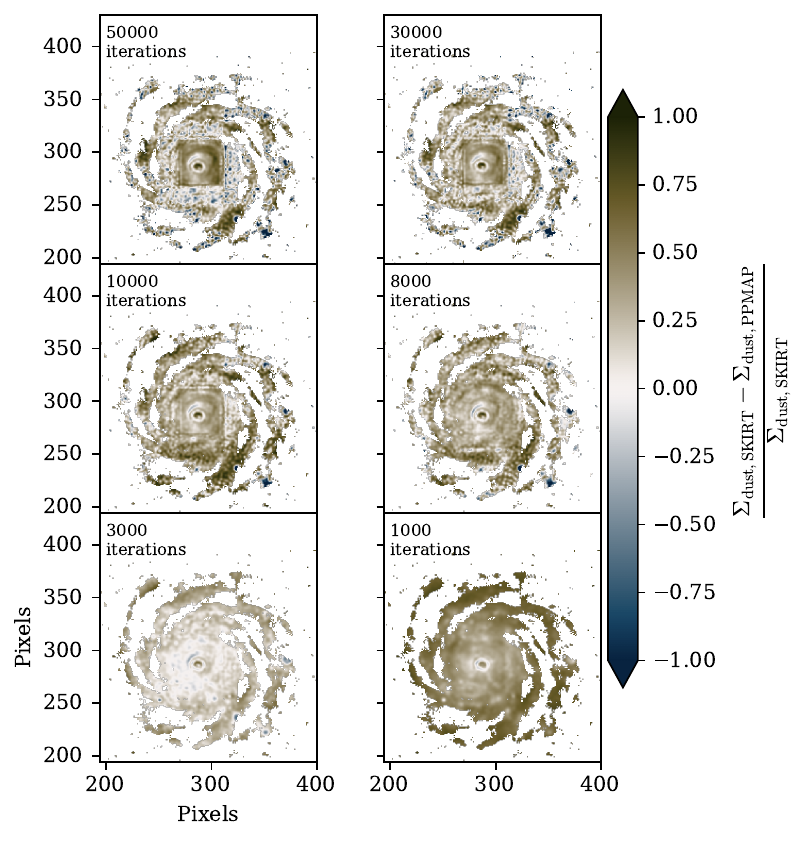} \caption{Fractional residual of SKIRT input dust mass surface density grid and \textsc{PPMAP}'s dust mass surface density values after offset correction.  Only pixels with signal at least 5 times the noise in the 250 $\mu$m band are shown.}
\label{fig:fracresi_iter}
\end{figure}

We also compare the fractional residuals for each iteration run after offset correction,  for pixels with signal at least 5 times the noise in the 250 $\mu$m observation (see Figure \ref{fig:fracresi_iter}).  Values close to zero indicate that the absolute difference between input and output dust mass surface density is only a small fraction of the original input value.  We find that,  after offset correction,  \textsc{PPMAP} generally underestimates the dust mass surface density.  The 3000 iteration run recovers the input values much better,  showing more regions of the galaxy with a fractional residual close to zero.

\begin{table*}
\centering
\caption{Spearman rank correlation coefficient and p-value for the input \textsc{SKIRT} dust distribution that went into \textsc{PPMAP} vs output dust mass surface density produced by \textsc{PPMAP} following offset correction.  Coefficient values between 0.40 and 0.69 indicate a strong correlation.  Coefficient values greater than or equal to 0.70 indicate a very strong correlation. }
\label{tab:spear_iter}
\begin{tabular}{c c c cl}
\hline \\
No. of iterations & Spearman rank coefficient & p-value  \\
\hline
50000 & 0.75 & $\ll 0.01$ \\
30000 & 0.78 & $\ll 0.01$ \\
10000 & 0.87 & $\ll 0.01$ \\
8000 & 0.88 & $\ll 0.01$ \\
3000 & 0.98 & $\ll 0.01$ \\
1000 & 0.97 & $\ll 0.01$ \\
\hline \\
\hline \\
No.  of iterations ($^{\circ}$) & Spearman rank coefficient & p-value  \\
\hline
30000 & 0.65 & $\ll 0.01$ & \hspace{-1em}\rdelim\}{3}{*}[Low SNR] \\
8000 & 0.87 & $\ll 0.01$  \\
3000 & 0.97 & $\ll 0.01$ \\
\hline
30000 & 0.71 & $\ll 0.01$ & \hspace{-1em}\rdelim\}{3}{*}[High SNR] \\
8000 & 0.93 & $\ll 0.01$  \\
3000 & 0.98 & $\ll 0.01$ \\
\hline \\
\hline \\
Inclination angle of simulated galaxy ($^{\circ}$) & Spearman rank coefficient & p-value  \\
\hline
45.57 & 0.98 & $\ll 0.01$  \\
72.54 & 0.99 & $\ll 0.01$ \\
90.0 & 0.99 & $\ll 0.01$ \\
\hline
\end{tabular}
\end{table*}

\newpage

\newpage
\begin{landscape}
\begin{figure}
\centering
\includegraphics[width=22cm]{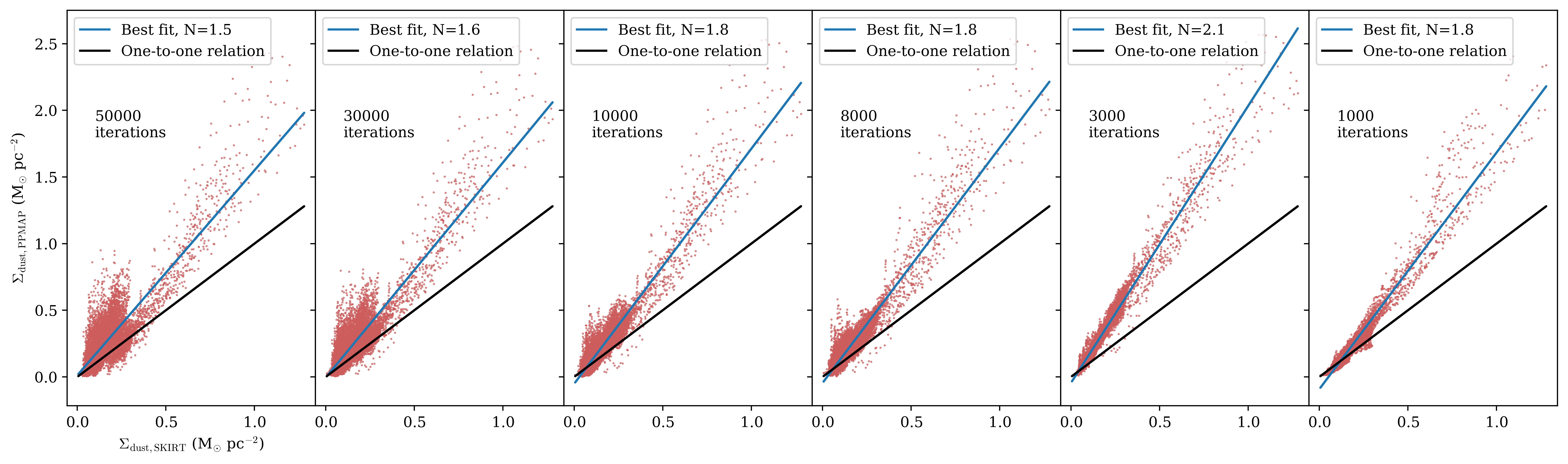} \caption{Scatter plot of SKIRT input dust mass surface density grid and \textsc{PPMAP}'s  dust mass surface density values after masking out all of the pixels below a 5$\sigma$ threshold in the 250 $\mu$m input image. The {\magi blue line} shows our best fit model and the {\magi black line} shows the one-to-one relation between the two quantities.}
\label{fig:scatter_5sig}
\end{figure}

\begin{figure}
\centering
\includegraphics[width=22cm]{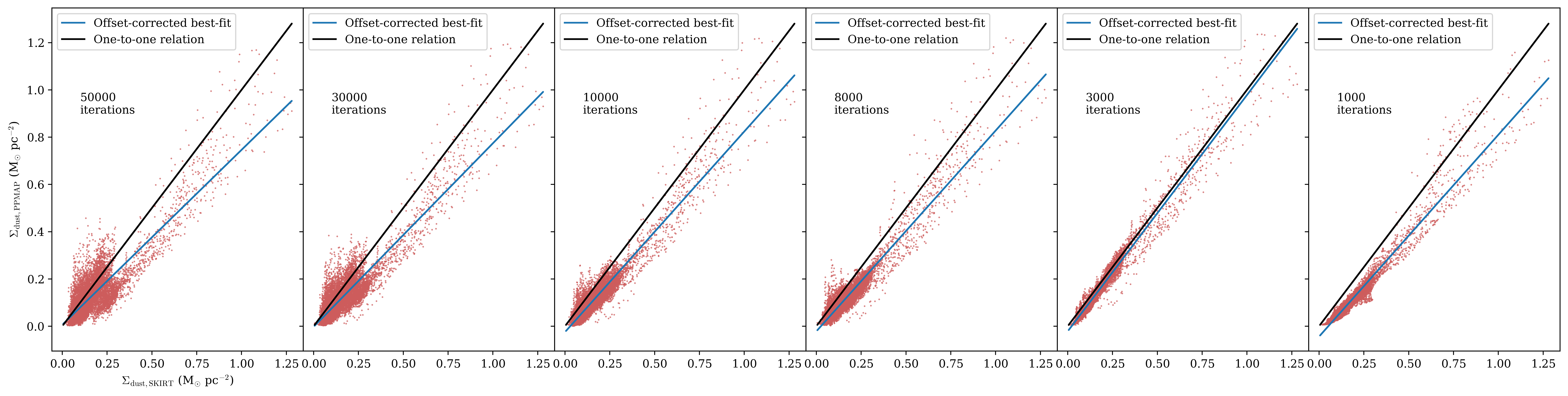} \caption{Scatter plot of SKIRT input dust mass surface density grid and \textsc{PPMAP}'s dust mass surface density values after offset correction. The {\magi blue line} shows our original best fit model from Figure \ref{fig:scatter_5sig} and the {\magi black line} shows the one-to-one relation between the two quantities.}
\label{fig:scatter_5sigoffcorr}
\end{figure}
\end{landscape}

\newpage 

\subsection{PPMAP performance with varying signal-to-noise levels}
\label{ssec:snr_vary}
\begin{figure}[h!]
\centering
\includegraphics[width=16cm]{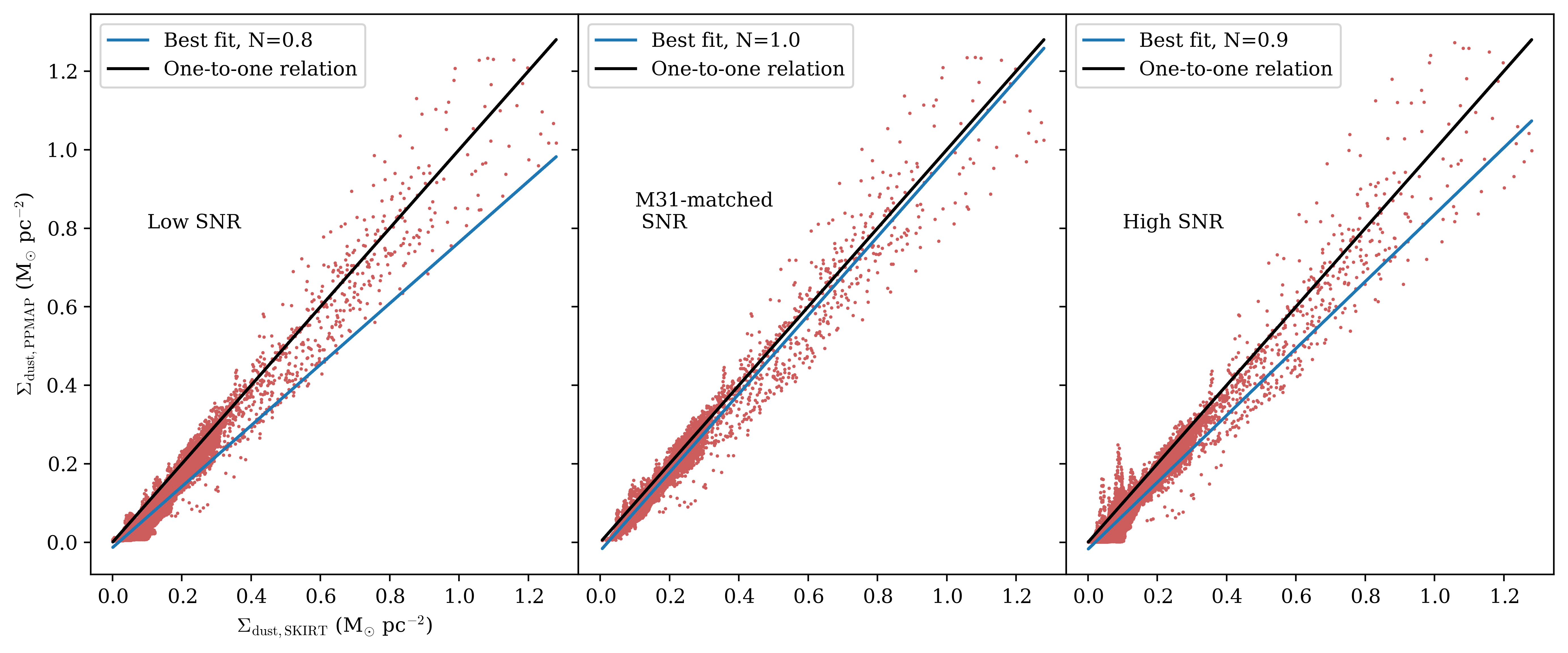} \caption{Scatter plot of SKIRT input dust mass surface density grid and \textsc{PPMAP}'s dust mass surface density values (after masking out all of the pixels below a 5$\sigma$ threshold in the 250 $\mu$m input image) for different SNR cases,  {\magi following offset correction using the 3000 iteration run}. The {\magi blue line} shows our best fit model and the {\magi black line} shows the one-to-one relation between the two quantities.}
\label{fig:scattersnr_5sigoffcorr}
\end{figure}

To check whether the signal-to-noise ratio of the input galaxy images affect \textsc{PPMAP}'s dust surface density estimation,  we also feed the algorithm with low SNR and high SNR images.   Figure \ref{fig:scattersnr_5sigoffcorr} shows a comparison of the dust mass surface density input vs output when \textsc{PPMAP} is given input galaxy images with low SNR,  SNR matched to M31 (same as the different iteration runs in Figure \ref{fig:scatter_5sigoffcorr}) and with high SNR.  We run \textsc{PPMAP} for 3000 iterations.  Once again,  \textsc{PPMAP} tends to generally underestimate the dust mass surface density for both the low and high SNR cases.

\begin{landscape}
\begin{figure}[h!]
\centering
\includegraphics[width=22cm]{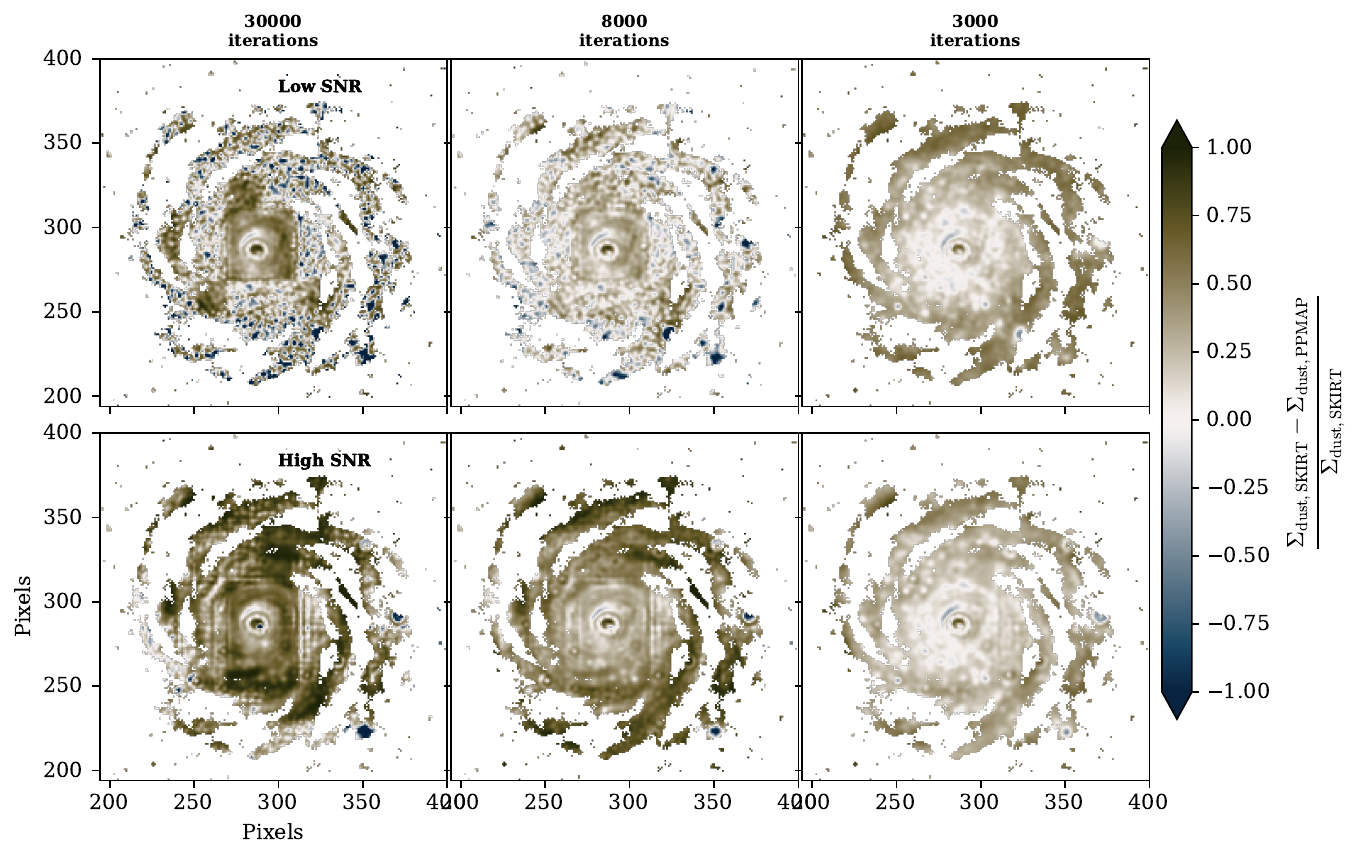} \caption{\textit{Top panel:} Fractional residual of SKIRT input dust mass surface density grid and \textsc{PPMAP}'s  dust mass surface density values with low signal-to-noise input images.  \textit{Bottom panel:} Fractional residual of SKIRT input dust mass surface density grid and \textsc{PPMAP}'s  dust mass surface density values with high signal-to-noise input images. \textit{From left to right:} The results of running 30000 iterations,  8000 iterations and 3000 iterations,  {\magi after offset correction using the 3000 iteration run.}}
\label{fig:fracresihighlow}
\end{figure}
\end{landscape}

To check how this effect changes with the choice of the number of iterations,  we also repeat this test for two more choices of iterations: 8000 and 30000.  Figure \ref{fig:fracresihighlow} shows fractional residual of the input minus output dust mass surface density values at these different iterations for both low and high SNR input galaxy images.  Once again,  it is evident that the 3000 iteration run best reproduces the input dust estimates as it shows fractional residual values close to zero across the galaxy and produces very little spurious artefacts.

\newpage

\subsection{PPMAP performance with different inclinations}
Since \textsc{PPMAP} has been applied to both our galaxy and external galaxies at different inclinations (e.g.  M31 with an inclination of 77.7$^{\circ}$),  we feed the algorithm multiwavelength images of the Auriga galaxy at different inclinations to see how it behaves with this input.  This time,  we fix the number of iterations at 3000.  Figure \ref{fig:diff_incl_scatter_5sigoffcorr} shows our result for pixels with at least a 5$\sigma$ signal-to-noise level in the input 250$\mu$m image.  The model described in equation \ref{eq:scatmodel} is fit to the scatter points after offset correction.  We find a mostly linear relationship between the input and output dust mass surface density values for 0,  45 and 72$^{\circ}$ inclination angles and the offset correction matches especially well for the 0 and 72$^{\circ}$ output values.  For these inclinations,  there is a small deviation from the linear relationship at very low surface densities ($\Sigma_{\mathrm{dust}} \lesssim 0.1$ M$_{\odot}$ pc$^{-2}$)

For the edge-on galaxy at 90$^{\circ}$ inclination,  there is a bigger deviation from the linear relationship at low surface densities (< 1 M$_{\odot}$ pc$^{-2}$) where \textsc{PPMAP} underestimates the dust content.  To evaluate what may be causing this,  we study the fractional residual of dust mass surface density values.  Figure \ref{fig:diffincl_fracresi} shows our result.  The largest underestimates (darker brown regions) by \textsc{PPMAP} seem to be on the outskirts of the galaxy in the edge-on image.  These are also likely to be regions with low dust mass surface densities, where the relation between the input and output values of the simulation become sub-linear ($\Sigma_{\mathrm{dust}} \lesssim 0.5$ M$_{\odot}$ pc$^{-2}$).  We are uncertain of the reason why \textsc{PPMAP} does not deal with these low $\Sigma_{\mathrm{dust}}$ values well. 
\newpage
\begin{landscape}

\begin{figure}
\centering
\includegraphics[width=23cm]{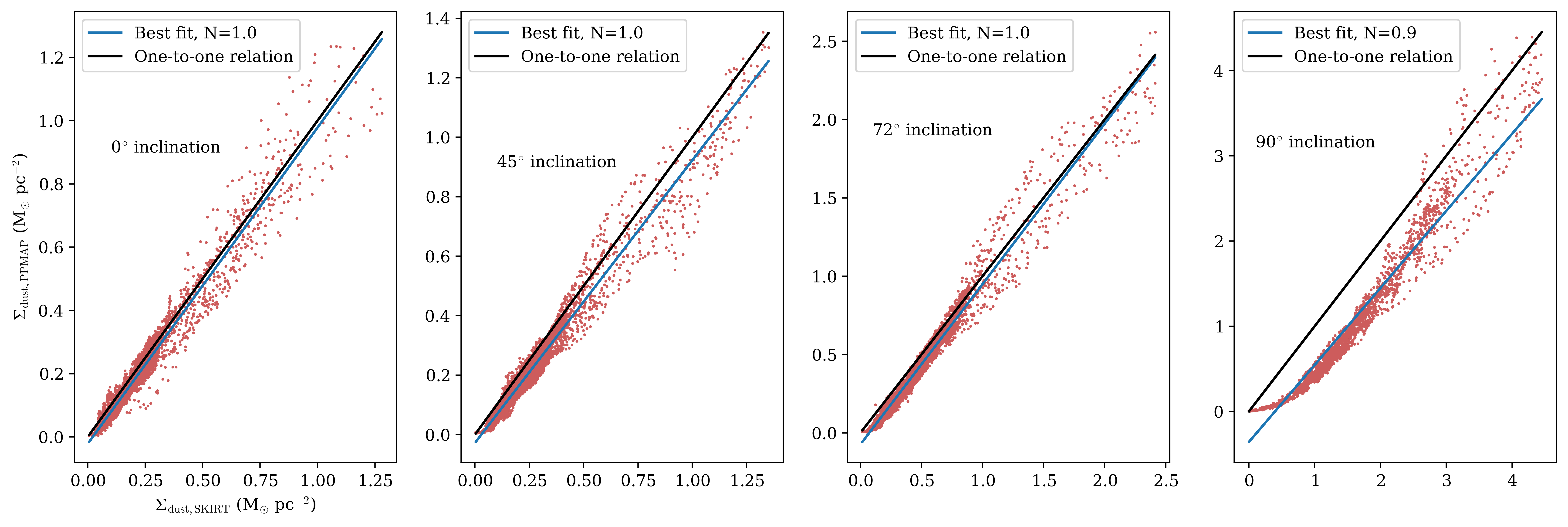} \caption{Scatter plot of SKIRT input dust mass surface density grid and \textsc{PPMAP}'s dust mass surface density values after masking out all of the pixels below a 5$\sigma$ threshold in the 250 $\mu$m input image,  for the galaxy at different inclinations.  The {\magi blue line} shows our best fit model and the {\magi black line} shows the one-to-one relation between the two quantities. {\magi The values at all inclinations have been offset-corrected using the 3000 iteration run for the 0$^{\circ}$ inclination image.}}
\label{fig:diff_incl_scatter_5sigoffcorr}
\end{figure}

\begin{figure}[h]
\centering
\includegraphics[width=23cm]{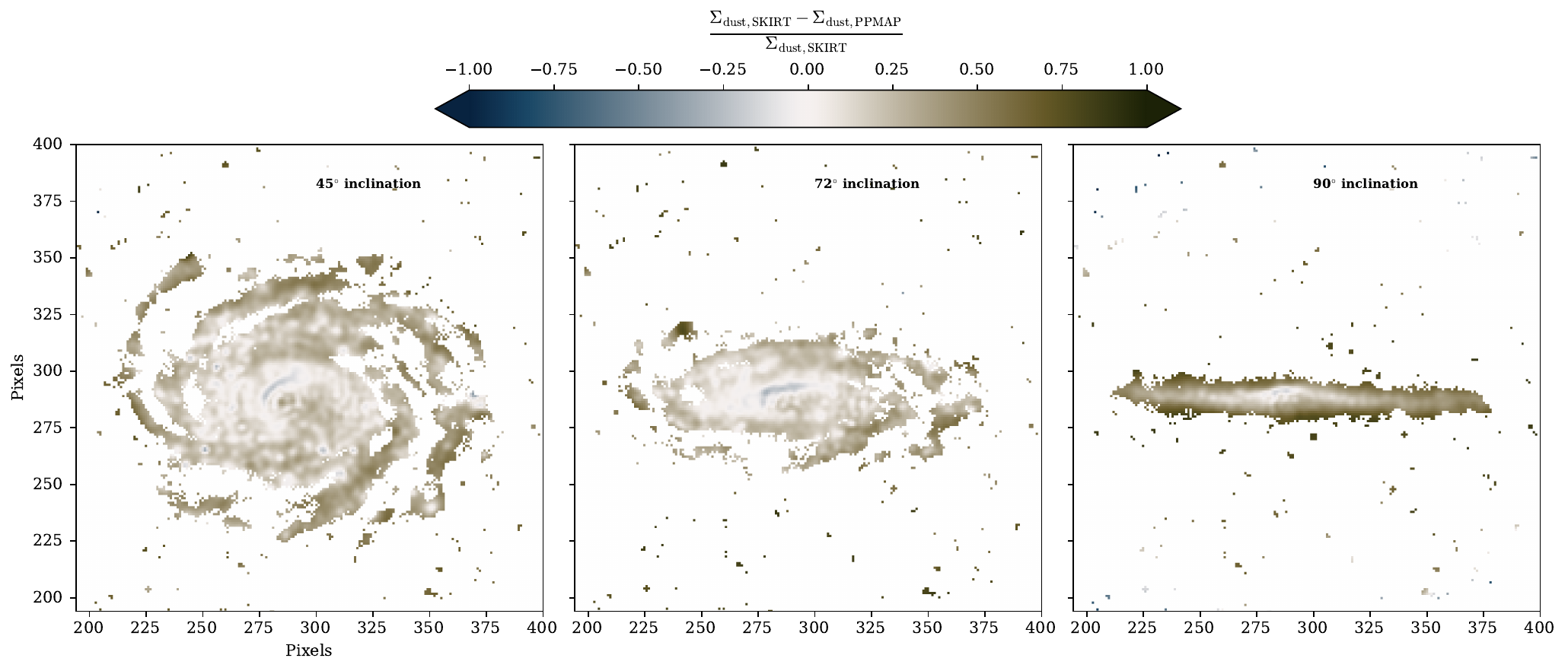} \caption{Fractional residual of SKIRT input dust mass surface density grid and \textsc{PPMAP}'s  dust mass surface density values with different inclination galaxy input images and the number of iterations set to 3000.  {\magi The values at all inclinations have been offset-corrected using the 3000 iteration run for the 0$^{\circ}$ inclination image.} \textit{From left to right:} 45.57 deg inclination,  72.54 deg inclination,  90.0 deg inclination.}
\label{fig:diffincl_fracresi}
\end{figure}

\end{landscape}

\subsection{Limitations of this work}
In this work,  we assume that dust grains emit with a constant dust emissivity index ($\beta$ = 1.8) and a range of temperatures between 10 and 50 K.  Since it is not possible to extract the dust temperature information from the input SKIRT dust grid,  we are unable to test \textsc{PPMAP}'s ability to produce the differential temperatures at this stage.  The assumption of a constant $\beta$,  although commonly used in simulations and observational analysis,  may also be an issue for galaxies on more local scales where there are observed variations in dust properties (e.g.  \citealt{Smith2012}).  Our tests are limited to the comparison of integrated column density, and hence dust mass surface density, produced by \textsc{PPMAP} and only the diffuse dust emission in the Auriga galaxy which has been modelled by SKIRT.

\section{Summary}
\label{sec:ppmapsummary}
\textsc{PPMAP} is a Bayesian algorithm (\citealt{Marsh2015}) which has been successfully applied to many Milky Way observations (\citealt{Marsh2017},  \citealt{Howard2019} , \citealt{Howard2021}) and to \textit{Herschel} observations of M31 (\citealt{Marsh2018}, \citealt{Whitworth2019}).  However,  the confidence in \textsc{PPMAP's} dust estimates thus far has come from tests performed using simulated galactic objects and reasonable estimates provided by the algorithm when compared with previous observations.  In this chapter,  we have attempted to test the capability of \textsc{PPMAP} in estimating the dust mass surface density of a simulated galaxy for the first time.  We use dust-infused synthetic observations of a galaxy (Au 3) from the Auriga simulation suite.  This was not possible to do before the analysis in Chapter 2 as these synthetic observations were only released publicly in late 2021 by \cite{Kapoor2021}.  We perform three tests to see if \textsc{PPMAP} recovers the input dust distribution that has gone into the simulated Auriga galaxy: varying the number of iterations that \textsc{PPMAP} runs,  varying the signal-to-noise levels in the input images taken in by the algorithm and, finally,  varying the inclination angle of the input galaxy image.
Our key findings are:
\begin{enumerate}
    \item For the input images of Au 3 across the 6 wavebands of 70, 100, 160, 250, 350 and 500 $\mu$m,  approximately 3000 iterations are required for \textsc{PPMAP} to produce the dust mass surface density map with the best visual resemblance to the input dust distribution.  {\magi Below this iteration number,  the algorithm does not always converge.  Above this value,  the algorithm overiterates and produces spurious artefacts in the output image. }
    \item For the input images of Au 3 across the 6 wavebands of 70, 100, 160, 250, 350 and 500 $\mu$m,  approximately 3000 iterations are required for \textsc{PPMAP} to produce the {\magi tightest} linear relation between the input and output dust mass surface density.  The input and output dust mass surface density values are very strongly correlated for the 3000 iteration run and most other runs (Spearman rank coefficient > 0.7,  p-value $\ll$ 0.01). There is an offset between \textsc{PPMAP}'s output dust mass surface density values and the input dust distribution, for all testing scenarios.  {\magi This is most likely due to the difference in $\kappa$ used by THEMIS model which feeds into SKIRT and the $\kappa$ used by \textsc{PPMAP}.} We correct for this offset using a correction factor from the 3000 iteration run. 
    \item For both the low and high SNR input images,  \textsc{PPMAP} underestimates the dust mass surface density,  following offset-correction.
    \item \textsc{PPMAP} produces a mostly linear relation between the input and output dust mass surface density for Au 3 at 0, 45 and 72$^{\circ}$ inclination, {\magi following offset correction},  but fails to do so for the edge-on case.
\end{enumerate}

%% file: Chapters/Conclusion.tex
\chapquote{``There is stardust in your veins. We are literally,  ultimately children of the stars."}{Jocelyn Bell Burnell}{}

\section{Thesis purpose \& overview}
Despite \textit{Herschel} surveys of our neighbouring galaxy,  Andromeda (M31),  revolutionising our view of dust continuum emission in the galaxy,  previous studies left some unanswered questions.  Existing studies failed to find the cause of radial variations in dust temperature and $\beta$ and did not test the effect of dense molecular gas on the radial variations in dust properties at sub-kpc scales.  Another unknown, which would help us to constrain the amount of very cold dust or give us clues about any different dust types,  was whether there is any excess emission in M31 at longer wavelengths like 850 $\mu$m. Moreover,  while a \textit{Herschel} study gave us the global star formation rate for M31,  we did not know if the star formation efficiency depends on position within the galaxy.  We also did not know to what extent observational dust properties like dust temperature and $\beta$ affect star formation in this galaxy.  We were left with the question of how much CO-dark gas is there in M31,  if any,  and where is it? Finding any CO-dark gas would have implications for the amount of molecular gas we are actually detecting using the CO tracer method and would raise questions about whether any CO is missing due to the photodissociation of this molecule.

Therefore,  the goal of this thesis has been to get a fuller picture of the effect of ISM interactions with star formation in our big neighbour using existing \textit{Herschel} observations and new JCMT observations from the HASHTAG large programme.
We study the interplay of dust with gas properties and star formation in M31 at the scale of individual molecular clouds (sub-kpc).  We test whether the variations in the dust emissivity index ($\beta$) previously found in M31 is due to an increase of $\beta$ in dense molecular gas regions (Chapter \ref{chapter:betavar}).  We search for a submillimetre excess in M31 at 450 and 850 $\mu$m wavelengths (Chapter \ref{chapter:sedfit}).  We investigate how well dust traces the gas of M31 when compared to CO and \textsc{HI} within clouds (Chapter \ref{chapter:dustmass}).  We investigate whether the star formation efficiency of clouds within M31 depends the galactocentric distance of the cloud{\magi ,  and} we check whether observational dust properties like dust temperature and $\beta$ influence the star formation efficiency of clouds (Chapter \ref{chapter:sfe}).  Finally,  as a side branch of work,  we test the Bayesian algorithm \textsc{PPMAP} with a simulated galaxy from the Auriga simulation suite (Chapter \ref{chapter:ppmaptest}) with the motivation to make it possible to apply \textsc{PPMAP} in the future to new observations of galaxies like M31.  In the process of the work conducted in this thesis,  we also produce three new cloud catalogues (Appendix \ref{ap:cloud_catscarmappmap} \& \ref{ap:hashtagcloudcat}).

\section{Key Results}
Our key results are as follows:
\begin{enumerate}
    \item Using CARMA CO observations and \textsc{PPMAP} dust emissivity index measurements,  we see a radial variation in $\beta$ in agreement with previous studies (e.g. \citealt{Smith2012}, \citealt{Draine2014}, \citealt{Whitworth2019}), with a decrease in $\beta$ going from the inner ring to outer dusty,  star-forming ring.  We find no evidence for radial variations in $\beta$ being caused by an increase of $\beta$ in dense molecular gas regions at radii between $5-7.5$ kpc and $9-15$ kpc.  We conclude that an increase of $\beta$ in dense molecular gas regions is not the prominent driver of the radial variations in $\beta$ in M31.  A speculative possibility is that the variations are driven by a genuine radial change in the C/Si ratio within dust grains. 
    \item Using CARMA CO observations and \textsc{PPMAP} dust mass surface density measurements,  we produce a CO-selected and dust-selected catalogue.  We find a population of clouds in our dust-selected catalogue with lower median CO-traced molecular GDR than in our CO-selected catalogue. These may be clouds containing CO-dark molecular gas, although we are unable to rule out the possibility that these structures are confused with features in the atomic phase of the ISM.
    \item Using new observations at 450 and 850 $\mu$m from the HASHTAG large programme and existing \textit{Herschel} observations,  we produce maps showing the spatial distribution of $\Sigma_{\mathrm{dust}}$,  $T_{\mathrm{dust}}$,  $\beta$ from SED fitting.  We find that a variable dust emissivity index ($\beta$) provides statistically better fits of the modified blackbody model to our data,  consistent with past studies showing radial variations in this parameter across M31.  The next best model is a single temperature modified blackbody model with a fixed $\beta$ of 2.
	\item Using the $\Sigma_{\mathrm{dust}}$ map obtained from SED fitting of HASHTAG and \textit{Herschel} observations, we search for a sub-mm excess in M31.  We find no strong evidence for excess emission at long wavelengths,  in particular at 850 $\mu$m suggesting that our observations are not sensitive to cold dust.  To see a real astrophysical excess that is 5 times above the noise level,  we would require dust mass at least one order of magnitude higher.  We do not rule out the possibility that there is still a large amount of cold dust in M31 masked by the emission from warmer dust.
	\item Using the $\Sigma_{\mathrm{dust}}$ map obtained from SED fitting of HASHTAG and \textit{Herschel} observations, we produce a catalogue of 422 dust-traced clouds, probing a spatial resolution of 68 pc.  We find a very strong positive correlation between the total dust-traced gas mass and the total molecular + atomic gas mass of the clouds,  with statistical significance of greater than 99\%.  We show that there is a discrepancy between the amount of gas mass in the clouds of M31 traced by dust vs combined CO and \textsc{HI}.  We propose that this discrepancy could be due to an incorrect assumption of the constant gas-to-dust ratio (GDR) or due to contributions from CO-dark gas.  From our CO-dark gas fraction calculation,  an infeasible amount of molecular gas would need to be CO-dark to reproduce the offset.  Knowing this and the information that a large number of our clouds are \textsc{HI}-dominated,  we suggest that variations in the GDR value is a more likely explanation for the offset.  We find a large variation in the GDR of clouds with galactocentric radius which could be due to M31's negative metallicity gradient.
	\item For the clouds extracted from the HASHTAG $\Sigma_{\mathrm{dust}}$ map,  we find long total dust-traced gas depletion times across our clouds with a median value of 12.2 Gyr.  We find an average CO-traced molecular gas depletion time for our clouds of 2.6 Gyr,  larger than values found for the Milky Way and nearby galaxies. 
	\item We find that the total star formation rate of clouds is strongly correlated with both CO-traced molecular gas and \textsc{HI} gas,  implying that both of these gas phases may be important for star formation in this galaxy.  We do not see any evidence of total or molecular gas depletion time being associated with distance from the galactic centre.
	\item We find a broad range of total dust-traced and CO-traced gas depletion times for clouds suggesting that we may be catching clouds at different evolutionary stages.
	\item  We find a very strong anti-correlation between dust temperature and total dust-traced gas depletion time and a moderate correlation between $\beta$ and depletion time,  suggesting that observational dust properties have a significant connection to star formation.  We find a weak correlation between the CO-traced molecular gas fraction of clouds and their average dust temperature,  suggesting that dust heating by stars is not dependent on the gas phase of the surrounding ISM. 
	\item We use a simulated galaxy from the Auriga simulation suite to test the algorithm \textsc{PPMAP}.  For the input images of the Auriga galaxy across the 6 wavebands of 70, 100, 160, 250, 350 and 500 $\mu$m,  approximately 3000 iterations are required for \textsc{PPMAP} to produce the dust mass surface density map with the best visual resemblance to the input dust distribution.  For both the low and high SNR input images,  \textsc{PPMAP} underestimates the dust mass surface density.  The input and output dust mass surface density values are very strongly correlated for most of our tests (Spearman rank coefficient > 0.7,  p-value $\ll$ 0.01).
  	\item \textsc{PPMAP} produces a mostly linear relation between the input and output dust mass surface density for the simulated galaxy at 0, 45 and 72$^{\circ}$ inclination but fails to do so for the edge-on case.
\end{enumerate}

\section{Future work}
\begin{figure}[h!]
\centering
\includegraphics[width=12cm]{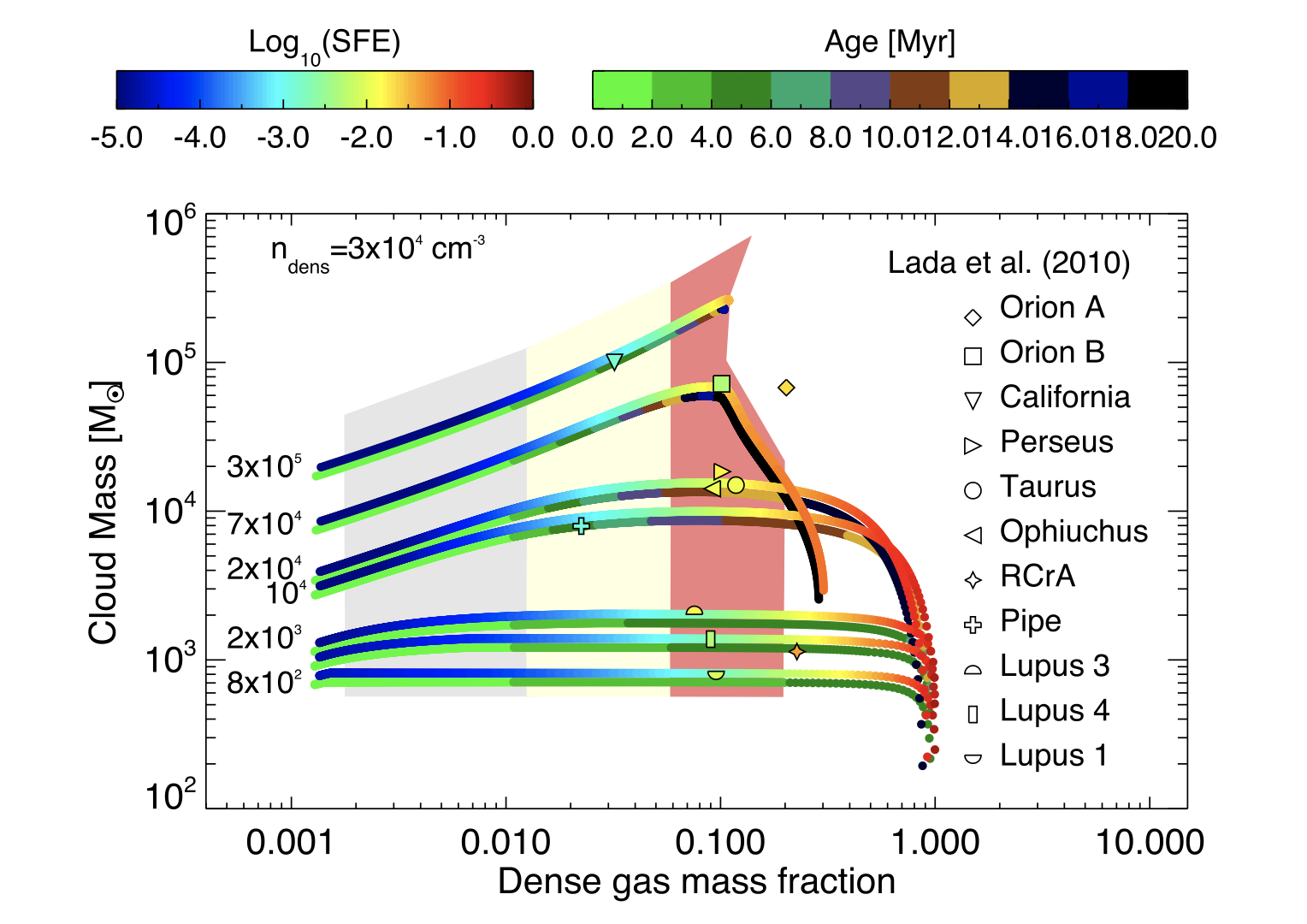} \caption{Figure 2 from \cite{Vazquez-Semadeni2018}. Their figure shows the instantaneous cloud mass vs instantaneous dense gas mass fraction (i.e.  mass at densities $n \geq 3 \times 10^4$ cm$^{-3}$) of model clouds. The total mass of the cloud is labelled next to each line. The tracks consist of two coloured lines. The lines using the colorbar at the top left indicate the instantaneous SFE of the model. The lines using the colorbar at the top right show the cloud age since becoming 25\% molecular.  The points show the clouds observed in the Milky Way by \cite{Lada2010} with colours representing their star formation efficiency. The gray,  cream and rose-coloured vertical bands,  respectively,  indicate the evolutionary periods during which 25, 50 and 75\% of the gas mass is between $10^{3}$ cm$^{-3}$ (left edge of each band) and $3 \times 10^{3}$ cm$^{-3}$ (right edge of the band).  \copyright{\citealt{Vazquez-Semadeni2018}.}}
\label{fig:vassem_cloudevolve}
\end{figure}

There are several interesting avenues through which we can exploit the new HASHTAG survey data and tackle existing questions about M31,  continuing on from this work.  Below we list a few projects for future work:
\begin{itemize}
\item \textbf{Understanding the cause of radial variations of dust properties} - One plausible cause for radial variations in the dust emissivity index in M31 could be due to changes in the C/Si ratio as we move from the centre of the galaxy to the outskirts. Studies have shown that M31 has a negative metallicity gradient as you move outwards from a galactocentric radius of $\sim$ 4 kpc (\citealt{Galarza1999},  \citealt{Gregersen2015}).  A more recent study by \cite{Gibson2023} looking at the inner 7 kpc has also found an increase in metallicity with radius near the centre of the galaxy.  Therefore,  a natural progression of our work would be to examine how the dust temperature and dust emissivity index within extracted clouds vary with metallicity. 

\item \textbf{Exploring dense {\magi and CO-dark gas} in Andromeda} - Hydrogen cyanide (HCN) is a tracer of dense gas.  When we look at the efficiency of star formation,  what we are really asking is how many high density regions can be formed from the diffuse molecular gas.  {\magi Therefore,  we can study quantities like the dense gas fraction traced by HCN/CO and check if this correlates with star formation efficiency. } This might provide some insight into regions of low vs high star formation efficiency (e.g.  \citealt{Usero2015}, \citealt{Heyer2022}). This work would require applying to observe HCN emission using the Northern Extended Millimeter Array (NOEMA) at $\sim$ 2.9 mm wavelength. 

{\magi C[II] line emission at 158 $\mu$m can be used to trace photodissociation regions and there is evidence from observations (\citealt{Pineda2013},  \citealt{Madden2020}) that a significant amount of C[II] can come from regions of CO-dark gas.  C[II] data exist for five fields within M31 taken by \textit{Herschel} instruments (\citealt{Kapala2014}) at $\sim$50 pc spatial resolution,  and a useful next step might be to calculate the levels of C[II] emission for clouds which fall within those five regions. }

\item \textbf{Studying molecular cloud evolution} - In this work,  we find a broad mix star formation efficiencies for clouds extracted from observations taken by the \textit{Herschel} and HASHTAG surveys.  This might be because we are catching the clouds at different evolutionary stages.  A cloud evolution framework has been previously suggested in observational studies by \cite{Kawamura2009} and \cite{Miura2012} for the LMC and M33,  whereby cloud evolution is classified into three categories: 1) showing no signatures of massive star formation,  2) showing signatures of embedded star formation traced by compact \textsc{HII} regions and 3) showing more active star formation measured by clouds containing stellar clusters.  Future work with an existing high resolution H$\alpha$ emission map and the PHAT optical map of M31 to classify the clouds according to this framework can help to distinguish any trends in star formation efficiency with evolutionary stage. 

A model presented by \cite{Vazquez-Semadeni2018},  has used a set of 11 molecular clouds within the MW with an observed scatter of SFEs,  to show that the mass of the cloud at the time it is observed might not be the maximum mass reached by the cloud,  but simply the mass of a cloud at a particular evolutionary stage.  As such,  two clouds of identical dense gas mass may follow a completely different evolutionary track (see Figure \ref{fig:vassem_cloudevolve}).  Their model does not assume any particular timescale (such as the freefall time) for cloud collapse but assumes that the entire cloud is under global collapse.  It is possible that we are seeing a similar phenomenon by simply catching our clouds at different times during their lifecycle.  It would be interesting to see if their model is applicable on extragalactic scales and where our clouds sit within their proposed evolutionary tracks.

\item \textbf{Applying the star formation uncertainty principle to the star formation law} - An issue which has been predicted at least for Galactic observations when moving from lower resolution studies to more resolved studies is the breakdown of star formation laws like the Kennicutt-Schmidt relationship due to star formation processes not being sampled well enough at smaller spatial scales (\citealt{Kruijssen2014}).  Our study is the first of its kind probing M31 at much finer spatial scales than previous work.  Motivated by the observed scatter in SFR for clouds within M31,  a possible avenue to try is applying the star formation uncertainty principle to our work to establish on what scales the star formation relations may start breaking down.

\item \textbf{Testing \textsc{PPMAP} with simulated galaxy emission matching JCMT wavelengths} - The next step to testing \textsc{PPMAP} using a simulated galaxy could be to apply \textsc{PPMAP} to the remaining Auriga galaxies to get a larger sample size for statistical analysis.  More interestingly,  it would be useful to test and optimise the algorithm for use at JCMT wavelengths.  The Auriga galaxy simulation data cube used in our work is comprised of 50 wavelengths,  including 450 and 850 $\mu$m.  Therefore,  a similar test to the one in Chapter 6 could be repeated to find the optimal scaling factor for the algorithm's output for use on the HASHTAG survey maps. 

\end{itemize}

\section{Concluding remarks}
While M31 is one of the popular testbeds for studying the nature of the interstellar medium in a galaxy like our own,  there is a gap in our understanding of dust properties and star formation efficiency on the scale of individual clouds within this galaxy.  Although many surveys and studies have conducted global and pixel-by-pixel analyses of M31's ISM,  we are lacking in our understanding of why dust properties vary with distance from the galactic centre,  and how clouds in our big neighbour evolve as stars begin to form and gradually eat up the gas.  With new observations of M31 allowing us to resolve individual clouds,  in this thesis we are able to apply common source extraction and fitting techniques,  already readily used on observations of the Milky Way,  on this external galaxy,  in order to first identify clouds and then probe their internal processes. 

We contribute to the understanding of whether variations in dust properties are caused by the interplay of dust with dense gas.  Our research also compares whether the intensity of emission from M31’s dust deviates from predicted models,  for the first time incorporating high resolution maps of the M31 at 450 and 850 $\mu$m wavelengths. The work in this thesis confirms that some properties of interstellar dust emission are directly correlated with regions in which stars form.  We learn that dusty clouds within this galaxy have a slow gas depletion time in the context of actively star-forming galaxies in the local universe,  raising interesting questions about the nature of cloud evolution in M31.

%% file: Chapters/An_appendix_betavar.tex
\chapter{Catalogue of clouds from CARMA and PPMAP}
{\magi In Chapter 2,  we use both CO and dust to trace molecular clouds. The catalogue of 140 clouds extracted from CARMA CO and 196 clouds extracted from \textsc{PPMAP} dust data are in tables A.1 and A.2 below\footnote{These catalogues are available in the Flexible Image Transport System (FITS) table format from the journal \textit{Monthly Notices of the Royal Astronomical Society}: \url{https://academic.oup.com/mnras/article/511/4/5287/6414546\#supplementary-data}}.  Our methods for calculating the quantities in the catalogues are below:

\begin{itemize}
\item \textbf{Cloud ID:} A random identification number assigned by the \texttt{astrodendro} algorithm.
\item \textbf{Mean $T_{\mathrm{dust}}$:} This is the dust mass surface density-weighted mean dust temperature of each cloud obtained using the method described in Chapter 2.
\item \textbf{Mean $\beta$:} This is the dust mass surface density-weighted mean dust emissivity index of each cloud obtained using the method described in Chapter 2.
\item \textbf{Total CO-traced gas mass:} This is the total CO-traced gas mass of each cloud obtained using the method described in Chapter 2.
\item \textbf{Total dust mass:} This is the total dust mass of each cloud obtained using the method described in Chapter 2.
\item \textbf{CO-traced GDR:} This is the CO-traced gas-to-dust ratio of each cloud obtained using the method described in Chapter 2.
\item \textbf{RA [J2000]:} This is the Right Ascension coordinate position of the pixel in the cloud with peak intensity,  given in units of degrees. 
\item \textbf{Dec [J2000]:} This is the Declination coordinate position of the pixel in the cloud with peak intensity, given in units of degrees. 
\item \textbf{Radius to peak:} This is the distance from the centre of the galaxy to the pixel with the peak CO intensity (dust mass surface density) in the CO-traced (dust-traced) cloud.  We calculate this by assuming the area of an ellipse and solving for the semi-major axis,  $a$,  by substituting in $b = a \; cos(i)$; where $b$ is the semi-minor axis and $i$ is the inclination angle of M31 taken as 77$^{\circ}$.  Assuming a distance to the cloud of $\sim$ 785 kpc and the galactic centre to be at the position mentioned in Chapter 2,  $a$ becomes the physical radius of the peak pixel given in units of kpc.
\item \textbf{Total gas mass:} This is a linear combination of the total CO-traced molecular gas mass and the HI gas mass of each cloud obtained using the method described in Chapter 2.
\item \textbf{Total GDR:} This is the total gas-to-dust ratio of each cloud obtained by dividing the cloud's total gas mass by the cloud's total dust mass.
\item \textbf{Equivalent radius:} This is the radius of the cloud assuming its physical area ($A_{\mathrm{cloud}}$) can be defined as the area of a circle and that the distance to the cloud is $\sim$ 785 kpc.  We calculate this by substituting the physical area of the cloud into the equation for the area of a circle and solving to find the radius. 
\begin{equation}
R_{\mathrm{eq}} =  \sqrt{\frac{A_{\mathrm{cloud}}}{\pi}}
\end{equation}
where $R_{\mathrm{eq}}$ is the equivalent radius.
\end{itemize}}
\begin{landscape}

\setlength\tabcolsep{2.3pt}

\sisetup{
    tight-spacing           = true,
    round-mode              = places,
    round-precision         = 1,
    }

\input{Chapters/Plots_Tables/CO_selected_catalogue.tex}

\input{Chapters/Plots_Tables/Dust_selected_catalogue.tex}

\end{landscape}

%% file: Chapters/Plots_Tables/CO_selected_catalogue.tex
\begin{table*}
\caption{Properties of 140 clouds in the CO-selected catalogue,  ordered by highest to lowest peak CO intensity.}
\scriptsize
\begin{tabular}{cccSSccccScc}
\hline
Cloud ID & Mean $T_{\mathrm{dust}}$ & Mean $\beta$ & {Total CO-traced gas mass} & {Total dust mass }& CO-traced GDR & RA [J2000] & Dec [J2000] & Radius to peak & {Total gas mass} & Total GDR & Equivalent radius \\
$\mathrm{}$ & ($\mathrm{K}$) & $\mathrm{}$ & {($\mathrm{M_{\odot}}$)} & {($\mathrm{M_{\odot}}$)} & $\mathrm{}$ & ($\mathrm{{}^{\circ}}$) & ($\mathrm{{}^{\circ}}$) & ($\mathrm{kpc}$) & {($\mathrm{M_{\odot}}$)} & $\mathrm{}$ & ($\mathrm{pc}$) \\
\hline
22 & 18.3 & 2.1 & 4.1E+05 & 1.1E+04 & 36.7 & 11.3 & 41.6 & 11.9 & 7.1E+05 & 100.6 & 67.1 \\
93 & 14.0 & 2.3 & 2.0E+05 & 6.0E+03 & 32.7 & 11.1 & 41.6 & 5.9 & 4.2E+04 & 39.6 & 32.1 \\
96 & 13.2 & 2.5 & 1.8E+05 & 6.5E+03 & 27.0 & 11.1 & 41.6 & 6.0 & 3.4E+04 & 32.2 & 31.0 \\
44 & 15.6 & 2.0 & 4.5E+05 & 7.0E+03 & 64.2 & 11.4 & 41.8 & 11.6 & 3.8E+05 & 118.2 & 63.7 \\
95 & 13.7 & 2.4 & 2.9E+05 & 8.1E+03 & 35.6 & 11.0 & 41.6 & 5.9 & 6.2E+04 & 43.2 & 40.3 \\
179 & 15.8 & 1.8 & 1.7E+05 & 2.4E+03 & 71.3 & 11.2 & 41.9 & 12.1 & 1.6E+05 & 138.1 & 41.2 \\
\end{tabular}
\end{table*}

\newpage

\begin{table*}
\scriptsize
\begin{tabular}{cccSSccccScc}
Cloud ID & Mean $T_{\mathrm{dust}}$ & Mean $\beta$ & {Total CO-traced gas mass} & {Total dust mass }& CO-traced GDR & RA [J2000] & Dec [J2000] & Radius to peak & {Total gas mass} & Total GDR & Equivalent radius \\
$\mathrm{}$ & ($\mathrm{K}$) & $\mathrm{}$ & {($\mathrm{M_{\odot}}$)} & {($\mathrm{M_{\odot}}$)} & $\mathrm{}$ & ($\mathrm{{}^{\circ}}$) & ($\mathrm{{}^{\circ}}$) & ($\mathrm{kpc}$) & {($\mathrm{M_{\odot}}$)} & $\mathrm{}$ & ($\mathrm{pc}$) \\
\hline
14 & 17.1 & 1.8 & 1.6E+05 & 3.6E+03 & 45.0 & 11.3 & 41.6 & 12.4 & 2.6E+05 & 117.6 & 39.4 \\
15 & 19.2 & 1.9 & 7.2E+04 & 1.4E+03 & 51.3 & 11.3 & 41.6 & 12.4 & 1.2E+05 & 134.5 & 27.2 \\
167 & 16.2 & 1.9 & 1.5E+05 & 2.4E+03 & 62.5 & 11.2 & 41.9 & 10.7 & 1.4E+05 & 119.8 & 42.9 \\
21 & 20.5 & 1.9 & 8.8E+04 & 2.1E+03 & 41.5 & 11.3 & 41.6 & 12.1 & 1.5E+05 & 114.3 & 32.1 \\
101 & 13.9 & 2.5 & 2.6E+05 & 9.3E+03 & 28.0 & 11.0 & 41.6 & 5.9 & 1.1E+05 & 40.0 & 58.2 \\
24 & 17.2 & 1.8 & 1.5E+05 & 4.0E+03 & 36.3 & 11.3 & 41.6 & 12.0 & 3.0E+05 & 109.9 & 44.6 \\
\end{tabular}
\end{table*}

\newpage

\begin{table*}
\scriptsize
\begin{tabular}{cccSSccccScc}
Cloud ID & Mean $T_{\mathrm{dust}}$ & Mean $\beta$ & {Total CO-traced gas mass} & {Total dust mass }& CO-traced GDR & RA [J2000] & Dec [J2000] & Radius to peak & {Total gas mass} & Total GDR & Equivalent radius \\
$\mathrm{}$ & ($\mathrm{K}$) & $\mathrm{}$ & {($\mathrm{M_{\odot}}$)} & {($\mathrm{M_{\odot}}$)} & $\mathrm{}$ & ($\mathrm{{}^{\circ}}$) & ($\mathrm{{}^{\circ}}$) & ($\mathrm{kpc}$) & {($\mathrm{M_{\odot}}$)} & $\mathrm{}$ & ($\mathrm{pc}$) \\
\hline
97 & 14.7 & 2.2 & 5.3E+04 & 2.2E+03 & 24.4 & 11.4 & 41.9 & 11.7 & 8.5E+04 & 63.8 & 27.2 \\
144 & 14.2 & 2.6 & 6.3E+04 & 1.5E+03 & 41.7 & 11.0 & 41.6 & 6.0 & 2.1E+04 & 55.9 & 28.5 \\
69 & 15.2 & 2.3 & 6.4E+04 & 2.1E+03 & 30.6 & 11.4 & 41.8 & 11.0 & 8.5E+04 & 71.6 & 31.0 \\
41 & 14.7 & 2.2 & 8.3E+04 & 2.7E+03 & 31.2 & 11.4 & 41.8 & 12.3 & 1.4E+05 & 84.9 & 36.4 \\
165 & 14.5 & 2.3 & 6.2E+04 & 2.2E+03 & 27.8 & 11.2 & 41.9 & 10.8 & 7.0E+04 & 59.1 & 31.0 \\
175 & 16.5 & 2.4 & 5.7E+04 & 2.0E+03 & 28.4 & 11.2 & 41.9 & 11.5 & 8.6E+04 & 71.5 & 31.0 \\
\end{tabular}
\end{table*}

\newpage

\begin{table*}
\scriptsize
\begin{tabular}{cccSSccccScc}
Cloud ID & Mean $T_{\mathrm{dust}}$ & Mean $\beta$ & {Total CO-traced gas mass} & {Total dust mass }& CO-traced GDR & RA [J2000] & Dec [J2000] & Radius to peak & {Total gas mass} & Total GDR & Equivalent radius \\
$\mathrm{}$ & ($\mathrm{K}$) & $\mathrm{}$ & {($\mathrm{M_{\odot}}$)} & {($\mathrm{M_{\odot}}$)} & $\mathrm{}$ & ($\mathrm{{}^{\circ}}$) & ($\mathrm{{}^{\circ}}$) & ($\mathrm{kpc}$) & {($\mathrm{M_{\odot}}$)} & $\mathrm{}$ & ($\mathrm{pc}$) \\
\hline
166 & 15.8 & 2.1 & 4.8E+04 & 1.1E+03 & 43.1 & 11.2 & 41.9 & 11.6 & 6.7E+04 & 102.4 & 29.7 \\
68 & 16.5 & 1.9 & 6.2E+04 & 1.6E+03 & 39.2 & 11.4 & 41.9 & 11.5 & 1.1E+05 & 107.4 & 33.3 \\
136 & 14.1 & 2.6 & 6.8E+04 & 3.0E+03 & 22.9 & 11.0 & 41.6 & 6.2 & 3.9E+04 & 36.0 & 35.4 \\
26 & 15.6 & 2.0 & 8.3E+04 & 2.7E+03 & 30.6 & 11.3 & 41.7 & 11.9 & 2.0E+05 & 106.1 & 40.3 \\
56 & 15.3 & 2.2 & 5.1E+04 & 1.2E+03 & 41.1 & 11.3 & 41.7 & 10.9 & 9.6E+04 & 117.6 & 31.0 \\
77 & 15.4 & 2.1 & 5.9E+04 & 2.1E+03 & 27.9 & 11.4 & 41.9 & 11.5 & 1.2E+05 & 84.5 & 33.3 \\
\hline
\end{tabular}
\end{table*}

%% file: Chapters/Plots_Tables/Dust_selected_catalogue.tex
\begin{table*}
\caption{Properties of 196 clouds in the dust-selected catalogue,  ordered by highest to lowest peak dust mass surface density.}
\scriptsize
\begin{tabular}{cccSSccccScc}
\hline
Cloud ID & Mean $T_{\mathrm{dust}}$ & Mean $\beta$ & {Total CO-traced gas mass} & {Total dust mass} & CO-traced GDR & RA [J2000] & Dec [J2000] & Radius to peak & {Total gas mass} & Total GDR & Equivalent radius \\
$\mathrm{}$ & ($\mathrm{K}$) & $\mathrm{}$ & {($\mathrm{M_{\odot}}$)} & {($\mathrm{M_{\odot}}$)} & $\mathrm{}$ & ($\mathrm{{}^{\circ}}$) & ($\mathrm{{}^{\circ}}$) & ($\mathrm{kpc}$) & {($\mathrm{M_{\odot}}$)} & $\mathrm{}$ & ($\mathrm{pc}$) \\
\hline
164 & 13.6 & 2.4 & 1.3E+06 & 5.2E+04 & 24.0 & 11.1 & 41.6 & 6.0 & 3.8E+05 & 31.3 & 100.5 \\
313 & 17.5 & 2.3 & 1.6E+05 & 6.4E+03 & 25.6 & 11.2 & 41.9 & 10.8 & 1.9E+05 & 55.4 & 43.8 \\
159 & 15.6 & 2.2 & 7.0E+05 & 3.2E+04 & 22.1 & 11.1 & 41.6 & 6.7 & 6.5E+05 & 42.6 & 103.1 \\
323 & 17.8 & 2.2 & 3.5E+05 & 1.2E+04 & 29.2 & 11.1 & 41.9 & 11.2 & 4.7E+05 & 68.4 & 72.9 \\
150 & 15.5 & 2.2 & 2.0E+05 & 7.3E+03 & 27.3 & 11.1 & 41.6 & 6.0 & 9.4E+04 & 40.1 & 42.1 \\
146 & 15.1 & 2.1 & 5.5E+05 & 3.5E+04 & 15.6 & 11.4 & 41.9 & 11.9 & 1.4E+06 & 54.1 & 104.8 \\
\end{tabular}
\end{table*}

\begin{table*}
\scriptsize
\begin{tabular}{cccSSccccScc}
Cloud ID & Mean $T_{\mathrm{dust}}$ & Mean $\beta$ & {Total CO-traced gas mass} & {Total dust mass} & CO-traced GDR & RA [J2000] & Dec [J2000] & Radius to peak & {Total gas mass} & Total GDR & Equivalent radius \\
$\mathrm{}$ & ($\mathrm{K}$) & $\mathrm{}$ & {($\mathrm{M_{\odot}}$)} & {($\mathrm{M_{\odot}}$)} & $\mathrm{}$ & ($\mathrm{{}^{\circ}}$) & ($\mathrm{{}^{\circ}}$) & ($\mathrm{kpc}$) & {($\mathrm{M_{\odot}}$)} & $\mathrm{}$ & ($\mathrm{pc}$) \\
\hline
52 & 16.3 & 2.1 & 5.9E+04 & 2.7E+03 & 21.5 & 11.3 & 41.7 & 11.9 & 1.4E+05 & 73.9 & 31.0 \\
27 & 20.4 & 1.9 & 6.6E+04 & 1.8E+03 & 36.1 & 11.3 & 41.6 & 12.1 & 1.1E+05 & 98.0 & 27.2 \\
50 & 14.6 & 2.1 & 9.3E+04 & 1.1E+04 & 8.2 & 11.4 & 41.8 & 13.3 & 5.3E+05 & 54.7 & 68.2 \\
297 & 14.9 & 2.1 & 4.6E+05 & 1.7E+04 & 28.0 & 11.3 & 41.9 & 11.3 & 6.6E+05 & 67.7 & 88.4 \\
302 & 15.7 & 2.1 & 1.0E+05 & 4.9E+03 & 21.0 & 11.2 & 41.9 & 11.5 & 1.5E+05 & 51.3 & 40.3 \\
168 & 14.0 & 2.3 & 8.5E+04 & 7.5E+03 & 11.4 & 11.4 & 41.9 & 10.9 & 3.0E+05 & 52.0 & 57.0 \\
\end{tabular}
\end{table*}

\begin{table*}
\scriptsize
\begin{tabular}{cccSSccccScc}
Cloud ID & Mean $T_{\mathrm{dust}}$ & Mean $\beta$ & {Total CO-traced gas mass} & {Total dust mass} & CO-traced GDR & RA [J2000] & Dec [J2000] & Radius to peak & {Total gas mass} & Total GDR & Equivalent radius \\
$\mathrm{}$ & ($\mathrm{K}$) & $\mathrm{}$ & {($\mathrm{M_{\odot}}$)} & {($\mathrm{M_{\odot}}$)} & $\mathrm{}$ & ($\mathrm{{}^{\circ}}$) & ($\mathrm{{}^{\circ}}$) & ($\mathrm{kpc}$) & {($\mathrm{M_{\odot}}$)} & $\mathrm{}$ & ($\mathrm{pc}$) \\
\hline
187 & 14.3 & 2.2 & 2.1E+04 & 1.6E+03 & 12.7 & 11.4 & 41.9 & 11.5 & 7.4E+04 & 57.5 & 27.2 \\
178 & 13.9 & 2.6 & 9.6E+04 & 1.0E+04 & 9.5 & 11.3 & 41.9 & 10.7 & 3.4E+05 & 43.6 & 70.8 \\
160 & 14.9 & 2.1 & 9.6E+04 & 7.9E+03 & 12.1 & 11.4 & 41.9 & 11.5 & 3.6E+05 & 57.7 & 56.3 \\
227 & 15.3 & 2.1 & 8.7E+04 & 2.6E+03 & 32.9 & 11.3 & 41.9 & 11.0 & 1.1E+05 & 75.3 & 34.4 \\
28 & 14.4 & 2.1 & 1.1E+05 & 1.6E+04 & 7.2 & 11.4 & 41.7 & 13.0 & 8.0E+05 & 57.5 & 82.8 \\
288 & 13.4 & 2.0 & 2.9E+04 & 4.3E+03 & 6.7 & 11.4 & 42.0 & 12.6 & 1.6E+05 & 43.8 & 43.8 \\
\end{tabular}
\end{table*}

\begin{table*}
\scriptsize
\begin{tabular}{cccSSccccScc}
Cloud ID & Mean $T_{\mathrm{dust}}$ & Mean $\beta$ & {Total CO-traced gas mass} & {Total dust mass} & CO-traced GDR & RA [J2000] & Dec [J2000] & Radius to peak & {Total gas mass} & Total GDR & Equivalent radius \\
$\mathrm{}$ & ($\mathrm{K}$) & $\mathrm{}$ & {($\mathrm{M_{\odot}}$)} & {($\mathrm{M_{\odot}}$)} & $\mathrm{}$ & ($\mathrm{{}^{\circ}}$) & ($\mathrm{{}^{\circ}}$) & ($\mathrm{kpc}$) & {($\mathrm{M_{\odot}}$)} & $\mathrm{}$ & ($\mathrm{pc}$) \\
\hline
279 & 13.3 & 2.3 & 4.3E+04 & 7.7E+03 & 5.6 & 11.2 & 41.8 & 9.8 & 2.2E+05 & 34.4 & 62.5 \\
244 & 14.4 & 2.6 & 5.6E+04 & 5.5E+03 & 10.2 & 11.0 & 41.6 & 6.3 & 1.2E+05 & 32.3 & 53.6 \\
188 & 14.1 & 2.4 & 2.4E+04 & 2.9E+03 & 8.5 & 11.4 & 41.9 & 11.1 & 1.2E+05 & 50.3 & 37.4 \\
35 & 14.2 & 2.2 & 2.1E+04 & 4.6E+03 & 4.5 & 11.4 & 41.7 & 12.5 & 2.8E+05 & 64.6 & 47.0 \\
205 & 14.6 & 2.4 & 3.1E+04 & 4.1E+03 & 7.5 & 11.1 & 41.6 & 6.6 & 8.1E+04 & 27.4 & 42.9 \\
172 & 14.5 & 2.0 & 3.5E+04 & 4.3E+03 & 8.2 & 11.4 & 42.0 & 12.3 & 1.9E+05 & 51.5 & 47.8 \\
\end{tabular}
\end{table*}

\begin{table*}
\scriptsize
\begin{tabular}{cccSSccccScc}
Cloud ID & Mean $T_{\mathrm{dust}}$ & Mean $\beta$ & {Total CO-traced gas mass} & {Total dust mass} & CO-traced GDR & RA [J2000] & Dec [J2000] & Radius to peak & {Total gas mass} & Total GDR & Equivalent radius \\
$\mathrm{}$ & ($\mathrm{K}$) & $\mathrm{}$ & {($\mathrm{M_{\odot}}$)} & {($\mathrm{M_{\odot}}$)} & $\mathrm{}$ & ($\mathrm{{}^{\circ}}$) & ($\mathrm{{}^{\circ}}$) & ($\mathrm{kpc}$) & {($\mathrm{M_{\odot}}$)} & $\mathrm{}$ & ($\mathrm{pc}$) \\
\hline
118 & 14.1 & 2.2 & 5.3E+04 & 2.2E+03 & 24.1 & 11.3 & 41.8 & 10.3 & 5.5E+04 & 49.2 & 34.4 \\
207 & 14.5 & 2.3 & 8.6E+04 & 3.9E+03 & 22.0 & 11.4 & 41.9 & 11.3 & 2.0E+05 & 72.3 & 47.0 \\
32 & 18.0 & 1.8 & 6.1E+04 & 2.7E+03 & 22.4 & 11.3 & 41.6 & 12.3 & 2.1E+05 & 101.4 & 39.4 \\
36 & 16.0 & 2.0 & 7.5E+04 & 2.9E+03 & 26.1 & 11.3 & 41.6 & 11.7 & 2.0E+05 & 95.2 & 41.2 \\
107 & 15.2 & 2.0 & 8.0E+04 & 7.3E+03 & 11.0 & 11.4 & 41.9 & 11.8 & 4.1E+05 & 67.4 & 64.8 \\
325 & 14.8 & 2.1 & 1.3E+04 & 1.7E+03 & 7.7 & 11.1 & 41.9 & 12.7 & 8.5E+04 & 57.8 & 29.7 \\
94 & 14.2 & 2.5 & 5.2E+04 & 2.9E+03 & 18.0 & 11.2 & 41.7 & 9.3 & 7.0E+04 & 42.2 & 40.3 \\
198 & 13.9 & 2.4 & 3.3E+04 & 3.7E+03 & 8.8 & 11.3 & 41.9 & 10.3 & 1.2E+05 & 41.8 & 46.2 \\
\hline
\end{tabular}
\end{table*}

%% file: Chapters/An_appendix_ppmaptest.tex
\chapter{PPMAP \& SKIRT additional details}
\label{ap:ppmapskirt}

\section{Scatter in SKIRT runs}
\label{sec:skirtscatter}
\begin{figure}[h]
\centering
\includegraphics[width=16cm]{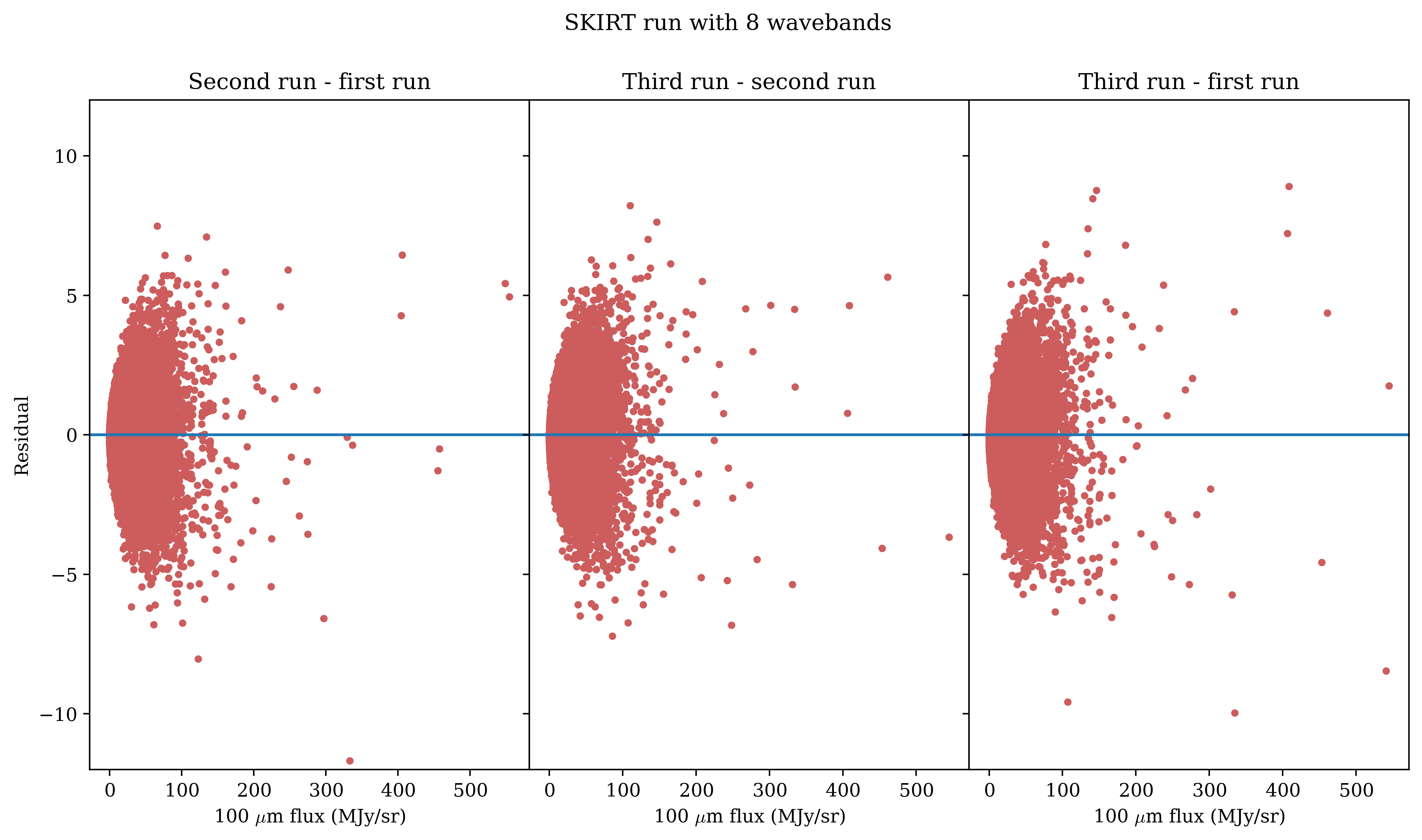}
\caption{Scatter in SKIRT radiative transfer code runs due to random sampling.  SKIRT was run with 8 `observers' corresponding to 8 different wavebands.  This figure shows only the 100 $\mu$m emission output.  \textit{Left panel:} Residual of the second run and the first run.  \textit{Middle panel:} Residual of the third run and the second run.  \textit{Right panel:} Residual of the third run and the first run. }
\label{fig:skirtscatter}
\end{figure}
{\magi \noindent In Chapter 6,  we remove the star-forming regions from the simulated galaxy images before feeding this synthetic dataset into the \textsc{PPMAP} algorithm.  Prior to doing this removal,  we run the \textsc{SKIRT} radiative transfer code multiple times to check if it reproduces the same emission from the star-forming regions each time.  The residuals of the 100 $\mu$m emission from three different runs can be seen in Figure \ref{fig:skirtscatter}.  We find that \textsc{SKIRT} is randomly drawing from a distribution for each run and that there is a scatter in the emission values. We have also checked if running \textsc{SKIRT} with different numbers of wavebands produces a scatter and find a similar result.   As such,  our removal of star-forming regions in Chapter 6 is not a perfect subtraction but is the best that we can do to reproduce the star-forming regions from the original simulated galaxy images created by \cite{Kapoor2021}.}

\section{Selected regions for SNR matching}
\label{sec:snrreg}
\begin{figure}[h]
\centering
\includegraphics[width=16cm]{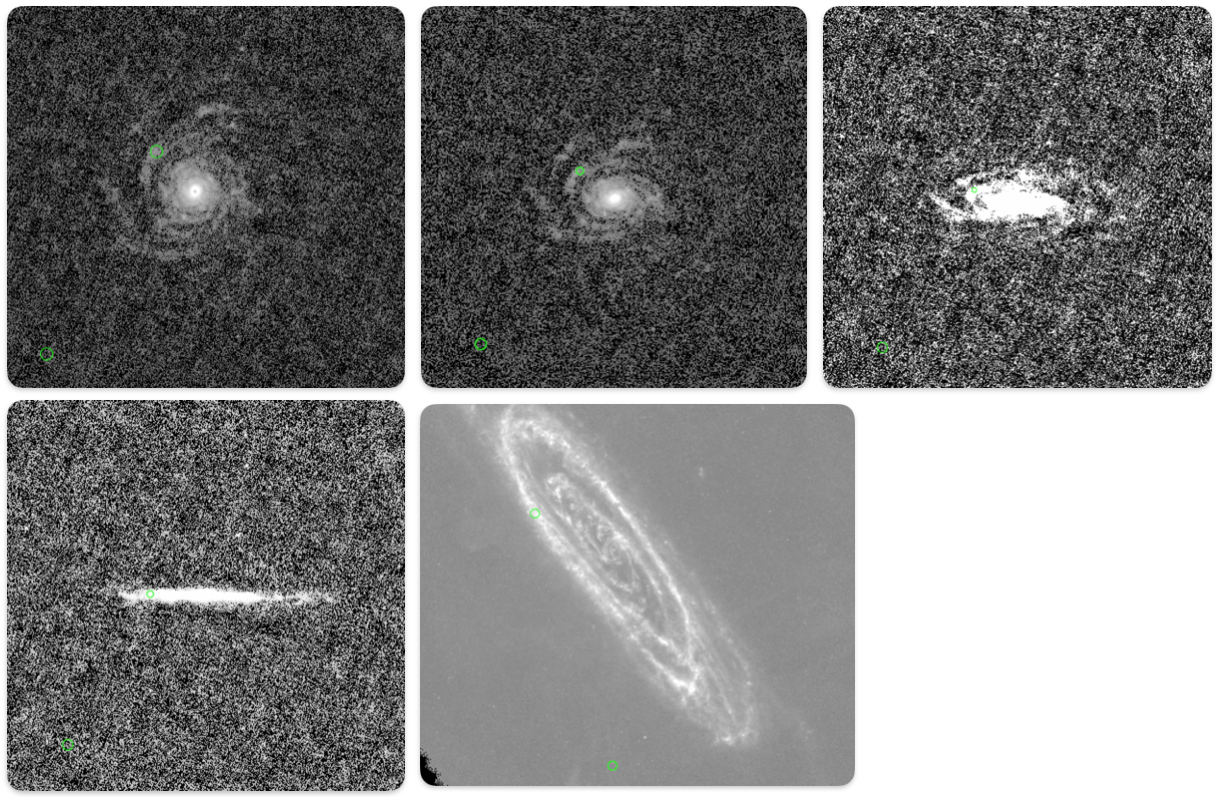}
\caption{Regions {\magi (\textit{green circle})} in the Auriga galaxy selected for signal-to-noise level matching.  \textit{Top row (left to right):} 0$^{\circ}$ inclination,  45$^{\circ}$ inclination,  72$^{\circ}$ inclination.  \textit{Bottom row (left to right):}  90$^{\circ}$ inclination,  example \textit{Herschel} M31 image.}
\label{fig:snrreg}
\end{figure}
{\magi In Chapter 6,  we add noise to the simulated galaxy images created by \cite{Kapoor2021}.  We match the noise,  in the first instance, to the noise level in \textit{Herschel} M31 image from the HELGA collaboration (\citealt{Fritz2012},  \citealt{Smith2012}).  Figure \ref{fig:snrreg} shows the regions selected to find a scaling factor for adjusting the signal-to-noise ratio of the simulated images at different inclinations. }

%% file: Chapters/An_appendix_dustmass.tex
\chapter{Cloud catalogue from the HASHTAG survey}
{\magi In Chapter 4,  we use dust to trace clouds. Table C.1 shows the catalogue of 422 clouds extracted from the dust mass surface density map produced by SED fitting of \textit{Herschel} and JCMT sub-mm/far-infrared observations at 100,  160,  250,  450 and 850 $\mu$m. The cloud properties listed in the catalogue have been calculated using the methods described in Chapter 4 and in a similar fashion to the methods used for producing the CARMA and PPMAP catalogues presented in Appendix A.}

\begin{landscape}

\setlength\tabcolsep{2.3pt}

\sisetup{
    tight-spacing           = true,
    round-mode              = places,
    round-precision         = 1,
    }

\input{Chapters/Plots_Tables/HASHTAGdustcloudtab.tex}

\end{landscape}

%% file: Chapters/Plots_Tables/HASHTAGdustcloudtab.tex
\begin{table*}
\caption{Properties of 422 clouds in the dust-selected catalogue created from \textit{Herschel} and JCMT observations.}
\scriptsize
\begin{tabular}{cccSSScccScc}
\hline
Cloud ID & Mean $T_{\mathrm{dust}}$ & Mean $\beta$ & {Total dust mass} & {Total CO-traced gas mass} & {Total HI gas mass} & RA [J2000] & Dec [J2000] & Radius to peak & {Total SFR} & Depletion time & Eq. radius \\
$\mathrm{}$ & ($\mathrm{K}$) & $\mathrm{}$ & {($\mathrm{M_{\odot}}$)} & {($\mathrm{M_{\odot}}$)} & {($\mathrm{M_{\odot}}$)} & ($\mathrm{{}^{\circ}}$) & ($\mathrm{{}^{\circ}}$) & ($\mathrm{kpc}$) & {($\mathrm{M_{\odot}\,yr^{-1}}$)} & ($\mathrm{Gyr}$) & ($\mathrm{pc}$) \\
\hline
12 & 14.6 & 2.4 & 1.2E+04 & 4.9E+04 & 6.4E+05 & 10.2 & 40.6 & 13.7 & 4.2E-05 & 29 & 81.5 \\
21 & 17.9 & 1.8 & 7.3E+03 & 7.6E+04 & 4.5E+05 & 10.2 & 40.6 & 12.8 & 4.0E-05 & 18 & 59.0 \\
36 & 17.5 & 1.9 & 2.9E+03 & 4.1E+04 & 1.3E+05 & 10.1 & 40.6 & 11.8 & 1.9E-05 & 15 & 42.7 \\
39 & 17.2 & 1.8 & 2.5E+03 & 2.3E+04 & 1.7E+05 & 10.3 & 40.6 & 13.6 & 1.1E-05 & 23 & 40.7 \\
40 & 19.2 & 1.5 & 2.9E+04 & 5.7E+05 & 1.3E+06 & 10.2 & 40.6 & 12.0 & 1.9E-04 & 15 & 92.9 \\
41 & 20.5 & 1.4 & 4.6E+03 & 8.7E+04 & 2.8E+05 & 10.2 & 40.6 & 12.2 & 4.2E-05 & 11 & 42.7 \\
\end{tabular}
\end{table*}

\begin{table*}
\tiny
\begin{tabular}{cccSSScccScc}
Cloud ID & Mean $T_{\mathrm{dust}}$ & Mean $\beta$ & {Total dust mass} & {Total CO-traced gas mass} & {Total HI gas mass} & RA [J2000] & Dec [J2000] & Radius to peak & {Total SFR} & Depletion time & Eq. radius \\
$\mathrm{}$ & ($\mathrm{K}$) & $\mathrm{}$ & {($\mathrm{M_{\odot}}$)} & {($\mathrm{M_{\odot}}$)} & {($\mathrm{M_{\odot}}$)} & ($\mathrm{{}^{\circ}}$) & ($\mathrm{{}^{\circ}}$) & ($\mathrm{kpc}$) & {($\mathrm{M_{\odot}\,yr^{-1}}$)} & ($\mathrm{Gyr}$) & ($\mathrm{pc}$) \\
\hline
175 & 16.0 & 2.1 & 9.4E+03 & 8.0E+04 & 5.2E+05 & 10.4 & 40.7 & 12.4 & 3.0E-05 & 32 & 77.3 \\
183 & 16.8 & 2.2 & 1.7E+04 & 2.4E+05 & 6.6E+05 & 10.2 & 40.8 & 8.9 & 7.4E-05 & 23 & 91.1 \\
187 & 17.3 & 1.9 & 1.5E+04 & 3.5E+05 & 4.8E+05 & 10.2 & 40.8 & 8.7 & 7.3E-05 & 20 & 84.5 \\
189 & 16.2 & 2.2 & 1.1E+04 & 1.1E+05 & 3.8E+05 & 10.2 & 40.8 & 8.5 & 4.5E-05 & 23 & 76.2 \\
194 & 16.8 & 1.9 & 5.3E+03 & 7.3E+04 & 2.7E+05 & 10.4 & 40.8 & 11.7 & 2.1E-05 & 25 & 56.2 \\
195 & 14.3 & 2.3 & 1.3E+04 & 1.4E+05 & 5.5E+05 & 10.4 & 40.8 & 11.9 & 2.7E-05 & 47 & 84.5 \\
\end{tabular}
\end{table*}

\begin{table*}
\tiny
\begin{tabular}{cccSSScccScc}
Cloud ID & Mean $T_{\mathrm{dust}}$ & Mean $\beta$ & {Total dust mass} & {Total CO-traced gas mass} & {Total HI gas mass} & RA [J2000] & Dec [J2000] & Radius to peak & {Total SFR} & Depletion time & Eq. radius \\
$\mathrm{}$ & ($\mathrm{K}$) & $\mathrm{}$ & {($\mathrm{M_{\odot}}$)} & {($\mathrm{M_{\odot}}$)} & {($\mathrm{M_{\odot}}$)} & ($\mathrm{{}^{\circ}}$) & ($\mathrm{{}^{\circ}}$) & ($\mathrm{kpc}$) & {($\mathrm{M_{\odot}\,yr^{-1}}$)} & ($\mathrm{Gyr}$) & ($\mathrm{pc}$) \\
\hline
322 & 18.3 & 1.8 & 2.5E+04 & 4.5E+05 & 1.5E+06 & 10.7 & 41.0 & 10.6 & 1.5E-04 & 16 & 109.3 \\
324 & 18.5 & 1.8 & 3.5E+03 & 8.7E+04 & 2.0E+05 & 10.2 & 41.0 & 9.3 & 2.4E-05 & 15 & 46.5 \\
325 & 17.1 & 1.9 & 2.4E+04 & 2.4E+05 & 1.5E+06 & 10.7 & 41.0 & 11.2 & 1.3E-04 & 18 & 119.5 \\
327 & 15.4 & 2.8 & 5.1E+03 & 5.8E+04 & 9.8E+04 & 10.6 & 41.0 & 5.6 & 1.6E-05 & 31 & 59.0 \\
328 & 16.2 & 2.2 & 3.0E+04 & 1.2E+05 & 1.5E+06 & 10.7 & 41.0 & 11.1 & 1.4E-04 & 21 & 139.4 \\
334 & 17.6 & 2.0 & 2.7E+04 & 6.1E+05 & 1.0E+06 & 10.7 & 41.0 & 8.1 & 1.5E-04 & 18 & 119.5 \\
\end{tabular}
\end{table*}

\begin{table*}
\tiny
\begin{tabular}{cccSSScccScc}
Cloud ID & Mean $T_{\mathrm{dust}}$ & Mean $\beta$ & {Total dust mass} & {Total CO-traced gas mass} & {Total HI gas mass} & RA [J2000] & Dec [J2000] & Radius to peak & {Total SFR} & Depletion time & Eq. radius \\
$\mathrm{}$ & ($\mathrm{K}$) & $\mathrm{}$ & {($\mathrm{M_{\odot}}$)} & {($\mathrm{M_{\odot}}$)} & {($\mathrm{M_{\odot}}$)} & ($\mathrm{{}^{\circ}}$) & ($\mathrm{{}^{\circ}}$) & ($\mathrm{kpc}$) & {($\mathrm{M_{\odot}\,yr^{-1}}$)} & ($\mathrm{Gyr}$) & ($\mathrm{pc}$) \\
\hline
467 & 16.2 & 2.3 & 7.6E+03 & 1.5E+05 & 3.2E+05 & 10.9 & 41.2 & 8.6 & 3.1E-05 & 25 & 61.8 \\
468 & 16.3 & 2.1 & 2.8E+04 & 6.3E+05 & 1.4E+06 & 10.4 & 41.3 & 9.9 & 1.0E-04 & 27 & 115.2 \\
470 & 16.9 & 2.4 & 5.0E+03 & 1.1E+04 & 1.4E+05 & 10.5 & 41.2 & 5.1 & 2.4E-05 & 21 & 57.6 \\
472 & 16.1 & 2.6 & 1.2E+04 & 1.0E+05 & 3.3E+05 & 10.5 & 41.3 & 5.8 & 5.3E-05 & 22 & 86.4 \\
474 & 17.9 & 1.9 & 2.1E+04 & 1.8E+05 & 1.5E+06 & 11.1 & 41.3 & 13.2 & 1.2E-04 & 18 & 113.0 \\
479 & 16.6 & 2.0 & 3.3E+03 & 1.0E+05 & 1.7E+05 & 10.4 & 41.3 & 10.0 & 1.1E-05 & 31 & 40.7 \\
\end{tabular}
\end{table*}

\begin{table*}
\tiny
\begin{tabular}{cccSSScccScc}
Cloud ID & Mean $T_{\mathrm{dust}}$ & Mean $\beta$ & {Total dust mass} & {Total CO-traced gas mass} & {Total HI gas mass} & RA [J2000] & Dec [J2000] & Radius to peak & {Total SFR} & Depletion time & Eq. radius \\
$\mathrm{}$ & ($\mathrm{K}$) & $\mathrm{}$ & {($\mathrm{M_{\odot}}$)} & {($\mathrm{M_{\odot}}$)} & {($\mathrm{M_{\odot}}$)} & ($\mathrm{{}^{\circ}}$) & ($\mathrm{{}^{\circ}}$) & ($\mathrm{kpc}$) & {($\mathrm{M_{\odot}\,yr^{-1}}$)} & ($\mathrm{Gyr}$) & ($\mathrm{pc}$) \\
\hline
606 & 16.8 & 2.5 & 1.4E+04 & 3.3E+05 & 2.4E+05 & 10.9 & 41.4 & 4.0 & 7.1E-05 & 20 & 92.0 \\
609 & 16.6 & 2.0 & 3.5E+04 & 9.5E+05 & 1.4E+06 & 10.6 & 41.4 & 9.8 & 1.6E-04 & 22 & 124.9 \\
610 & 15.6 & 2.7 & 3.0E+03 & 1.1E+05 & 3.7E+04 & 10.8 & 41.4 & 3.0 & 7.5E-06 & 40 & 44.6 \\
613 & 22.2 & 1.7 & 3.2E+03 & 4.0E+04 & 2.1E+05 & 11.2 & 41.4 & 12.5 & 1.2E-04 & 3 & 42.7 \\
615 & 18.4 & 2.0 & 1.2E+04 & 4.5E+05 & 2.0E+05 & 11.0 & 41.4 & 5.8 & 9.5E-05 & 13 & 78.4 \\
617 & 16.0 & 2.7 & 4.1E+03 & 5.7E+04 & 8.3E+04 & 10.7 & 41.4 & 5.1 & 3.3E-05 & 12 & 44.6 \\
\end{tabular}
\end{table*}

\begin{table*}
\tiny
\begin{tabular}{cccSSScccScc}
Cloud ID & Mean $T_{\mathrm{dust}}$ & Mean $\beta$ & {Total dust mass} & {Total CO-traced gas mass} & {Total HI gas mass} & RA [J2000] & Dec [J2000] & Radius to peak & {Total SFR} & Depletion time & Eq. radius \\
$\mathrm{}$ & ($\mathrm{K}$) & $\mathrm{}$ & {($\mathrm{M_{\odot}}$)} & {($\mathrm{M_{\odot}}$)} & {($\mathrm{M_{\odot}}$)} & ($\mathrm{{}^{\circ}}$) & ($\mathrm{{}^{\circ}}$) & ($\mathrm{kpc}$) & {($\mathrm{M_{\odot}\,yr^{-1}}$)} & ($\mathrm{Gyr}$) & ($\mathrm{pc}$) \\
\hline
733 & 19.4 & 2.2 & 3.4E+03 & 7.8E+04 & 5.0E+04 & 11.1 & 41.6 & 6.1 & 5.7E-05 & 6 & 40.7 \\
735 & 14.6 & 2.7 & 1.5E+04 & 3.4E+05 & 1.8E+05 & 11.0 & 41.6 & 5.0 & 2.6E-05 & 60 & 92.0 \\
737 & 18.7 & 1.8 & 3.4E+03 & 3.7E+04 & 2.3E+05 & 11.3 & 41.6 & 12.7 & 3.4E-05 & 10 & 40.7 \\
745 & 16.7 & 2.4 & 6.9E+03 & 1.7E+05 & 1.0E+05 & 11.1 & 41.6 & 6.5 & 4.9E-05 & 14 & 64.4 \\
746 & 19.4 & 1.6 & 5.0E+04 & 1.3E+06 & 2.9E+06 & 11.3 & 41.6 & 12.8 & 4.3E-04 & 12 & 142.9 \\
747 & 16.9 & 2.2 & 5.6E+03 & 1.4E+05 & 6.6E+04 & 11.1 & 41.6 & 6.1 & 2.9E-05 & 19 & 40.7 \\
\end{tabular}
\end{table*}

\begin{table*}
\tiny
\begin{tabular}{cccSSScccScc}
Cloud ID & Mean $T_{\mathrm{dust}}$ & Mean $\beta$ & {Total dust mass} & {Total CO-traced gas mass} & {Total HI gas mass} & RA [J2000] & Dec [J2000] & Radius to peak & {Total SFR} & Depletion time & Eq. radius \\
$\mathrm{}$ & ($\mathrm{K}$) & $\mathrm{}$ & {($\mathrm{M_{\odot}}$)} & {($\mathrm{M_{\odot}}$)} & {($\mathrm{M_{\odot}}$)} & ($\mathrm{{}^{\circ}}$) & ($\mathrm{{}^{\circ}}$) & ($\mathrm{kpc}$) & {($\mathrm{M_{\odot}\,yr^{-1}}$)} & ($\mathrm{Gyr}$) & ($\mathrm{pc}$) \\
\hline
898 & 15.4 & 2.2 & 5.6E+03 & 1.3E+04 & 2.8E+05 & 11.0 & 41.8 & 11.4 & 1.9E-05 & 29 & 56.2 \\
899 & 15.4 & 2.0 & 3.0E+03 & 1.7E+04 & 1.3E+05 & 11.5 & 41.8 & 14.1 & 8.1E-06 & 38 & 44.6 \\
902 & 16.8 & 2.0 & 1.2E+04 & 5.7E+04 & 6.1E+05 & 11.0 & 41.8 & 11.4 & 1.1E-04 & 10 & 81.5 \\
905 & 17.0 & 1.9 & 5.8E+03 & 3.5E+04 & 3.0E+05 & 11.0 & 41.8 & 10.8 & 2.6E-05 & 23 & 54.7 \\
907 & 15.1 & 2.3 & 1.5E+04 & 7.5E+04 & 5.4E+05 & 11.0 & 41.8 & 10.5 & 4.7E-05 & 33 & 93.8 \\
910 & 14.1 & 2.3 & 1.0E+04 & 2.2E+04 & 4.3E+05 & 10.9 & 41.8 & 12.3 & 2.1E-05 & 48 & 77.3 \\
\end{tabular}
\end{table*}

\begin{table*}
\tiny
\begin{tabular}{cccSSScccScc}
Cloud ID & Mean $T_{\mathrm{dust}}$ & Mean $\beta$ & {Total dust mass} & {Total CO-traced gas mass} & {Total HI gas mass} & RA [J2000] & Dec [J2000] & Radius to peak & {Total SFR} & Depletion time & Eq. radius \\
$\mathrm{}$ & ($\mathrm{K}$) & $\mathrm{}$ & {($\mathrm{M_{\odot}}$)} & {($\mathrm{M_{\odot}}$)} & {($\mathrm{M_{\odot}}$)} & ($\mathrm{{}^{\circ}}$) & ($\mathrm{{}^{\circ}}$) & ($\mathrm{kpc}$) & {($\mathrm{M_{\odot}\,yr^{-1}}$)} & ($\mathrm{Gyr}$) & ($\mathrm{pc}$) \\
\hline
1084 & 12.9 & 2.5 & 4.6E+03 & 2.9E+03 & 1.6E+05 & 11.5 & 41.9 & 13.2 & 6.2E-06 & 74 & 51.5 \\
1099 & 17.1 & 1.8 & 7.0E+03 & 1.5E+05 & 2.8E+05 & 11.3 & 41.9 & 11.2 & 2.4E-05 & 29 & 59.0 \\
1104 & 15.5 & 2.2 & 6.0E+03 & 8.0E+04 & 2.3E+05 & 11.4 & 41.9 & 11.3 & 1.7E-05 & 36 & 54.7 \\
1105 & 17.2 & 1.7 & 3.6E+03 & 6.8E+04 & 1.4E+05 & 11.3 & 41.9 & 11.0 & 1.1E-05 & 33 & 40.7 \\
1108 & 17.3 & 2.0 & 1.3E+04 & 2.9E+05 & 4.4E+05 & 11.2 & 41.9 & 11.3 & 6.8E-05 & 19 & 81.5 \\
1109 & 17.0 & 2.1 & 3.0E+03 & 3.0E+04 & 1.5E+05 & 11.2 & 41.9 & 11.6 & 1.4E-05 & 22 & 44.6 \\
\hline
\end{tabular}
\end{table*}